\documentclass[twoside]{cernyrep}

\usepackage[utf8]{inputenc}
\DeclareUnicodeCharacter{2212}{-}

\usepackage[onehalfspacing]{setspace}
\newcommand{\chapabstract}[1]{
    \begin{quote}
        \singlespacing\small
        \rule{14cm}{1pt}
        #1
        \vskip-4mm
        \rule{14cm}{1pt}
\end{quote}}

\usepackage{graphicx}
\usepackage{caption}
\usepackage{epstopdf}

\usepackage{booktabs}
\usepackage{subfloat}
\usepackage{acronym}
\usepackage{subfig}
\usepackage{amsmath}
\usepackage{color}
\usepackage{balance}
\usepackage{authblk}
\usepackage{fancyhdr}
\usepackage{multirow}
\usepackage{afterpage}

\usepackage[hang,flushmargin]{footmisc}
\fancypagestyle{plain}{}
\usepackage{emptypage}
\usepackage{lipsum}
\usepackage{etoc}
\etocsettocstyle{}{}

\newcommand\blfootnote[1]{%
  \begingroup
  \renewcommand\thefootnote{}\footnote{#1}%
  \addtocounter{footnote}{-1}%
  \endgroup
}
\usepackage{texnames}
\usepackage{ctable}
\usepackage[T1]{fontenc}

\usepackage{blindtext}

\usepackage[bookmarks, colorlinks=true, linktoc=page, pdftex, linkcolor=blue, citecolor=blue, urlcolor=blue]{hyperref}
\usepackage{float}
\usepackage{xcolor}
\usepackage[toc,page]{appendix}
\usepackage[bookmarks, colorlinks=true, pdftex, linkcolor=blue, citecolor=blue, urlcolor=blue]{hyperref}
\newlength{\oddmarginwidth}
\newlength{\evenmarginwidth}
\fancyhfoffset[LO,RE]{\oddmarginwidth}
\fancyhfoffset[LE,RO]{\evenmarginwidth}
\begin{document}


\title{Quantum Field Theory and the Electroweak Standard Model}

\pagenumbering{arabic}
\setcounter{page}{1}



\chapter*{Quantum field theory and the electroweak Standard Model}

\renewcommand{\theequation}{\thesection.\arabic{equation}}
\newcommand{\bd}{\begin{displaymath}}
\newcommand{\ed}{\end{displaymath}}
\newcommand{\be}{\begin{equation}}
\newcommand{\ee}{\end{equation}}
\newcommand{\bear}{\begin{eqnarray}}
\newcommand{\eear}{\end{eqnarray}} 
\newcommand{\ba}{\begin{array}}
\newcommand{\ea}{\end{array}}
\newcommand{\del}{\partial}
\newcommand{\nin}{\noindent}
\newcommand{\vs}{\vskip0.2cm}
\newcommand{\cL}{{\cal L}}
\newcommand{\cO}{{\cal O}}
\newcommand{\cA}{\cal A}
\newcommand{\mbf}{\mathbf}
\newcommand{\Tr}{\rm Tr}
\newcommand{\dsl}{\hskip-0.1cm\not\hskip-0.09cm}
\newcommand{\dslcv}{\hskip-0.1cm\not\hskip-0.13cm}
\newcommand{\dslcvex}{\hskip-0.01cm\not\hskip-0.025cm}
\newcommand{\grad}{\vec{\bigtriangledown}}
\newcommand{\vf}{\mathbf}
\vspace{-11mm}


\begin{flushleft}
\blfootnote{
This article should be cited as:
Quantum Field theory and the electroweak Standard Model, Gustavo Burdman,
DOI:~\href{https://doi.org/10.23730/CYRSP-2024-XXX.\thepage}{10.23730/CYRSP-2023-XXX.\thepage}, in:
Proceedings of the 2023 CERN Latin-American School of High-Energy Physics,
\\CERN Yellow Reports: School Proceedings, CERN-2023-XXX,
DOI:~\href{https://doi.org/10.23730/CYRSP-2023-XXX}{10.23730/CYRSP-2023-XXX}, p. \thepage.
\\ \copyright\space CERN, 2024. Published by CERN under the
\href{http://creativecommons.org/licenses/by/4.0/}{Creative Commons Attribution 4.0 license}.
}
\end{flushleft}


\label{sec:Burdman}

\noindent
{\it Gustavo Burdman\\}
{ {Institute of Physics - University of São Paulo, Brazil\\}}
\vspace{-11mm}
\chapabstract{



In these lectures we  give an introduction and overview of
the electroweak standard model (EWSM) of particle physics.
We first introduce the basic concepts of quantum field theory necessary
to build the EWSM: abelian and non-abelian gauge theories,
spontaneous symmetry breaking and the Higgs mechanism.  
We also introduce some basic concepts of renormalization, so as to
be able to understand the full power of electroweak precision tests
and their impact on our understanding the EWSM and its possible
extensions. We discuss the current status of experimental tests and
conclude by pointing the problems still existing in particle physics
not solved by the EWSM and how these impact the future of the field.

}

\tableofcontents

\newpage
\noindent
\section{Quantum field theory basics and  gauge theories}
\label{sec:lecture1}


\subsection{Quantum field theory basics}
\subsubsection{Why quantum field theory}
\label{sec:whyqft}

Quantum field theory (QFT)~\cite{gabqft} is, at least  in its origin, the result of
trying to work with both quantum mechanics and special relativity. 
Loosely speaking, the uncertainty principle tells us that we can
violate energy conservation by $\Delta E$ as long as it is for a
small $\Delta t$. But on the other hand, special relativity tells us
that energy can be converted into matter. So if we get a large energy
fluctuation $\Delta E$ (for a short $\Delta t$) this energy might be
large enough to produce new particles, at least for that short period
of time. However, quantum mechanics does not allow for such
process. For instance, the Schr\"{o}dinger equation for an electron
describes the evolution of just this one electron, independently of
how strongly it interacts with a  given potential. The same continues
to be true of its relativistic counterpart, the Dirac equation.  We need a framework
that allows for the creation (and annihilation) of quanta. This is
QFT. 

We can say the same thing by being a bit more precise so that we can
start to see how we are going to tackle this problem. Let us consider
a {\em classical} source  that emits particles with an amplitude
$J_E(x)$, where $x\equiv x_\mu$ is the space-time position. We also
consider an absorption source of amplitude $J_A(x)$.
We assume that a particle of mass $m$ that is emitted at $y$
propagates freely before being absorbed at $x$~\cite{Banks:2014twn}.

\vs
\begin{figure}[h]
  \begin{center}
\includegraphics[width=0.3\textwidth]{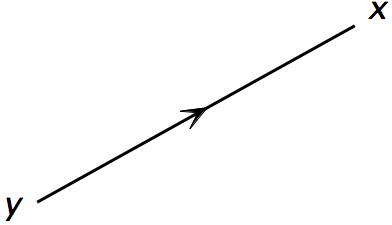}
\caption{Emission, propagation and absorption of a particle.}
\end{center}
\end{figure}
\vs\nin
The quantum mechanical amplitude is given by 
\vs
\be
{\cal A} = \int d^4x\, d^4y \,\langle x| e^{-iH\Delta t} |y\rangle\,
J_A(x)\,J_E(y)~,
\label{qmamplitude}
\ee
\vs\nin
 where $\Delta t=x_0-y_0$. Here we have used the notation 
\vs\be
d^4x\equiv dt\,d^3x~,
\ee\vs\nin
to denote the Minkowski space four-volume, i.e. we are integrating over
time and all space. 
We want to check if the amplitude in (\ref{qmamplitude}) is
Lorentz invariant, i.e. if it is compatible with special relativity.
Writing
\vs
\be
H = \sqrt{p^2 + m^2}\equiv \omega_p~,\label{hamiltonian2}
\ee\vs\nin
as the frequency associated with momentum $p$, then the amplitude is
\vs
\be
{\cal A } = \int d^4x\, d^4y \,\langle x| e^{-i\omega_p(x_0-y_o)} |y\rangle\,
J_A(x)\,J_E(y)~.
\ee\vs\nin
If we go to momentum space using
\vs
\be
|x\rangle = \int \frac{d^3p}{(2\pi)^{3/2}}\,
|p\rangle\,e^{-i\vec{p}\cdot\vec{x}}~,
\label{x2momentum}
\ee\vs\nin
 and analogously for $|y\rangle$, we obtain
\vs
\be
{\cal A } = \int d^4x\, d^4y\, \int \frac{d^3p}{(2\pi)^{3/2}}\,\,\langle p|\,e^{i\vec{p}\cdot\vec{x}} \,e^{-i\omega_p(x_0-y_o)}\,\int \frac{d^3p'}{(2\pi)^{3/2}}\, |p'\rangle\,
e^{-i\vec{p'}\cdot\vec{y}} \, J_A(x)\,J_E(y)~.
\ee
\vs\nin
Using that
\vs
\be
\langle p|p'\rangle = \delta^3(\vec{p}-\vec{p'})\,N^2_p~,
\label{norma}
\ee
\vs\nin
where $N_p$ is the momentum dependent normalization, we now have
\vs\be
{\cal A } = \int d^4x\, d^4y \, J_A(x)\,J_E(y)\,
\int\frac{d^3p}{(2\pi)^3}\,N^2_p\,\,e^{-ip^\mu\,(x_\mu-y_\mu)}~,
\label{amp6}
\ee\vs\nin
In the last exponential factor in (\ref{amp6}) we use covariant
notation, i.e.
\vs
\be
p^\mu\,(x_\mu-y_\mu) = p_0(x_0-y_0) - \vec{p}\cdot(\vec{x}-\vec{y}) =
\omega_p\Delta t - \vec{p}\cdot(\vec{x}-\vec{y})~.
\ee\vs\nin
To check if ${\cal A}$ is Lorentz invariant we are going to define the
four-momentum integration with a Lorentz invariant measure. Defining
\vs
\be
d^4p = dp_0\,\,d^3p~,
\ee
\vs\nin
we now can compute the Lorentz invariant combination
\vs
\be
d^4p\,\,\delta(p^2-m^2)~,
\ee
\vs\nin
where the delta function ensures that $p^2=p_\mu\,p^\mu=m^2$. 
Then we do the integral on $p_0$ as in
\vs
\be
\int dp_0\,\delta(p^2-m^2) = \int dp_0\,\delta(p_0^2 - |\vec{p}|^2
  -m^2) = \int dp_0 \, \frac{\delta(p_0-\omega_p)}{|2p_0|} = \int dp_0\,\frac{\delta(p_0-\omega_p)}{2w_p}~,
\label{deltapzero}
\ee
\vs\nin
remembering that $\omega_p=+\sqrt{p^2+m^2}$ positive. Only the
positive root contributes in (\ref{deltapzero}) since the fact that
$p^\mu$ is always time-like means that the {\em sign} of $p_0$ is
invariant. This, in turn, means that the $p_0$ integration interval is
$(0,\infty)$, and the negative root is outside the integration
region. 

\nin
This allows us to rewrite the amplitude as 
\vs
\be
{\cal A } = \int d^4x\, d^4y \, J_A(x)\,J_E(y)\,
\int\frac{d^4p}{(2\pi)^3}\,\delta(p^2-m^2)\,2\omega_p\,\,N^2_p\,\,e^{-ip^\mu\,(x_\mu-y_\mu)}~.
\label{almostLI}
\ee
\vs\nin
The expression above appears Lorentz invariant other than for the
momentum dependent factor 
\vs
\be
2\omega_p\,N_p^2~.
\ee\vs\nin
 Thus, the choice (up to an irrelevant constant) 
\vs
\be
N_p^2 = \frac{1}{2\omega_p}~,
\label{relnorma}
\ee
\vs\nin
results in the Lorentz invariant amplitude
\vs
\be
{\cal A } = \int d^4x\, d^4y \, J_A(x)\,J_E(y)\,
\int\frac{d^4p}{(2\pi)^3}\,\delta(p^2-m^2)\,\,e^{-ip^\mu\,(x_\mu-y_\mu)}~.
\label{ampLI}
\ee\vs\nin
Although the quantum mechanical amplitude in (\ref{ampLI}) is
manifestly Lorentz invariant, there remains a problem: this expression
is valid even if the interval separating $x$ from $y$ is spatial,
i.e. even if the separation is non-causal. This is obviously wrong,
since we started from the assumption that there is an {\em emitting} source
at $y$  and an {\em absorbing} source at $x$, for which the causal order is
crucial, which means that the way it is now the separation should not
be spatial. 

\nin
In order to solve this problem, we are going to  allow {\em all} sources to
both emit and absorb, i.e. at any point $x$ we have 
\vs
\be
J(x) = J_E(x) + J_A(x)~.
\ee  
\vs\nin
The amplitude then reads
\vs
\be
{\cal A } = \int d^4x\, d^4y \, J(x)\,J(y)\,
\int\frac{d^3p}{(2\pi)^3\,2\omega_p}\,\left\{ \theta(x_0-y_0)\,
    e^{-ip^\mu\,(x_\mu-y_\mu)} + \theta(y_0-x_0)  \, e^{+ip^\mu\,(x_\mu-y_\mu)} \right\}~.
\label{ampcausal}
\ee\vs\nin
The first term in (\ref{ampcausal}) corresponds to the emission in $y$
and absorption in $x$, since the function $\theta(x_0-y_0)\not=0$ for
$x_0>y_0$. For the opposite time order, this term is zero and then
only the second term contributes. The sign inversion in the exponential of the second term in
(\ref{ampcausal}) needs some explaining. Surely, the time component
$p_0 \,(y_0-x_0)=-p_0\,(x_0-y_0)$ comes from just the inversion of the
causal order. However, the inversion of the space component from
$\vec{p}\cdot (\vec{x}-\vec{y})$ to  $-\,\vec{p}\cdot
(\vec{x}-\vec{y})$ is possible by changing $d^3p$ to $-d^3p$ and
switching the limits of the spatial momentum integration to preserve
the overall sign. 

\nin
So for time like separations, when
the order of the events is an observable, only one of these terms
contributes. On the other hand, for space like separations {\em both}
terms contribute. Different observers would disagree on the temporal
order of the event, however all of them would write the same
amplitude. So this amplitude is both Lorentz invariant and causal. It
is typically written as 
\vs
\be
{\cal A } = \int d^4x\, d^4y \, J(x)\,J(y)\,D_F(x-y)~,
\label{ampdf}
\ee\vs\nin
where we defined 
\vs
\be
D_F(x-y) \equiv \int\frac{d^3p}{(2\pi)^3\,2\omega_p}\,\left\{ \theta(x_0-y_0)\,
    e^{-ip^\mu\,(x_\mu-y_\mu)} + \theta(y_0-x_0)  \,
    e^{+ip^\mu\,(x_\mu-y_\mu)} \right\}~.
  \label{feynmanprop}
\ee
\vs\nin
The two-point function above is what is called a Feynman propagator. 
To summarize so far, in order to obtain a Lorentz invariant and causal
quantum mechanical amplitude for the emission, propagation and
absorption of a particle we had to allow for all points in spacetime
to both emit and absorb, and we needed to allow for all possible time
orders. There is still one more thing we need to introduce.

\subsubsection{Charged particles}

Here is the problem: if the particle propagating between $y$ and $x$
is charged, for instance under standard electromagnetism,
i.e. electrically charged, then because the amplitude (\ref{ampdf})
does not tell us the order of events in the case of space like
separation, we do not know the sign of the current. For instance,
suppose a negatively charged particle. Is it being absorbed or emitted
? We concluded above that this absolute statement should not be
allowed. But this means that we cannot know the direction of the
current.  
\vs\vs\vs\vs
\begin{figure}[h]
  \begin{center}
\includegraphics[width=0.6\textwidth]{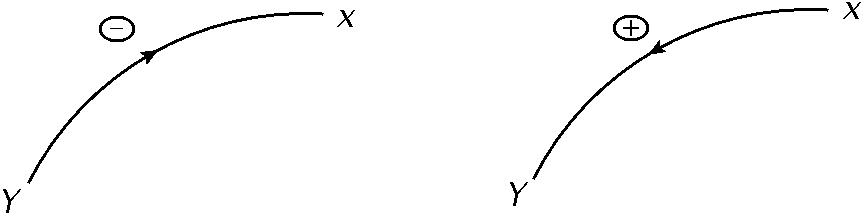}
\caption{Emission, propagation and absorption of a charged
  particle. Consistency with either temporal order is restored by
  having anti-particles. Emission of a negatively charged particle at
  $y$ followed by absorption at $x$ is equivalent to emission of the
  positively-charged anti-particle at $x$, followed by absorption at $y$. }
\end{center}
\end{figure}
\vs\nin
The solution to this problem is that for each negatively charged
particle, there must be a positively charged particle with the same
mass, its anti-particle. With this addition, it will not be possible
to distinguish between say the emission of a negatively charged
particle or the absorption of its positively charged anti-particle. 

\nin
In general, any time a particle has an internal quantum number that
may distinguish emission from absorption it should have a distinct
anti-particle that would restore the desired indistinguishability. For
instance, neutral kaons have no electric charge, but they carry a
quantum number called ``strangeness'' which distinguishes the neutral kaon
from the neutral anti-kaon. In the absence of any distinguishing
internal quantum number, a particle can be its own anti-particle. 

\nin
Finally, to illustrate the relationship between propagation and
particle or anti-particle identity, we consider the scattering of a
particle off a localized potential. We first consider the situation 
with emission at $y$, followed by interaction at $z$ and finally
absorption at $x$, i.e. the time order is $x_0>z_0>y_0$. 
\vs
\begin{figure}[h]
\begin{center}
\includegraphics[width=0.3\textwidth]{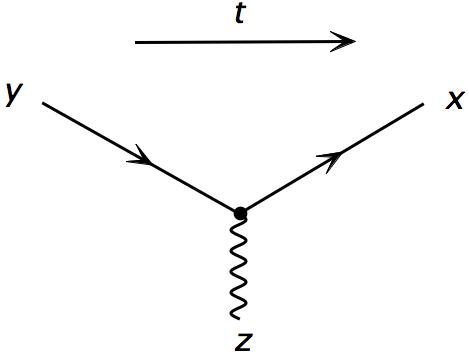}
\caption{Scattering off a localized potential.}
\end{center}
\end{figure}

\begin{figure}[h]
\begin{center}
\includegraphics[width=0.3\textwidth]{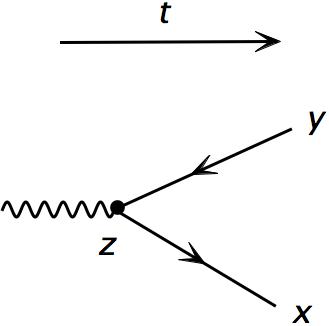}
\caption{Particle -- anti-particle pair creation.}
\label{fig:1.4}
\end{center}
\end{figure}

\nin
The amplitude for this is 
\vs
\be
{\cal A}_{\rm scatt.} = \int d^4x\,d^4y\,J(y)\,D_F(z-y)\,{\cal A}_{\rm
  int.}(z)\,D_F(z-x)\,J(x)~,
\label{ampscat}
\ee\vs\nin
where ${\cal A}_{\rm int.}(z)$ is the amplitude for the local interaction
with the potential at $z$.  
But we know that the  amplitude is non-zero even if events are
spatially separated. In this case then, it is possible to have a
non-zero amplitude corresponding to the following time order: $y_0,x_0
> z_0$. This now would correspond to the diagram in the
Figure~\ref{fig:1.4}.
\vs
\vs\nin
In this time order, a pair is created from the ``vacuum'' at $z$. The
arrows indicate that a particle propagates between $z$ and $x$, where
it is absorbed, whereas an anti-particle travels from $z$ to $y$. 
Thus, the creation of a pair particle--anti-particle, assuming there
is enough energy, is an unavoidable consequence of the marriage
between quantum mechanics and special relativity. 
All of the arguments above lead us to the fact that relativistic
quantum mechanics, compatible with causality, must be a theory of
quantized local fields. That is to say, we must be able to create or
annihilate quanta of the fields locally, including particles and
anti-particles. We will define what we really mean by this below.


\subsubsection{Some classical field theory}

Here we start by considering a field or set of fields $\phi(x)$, where
$x$ is the spacetime position. The Lagrangian is a functional of
$\phi(x)$ and its derivatives
\vskip0.2cm
\be
\frac{\del\phi(x)}{\del x^\mu}=\del_\mu\phi(x)~.
\ee
\vs\nin
Here $\phi(x)$ can be a set of fields with an internal index $i$, such
that
\vs
\be
\phi(x)=\left\{\phi_i(x)\right\}~.
\ee 
\vs\nin
We will start with the \underline{Lagrangian formulation}. 
We define the Lagrangian density ${\cal L}(\phi(x),\del_\mu\phi(x))$
by 
\vs
\be
L= \int d^3x\, {\cal L}(\phi(x),\del_\mu\phi(x)). 
\ee
\vs\nin
In this way the action is 
\vs
\be
S = \int dt\,L = \int d^4x\,{\cal L}(\phi(x),\del_\mu\phi(x))~,
\label{actionrel}
\ee
\vs\nin
where we are again using the Lorentz invariant spacetime volume element
\vs
\be
d^4x = dt \,d^3x~.
\ee
\vs\nin
From (\ref{actionrel}) is clear that ${\cal L}$ must be Lorentz
invariant. In addition, $\cal L$ might also be invariant under other
symmetries of the particular theory we are studying. These are
generally called internal symmetries and we will study them in more
detail later in the rest of the course.

\nin We vary the action in (\ref{actionrel}) in order to
find the extremal solutions (i.e. $\delta S=0$) and obtain the {\em classical } equations of motion, just as we obtain classical mechanics from extremizing the action of a system of particles. 
We get
\vs
\be
\delta S = \int d^4x\,\left\{\frac{\del\cL}{\del\phi}\,\delta\phi +
  \frac{\del\cL}{\del(\del_\mu\phi)}\,\delta(\del_\mu\phi) \right\}
\ee
\vs\nin 
But we have that
\vs
\be
\delta(\del_\mu\phi) = \del_\mu(\delta\phi)~,
\ee
\vs\nin
so the variation of the action is
\vs
\bear
\delta S &=& \int d^4x\,\left\{\frac{\del\cL}{\del\phi}\,\delta\phi +
  \frac{\del\cL}{\del(\del_\mu\phi)}\,\del_\mu(\delta\phi)\right\}~,\nonumber\\
&=&\int d^4x \,\left\{\left(\frac{\del\cL}{\del\phi} -
    \del_\mu\left(\frac{\del\cL}{\del(\del_\mu\phi)}\right)\right)\,\delta\phi
+
\del_\mu\left(\frac{\del\cL}{\del(\del_\mu\phi)}\,\delta\phi\right)\right\}~.\label{deltaSfields} 
\eear
\vs\nin
In the second line in (\ref{deltaSfields}) we have integrated by
parts. The last term is a four-divergence, i.e.  a total
derivative. Since the integral is over the volume of all of spacetime,
the resulting (hyper-)surface term must be evaluated at infinity. But the value of the field
variation at these extremes is 
$\delta\phi=0$. Thus, the (hyper-)surface term in (\ref{deltaSfields})
does not contribute. 

\nin
Then imposing $\delta S=0$, we see that the first term in
(\ref{deltaSfields})  multiplying $\delta\phi$ must vanish for all
possible values of $\delta\phi$. We obtain 
\vs
\be
\boxed{~\frac{\del\cL}{\del\phi} -
\del_\mu\left(\frac{\del\cL}{\del(\del_\mu\phi)}\right) =0~}~, 
\label{euler-lagrange}
\ee
\vs\nin
which are the E\"{u}ler-Lagrange equations, one for each of the
$\phi_i(x)$, also known as equations of motion.

\nin If now we want to go to the \underline{Hamiltonian formulation},
we start by defining the canonically conjugated momentum by
\vs
\be
p(x) = \frac{\del L}{\del\dot{\phi}(x)} =
\frac{\del}{\del\dot{\phi}(x)}\,\int d^3y\,\cL(\phi(y),
\del_\mu{\phi}(y))~,
\ee
\vs\nin
which results in the momentum density
\vs
\be
\pi(x) = \frac{\del\cL}{\del\dot{\phi}(x)}~.
\ee
\vs\nin
Here $\pi(x)$ is the momentum density canonically conjugated to
$\phi(x)$. Then the Hamiltonian is given by 
\vs
\be
H = \int d^3x\,\pi(x)\,\dot{\phi}(x) - L~,
\ee
\vs\nin
which leads to the Hamiltonian density  
\vs
\be
{\cal H}(x) = \pi(x)\,\dot{\phi}(x) - \cL(x)~,
\label{hamiltonian3}
\ee
\vs\nin
where we must remember that we evaluate at a fixed time $t$,
i.e. $x=(t,\mbf{x})$ for fixed $t$.  
The Lagrangian formulation allows for a Lorentz invariant
treatment. On the other hand, the Hamiltonian formulation might have
some advantages. For instance, it allows us to impose canonical
quantization rules. 

\nin
\underline{Example}:
We start with a simple example: the non-interacting theory of real
scalar field. The Lagrangian density is given by 
\vs
\be
\cL = \frac{1}{2}\,\del_\mu\phi\,\del^\mu\phi -
\frac{1}{2}\,m^2\,\phi^2~,\label{LforRealScalar}
\ee
\vs\nin
We will call the first term in (\ref{LforRealScalar}) the kinetic
term. In the second term $m$ is  the mass parameter, so this we will call
the mass term.
We first obtain the equations of motion by using the
E\"{u}ler-Lagrange equations (\ref{euler-lagrange}). We have
\vs
\be
\frac{\del\cL}{\del\phi}=-m^2\,\phi~,\qquad
\frac{\del\cL}{\del(\del_\mu\phi)} = \del_\mu\phi~, 
\ee
\vs\nin 
giving us
\vs
\be
\boxed
{(~\del^2 + m^2)\,\phi = 0~}~,
\label{klein-gordon}
\ee
\vs\nin
where the  D'Alembertian  operator is defined by
$\del^2=\del_\mu\del^\mu$. The equation of motion (\ref{klein-gordon})
is called the Klein-Gordon equation. 
\nin
This might be a good point for a comment. In ``deriving'' the equations
of motion (\ref{klein-gordon}) , we started with the ``given''
Lagrangian density (\ref{LforRealScalar}).  But in general this is not
how it works. Many times we have information that leads to the
equations of motion, so we can guess the Lagrangian that would
correspond to them. This would be a bottom up construction of the
theory. In this case, the Klein-Gordon equation is just the
relativistic dispersion relation $p^2=m^2$, noting that
$-i\del_\mu=p_\mu$. So we could have guessed (\ref{klein-gordon}), and
then derive $\cL$. However, we can invert the argument: the Lagrangian
density (\ref{LforRealScalar}) is the most general non-interacting
Lagrangian for a real scalar field of mass $m$ that respects Lorentz
invariance. So imposing the symmetry restriction on $\cL$ we can build
it and then really derive the equations of motion. In general, this
procedure of writing down the most general Lagrangian density
consistent with all the symmetries of the theory will be limiting
enough to get the right dynamics\footnote{Actually, in the presence of
interactions we need to add one more restriction called
renormalizability. Otherwise, in general there will be infinite terms
compatible with the symmetries.}. 

\nin
Now we want to  derive the form of the  Hamiltonian in this
example. 
It is convenient to first write the Lagrangian density
(\ref{LforRealScalar}) as
\vs
\be
\cL = \frac{1}{2}\,\dot{\phi}^2 - \frac{1}{2}\,(\vec{\nabla}\phi)^2
-\frac{1}{2}\,m^2\phi^2~.
\ee
\vs\nin
The canonically conjugated momentum density is now
\vs
\be
\pi(x) = \frac{\del\cL}{\del\dot{\phi}} = \dot{\phi}~.
\ee
\vs\nin
Then, using (\ref{hamiltonian3}) we obtain the Hamiltonian 
\vs
\be
H =  \int d^3x \, \left\{\pi^2 - \frac{1}{2}\,\pi^2\, +
\frac{1}{2}\,(\vec{\nabla}\phi)^2 + \frac{1}{2}\,m^2\,\phi^2\right\}
\ee
\vs\nin
which results in 
\vs
\be
\boxed{
~H =  \int d^3x \, \left\{ \frac{1}{2}\,\pi^2\, +
\frac{1}{2}\,(\vec{\nabla}\phi)^2 + \frac{1}{2}\,m^2\,\phi^2\right\}~}~.
\label{HforRealScalar}
\ee
\vs\nin
We clearly identify the first term in (\ref{HforRealScalar}) as the
kinetic energy, the second term as the energy associated with spatial
variations of the field, and finally the third term as the energy
associated with the mass. 

\subsubsection{Continuous symmetries and Noether's theorem} 

In addition to being invariant under Lorentz transformations, the
Lagrangian density $\cL$ can be a scalar under other symmetry
transformations. In particular, when the symmetry transformation is
continuous, we can express it  as an infinitesimal variation of the
field $\phi(x)$ that leaves the equations of motion invariant. 
Let us consider the infinitesimal transformation
\vs
\be
\phi(x)\longrightarrow  \phi'(x)=\phi(x) + \epsilon\,\Delta\phi~,
\label{symtrans}
\ee
\vs\nin
where $\epsilon$ is an infinitesimal parameter.   The change induced
in the Lagrangian density is 
\vs
\be
\cL\longrightarrow \cL + \epsilon\,\Delta\cL~,
\ee
\vs\nin
where we factorized $\epsilon$ for convenience in the second term. 
This term can be written as 
\vs
\bear
\epsilon\,\Delta\cL &=& \frac{\del\cL}{\del\phi}\,(\epsilon\,\Delta\phi) +
\frac{\del\cL}{\del(\del_\mu\phi)}\,\del_\mu(\epsilon\,\Delta\phi)\nonumber\\
&=& \epsilon\,\Delta\phi\left\{\frac{\del\cL}{\del\phi}
  -\del_\mu\left(\frac{\del\cL}{\del(\del_\mu\phi)}\right)\right\}
+
\epsilon\,\del_\mu\left(\frac{\del\cL}{\del(\del_\mu\phi)}\,\Delta\phi\right)~.
\label{DeltaL}
\eear
\vs\nin
The first term in (\ref{DeltaL}) vanishes when we use the equations of
motion. The last term is a total derivative so it does not affect the
equations of motion when we minimize the action. We can take advantage
of this fact and define
\vs
\be
j^\mu\equiv \frac{\del\cL}{\del(\del_\mu\phi)}\,\Delta\phi~
\ee
\vs\nin
such that its four-divergence 
\vs
\be
\del_\mu\,j^\mu = 0~,
\ee
\vs\nin
up to terms that are total derivatives in the action, and therefore
do not contribute if we use the equations of motion. We call this
object the conserved current associated with the symmetry
transformation (\ref{symtrans}). We will illustrate this with the
following example.

\nin \underline{Example}:

\nin We consider a complex scalar field. That is, there is a real part
of $\phi(x)$ and an imaginary part, such that $\phi(x)$ and $\phi^*(x)$
are distinct. The Lagrangian density can be written as
\vs
\be
\cL = \del_\mu\phi^*\,\del^\mu\phi - m^2\,\phi^* \phi~.
\label{LComplexScalar}
\ee
 \vs\nin
The Lagrangian density in (\ref{LComplexScalar}) is invariant under
the following transformations
\vs
\bear
\phi(x) &\longrightarrow & e^{i\alpha}\,\phi(x)\nonumber\\
&~&\label{complextrans}\\
\phi^*(x) &\longrightarrow & e^{-i\alpha}\,\phi^*(x)\nonumber~,
\eear
\vs\nin
where $\alpha$ is an arbitrary  constant real parameter.  If we
consider the case when $\alpha$ is infinitesimal  ($\alpha \ll 1$), 
\vs
\bear
\phi(x) &\longrightarrow& \phi'(x)\simeq\phi(x) + i\alpha\phi(x) \\
\phi^*(x) &\longrightarrow& \phi^{*'}(x)\simeq\phi^*(x) -
i\alpha\phi^*(x)~,
\eear
\vs\nin 
which tells us that we can make the identifications
\vs
\bear
\epsilon\,\Delta\phi &=& i\alpha\,\phi\nonumber\\
\epsilon\,\Delta\phi^* &=& -i\alpha\,\phi^*~,
\eear
\vs\nin
with  $\epsilon=\alpha$. In other words we have 
\vs
\be
\Delta\phi = i\phi~,\qquad\Delta\phi^* = -i\phi^*~.
\ee
\vs\nin 
Armed with all these we can now build the current $j^\mu$ associated
with the symmetry transformations (\ref{complextrans}). In particular,
since there are two independent degrees of freedom, $\phi$ and
$\phi^*$, we will have two terms in $j^\mu$ 
\vs
\be
j^\mu = \frac{\del\cL}{\del(\del_\mu\phi)}\,\Delta\phi +
\frac{\del\cL}{\del(\del_\mu\phi^*)}\,\Delta\phi^*~,
\ee
\vs\nin
From (\ref{LComplexScalar}) we obtain
\vs
\be
\frac{\del\cL}{\del(\del_\mu\phi)} = \del^\mu\phi^*~,\qquad \frac{\del\cL}{\del(\del_\mu\phi^*)} = \del^\mu\phi
\ee
\vs\nin
which results in 
\vs
\be
j^\mu = i\left\{(\del^\mu\phi^*)\,\phi -
  (\del^\mu\phi)\,\phi^*\right\}~.
\label{current}
\ee
\vs\nin
We would like to check current conservation, i.e. check that
$\del_\mu\,j^\mu=0$. However, as we discussed above, this is only true
up to total divergences that do not affect the equations of motion. So
the strategy is to compute the four-divergence of the current and then
use the equations of motion to see if the result vanishes.
The equations of motion are easily obtained from the
E\"{u}ler-Lagrange equations applied to $\cL$ in
(\ref{LComplexScalar}). This results in 
\vs
\be
 (\del^2 +m^2)\phi^* = 0~,\qquad (\del^2+m^2)\phi = 0~,
\label{eomcomplex}
\ee
\vs\nin
i.e. both $\phi$ and $\phi^*$ obey the Klein-Gordon equation. 
Taking the four-divergence in  (\ref{current}) we obtain
\vs
\be
\del_\mu  j^\mu = i\left\{(\del^2\phi^*)\phi -
  (\del^2\phi)\phi^*\right\}~,
\ee
\vs\nin
Thus, this is not zero in general. But applying the equations of
motion in (\ref{eomcomplex}) we get
\vs
\be
\del_\mu j^\mu = i\left\{(-m^2\phi^*)\phi - (-m^2\phi)\phi^*\right\} =
0~,
\ee
\vs\nin
which then verifies current conservation. We conclude that, at least
at the classical level, as long as the equations of motion are valid,
the current is conserved.

\subsubsection{Field quantization}

Since, as we saw before, quantum field theory (QFT) emerges as we attempt to
combine quantum mechanics with special relativity it is natural to
start with quantum mechanics of a single particle.  We will see that
when trying to make this conform with relativistic dynamics, we will
naturally develop a way of thinking of the solution to this problem
that goes by the name of canonical quantization. Besides being
conceptually natural, this formalism will be useful when trying to
understand the statistics of different states.

\subsubsubsection{Quantum mechanics}

The Schr\"{o}dinger equation for the  wave-function of a free particle is 
\vs
\be
i\frac{\del}{\del t}\,\psi(\mathbf{x},t) = -\frac{1}{2m}\,
\nabla^2\psi(\mathbf{x},t)~,
\label{Schr}
\ee
\vs\nin
where we set $\hbar =1$. In terms of states and operators, we can
define the wave function as $\psi(\mathbf{x},t) = \langle
\mathbf{x}|\psi,t\rangle$, i.e. in term of the state $|\psi,t\rangle$
projected onto the position state $|\mathbf{x}\rangle$. More generally,
eq.~(\ref{Schr}) can be written as 
\vs
\be
i\frac{\del}{\del t}\,|\psi,t\rangle = H\,|\psi,t\rangle~,
\ee\vs\nin
where $H$ is the Hamiltonian which in the non-relativistic free-particle case is just
\vs\be
H=\frac{\mathbf{p}^2}{2m}~,
\ee\vs\nin
 resulting in (\ref{Schr}). We would like to generalize this for the
 relativistic case, i.e. choosing 
\vs\be
H = + \sqrt{\mathbf{p}^2 + m^2}~,
~\label{relham}
\ee\vs\nin
where again we use $c=1$. If one uses this Hamiltonian in the
Schr\"{o}dinger equation one gets
\vs\be
i\,\frac{\del}{\del t} \psi(\mathbf{x},t) = \sqrt{-\nabla^2 + m^2}\,\psi(\mathbf{x},t) ~.
\ee\vs\nin
But this is problematic for a relativistic equation since time and
space derivatives are of different order. If the equation has to have
any chance of being Lorentz invariant, it needs to have the same
number of time and space derivatives. One simple way to do this is to
apply the time derivative operator twice on both sides. This results in
\vs
\be
-\frac{\del^2}{\del t^2}\,\psi(\mbf{x},t) = \left(-\nabla^2 + m^2\right)\,\psi(\mbf{x},t)~.
\ee\vs\nin
This is the Klein-Gordon equation for the wave function $\psi(\mbf{x},t)$, and is clearly consistent
with the relativistic dispersion relation (\ref{relham}), once we make
the identifications
\vs\be
i\frac{\del}{\del t} \leftrightarrow H~\qquad\qquad -i\nabla
\leftrightarrow \mbf{p}~,
\ee\vs\nin
where $H$ and $\mbf{p}$ are the Hamiltonian and momentum operators. 
In covariant notation, and using 
\vs\be
\frac{\del}{\del x^\mu} \equiv \del_\mu = (\frac{\del}{\del t},
  \mbf{\nabla})~,\qquad
\frac{\del}{\del x_\mu} \equiv \del^\mu = (\frac{\del}{\del t},
  -\mbf{\nabla})~,
\ee\vs\nin
we can write the Klein-Gordon equation as
\vs\be
(\del_\mu\,\del^\mu +m^2)\,\psi(\mbf{x},t)~ =0~.
\ee\vs\nin
This is manifestly Lorentz invariant. However it has several
problems. The fact that this equation has two time derivatives implies
for instance that $|\psi(\mbf{x},t)|^2$ is not generally time
independent, so we cannot interpret it as a conserved probability,
as it is in the case of the Scr\"{o}dinger equation. This issue is tackled
by Dirac, which derives a relativistic equation for the wave-function
that is first order in both time and space derivatives. But this
equation will be valid for spinors, not scalar wave-functions. We will
study it in more detail later. But it  does not resolve the
central issue, as we see below. 

\nin
 
Both the Klein-Gordon and the Dirac equations admit solutions with
negative energies. This would imply that the system does not have a
ground state, since it would be always energetically favorable to go
to the negative energy states. 
Since  the Dirac equation describes fermions, one can use
Pauli's exclusion principle and argue, as Dirac did, that all the
negative energy states are already occupied. This is the so-called
Dirac sea. According to this picture, an electron would not be able to
drop to  negative energy states since these  are already
filled. Interestingly, this predicts that in principle it should be
possible to kick one of the negative energy states to a positive
energy state. Then, one would see an electron appear. But this would
leave a hole in the sea, which would appear as a positively charged
state. This is Dirac's prediction of the existence of the positron. Is
really nice, but now we need an infinite number of particles in the
sea, whereas we were supposed to be describing the wave-function of
{\em one} particle. Besides, this only works for wave-functions
describing fermions. What about bosons ?

What we are seeing is the inadequacy of the relativistic description
of the one-particle wave-function. At best, as in the case of
fermions, we were driven from a
one-particle description to one with an infinite number of particles. 
At the heart of the problem is the fact that, although now we have the
same number of time and space derivatives, position and time are
not treated on the same footing in quantum mechanics. There is in fact
a position operator, whereas time is
just a parameter labeling the states.

\nin
On the other hand, we can consider 
operators labeled by the {\em spacetime} position $x^\mu=(t,\mbf{x})$,
such as in
\nin\be
\phi(t,\mbf{x}) = \phi(x)~.
\ee
These objects are called 
 quantum fields.  They are clearly in the
Heisenberg picture, whereas if we choose the time-independent
Schr\"{o}dinger picture quantum fields they are only labeled by the
spatial component of the position as in $\phi(\mbf{x})$. These quantum
fields will be our dynamical degrees of freedom. All spacetime
positions have a value of $\phi(x)$ assigned. As we will see in more
detail below, the quantization of these fields will result in infinitely
many states. So we will abandon the idea of trying to describe the
quantum dynamics of {\em one } particle. This formulation will allow
us to include {\em antiparticles} and  (in the presence of
interactions) also other particles associated with other quantum
fields. It solves one of the problems mentioned earlier, the fact that
relativity and quantum mechanics should allow the presence of these
extra particles as long as there is enough energy, and/or the
intermediate process that {\em violates } energy conservation by
$\Delta E$ lasts a time $\Delta t$ such that $\Delta E \Delta t \sim
\hbar$.

The behavior of quantum fields under Lorentz transformations will
define their properties. We can have scalar fields $\phi(x)$, i.e. no
Lorentz indices; fields that transform as four-vectors: $\phi^\mu(x)$;
as spinors: $\phi_a(x)$, with $a$ a spinorial index; as tensors, as in
the rank 2 tensor $\phi^{\mu\nu}(x)$; etc. We will start with the
simplest kind, the scalar field.

\subsubsubsection{Canonical description of quantum fields} 

First, let us assume a  scalar field $\phi(x)$ that obeys the
Klein-Gordon equation. The exact meaning of this will become clearer
below. But for now it suffices to  assume that our dynamical
variable obeys a relativistic equation relating space and time
derivatives:
\vs\be
(\del^2 + m^2)\phi(x) = 0~,
\label{kleingordon}
\ee\vs\nin
where we defined the D'Alembertian as $\del^2\equiv \del_\mu\del^\mu$,
and $m$ is the mass of the particle states associated with the field
$\phi(x)$. We also assume the scalar field in question is real. That
is 
\vs\be
\phi(x)=\phi^\dagger(x)~, 
\label{phireal}
\ee\vs\nin
where we already anticipate to elevate the field to an operator, hence
the $\dagger$. 
It is interesting to solve the Klein-Gordon equation for the
classical field in momentum space. The most general solution has the
following form
\vs\be
\phi(\mbf{x},t) = \int \, \frac{d^3p}{(2\pi)^3}\, N_p\,\left\{
a_p\,e^{-i(\omega_p t - \mbf{p}\cdot \mbf{x})} +
  b^\dagger_p\,e^{i(\omega_p t-\mbf{p}\cdot\mbf{x})}\right\}~,
\label{realscalar}
\ee\vs\nin
where, as defined earlier, $\omega_p=+\sqrt{\mbf{p}^2 + m^2}$.
Here, $N_p$ is a momentum-dependent normalization to be determined
later, and the momentum-dependent coefficients $a_p$ and $b^\dagger_P$
will eventually be elevated to operators. 
In general $a_p$ and $b_p^\dagger$ are independent. However, when we
impose (\ref{phireal}), this results in
\vs\be
a_p =b_p~.
\label{aeqb}
\ee\vs\nin
This is not the case, for instance, if $\phi(x)$ is a complex scalar field.

\nin
At this point and before we quantize the system, we remind  ourselves
of the fact that the Klein-Gordon equation (\ref{kleingordon}) is obtained from the
Lagrangian density 
\vs\be
{\cal L} = \frac{1}{2}\del_\mu\phi\,\del^\mu\phi -\frac{1}{2}\,m^2\,\phi^2~.
\label{kglagrangian}
\ee\vs\nin
To convince yourself of this just use the E\"{u}ler-Lagrange equations
from the previous lecture to derive (\ref{kleingordon}) from
(\ref{kglagrangian}). Then, since $\phi(x)$ is our dynamical variable,
the canonically conjugated momentum is
\vs\be
\pi(x) = \frac{\del {\cal L}}{\del \dot{\phi}(x)}=\dot{\phi}~, 
\ee
\vs\nin
which, using (\ref{realscalar}),  results in 
\vs\be
\pi(\mbf{x},t) = \int\,\frac{d^3x}{(2\pi)^3}\,N_p\,\left\{ -i\omega_p\,a_p\,e^{-i(\omega_p t - \mbf{p}\cdot \mbf{x})} +
  i\omega_p\,a^\dagger_p\,e^{i(\omega_p
    t-\mbf{p}\cdot\mbf{x})}\right\}~.
\label{pi}
\ee\vs\nin
Having the field and its conjugate momentum defined we can then impose
quantization conditions. It is useful first to refresh our memory  on how
this is done in quantum mechanics. 

\subsubsubsection{Canonical quantization in quantum mechanics}

Let us consider a particle of mass $m=1$ in some units. 
Its Lagrangian is 
\vs\be
L = \frac{1}{2}\,\dot{q}^2 -V(q)~~,
\ee\vs\nin
where $V(q)$ is some still unspecified potential, which we assume it
does not depend on the velocities. The associated
Hamiltonian is 
\vs
\be
H = \frac{1}{2}\,p^2 + V(q)~~,
\ee\vs\nin
where the conjugate momentum is $p=\del L/\del\dot{q}=\dot{q}$.
 To quantize the system we elevate $p$ and $q$ to operators and impose
 the commutation relations
\vs\be
[q,p] = i, \qquad [q,q]=0=[p,p]~. 
\label{pqcom}
\ee\vs\nin
Notice that if we are in the Heisenberg description, the commutators
should be evaluated at equal time, i.e. $[q(t),p(t)]=i$, etc. 
We change to a description in terms
of the operators
\vs\bear
a &\equiv &\frac{1}{2\omega} \left(\omega\,q+ ip\right)~\nonumber\\
a^\dagger &\equiv &\frac{1}{2\omega} \left(\omega\,q- ip\right)~,
\label{aandadagger}
\eear\vs\nin
where $\omega$ is a constant with units of energy. 
It is straightforward, using the commutators in (\ref{pqcom}), to prove
that these operators satisfy the following commutation relations
\vs\be
[a,a^\dagger]=1, \qquad [a,a]=0=[a^\dagger,a^\dagger]~.
\label{adaggeracom}
\ee\vs\nin
We define the ground state of the system by the following relation
\vs\be
a |0\rangle =0 ~,
\label{vacdef}
\ee\vs\nin
where the $0$ in the state refers to the absence of quanta. Then,
assuming the ground state (or vacuum) is a normalized state, we have
\vs\be
1 = \langle 0 | 0\rangle = \langle 0 | [a,a^\dagger] | 0\rangle =
\langle 0 |a a^\dagger |0\rangle - \langle 0| a^\dagger a | 0\rangle~.
\ee\vs\nin
Since the last term vanishes when using (\ref{vacdef}), we arrive at 
\vs\be
\langle 0 |0\rangle = \langle 0 |a a^\dagger |0\rangle~.
\ee\vs\nin
This is achieved only if we have 
\vs\bear
&&a^\dagger|0\rangle =|1\rangle~,\nonumber\\
&&a| 1\rangle = |0\rangle~,
\eear
\vs\nin
which means that $a$ and $a^\dagger$ are ladder operators. We
interpret the state $|1\rangle$ as a state with one particle. In this
way $a$ and $a^\dagger$ can also we called annihilation and creation
operators. The simplest example is, of course, the simple harmonic
oscillator, with $V(q) = \omega^2\,q^2/2$. 

\subsubsubsection{Quantizing fields}

We are now ready to generalize the canonical quantization procedure
for fields. We will impose commutation relations for the field in
(\ref{realscalar}) and its conjugate momentum in (\ref{pi}), which
means that we elevated them to operators, specifically in the
Heisenberg representation. The
quantization condition is 
\vs\be
[\phi(\mbf{x},t), \pi(\mbf{x'},t)] = i \delta^{(3)}
(\mbf{x}-\mbf{x'})~.
\label{phipicom}
\ee\vs\nin
Here we see that the commutator is defined at equal times, as it
should for Heisenberg operators. All 
other possible commutators vanish, i.e.
\vs\be
[\phi(\mbf{x},t), \phi(\mbf{x'},t)] = 0 = [\pi(\mbf{x},t), \pi(\mbf{x'},t)] 
\label{zerocom}
\ee\vs\nin
Now, when we turned $\phi(x)$ into an operator, so did  $a_p$ and $a_p^\dagger$.
In order to see what the imposition of (\ref{phipicom}) and (\ref{zerocom}) implies for
the commutators of the operators $a_p$ and $a^\dagger_p$, we write out
(\ref{phipicom}) using the explicit expressions (\ref{realscalar}) and
(\ref{pi}) for the field and its momentum in terms of them. We obtain
\vs\bear
[\phi(\mbf{x},t),\pi(\mbf{x'},t)] &=&
\int\,\frac{d^3p}{(2\pi)^3}\,N_p\,\int\,\frac{d^3p'}{(2\pi)^3}\,N_{p'}\,\left\{
  i\omega_{p'} e^{-i(\omega_p -\omega_{p'})t} e^{i\mbf{p}\cdot \mbf{x} -
      i\mbf{p'}\cdot \mbf{x'}}\,[a_p,a^\dagger_{p'} ]
\right.\nonumber \\
&&\left.
-i\omega_{p'} e^{i(\omega_p -\omega_{p'})t} e^{-i\mbf{p}\cdot \mbf{x} +
      i\mbf{p'}\cdot \mbf{x'}}\,[a^\dagger_p,a_{p'} ] \right\}~,
\label{phipi2acom}
\eear\vs\nin
where we have already assumed that 
\vs\be
[a_p,a_{p'}] = 0 = [a^\dagger_p,a^\dagger_{p'}]~.
\ee\vs\nin
The question is what are the commutation rules for
$[a_p,a^\dagger_{p'}]$. 
Now we will show that in order for (\ref{phipicom}) to be satisfied,
we need to impose 
\vs\be
[a_p, a^\dagger_{p'}] = (2\pi)^3\,\delta^{(3)}(\mbf{p}-\mbf{p'})~.
\ee\vs\nin 
If we do this in (\ref{phipicom}) we see that $\omega_p=\omega_{p'}$,
$N_p=N_{p'}$, and we obtain 
\vs\be
[\phi(\mbf{x},t),\pi(\mbf{x'},t)]  = i \int\,\frac{d^3p}{(2\pi)^3}\,
N^2_p\,\omega_p\,\left\{ e^{i\mbf{p}\cdot (\mbf{x}-\mbf{x'})} +
  e^{-i\mbf{p}\cdot (\mbf{x}-\mbf{x'})} \right\}~.
\ee\vs\nin
But we notice that since 
\vs\be
\delta^{(3)}(\mbf{x}-\mbf{x'}) = \int
\frac{d^3p}{(2\pi)^3}\,e^{i\mbf{p}\cdot (\mbf{x}-\mbf{x'})} = \int
\frac{d^3p}{(2\pi)^3}\,e^{-i\mbf{p}\cdot (\mbf{x}-\mbf{x'})}~,
\ee\vs\nin
then if 
\vs\be
N^2_p\,\omega_p = \frac{1}{2} ~,
\ee\vs\nin
we recover the result of ({\ref{phipicom}). In other words 
\vs\be
\boxed{[\phi(\mbf{x},t),\pi(\mbf{x'},t)] = i\delta^{(3)}(\mbf{x}-\mbf{x'})}
\longleftrightarrow \boxed{ [a_p,a^\dagger_{p'}] = (2\pi)^3
  \delta^{(3)}(\mbf{p}-\mbf{p'}) }~,
\ee\vs\nin
as long as
\vs\be
N_p =\frac{1}{\sqrt{2\omega_p}}~.
\ee\vs\nin
We can now go back to the expression (\ref{realscalar}) for the real
scalar field, and rewrite it in covariant form as 
\vs\be
\phi(x) = \int\,\frac{d^3p}{(2\pi)^3\sqrt{2\omega_p}}\,\left\{ a_p
e^{-ip_\mu x^\mu} + a^\dagger_p e^{ip_\mu x^\mu}\right\}~,
\label{phicovariant}
\ee\vs\nin
where we used that 
\vs\be
p_\mu x^\mu = p_0x_0 - \mbf{p}\cdot \mbf{x}= \omega_p t - \mbf{p}\cdot \mbf{x}
\ee\vs\nin
Once again, since we define the vacuum state by
\vs\be
a_p|0\rangle =0~,
\ee\vs\nin
we conclude that $a_p$ and $a^\dagger_p$ are ladder operators, just as
in the quantum mechanical case seen above. In other words we have 
\vs\be
a^\dagger_p |0\rangle = |1_p\rangle~,
\ee\vs\nin
where $|1_p\rangle$ corresponds to the state containing one particle
of momentum $\mbf{p}$. 
\nin Conversely, and analogously to the quantum
mechanical case, we have 
\vs\be
a_p|1_p\rangle =|0\rangle~.
\ee\vs\nin
This allows us to interpret the operators $\phi(x)$ and
$\phi^\dagger(x)$ in the following form:

\vs
\nin
\underline{The operator $\phi(x)$}: 
\begin{itemize}
\item Annihilates a {\em particle} of momentum $\mbf{p}$
\item Creates an {\em anti-particle} of momentum $\mbf{p}$ 
\end{itemize}
 \nin
On the other hand, 

\vs
\nin
\underline{The operator $\phi^\dagger(x)$}: 
\begin{itemize}
\item Annihilates an {\em anti-particle} of momentum $\mbf{p}$
\item Creates a {\em particle} of momentum $\mbf{p}$ 
\end{itemize}
 \nin
Of course in our case, a real scalar field, particles and
anti-particles are the same due to (\ref{aeqb}). On the other hand,
if $\phi$ was for instance complex, particles and anti-particles would
be created and annihilated by different operators, and they would
carry different ``charges'' under the global $U(1)$ symmetry of the
Lagrangian.

\subsubsection{Quantization of fermion fields}
\nin
We will consider the spinor $\psi(\mbf{x},t)$ as a field and use to
quantize the fermion field theory. For this we need 
to know its conjugate momentum. So it will be  helpful to have the
Dirac Lagrangian. We will first insist in imposing {\em commutation} rules just as for the
scalar field. But this will result in a disastrous Hamiltonian. Fixing
this problem will require a drastic modification of the commutation relations for the ladder
operators. 

The first step for the quantization procedure is to have the Dirac Lagrangian. 
 Starting from the Dirac equation
\vs\be
(i\gamma^\mu\del_\mu-m)\psi(x)=0~, 
\label{diracpsi}
\ee\vs\nin
we can obtain the conjugate equation
\vs\be
\bar{\psi}(x)\,(i\gamma^\mu\del_\mu + m) =0 ~,
\label{diracpsibar}
\ee\vs\nin
where in this equation the derivatives act to their left on
$\bar{\psi}(x)$.
From these two equations for $\psi$ and $\bar{\psi}$ is clear that the
Dirac Lagrangian  must be 
\vs\be
{\cal L} = \bar{\psi}(x)\,(i\gamma^\mu\del_\mu -m)\,\psi(x)~.
\label{diraclag}
\ee\vs\nin
It is straightforward to check the the E\"{u}ler-Lagrange equations
result in (\ref{diracpsi}) and (\ref{diracpsibar}). For instance,
\vs\be
\frac{\del\cL}{\del\bar{\psi}} -
\del_\mu\left(\frac{\del\cL}{\del(\del_\mu\bar{\psi})}\right) = 0~. 
\ee\vs\nin
But the second term above is zero since $\cL$ does not depend (as
written) on $\del_\mu\bar{\psi}$.  Thus, we obtain the Dirac equation (\ref{diracpsi})
for $\psi$. Similarly, if we use $\psi$ and $\del_\mu\psi$ as the
variables to put together  the  E\"{u}ler-Lagrange equations, we
obtain (\ref{diracpsibar}).

\nin
From the Dirac Lagrangian we can obtain the conjugate momentum density defined by
\vs\be
\pi(x) = \frac{\del\cL}{\del(\del_0\psi)}= i\bar{\psi}\gamma^0 = i\psi^\dagger~.
\label{diracmomentum}
\ee\vs\nin
This way, if we follow the quantization playbook we used for the scalar
field, we should impose
\vs\be
[\psi_a(\mbf{x},t),\pi_b(\mbf{x'},t)] =
[\psi_a(\mbf{x},t),i\psi^\dagger_b(\mbf{x'},t)]= i \delta^{(3)}(\mbf{x}-\mbf{x'})\,\delta_{ab}~,
\ee\vs\nin
or just
\vs\be
[\psi_a(\mbf{x},t),\psi^\dagger_b(\mbf{x'},t)] =
 \delta^{(3)}(\mbf{x}-\mbf{x'})\,\delta_{ab}~,
\label{comutator}
\ee\vs\nin
Following the same steps as in the case of the scalar field, we now
expand $\psi(x)$ and $\psi^\dagger(x)$ in terms of solutions of the
Dirac equation in momentum space. 
As we will see later, this will not work. But it is interesting to see
why, because this will point directly to the correct quantization
procedure. 
The most general expression for the fermion field in terms of the
solutions of the Dirac equation in momentum space is 
\vs\bear
\psi(x) &=&
\int\frac{d^3p}{(2\pi)^3}\,\frac{1}{\sqrt{2E_p}}\,\sum_s\left(
    a_p^s\,u^s(\mbf{p}) \,e^{-iP\cdot x} + b^{s\dagger}_p\,v^s(\mbf{p})\,
    e^{+iP\cdot x}\right)~,\label{psi}\\
\psi^\dagger(x) &=&
\int\frac{d^3p}{(2\pi)^3}\,\frac{1}{\sqrt{2E_p}}\,\sum_s\left(
    a_p^{s\dagger}\,u^{s\dagger}(\mbf{p}) \,e^{iP\cdot x} + b^{s}_p\,v^{s\dagger}(\mbf{p})\,
    e^{+iP\cdot x}\right)~,\label{psidagger}\
\eear\vs\nin
The imposition of the quantization rule (\ref{comutator}) on the field
and its conjugate momentum in (\ref{psi}) and (\ref{psidagger})
would imply that the coefficients $a^s_p$, $a^{s\dagger}_p$, $b^s_p$ and
$b^{s\dagger}_p$ are ladder operators associated to the $u$-type and
$v$-type ``particles''. But before we impose commutation rules on them
we are going to compute the Hamiltonian in terms of these operators. 

\nin
Remember that the Hamiltonian is defined by
\vs\bear
H &=& \int d^3x \,\left\{ \pi(x)\,\del_0\psi(x) - \cL \right\} ~,\nonumber\\
&=& \int d^3x\, \left\{ i\psi^\dagger(x)\,\del_0\psi(x)
  -\bar{\psi}(x)\left(i\gamma^\mu\del_\mu -
    m\right)\psi(x)\right\}~,\label{hamil1}
\eear\vs\nin
which results in 
\vs\be
\boxed{
H = \int d^3x\, \bar{\psi}(x)\left( -i\mbf{\gamma}\cdot \mbf{\nabla} + m
\right)\psi(x) ~,\label{hamil2}
}
\ee\vs\nin
Inserting (\ref{psi}) and (\ref{psidagger}) into (\ref{hamil2}) we
have
\vs\bear
H &=& \int d^3x\, \left\{ \int
  \frac{d^3k}{(2\pi)^3}\,\frac{1}{\sqrt{2E_k}}\,\sum_r\left( 
a^{r\dagger}_k\,\bar{u}^{r\dagger}(\mbf{k})\,e^{iK\cdot x} +
b^r_k\,\bar{v}^r(\mbf{k})\,e^{-iK\cdot x}\right) \right. \label{hamil3}\\
 && \left. \times \int
\frac{d^3p}{(2\pi)^3}\,\frac{1}{\sqrt{2E_p}}\,\sum_s\left( a^s_p
  \,e^{-iP\cdot x}\,(\mbf{\gamma}\cdot \mbf{p} +m)\,u^s(\mbf{p}) 
+ b^{s\dagger}_p\,e^{+iP\cdot x}\,(-\mbf{\gamma}\cdot \mbf{p} +
m)\,v^s(\mbf{p}) \right) \right\}~,\nonumber
\eear\vs\nin
In the second line of (\ref{hamil3}) the Hamiltonian operator was
applied to the exponentials. Since $P\cdot x = E x_0 -
\mbf{p}\cdot\mbf{x}$, the $-i$ in the operator cancels with the 
$+i\mbf{p}\cdot\mbf{x}$ in when the derivative acts on the $-P\cdot x$ 
exponential.   The opposite sign is picked up when acting on the
$+P\cdot x$ exponential. Furthermore, since 
\vs\be
(\dsl p -m)\,u^s(\mbf{p}) =0 \quad\Longrightarrow \quad(E_p\gamma^0
-\mbf{\gamma}\cdot\mbf{p} - m)\,u^s(\mbf{p}) =0~,
\ee\vs\nin 
which results in
\vs\be\boxed{
(\mbf{\gamma}\cdot\mbf{p} + m)\,u^s(\mbf{p}) =
E_p\,\gamma^0\,u^s(\mbf{p})~.}
\label{oponu}
\ee\vs\nin
Similarly, applying 
\vs\be
(\dsl p +m)\,v^s(\mbf{p}) =0 \quad\Longrightarrow \quad(E_p\gamma^0
-\mbf{\gamma}\cdot\mbf{p} + m)\,u^s(\mbf{p}) =0~,
\ee\vs\nin 
which gives us
\vs\be\boxed{
(-\mbf{\gamma}\cdot\mbf{p} + m)\,v^s(\mbf{p}) =
-E_p\,\gamma^0\,v^s(\mbf{p})~.}
\label{oponv}
\ee\vs\nin
Using (\ref{oponu}) and (\ref{oponv}) and that 
\vs\be
\int d^3x e^{\pm i(\mbf{k}-\mbf{p})\cdot\mbf{x}}  =
(2\pi)^3\,\delta^{(3)} (\mbf{k}-\mbf{p})~,
\ee\vs\nin
in (\ref{hamil3}) we can get rid of 2 of the 3 integrals. Then we have
\vs\bear
H &=& \int \frac{d^3p}{(2\pi)^3}\,\frac{1}{2E_p}\,\sum_{r,s}\,\left\{
  a^{r\dagger}_p\,a^s_p\,u^{r\dagger}(\mbf{p})\,\gamma^0\,E_p\gamma^0\,u^s(\mbf{p})
\right.\nonumber\\
&&\qquad\qquad\qquad\quad -\left. b^r_p\,b^{s\dagger}_p\,v^{r\dagger}(\mbf{p})\,\gamma^0\,E_p\,\gamma^0\,v^s(\mbf{p})\right\}~,
\eear\vs\nin
where we have also use the orthogonality of the $u^s(\mbf{p})$ and
$v^s(\mbf{p})$ solutions. Finally, using the normalization os spinors
\vs\bear
u^{r\dagger}(\mbf{p})\,u^s(\mbf{p}) &=& 2E_p\,\delta^{rs}~, \nonumber\\
v^{r\dagger}(\mbf{p})\,v^s(\mbf{p}) &=& 2E_p\,\delta^{rs}~, 
\eear\vs\nin
 we obtain
\vs\be
H =\int \frac{d^3p}{(2\pi)^3}\,\sum_s\,\left\{ E_p\,a^{s\dagger}_p\,a^s_p
  - E_p\, b^s_p\,b^{s\dagger}_p\right\}~.
\label{hamil4}
\ee\vs\nin
In order to have a correct form of the Hamiltonian, we must rearrange
the second term in (\ref{hamil4}) into a number operator, such as the
first term. For this purpose, we need to apply the commutation rules
on $b^s_p$ and $b^{s\dagger}_p$. If we were to impose the same
commutation rules we used for scalar fields, and also in
(\ref{comutator}), we would have
\vs\be
[a^r_p,a^{s\dagger}_k] =
(2\pi)^3\,\delta^{(3)}(\mbf{p}-\mbf{k})\,\delta^{rs}, \qquad 
[b^r_p,b^{s\dagger}_k] =
(2\pi)^3\,\delta^{(3)}(\mbf{p}-\mbf{k})\,\delta^{rs}, 
\label{wrongcomute}
\ee\vs\nin
and zero otherwise. This would result in a Hamiltonian
\vs\be
H= \int \frac{d^3p}{(2\pi)^3}\,E_p\,\sum_s\,\left\{
a^{s\dagger}_p\,a^s_p - b^{s\dagger}_p\,b^s_p\right\} - \int\,E_p
d^3p\,\delta^{(3)}(\mbf{0}) ~.\label{wronghamil}
\ee\vs\nin
The last term in (\ref{wronghamil}) is an infinite constant. It
corresponds to the sum over all the zero-point energies of the infinite
harmonic oscillators each with a ``frequency'' $E_p$. This will always
be present in quantum field theory (just as the zero-point energy is
present in the harmonic oscillator! ) and we will deal with it throughout
the course. However, since it is a constant, we can always shift the
origin of the energy in order to cancel it\footnote{The fact that this
  constant is negative will remain true and is an important fact. For
  instant, for scalar fields is positive.}. So this is not what is
wrong with this Hamiltonian. The problem is in the first term,
particularly the negative term. The presence of this negative term
tells us that we can lower the energy by producing additional $v$-type
particles. For instance, the state $|\bar{1}_p\rangle $ with one such
particle would have an energy 
\vs\be
\langle \bar{1}_p | H|\bar{1}_p\rangle = -E_p < \langle 0|H|0\rangle ~,
\ee\vs\nin
 smaller than the vacuum. This means that we have a runaway
 Hamiltonian, i.e. its ground state corresponds to the state with
 infinite such particles. This is of course non-sense. The problem
 comes from the use of the commutation relations
 (\ref{wrongcomute}). On the other hand if we used    anti-commutation
 relations such as 
\vs\be
\{a^r_p,a^{s\dagger}_k\} =
(2\pi)^3\,\delta^{(3)}(\mbf{p}-\mbf{k})\,\delta^{rs}, \qquad 
\{b^r_p,b^{s\dagger}_k\} =
(2\pi)^3\,\delta^{(3)}(\mbf{p}-\mbf{k})\,\delta^{rs}, 
\label{anticomute}
\ee\vs\nin
together with 
\vs\be
\{a^r_p,a^{s}_k\} =0=\{a^{r\dagger}_p,a^{s\dagger}_k\}~,\qquad
\{b^r_p,b^{s}_k\} = 0= 
\{b^{r\dagger}_p,b^{s\dagger}_k\}~,
\label{zeroanticomute}
\ee\vs\nin
and we go  back to  (\ref{hamil4}),  using (\ref{zeroanticomute})
instead of (\ref{wrongcomute}) we obtain
\vs\be\boxed{
H= \int \frac{d^3p}{(2\pi)^3}\,E_p\,\sum_s\,\left\{
a^{s\dagger}_p\,a^s_p + b^{s\dagger}_p\,b^s_p\right\} +
{\rm constant} ~.}\label{hamil5}
\ee\vs\nin
This is now a well behaved Hamiltonian, where for each fixed value of
the momentum we have a contribution to the energy of $a^{s\dagger}_p\,a^s_p$ number
of particles of type $u$, and $b^{s\dagger}_p\,b^s_p$ number of
particles of type $v$. This is the expected form of the Hamiltonian,
and we arrived at it by using the anti-commutation relations
(\ref{anticomute}) and (\ref{zeroanticomute}) for the ladder
operators. It is straightforward to show that they imply
anti-commutation rules also for the fermion field and its conjugate
momentum. That is
\vs\be\boxed{
\{\psi_a(\mbf{x},t),\psi_b^\dagger(\mbf{x'},t)\} =
\delta^{(3)}(\mbf{x}-\mbf{x'})\,\delta_{ab}~,}
\label{anticomutepsi} 
\ee\vs\nin
and zero otherwise, instead of (\ref{comutator}).

\subsubsubsection{Charge operator and fermion number}

\nin
In order to better understand the meaning of the $u$ and $v$
solutions it is useful to build another operator other than the
Hamiltonian. We start with the Dirac current. We know that it is given
by
\vs\be
j^\mu = \bar{\psi}\gamma^\mu\psi~,
\label{diraccurrent}
\ee\vs\nin
satisfying current conservation
\vs\be
\del_\mu\,j^\mu = 0~.
\label{jconserv}
\ee\vs\nin
Noether's theorem tells us that the conserved current is associated
with a conserved charge defined by 
\vs\be
Q = \int d^3x\,j^0(x) = \int d^3x\,\bar{\psi}(x)\gamma^0 \psi(x) =
\int d^3x\,\psi^\dagger(x)\,\psi(x)~. 
\ee\vs\nin
We have seen this before: it is the probability density obeying a
continuity equation (\ref{jconserv}). The fact that the  charge $Q$ is
time independent is a direct consequence of (\ref{jconserv}). We build
this operator in terms of ladder operators in momentum space just as
we did for the Hamiltonian. Using (\ref{psi}) and (\ref{psidagger})
and following the same steps that lead to (\ref{hamil3}) we obtain
\vs\be
Q = \int \frac{d^3p}{(2\pi)^3}\, \sum_s\left\{ a^{s\dagger}_p\,a^s_p + b^s_p\,b^{s\dagger}_p\right\}~,
\ee\vs\nin
Using the anti-commutation relations (\ref{zeroanticomute}) on the second term we arrive at
\vs\be
Q = \int \frac{d^3p}{(2\pi)^3}\, \sum_s\left\{ a^{s\dagger}_p\,a^s_p -
  b^{s\dagger}_p\,b^s_p\right\}~,
\label{chargeop}
\ee\vs\nin
where we have omitted the $a$ and $b$-independent, infinite constant. 
We see clearly that each $u$-type  particle contributes to $Q$ with
$+1$, whereas each $v$-type particle contributes with $-1$. The
continuous symmetry associated with the current $j^\mu$ is just the
global fermion number. That is the Lagrangian is invariant under 
\vs\bear
&&\psi(x)\longrightarrow e^{i\alpha}\,\psi(x) ~,\nonumber\\
&&\psi^\dagger(x) \longrightarrow e^{-i\alpha}\,\psi^\dagger(x)~.
\eear\vs\nin
with $\alpha$ a real constant. This just says that the Dirac
Lagrangian conserves fermion number, meaning that there are fermions
with charge $+1$ and anti-fermions (the $v$-type states) with
charge $-1$. To summarize:
\vs\nin
\begin{itemize}
\item $a^s_p$: annihilates fermions 
\item $b^{s\dagger}_p$  creates anti-fermions
\item $a^{s\dagger}_p$ creates fermions
\item $b^s_p$ annihilates anti-fermions
\end{itemize}
Or, in other words
\begin{itemize}
\item $\psi(x)$ annihilates fermions or creates anti-fermions
\item $\psi^\dagger(x)$ creates fermions or annihilates anti-fermions
\end{itemize}

\subsubsubsection{Pauli exclusion principle and statistics}

One of the most important consequences of having anti-commutation
rules for the ladder and field operators is that fermions obey
Fermi-Dirac statistics and the Pauli exclusion principle. 
To see this, consider a two fermion state. It is built out of creation
operators as 
\vs\be
|1^s_p \, 1^r_k\rangle = a^{s\dagger}_p\,a^{r\dagger}_k |0\rangle~.
\label{twofstate}
\ee\vs\nin
The anti-commutation rules (\ref{zeroanticomute})  imply
\vs\be
a^{s\dagger}_p \,a^{r\dagger}_k = - \, a^{r\dagger}_k\,a^{s\dagger}_p~. 
\ee\vs\nin
which means that the state is odd under the exchange of two particles (for
instance switching positions), or 
\vs\be
|1^s_p \, 1^r_k\rangle  = - |1^r_k \, 1^s_p\rangle 
\ee\vs\nin
In particular if both fermions have the same exact quantum numbers,
here in our example the helicity $s$ and the momentum $\mbf{p}$,  we
have
\vs\be
|1^s_p \, 1^s_p\rangle  = - |1^s_p \, 1^s_p\rangle  = 0~,
\label{pauli}
\ee\vs\nin
which means that this state is forbidden. As a result, we cannot put
two fermions (or two anti-fermions) with the exact same quantum
numbers in the same state. So the occupation numbers in states made of fermions 
 are either $0$ or $1$ for a given set of quantum numbers. This is
 what is called Fermi-Dirac statistics. Equation (\ref{pauli}) is an
 expression of the Pauli exclusion principle.  

 \subsubsection{Interactions and Feynman rules}
 \label{sec:feynrules}

 In Section~\ref{sec:whyqft} we derived an expression for the amplitude
 for a particle to be produce in one pont of space time, propagate and
 be annihilated in another point. The kernel of the amplitude defined
 in (\ref{ampdf}) is the two-point function $D_F(x-y)$ in
 (\ref{feynmanprop}). But since we now know that quantum fields act as
 creation and annihilation operators for quanta of the fields, we can
 write 
 \begin{eqnarray}
D_F(x-y)  &=&  \int\frac{d^3p}{(2\pi)^3\,2\omega_p}\,\left\{ \theta(x_0-y_0)\,
    e^{-ip^\mu\,(x_\mu-y_\mu)} + \theta(y_0-x_0)  \,
              e^{+ip^\mu\,(x_\mu-y_\mu)} \right\} \nonumber\\
   & & \nonumber\\
          &=& \langle 0|T \phi(x) \,\phi(y) |0\rangle~.
              \label{twopoint}
\end{eqnarray}
 In the second line in (\ref{twopoint}) we see that $D_F(x-y)$ is the
 ground state (or vacuum) expectation value of the product of two
 field operators evaluated at different points in spacetime in a {\em
   time ordered} form bu the application of the time order operator
 $T$. The two-point function above is called the Feynman
 propagator. It isa {\em causal} propagator, in the sense that both
 possible time orderings ($x_0>y_0$ and $x_0<y_0$) are taking into
 account in it. But there are other correlation functions we can be
 interested in. For instance, we could want to know the four-point
 correlation function
 \begin{equation}
G^{(4)}(x_1,x_2,x_3,x_4) \equiv   \langle
0|T\phi(x_1)\,\phi(x_2)\,\phi(x_3)\,\phi(x_4)|0\rangle~.
\label{fourpoint}
\end{equation}
But since this is a free theory and there are no interactions, the
only thing that a particle created somewhere can do is propagate and
be annihilated somewhere else. So this four-point function can be
diagrammatically  expressed as seen in Figure~\ref{fig:fourpoint}.
\begin{figure}[h]
  \begin{center}
    \includegraphics[width=4in]{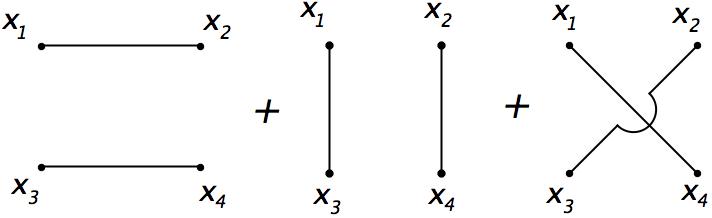}
    \caption{Four-point correlation function in the free scalar
      theory. It is the sum over the products of all possible pairs of propagators.} 
    \label{fig:fourpoint}
\end{center}
\end{figure}
\vs\nin
The result is the sum over the product of all possible combinations of
two propagators: 
\begin{eqnarray}
  G^{(4)}(x_1,x_2,x_3,x_4) &=& D_F(x_1-x_2)\,D_F(x_3-x_4) +
D_F(x_1-x_3)\,D_F(x_2-x_4) \nonumber\\
                           &&+ D_F(x_1-x_4)\,D_F(x_2-x_3)~.
\end{eqnarray}
We can generalize this result for the $n$-point correlation function
as
\vs\be
G^{(n)}(x_1,\dots,x_{n}) = \sum_{\rm all~pairings}
D_F(x_{i_1}-x_{i_2})\dots D_F(x_{n-1}-x_{n})~.
\label{wicsth}
\ee\vs\nin
That is, in the free scalar theory the $n$-point correlation function
is given by the product of all possible products of pairings of two
points into propagators
(2-point functions). For instance, for the 6-point correlation
function we would need products of three propagators, etc. This result
reflects something called Wick's theorem. And although it looks that it
would be useful only in free theories, we will see below how we can
still use it in the presence of interactions, as long as we make use
of perturbation theory.

\subsubsubsection{Perturbation theory}

In the presence of interactions the correlation functions will
change. But in general the solution of the problem is better approached
by using a controlled approximation, typically in powers of the
interaction's strength, i.e. its coupling.
The lagrangian now is given by
\begin{equation}
  {\cal L} = {\cal L}_0 + {\cal L}_{\rm int.} ~,
  \label{realscalagrangian}
\end{equation}
where ${\cal }_0$ is the free theory Lagrangian and ${\cal L}_{\rm
  int.}$ denotes the interaction Lagrangian. The latter involves more
than two fields and must respect not just Lorentz invariance, but also
any other symmetry we impose. For instance for real scalar field the
interaction
\begin{equation}
{\cal L}_{\rm int.} = -\frac{\lambda}{4!}\,\phi^4~,
\end{equation}
is invariant under the discrete symmetry $\phi(x) \to -\phi(x)$,
whereas for a complex scalar field the interaction
\begin{equation}
{\cal L}_{\rm int.} = -\frac{\lambda}{2}\,(\phi\phi^*)^2~,
\end{equation}
respects a {\em global} $U(1)$ transformation, i.e. the Lagrangian is
invariant under $\phi(x)\to e^{i\alpha}\phi(x)$ with $\alpha$ a real
constant. Since this is a continuous symmetry, there is a conserved
current associated with it.\footnote{A {\em local} continuous
  transformation (basically with $alpha = \alpha(x)$) is the case of
  gauge theories. We will discuss them  later below.}   For
simplicity, let us consider the case of a real scalar field with
Lagrangian
\begin{equation}
{\cal L} = \frac{1}{2}\partial_\mu\phi\partial^\mu\phi
-\frac{m^2}{2}\phi^2-\frac{\lambda}{4!}\,\phi^4~.
\label{eq:lagrealscalar}
\end{equation}
In general, the n-point correlation functions of the theory can be
written in the functional integral approach as
\begin{equation}
\langle 0 |T\phi(x_1)\dots \phi(x_n)|0\rangle = G^{(n)}(x_1,\dots x_n)
= \frac{\int {\cal D}\phi
  \, e^{i\int d^4x \cal{L}}\,\phi(x_1)\dots\phi(x_n) }{\int
  {\cal D}\phi e^{i\int d^4x {\cal L} }}~.
\label{eq:npointcorr}
 \end{equation} 
But since the Lagrangian in (\ref{eq:npointcorr}) and
(\ref{eq:lagrealscalar}) contains term that are non-quadratic in the
field $\phi(x)$ we cannot perform the functional integrals as easily
as in the free theory, where they can be turned into basically Gaussian
integrals.  As a result, we make use of perturbation theory in the
interaction coupling $\lambda$. To implement this in the functional
integral we must expand the exponential in powers of ${\cal L}_{\rm
  int.}$. We start with the denominator in (\ref{eq:npointcorr})
above. It expansion reads
\begin{eqnarray}
\int  {\cal D}\phi e^{i\int d^4x\{ {\cal L}_0+{\cal L}_{\rm int.}\}}  &=&
\int  {\cal D}\phi e^{i\int d^4x {\cal L}_0} +\int {\cal D}\phi \, e^{i\int d^4z
  \,{\cal L}_0}\,i\left(-\frac{\lambda}{4!}\right) \int  d^4x\,\phi(x)^4~\nonumber\\
  &~& ~\label{eq:intervac}\\
& & + \int {\cal  D}\phi \,e^{i\int d^4z
  {\cal L}_0}\,\frac{i^2}{2!}\left(\frac{-\lambda}{4!}\right)^2\int
d^4x \phi^4(x)\int d^4y \phi^4(y) +\dots~\nonumber
\end{eqnarray}  
 We interpret the first term in the right hand side of
 (\ref{eq:intervac}) as the vacuum-to-vacuum amplitude in the free
 theory, whereas the terms or order $\lambda$ and higher can be seen
 as corrections to this ``vacuum persistence'' due to the presence of
 interactions. Then, we see that the left hand side can be thought of
 as the corrected vacuum persistence in the presence of the
 interactions
 \begin{equation}
\langle\tilde 0|\tilde 0\rangle = \langle 0| 0\rangle +
\dots~,
\label{eq:newvacuum}
\end{equation}
where we denoted $|\tilde 0\rangle$ as the corrected vacuum state.
\begin{figure}[h]
\begin{center}
\includegraphics[width=4in]{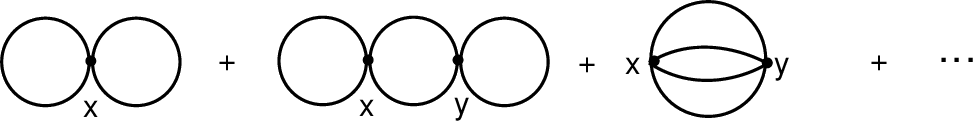}
\caption{Corrections to the vacuum state coming from the
  interactions. 
The first two bubbles are the order $\lambda$, whereas the third and
fourth diagrams are the $\lambda^2$ corrections appearing in
$\langle\tilde 0|\tilde 0\rangle - \langle 0| 0\rangle $.
}
\label{fig:vacuumbubbles}
\end{center}
\end{figure}
We can see this diagrammatically in Figure~\ref{fig:vacuumbubbles}. The
fact that the Lagrangian appearing in the exponent in the expressions
in (\ref{eq:intervac}) is the free theory one, allows us to apply
Wick's theorem also here. For instance, the contribution of order
$\lambda$ can be written as
\begin{equation}
-i\frac{\lambda}{4!}\,\int d^4x \langle 0|T
\phi(x)\phi(x)\phi(x)\phi(x)|0\rangle =  -i\frac{\lambda}{4!}\,\int
d^4x D_F(x-x) D_F(x-x)~.
\end{equation}

The term of order $\lambda^2$ in the second line of (\ref{eq:intervac}) will result in the products of 
four propagators giving terms such as
\begin{equation}
\int d^4x\,\int d^4y \,D_F(x-x)\,D_F(x-y)\,D_F(x-y)\,D_F(y-y)~,
\end{equation}
as represented in the third diagram in Figure~\ref{fig:vacuumbubbles}, or  in
the following combination 
\begin{equation}
\int d^4x\,\int d^4y \,D_F(x-y)\,D_F(x-y)\,D_F(x-y)\,D_F(x-y)~,
\end{equation}
as represented by the last diagram of
Figure~\ref{fig:vacuumbubbles}.The vacuum bubbles in Figure~\ref{fig:vacuumbubbles}
of the {\em denominator} in (\ref{eq:npointcorr}) are just corrections
to the vacuum state and will cancel with corresponding vacuum bubbles
in the {\em numerator } of the correlation
functions. So we do not need to concern ourselves with these vacuum
bubbles since we are interested in diagrams with connection to
external points and their {\em connected } corrections.

For example, let us consider the order $\lambda$ corrections to the
two point correlation function.
\vs
\begin{figure}[h]
\begin{center}
\includegraphics[width=4in]{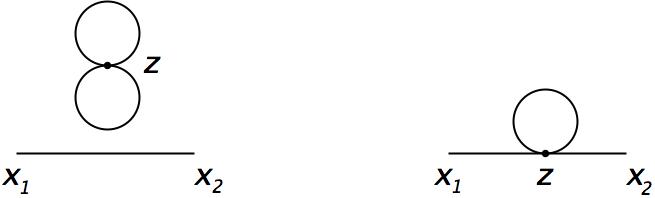}
\caption{
Order $\lambda$ corrections to the  two-point function in the theory
described in the text. 
}
\label{fig:olcorr2pf}
\end{center}
\end{figure}
\vs\nin
The two point function to this order comes from the perturbative expansion
\vs\begin{equation}
 G^{(2)}(x_1,x_2) =
\frac{1}{\int {\cal D}\phi e^{\int d^4x {\cal L}_0}}\,\int {\cal D}\phi\,e^{i\int d^4x\,{\cal L}_0}\,\phi(x_1)\,\phi(x_2)~
\times\left(1-i\frac{\lambda}{4!}\, \int d^4x\,\phi^4(x) +
  \dots\right)~,
\label{eq:twopointol}
\end{equation}
where we are already omitting the corrections in the denominator
since, as mentioned earlier,
they will be cancelled by vacuum bubbles in the numerator. Thus, the functional
integrals can be performed using Wick's theorem, since they  only
depend  on the free Lagrangian ${\cal L}_0$.
For instance, the first term is clearly the free propagator
$D_F(x_1-x_2)$, the zeroth order in $\lambda$. The second term, the
contribution to order $\lambda$, is given by 
\begin{equation}
-i\frac{\lambda}{4!} \, \frac{1}{\int {\cal D}\phi e^{\int d^4x {\cal L}_0}}\,\,\int d^4y\int {\cal D}\phi\,e^{i\int
  d^4x\,\cL_0}\,\phi(x_1)\,\phi(x_2) \,\phi^4(y)~. 
\label{eq:olterm}
\end{equation}
The application of Wick's theorem to the expression above in
(\ref{eq:olterm}) results in two terms, corresponding to the two ways
to pair ( sometimes called contraction) the two fields evaluated in
the {\em external} points with the four fields in the local
interaction. These are given by
\begin{equation}
-i\frac{\lambda}{4! } \int d^4y\,\left\{3\,
  D_F(x_1-x_2)\,D_F(y-y)\,D_F(y-y) +
  12\,D_F(x_1-y)\,D_F(x_2-y)\,D_F(y-y)\right\} ~,
\label{eq:twotopol}
\end{equation} 
where the factors of $3$ and $12$ are the combinatoric factors of the
two types of diagrams: free propagation from $x_1$ to $x_2$ plus
vacuum correction, and correction of the propagator to order
$\lambda$. These terms 
correspond to the two topologies shown in
Figure~\ref{fig:olcorr2pf}. The disconnected diagram on the left is
just the free propagator plus an order $\lambda$ correction of the
vacuum. It will be cancelled by the corresponding vacuum correction in
the denominator. The diagram on the right of
Figure~\ref{fig:olcorr2pf} is more interesting: represents a genuine order
$\lambda$ correction to the propagator. 

Let us now  consider the four point function. Up to order $\lambda$ in
perturbation theory we can write as
\begin{equation}
G^{(4)}(x_1,x_2,x_3,x_4) =\frac{1}{\int {\cal D}\phi e^{\int d^4x
    {\cal L}_0}} \,\int {\cal D}\phi\,\,e^{i\int
  d^4x\,\cL_0}\,\phi(x_1)\phi(x_2)\phi(x_3)\phi(x_4) \left(1
  -i\frac{\lambda}{4!}\,\int d^4y \phi^4(y) + \dots\right)~.
\label{eq:g4tol}
\end{equation}
Of course, the order zero is the four point function of free theory,
where there are no interactions, just propagation from one point to
another. The order $\lambda$ term leads to several diagrams. However,
we want to focus on a special diagram where each of the external
points is connected via a propagator to the interaction point, here
denoted by the coordinate $y$. This {\em fully connected}
contributions to the four point function can be written as
\begin{equation}
(-i\lambda)\,\int
d^4z\,D_F(x_1-z)\,D_F(x_2-z)\,D_F(x_3-z)\,D_F(x_4-z)~,
\label{eq:conn4p}
\end{equation}  
where we have used Wick's theorem. Inspecting (\ref{eq:g4tol}), we see
that there are $4!$ ways to obtain the result above, also represented
by Figure~\ref{fig:ol4p}. 
\begin{figure}[h]
\begin{center}
  \includegraphics[width=2.5in]{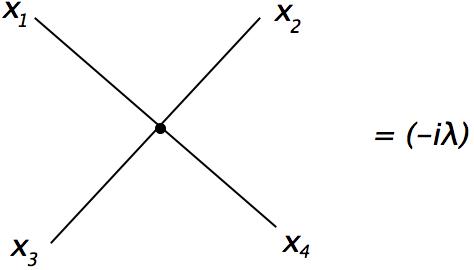}
\caption{
Connected diagram contribution to the four-[point function to order $\lambda$. 
}
\label{fig:ol4p}
\end{center}
\end{figure}
This fully connected diagram will be used below in order to define the
basic rules of the interacting theory in question. Other,
non-connected diagrams contributing to the four point correlation
function to order  $\lambda$  can be seen in Figure~\ref{fig:disl4p}.
\begin{figure}[h]
\begin{center}
  \includegraphics[width=2.5in]{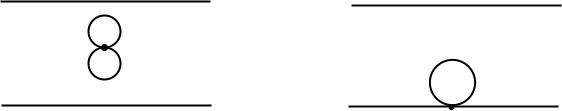}
\caption{
Disconnected  diagram contributions to the four-point function to order $\lambda$. 
}
\label{fig:disl4p}
\end{center}
\end{figure}
The diagram on the left is just the free theory contribution corrected
by a vacuum bubble. On the other hand, the disconnected contribution
on the right is an order $\lambda$ correction of one of the two
disconnected propagators. Unlike the previous one, this contribution is not cancelled by the
corrections in the denominator. However, in order to compute the
physical amplitudes of interest we will need only {\em connected}
diagrams. We will discuss the reason for this next to derive the
Feynman rules in momentum space.

\subsubsubsection{From correlation functions to amplitudes: Feynman
  rules in momentum space}

Although the correlation functions we have obtained using perturbation
theory are physically meaningful objects, they are not as useful to
compare with experimental observables. For this purpose it is
necessary to compute transition amplitudes, typically in momentum
space. For instance, we may want to compute the amplitude for the
scattering of two particles of given momenta $\mbf p_1$ and $\mbf p_2$  in the initial state
going into a final state with several particles. I.e. we want to
compute the 
\begin{equation}
  \langle {\mbf p_1 \mbf p_2}|{\mbf p_3}\dots{\mbf p_n}\rangle ~,
  \label{scatam1}
 \end{equation}
from our knowledge of a given quantum field theory's correlation
functions. In order to do this, we start by defining the initial state
as {\em asymptotic} states in the far past, i.e. for $t\to-\infty$, and
the final states as asymptotic states in the far future, i.e. for
$t\to+\infty$. In these $n$ particle amplitude, we assume that
asymptotic states are well defined momentum states, well separated
from each other, i.e. without appreciable superposition between any
two states. So in the far past or in the far future, these states are
 {\em not interacting with each other}. On the other hand, this does
 not mean that asymptotic states do not feel the effects of the
 interactions. They do not feel the interactions with the other 
 real particles in the amplitude, but they still feel the virtual effects of
 the interactions as they propagate. So the asymptotic states are not
 free states. We will clarify these important different later on. For
 now, the aim is to write the scattering amplitude in (\ref{scatam1})
 in terms of the correlation functions of our quantum field theory.

 We then start by defining the asymptotic states in the far past as
 satisfying
 \begin{equation}
|p\rangle =\sqrt{2\omega_p}\,a^\dagger_p(-\infty) |0\rangle~,
 \end{equation}
where $a^\dagger_p(-\infty)$ creates a particle of momentum  $\mbf{p}$
at $t\to -\infty$. Analogously,
\begin{equation}
  |p\rangle =\sqrt{2\omega_p}\,a^\dagger_p(+\infty) |0\rangle~,
 \end{equation}
creates a particle of momentum $\mbf{p}$ in the far future 
at $t\to +\infty$. If we consider the $2\to n$ scattering amplitude,
we want to compute the momentum space amplitude 
\vs\bear
\langle f| i\rangle = \langle p_3\dots p_n|p_1 p_2\rangle &=&
\sqrt{2\omega_{p_1}}\sqrt{2\omega_{p_2}}\sqrt{2\omega_{p_3}}\dots\sqrt{2\omega_{p_n}}\nonumber\\
&\times&\langle 0 | a_{p_3}(+\infty)\dots
a_{p_n}(+\infty)\,a^\dagger_{p_1}(-\infty)\,a^\dagger_{p_2}(-\infty) |
0\rangle~. 
\label{amplitude1}
\eear\vs\nin
Then, in order to obtain this observable  from the correlation
functions written in terms of fields, we need to invert the expansion
of fields in momentum space. 
In the case of a free scalar field, the momentum
expansion that needs to be inverted is
\vs\be
\phi(x) = \int\frac{d^3k}{(2\pi)^3}\frac{1}{\sqrt{2\omega_k}}\,\left(a_k
    e^{-ik\cdot x} + a^\dagger_k e^{+ik\cdot x}\right)~,
\label{freephi}
\ee\vs\nin
From this, it is straightforward to prove that
\vs\be
\langle0|\phi(x)|0\rangle=0, \qquad\qquad\langle 0| \phi(x) |p\rangle
=e^{-ip\cdot x}= e^{-i\omega_p t} e^{i\mbf{p}\cdot \mbf{x}}~,
\label{mincond1}
\ee\vs\nin
These are in fact the two conditions that we will need to maintain
once we consider an interacting theory. The first one tells us that in
fact $a_p$ annihilates the vacuum. The second condition ensures that
the creation operators $a^\dagger_p$ does create a single particle
state with momentum $\mbf{p}$. In the presence of interactions, the
main difference regarding creation and annihilation operators is that
they acquire time dependence. This is implicit in (\ref{amplitude1})
where we have $t\to\pm \infty$ to the well separated asymptotic
states.   So if we use the free field expansion (\ref{freephi}) to
invert it an obtain expression for the annihilation and creation
operators of asymptotic states in the presence of interactions, all we
need to guarantee is that (\ref{mincond1}) are still satisfied. We
will comment on this point below.

Making use of the free field expansion (\ref{freephi}) it is possible
to arrive at
vs\be\boxed{
i\int d^4x \,e^{ip\cdot x} \left(\del^2+m^2\right)\phi(x) =
\sqrt{2\omega_p}\left[a_p(+\infty) - a_p(-\infty)\right]}~,
\label{lsz1}
\ee\vs\nin
for annihilation operators and 
\vs\be\boxed{
-i\int d^4x \,e^{-ip\cdot x} \left(\del^2+m^2\right)\phi(x) =
\sqrt{2\omega_p}\left[a^\dagger_p(+\infty) - a^\dagger_p(-\infty)\right]}~,
\label{lsz2}
\ee\vs\nin
for the creation operators.
Then, the amplitude of interest in (\ref{amplitude1}) can be rewritten
as 
\vs\bear
\langle f|i\rangle &=& \sqrt{2\omega_{p_1}}\sqrt{2\omega_{p_2}}\sqrt{2\omega_{p_3}}\dots\sqrt{2\omega_{p_n}}\nonumber\\
&\times&\langle 0 | a_{p_3}(+\infty)\dots
a_{p_n}(+\infty)\,a^\dagger_{p_1}(-\infty)\,a^\dagger_{p_2}(-\infty) |0\rangle\nonumber\\
&=&\sqrt{2\omega_{p_1}}\dots\sqrt{2\omega_{p_n}} \quad \langle 0 | T\left(\left[a_{p_3}(+\infty)-a_{p_3}(-\infty)\right]\dots\right.\label{timeorder}\\
&&\left.\left[a_{p_n}(+\infty)-a_{p_n}(-\infty)\right]
\left[a^\dagger_{p_1}(+\infty)-a^\dagger_{p_1}(-\infty)\right]\left[a^\dagger_{p_2}(+\infty)-a^\dagger_{p_2}(-\infty)\right]\right)
| 0\rangle\nonumber~,
\eear\vs\nin
where in the last equality we used the fact that the time-ordering
operator $T$ tells us to put all earlier time operators (here, those
evaluated at $t\to-\infty$) to the right, whereas the later time
operators should be going on the left. 
Since 
\vs\be
a_p(-\infty)|0\rangle = 0~,\qquad \langle 0| a^\dagger_p(+\infty) = 0 ~,
\ee\vs\nin 
then the equality between the first and second line in
(\ref{timeorder}) holds. We can finally obtain the Lehmann, Symanzik
and Zimmermann  (LSZ) reduction formula by
using (\ref{lsz1}) and (\ref{lsz2}) above, which results in
\vs\bear
\langle f|i\rangle &=& i\int d^4x_3\, e^{ip_3\cdot x_3}\left(\del^2_{x_3} +
  m^2\right) \dotsi\int d^4x_n\, e^{ip_n\cdot x_n}\left(\del^2_{x_n} +
  m^2\right)\nonumber\\
&\times& \,i\int d^4x_1 \,e^{-ip_1\cdot x_1}\left(\del^2_{x_1} +
  m^2\right) i\int d^4x_2 \,e^{-ip_2\cdot x_2}\left(\del^2_{x_2} +
  m^2\right) \nonumber\\
&\times&\langle 0|
T\left(\phi(x_1)\,\phi(x_2)\,\phi(x_3)\dots\phi(x_n)\right)|0\rangle~.
\label{lszfinal}
\eear\vs\nin
The equation above gives the desired relation between the $2\to n-2$
amplitude in momentum space on the left, and the $n$-point  correlation
function on the right. Although the LSZ reduction formula in (\ref{lszfinal}) is
not the most convenient way to obtain the momentum space amplitudes,
we will make use of it to derive a set of rules, the Feynman rules,
that will greatly speed up the procedure.

But before we derive the Feynman rules from (\ref{lszfinal}) we must
comment on its validity in the presence of interactions. We derived
the LSZ formula from the simple assumption of the free field momentum expansion 
in (\ref{freephi}). In the presence of interactions, on the other hand, we need
to make sure that the asymptotic states created and/or annihilated at
$t\to\pm\infty$ are single-particle well separated momentum
eigenstates. For this to be the case we need to guarantee that
\vs\be
\langle 0| \phi(x)|0\rangle =0~,
\label{veviszero}
\ee\vs\nin
still holds. This is not always the case. In the presence of
interactions we could have  $\langle
0|\phi(x)|0\rangle = v \not=0$. However, in this case we can
{\em additively} shift the definition of the field as in $\phi(x)\to \phi(x)
+ v$, such that the new field $\phi(x)$ satisfies
(\ref{veviszero}). The other condition we should worry about is
\vs\be
\langle 0|\phi(x)|p\rangle = e^{-ip\cdot x}~,
\label{zerotop}
\ee\vs\nin
which, in the presence of interactions, would still guarantee that
$a_p(\pm\infty)$ still annihilates a single-particle state of momentum
$\mbf{p}$. But in the interaction theory the coefficient of the
exponential in (\ref{zerotop}) need not be equal to one. This requires
that we redefine (renormalize) the field $\phi(x)$ {\em
  multiplicatively} by a factor in such a way as to ensure that the
coefficient in front of the exponential is in fact one. Then, we see
that with the necessary redefinitions of the field in the presence of
interactions, the LSZ reduction formula is valid.

We are finally ready to derive the Feynman rules in momentum space
from the LSZ reduction formula. Since the correlation
function on the right side of (\ref{lszfinal}) will be expressed, in
perturbation theory, by sums of products of free propagators (Wick's
theorem) the action of the Klein-Gordon operators on them will result
in delta functions as in
\vs\be
(\del^2_{x_i} + m^2)\,D_F(x_i - y) = -i\delta^{(4)}(x_i-y)~,
\label{dfisgreen1}
\ee\vs\nin
where $x_i$ and $y$ are external points. This removal of the external
propagators, will result in only
{\em connected} correlation functions contributing to the
amplitudes. The reason for this is that in the LSZ reduction formula
there is a Klein-Gordon operator for each external line. Disconnected
diagrams contributing to correlation functions will have {\em less}
external propagators, resulting in the KG operators acting on delta
functions and finally a vanishing  contribution.

\nin
As an example, let us consider the fully connected diagram of
Figure~\ref{fig:ol4p}. The correlation function is given by
\vs\be
G^{(4)}_\lambda(x_1,x_2,x_3,x_4) = (-i\lambda)   \int
d^4y\,D_F(x_1-y)\,D_F(x_2-y)\,D_F(x_3-y)\,D_F(x_4-y)~.
\label{g4lambda}
\ee\vs\nin
The application of the LSZ reduction formula (\ref{lszfinal}) to the
expression above results in 
\vs\bear
\langle p_3 p_4 | p_1 p_2\rangle &=&  i\int d^4x_1\, e^{-ip_1\cdot
  x_1}\left(\del^2_{x_1}+m^2\right) i\int d^4x_2\, e^{-ip_2\cdot
  x_2}\left(\del^2_{x_2} +m^2\right) \label{fi1}\\
&&  i\int d^4x_3\, e^{ip_3\cdot
  x_3}\left(\del^2_{x_3}+m^2\right)  i\int d^4x_4\, e^{ip_4\cdot
  x_4}\left(\del^2_{x_4}+m^2\right) \,G^{(4)}_\lambda(x_1,x_2,x_3,x_4) \nonumber~.
\eear\vs\nin
Applying (\ref{fi1}) to (\ref{g4lambda}) we obtain
\vs\bear
\langle p_3 p_4 | p_1 p_2\rangle &=& (-i\lambda) \int d^4y\,\int
d^4x_1 \,e^{-ip_1\cdot x_1}\delta^{(4)}(x_1-y)\,\int d^4x_2\,e^{-ip_2\cdot
  x_2}\delta^{(4)}(x_2-y) \nonumber\\
&\times&\,\int d^4x_3\,e^{ip_3\cdot  x_3}\delta^{(4)}(x_3-y)
\,\int d^4x_4\,e^{ip_4\cdot  x_4}\delta^{(4)}(x_4-y)\nonumber\\
&=& (-i\lambda) \int d^4y\,e^{-i(p_1+p_2-p_3-p_4)\cdot y} \nonumber\\
&=& (-i\lambda)\,(2\pi)^4\,\delta^{(4)}(p_1+p_2-p_3-p_4)\label{fi2}
\eear\vs\nin
From this expression, we see that the amplitude is just the
insertion of the vertex factor $(-i\lambda)$ times a momentum
conservation delta function. The appearance of this delta function is
associated to the fact that all external points are connected to the
same internal point $y$ where the interaction takes place. That is, it
comes from the fact that the interaction is local. Another important
point is that, unlike for the order $\lambda^0$ above, the
singularities of the contribution to the four-point function
$G^{(4)}_\lambda(x_1,x_2,x_3,x_4) $ exactly match the action of the
Klein-Gordon operators in (\ref{fi1}). The result above is a first
example of a Feynman rule in momentum space. Insert the interaction
factor $(-i\lambda)$ and a momentum conservation delta function in each
vertex. Strip all external propagators (which is the result of
applying the LSZ reduction formula). This is schematically shown in
Figure~\ref{fig:fplfig1}.
\vs\vs
\begin{figure}[h]
\begin{center}
\includegraphics[width=1.7in]{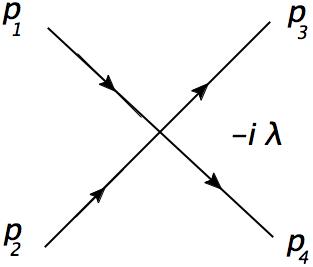}
\caption{
Momentum-space Feynman rule for the four-point amplitude to order $\lambda$ in
$\phi^4$ theory.  
}
\label{fig:fplfig1}
\end{center}
\end{figure}

Another important case to consider is that of going beyond leading
order. In the case of the four-point function we just computed, this
means going to order $\lambda^2$.
The $\lambda^2$ contribution to the four-point function can be
obtained from 
\vs\bear
G^{(4)}_{\lambda^2}(x_1,x_2,x_3,x_4) = \frac{1}{Z[0]}\,\int {\cal D}\phi
\,\phi(x_1)\phi(x_2)\phi(x_3)\phi(x_4)
\times \frac{1}{2!}\,\frac{(-i\lambda)^2}{(4!)^2}\,\int
d^4y\,\phi^4(y)\,\int d^4z\,\phi^4(z)~.\nonumber\\
&&~\label{fpl2}
\eear\vs\nin
In (\ref{fpl2}), the factor of $1/2!$ coming from the exponential expansion cancelled by the exchange
$y\leftrightarrow z$. We will concentrate on connected diagrams. There
are three ways of connecting the external fields to the eight fields
at points $y$ and $z$ of the interactions. They are depicted in
Figure~\ref{fig:threeloops}.
\vs\vs
\begin{figure}[h]
\begin{center}
  \includegraphics[width=1.7in]{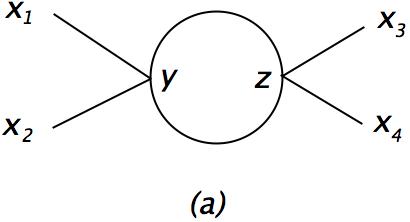}
  \qquad
  \includegraphics[width=1.2in]{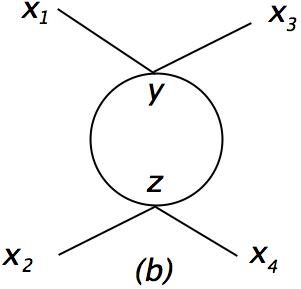}
  \qquad
  \includegraphics[width=1.2in]{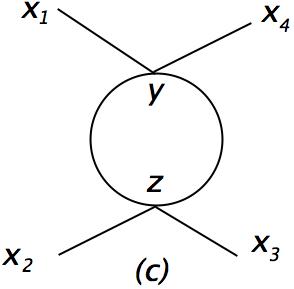}
\caption{
Connected diagrams contributing to  the four-point function to order $\lambda^2$ in
$\phi^4$ theory.  
}
\label{fig:threeloops}
\end{center}
\end{figure}
Let us focus on the first diagram $(a)$ since the other two will be
analogous with the obvious replacements. The combinatoric factor in
front of it can be obtained by counting the ways to match $\phi(x_1)$
with $\phi(y)$ (4), times the ways of matching $\phi(x_2)$ with the
remaining $\phi(y)$ (3), times the 4 ways of matching $\phi(x_3)$ with
$\phi(z)$, times the 3 ways to match $\phi(x_4)$ with
$\phi(z)$. Finally, we need to contract the remaining $\phi(y)$ and
$\phi(z)$, which brings an extra factor of 2. All in all, the
combinatoric factor times $1/(4!)^2$ results in an overall factor of 
\vs\be
(-i\lambda)^2\,\frac{1}{2}~.
\ee\vs\nin  
We can understand the factor of $1/2$ above in this diagram as a
{\em symmetry factor}. It is the factor we need to divide by if we
assume that at each vertex of the diagram we insert a factor of
$-i\lambda$, which is the coefficient for the four-point function at
order $\lambda$. In this diagram, using $-i\lambda$ at each vertex is
overcounting the combinatoric factor since it is tantamount to assuming that all the lines at
the two vertices are {\em un-contracted} fields. But we know that the internal
lines coming from the vertices result in contractions into two
propagators. To obtain the symmetry factors we see that the use of $-i\lambda$
will result in counting diagrams interchanging the internal
integration points $y$ and $z$ as distinct contributions. But this is
not the case. So we can think of this factor of $2$ as obtained by
exchanging the two internal propagators, resulting in 
undistinguishable contributions.
The result for the contribution to the four-point function is
\vs\bear
G^{(4)}_{(a)}(x_1,x_2,x_3,x_4) &=& \frac{(-i\lambda)^2}{2}\,\int
d^4y \,d^4z\, D_F(x_1-y) \,D_F(x_2-y) \,D_F(x_3-z)\,
D_F(x_4-z)\nonumber\\
&&\qquad\qquad D_F(y-z) D_F(y-z)~.
\label{fpla}
\eear\vs\nin
We want to obtain the $\cO(\lambda^2)$ contributions to the scattering amplitude for two particles of
initial fixed momenta to go to other two particles of known
final momenta. Applying (\ref{lszfinal}) on (\ref{fpla})  we get
\vs\be
\langle p_3 p_4| p_1 p_2\rangle_{(a)} = \frac{(-i\lambda)^2}{2}\,\int
d^4y \, d^4z\, e^{-i(p_1+p_2)\cdot y} \, e^{i(p_3+p_4)\cdot z} \,D_F(y-z)\,D_F(y-z)~,
\label{scatampa}
\ee\vs\nin
where the action of each Klein-Gordon operator $(\del^2_{x_i}+m^2)$ on
the propagators containing an external point $x_i$ in the argument
resulted in factors of  $-i\delta^{(4)}(x_i-y)$ and
$-i\delta^{(4)}(x_i-z)$ which we used to integrate over the
$x_i$'s. Since the two internal propagators do not have external
positions in their arguments they remain in (\ref{scatampa}). In order
to make further progress we are going to express these propagators in
momentum space by making use of 
\vs\be
D_F(y-z) = \int\frac{d^4q}{(2\pi)^4}\,e^{-iq\cdot
  (y-z)}\,\frac{i}{q^2-m^2+i\epsilon}~,
\label{dfinqspace}
\ee\vs\nin
in (\ref{scatampa}).
The final expression fot the amplitude in momentum space is
\vs\bear
\langle p_3 p_4| p_1 p_2\rangle_{(a)} &=&  \frac{(-i\lambda)^2}{2}\,
(2\pi)^4\,\delta^{(4)}(p_1+p_2-p_3-p_4)\nonumber\\
&&\times \int\frac{d^4q}{(2\pi)^4}\,
\frac{i}{q^2-m^2+i\epsilon}\,\,\frac{i}{k^2-m^2+i\epsilon}~,
\label{scatampafinal}
\eear\vs\nin
where the value of $k$ is fixed by delta functions at 
\nin\be
k=-p_1-p_2-q=-p_3-p_4+q~.
\ee\nin
As we can see from (\ref{scatampafinal}), there remains an
undetermined momentum $q$ which must be integrated over. 
This can be easily understood by looking at the diagram once again,
now in momentum space and with all these momenta drawn explicitly as
seen in Figure~\ref{fig:fplfig3}. 
\begin{figure}[h]
\begin{center}
  \includegraphics[width=1.7in]{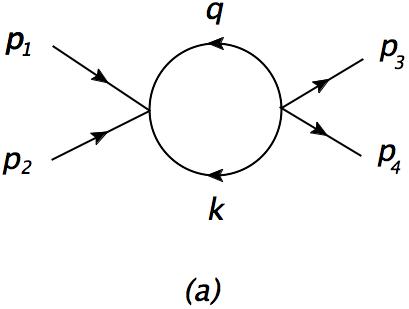}
\caption{
One of the Feynman diagrams in momentum space  for the four-point amplitude to order $\lambda^2$ in
$\phi^4$ theory.  
}
\label{fig:fplfig3}
\end{center}
\end{figure}
\nin 
It is clear that, although we have two internal lines, only one of the
two internal momenta are independent: there is a delta function
forcing momentum conservation at each vertex, but the overall momentum
conservation is not a constraint so there is one undetermined
momentum we still have to integrate over.
\begin{figure}[h]
\begin{center}
  \includegraphics[width=1.7in]{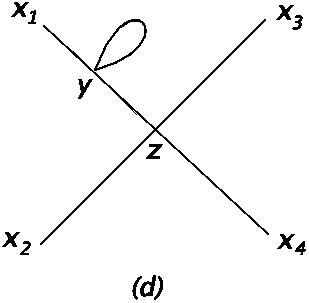}
\caption{
One of the Feynman diagrams in position space  for the four-point amplitude to order $\lambda^2$ in
$\phi^4$ theory.  It corresponds to an $O(\lambda)$ correction to one
of the external lines. 
}
\label{fig:renexleg}
\end{center}
\end{figure}
\nin 
\nin
Finally, before summarizing the Feynman rules, we comment on another
type of  order
$\lambda^2$ diagram, depicted in  Figure~\ref{fig:renexleg}. 
Its
contribution to the four-point correlation function is
\vs\vs\vs\vs
\bear
G^{(4)}(x_1,\dots,x_4)_{(d)} &=& \frac{(-i\lambda)^2}{2} \int d^4y\,
\,d^4z \, D_F(x_1-y) \,D_F(y-z)\, D_F(y-y)\, D_F(x_2-z) \nonumber\\
&&\qquad\qquad\times D_F(x_3-z)\, D_F(x_4-z)~.
\label{g4l2d}
\eear\vs\nin
We can see from (\ref{g4l2d}) above that applying the LSZ reduction
formula will result in two remaining propagators, but only one
undetermined momentum. This is because, when doing to momentum space,
the propagator $D_F(y-z)$ will be {\em on shell}, resulting in an
overall divergent contribution. These are in fact
part of the {\em renormalization} of the external legs of any diagram
and should not be considered when computing an amplitude.
These diagrams should be excluded from the calculation, since they are
going to be included by the renormalization process, which redefines
the fields (leading in this case to the redefinition of the
propagators) in the presence of interactions, as it was briefly
mentioned at the end of the derivation of the LSZ reduction
formula. In order to avoid including these diagrams, a rule can be
imposed: only consider diagrams without on shell propagators
(amputated diagrams).

We are finally ready to enumerate a set of rules to compute the
amplitudes  in momentum space without the need to apply the LSZ
reduction formula every time we need to compute one. These are the
Feynman rules of the theory.
\begin{enumerate}
\item \underline{Vertex}: Insert a factor of $-i\lambda$ for each
  vertex in the diagram. It is clear from (\ref{scatampafinal}) that
  this will get us the factor in front up to symmetry factors. Notice
  that this is the Feynman rule of the diagram at order $\lambda$
  (i.e. at ``tree'' level and without ``loops''). In general, deriving
  the tree-level interaction vertex is one of the first things we need
  to do in a theory in order to be able to obtain its Feynman rules.
\item \underline{Momentum conservation at each vertex}: The presence of the delta
  functions at each vertex that appear integrating  (\ref{scatampa}), tells us that
  momentum conservation must be enforced at each vertex in the
  diagram. This always results in an overall delta function enforcing
  total momentum conservation. For our case is the factor
  $(2\pi)^4\delta^{(4)}(p_1+p_2+p_3+p_4)$. 
\item \underline{Loop momentum integration}: Integrate over all the
  undetermined momenta. 
In our example from Figure~\ref{fig:fplfig3} , the product of the two
  delta functions after integrating  (\ref{scatampa}) is equivalent to the overall
  momentum conservation. So one of the two internal momenta remains
  free and must be integrated over.  
\item \underline{Symmetry factors}: We must divide by the symmetry
  factor of the diagram. In Figure~\ref{fig:fplfig3} this is $2$, since
  the internal propagators can be exchanged without consequence. The
  need to divide by the symmetry factors stems from the fact that in
  any generic diagram we use the vertex Feynman rule (here
  $-i\lambda$) for each
  interaction. But generally this has the correct combinatoric factor only in
  the tree-level interaction vertex. This ``mistake'' must be corrected by the symmetry factor.   
\end{enumerate}

These Feynman rules are specific to the example of the real scalar
field theory of (\ref{realscalagrangian}). However, the procedure to
derive the Feynman rules for any other quantum field theory is always
the same. The one difference is in the derivation of the vertex
Feynamn rule (Rule 1). The rest are analogous in all cases, although
in the presence of different kinds of fields, there may or may not be
symmetry factor to worry about. Using the Feynman rules of an
interacting theory, we can compute the amplitude of a desired process
in momentum space, and to the desired order in perturbation
theory. For instance, to leading order in perturbation theory, i.e. to
leading order in the coupling $\lambda$, the momentum space scattering amplitude of
two real scalar field going to two real scalar fields is given by
(\ref{fi2}). But if order $\lambda^2$ accuracy is required one needs
to add the diagrams such as that of Figure~\ref{fig:fplfig3}.
Once the momentum space amplitude is obtained, the next step os to
compute the actual physical observable, the cross section.

\subsubsection{Cross sections} 

\nin
Now that we know how to compute amplitudes for given processes, we
would like to make contact with observables such as cross sections and
decay rates based on those amplitudes. This will complete the path
from computing correlation functions and then amplitudes, which can be
easily obtained by using the derived Feynman rules of a given
theory.

\nin
 We will state the amplitude in the language of the S
matrix. Let us consider a scattering process with a given initial
state and a final state. We define the asymptotic states by 
\vs\bear
&&|i,{\rm in}\rangle \qquad\qquad {\rm for} \quad t\to -\infty\nonumber\\
&&|f,{\rm out}\rangle \qquad\qquad {\rm for} \quad t\to +\infty\label{inoutstates}~,
\eear\vs\nin
where the states labeled ``in'' are those asymptotic states created by
creation operators evaluated at times $-\infty$,
e.g. $a^\dagger(-\infty)$, etc; and the states labeled ``out'' are
those created by creation operators  evaluated at times $+\infty$,
such as $a^\dagger(+\infty)$. These two distinct sets of asymptotic
states are the ones we have used up until now to write down the
desired amplitude
\vs\be
\langle f,{\rm out}|i, {\rm in}\rangle~.
\label{amp1}
\ee\vs\nin
The ``in'' and ``out'' asymptotic states are however isomorphic,
i.e. there are the same set of states but labeled differently. We
can define a unitary transformation $\mbf{S}$ such that 
\vs\be
|i,{\rm in}\rangle =\mbf{S} |i,{\rm out}  \rangle~,
\label{strans} 
\ee\vs\nin
in such a way that we can rewrite  (\ref{amp1}) in terms of either
both ``in'' or``out'' states. 
\vs\be
\langle f,{\rm out}|i, {\rm in}\rangle = \langle f, {\rm in} | \mbf{S}
| i, {\rm in}\rangle =  \langle f, {\rm out} | \mbf{S}
| i, {\rm out}\rangle \equiv \langle f | \mbf{S} | i\rangle~.
\label{amp2} 
\ee\vs\nin
The last equality stems from the fact that we can equally express the
amplitude in terms of the ``in'' or the ``out'' states as long as is
an element of the $\mbf{S}$ matrix. 
The $\mbf{S}$ operator can be written as 
\vs\be
\mbf{S} \equiv \mbf{1} + i\mbf{T} ~,
\label{tmatrixdef} 
\ee\vs\nin
where we defined the T matrix elements. The identity in the first term
in (\ref{tmatrixdef}) reflects the fact that the amplitude must
include the possibility of no interaction. But in order to compute a
cross section we are only concerned with the part of the amplitude
that allows for interactions, i.e. the second term in
(\ref{tmatrixdef}). Schematically, we can express this as 
\vs\be
\langle f |\mbf{S}|i\rangle = {\rm disconnected ~~diagrams} \qquad
+\qquad{\rm LSZ ~~formula}~,
\ee\vs\nin
where the contributions of disconnected diagrams comes from the
identity in (\ref{tmatrixdef}). Thus, the LSZ formula will give the
contribution of the T matrix to a given amplitude. 

\nin
In general we want to compute the transition probability from an
initial state to a final state. In practice, we are mainly interested
in two cases: the decay of a particle to two or more particles, and
the scattering of two particles in the initial state into two or more
particles in the final state. 

\nin
We start with the scattering process $2\to n$. The transition
amplitude is given by
\vs\be
\langle \mbf{p_1}\dots\mbf{p_n}|\,i\mbf{T}\,|\mbf{p_A}\mbf{p_B}\rangle
\equiv (2\pi)^4\delta^{(4)}(P_A+P_B-P_1-\dots-P_n)\,i\,\cA~,
\label{ampdef}
\ee\vs\nin  
where we have defined the amplitude $\cA$ as the transition amplitude
with the overall momentum conservation delta function already factored
out. In order to obtain a probability, we will define it as the
squared of the transition amplitude appropriately normalized. 
\vs\be
P\equiv \frac{|\langle
  \mbf{p_1}\dots\mbf{p_n}|\,i\mbf{T}\,|\mbf{p_A}\mbf{p_B}\rangle|^2}{\langle
  \mbf{p_1}\dots\mbf{p_n}|\mbf{p_1}\dots\mbf{p_n}\rangle\langle
  \mbf{p_A}\mbf{p_B}|\mbf{p_A}\mbf{p_B}\rangle}~,
\label{probdef}
\ee\vs\nin
where the denominator corresponds to the normalization of the initial
and final states. 

\nin
We  start  by considering the numerator of (\ref{probdef}). This is 
\vs\bear
|\langle
\mbf{p_1}\dots\mbf{p_n}|\,i\mbf{T}\,|\mbf{p_A}\mbf{p_B}\rangle|^2 &=&
\left((2\pi)^4\delta^{(4)}(P_A+P_B-\sum_{f=1}^n P_f)\right)^2\,\left|
  \cA\right|^2\nonumber\\
&=&(2\pi)^4 \delta^{(4)}(P_A+P_B-\sum_{f=1}^nP_f) \,(2\pi)^4 \delta^{(4)}(0)\,\left|\cA\right|^2~,\label{num1} 
\eear\vs\nin
where $f=1,\dots,n$ labels the final state momenta. However, we can
write
\vs\be
\delta^{(4)}(0) = \delta(0) \delta^{(3)}(0) = \frac{1}{(2\pi)^4} \int
d^4x\,e^{i0\cdot x}~.
\label{deltaofzero}
\ee\vs\nin 
If we consider for a moment a finite volume $V$ and a finite time $T$,
the integral in (\ref{deltaofzero}) results in 
\vs\be
(2\pi)^4 \delta^{(4)}(0) = V T~.
\ee\vs\nin
For the denominator, we consider the asymptotic momentum eigenstates
normalized according to 
\vs\be
|\mbf{p}\rangle = \sqrt{2E_p}\,a^\dagger_p |0\rangle~,
\ee\vs\nin
such that the normalization of an eigenstate of momentum $\mbf{p}$ is
given by
\vs\bear
\langle \mbf{p}|\mbf{p}\rangle &=& 2E_p\,\langle 0|a_p\,a^\dagger_p
|0\rangle\nonumber\\
&=& 2E_p\,(2\pi)^3\,\delta^{(3)}(\mbf{p}-\mbf{p}) = 2\,E_p\,V~,
\label{den1}
\eear\vs\nin
where in the last equality we used (\ref{deltaofzero}).
Then, the two factors in the denominator of (\ref{probdef}) are
\vs\bear
\langle\mbf{p_A}\,\mbf{p_B}| \mbf{p_A}\,\mbf{p_B}\rangle &=&
2E_A\,2E_B\,V^2~\nonumber\\
\langle\mbf{p_1}\dots\mbf{p_n}| \mbf{p_1}\dots\mbf{p_n}\rangle &=&
2E_1\dots 2E_n V^n = \prod_f (2E_f V)~.\label{den2}
\eear\vs\nin
Replacing (\ref{deltaofzero}) and (\ref{den2}) into (\ref{probdef})
and dividing by $T$, we obtain the probability of transition for unit
time
\vs\be
\frac{P}{T} = \frac{(2\pi)^2\delta^{(4)}(P_A-P_B-\sum_f P_f)
  \,V\,\left|\cA\right|^2}{2 E_A \,2E_B\,V^2\,\prod_f (2E_f V)}~.
\label{probpert}
\ee\vs\nin
But this probability requires that we have precise knowledge of all
final state momenta. Often times we will need to either partially or
totally integrate over the phase space of the final states. For this
we need to know the probability that a given final state particle has 
momentum in the interval  
\vs\be
(\mbf{p_f}, \mbf{p_f} + d^3p_f)~, 
\label{fsinterval}
\ee\vs\nin
where $d^3p_f$ contains information about the momentum vector. We
would like then to convert (\ref{probpert}) into the differential
probability that the final states are in a region of the final state
phase space defined by (\ref{fsinterval}). In order to obtain this we
need to multiply (\ref{probpert}) by the number of states in each
interval defined by (\ref{fsinterval}) for each final state
particle. Given that we are using a finite volume $V$, the momentum of
each final state particle obeys the quantization rule
\vs\be
\mbf{p} = \frac{2\pi}{L} \,(n_1,n_2,n_3)~,
\label{pquantized}
\ee\vs\nin
where $L^3=V$, and the $n_i$ with $i=1,2,3$ refer to the number of
states in each spatial direction. Then, the number of states inside
the interval (\ref{fsinterval}) of size $d^3p$ is
\vs\bear
n_1\,n_2\,n_3 &=& \frac{L dp_x}{2\pi}\frac{L dp_y}{2\pi}\frac{L
  dp_z}{2\pi}\nonumber\\
&=&\frac{V\,d^3p}{(2\pi)^3}
\eear\vs\nin
Putting all these together we obtain the differential probability per
unit time
\vs\be
\frac{dP}{T} = \frac{(2\pi)^2\,\delta^{(4)}(P_A+P_B-\sum_f
  P_f)\,|\cA|^2}{2E_A\,2E_B\,V}\,\prod_{f=1}^n
\left(\frac{d^3p_f}{(2\pi)^3\,2E_f}\right)~,
\label{diffprob}
\ee\vs\nin  
Finally, in order to convert this into a differential cross section we
need to account for the incident flux. In other words, we are
interested in the differential probability per unit time {\em and} per
unit of initial flux so that we obtain a probability that depends
intrinsically on the amplitude $\cA$ and the final state phase space,
not on how intense our beams of $A$ and $B$ particles were. The flux
is the number of particles per unit volume times the relative velocity
of the particles. For instance, for  a typical head on collision 
\vs
\begin{figure}[h]
\begin{center}
\includegraphics[width=3in]{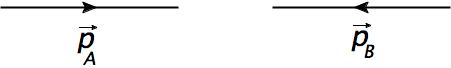}
\caption{
Head on collision. $\mbf{p_B} = -\mbf{p_A}$. 
}
\label{fig:headon}
\end{center}
\end{figure}
\vs\nin 
the initial flux ``seen'' by either the $A$ or the $B$ particle is given by   
\vs\be
\frac{|v_A^z-v_B^z|}{V}~.
\label{flux}
\ee\vs\nin
So dividing (\ref{diffprob}) by the flux in (\ref{flux}) we obtain
\vs\be
d\sigma = \frac{1}{2E_A\,2E_B}\frac{1}{|\mbf{v_A}-\mbf{v_B}|}(2\pi)^4\,\delta^{(4)}(P_A+P_B-\sum_f P_f)\,|{\cal A}|^2\,\prod_f \left(\frac{d^3p_f}{(2\pi)^3\,2E_f}\right)~,
\label{dsigma}
\ee\vs\nin
which is the differential cross section for the scattering of the two
initial particles with momenta $P_A$ and $P_B$ going into an
$n$-particle final state.

\nin
At this point we will make some comments:
\begin{itemize}
\item We can define the final state phase space by 
\vs\be
\int d\Pi_n\equiv
\int\prod_{f=1}^n\,\left(\frac{d^3p_f}{(2\pi)^3\,2E_f}\right)\,(2\pi)^4\,\delta^{(4)}(P_A+P_B-\sum_f
P_f)~.
\label{nfps}
\ee\vs\nin
It is separately Lorentz invariant. 
\item The amplitude squared $|{\cal A}|^2$ is also Lorentz invariant by
  itself.
\item The factor 
\vs\be
\frac{1}{E_A\,E_B|v_A^z-v_B^z|}~,
\ee\vs\nin
is not Lorentz invariant, but it is invariant under boosts in the $z$
direction. 
\end{itemize}

\subsubsubsection{Two-particle final state}

\nin
A very paradigmatic example is the scattering of two particles in the
initial state into two particles in the final state. We first compute
the two-particle phase space for $A+B\to 1+2$. We will use the center
of momentum frame. 
From (\ref{nfps}) we have 
\vs\bear
\int d\Pi_2 &=& \int \frac{d^3p_1}{(2\pi)^3}\,\frac{1}{2E_1}\,\int
\frac{d^3p_2}{(2\pi)^3}\,\frac{1}{2E_2}\,(2\pi)^4\,\delta^{(4)}(P_A+P_B-P_1-P_2)\nonumber\\
&=& \int \frac{d^3p_1}{(2\pi)^3}\,\frac{1}{4E_1E_2}\,2\pi\,\delta(E_A+E_B-E_1-E_2)~,
\label{tfps1}
\eear\vs\nin
where the second line is obtained by using the spatial delta function
to perform the  $d^3p_2$ integral. The final momentum differential is 
\vs\be
d^3p_1 = p_1^2\,dp_1\,d \Omega_1 = p_1^2\,dp_1\,d\cos\theta_1\,d\phi_1~,
\ee\vs\nin
with $\theta_1$ the angle of $\mbf{p_1}$ with respect to the direction
of the incoming momentum $\mbf{p_A}$, and $\phi_1$ the corresponding azimuthal
angle.  There is typically no azimuthal angle dependence in $|\cA|^2$,
so we can integrate over $\phi_1$ obtaining a factor of $2\pi$.  Then
(\ref{tfps1}) now reads
\vs\be
\int d\Pi_2 =\int \frac{p_1^2\,dp_1}{(2\pi)^3\,4\,E_1\,E_2}\,(2\pi\,d\cos\theta_1)\,\,2\pi\,\delta\left(E_A+E_B-\sqrt{p_1^2+m_1^2}-\sqrt{p_1^2+m_2^2}\right)~,
\label{tfps2}
\ee\vs\nin
where we have used that $\mbf{p_1}=-\mbf{p_2}$ in the delta function,
which stems from the fact that we have used the spatial delta function
in the center of momentum frame. We are now in a position to perform
the integral in the absolute value of the spatial momentum of the
particle $1$, $p_1$, by using the delta function. Restoring the
differential solid angle to have a more general expression, we have
\vs\bear
\int d\Pi_2 &=& \int \frac{p_1^2}{(2\pi)^2\,4E_1\,E_2}
\frac{d\Omega_1}{|\frac{p_1}{E_1} + \frac{p_1}{E_2}|}\nonumber\\
&=&\int \frac{1}{16\pi^2}\,\frac{p_1}{E_1+E_2}\,d\Omega_1~.
\eear\vs\nin
But, since $E_1+E_2=E_{\rm CM}$ then we obtain
\vs\be\boxed{
\int d\Pi_2 = \int \frac{1}{16\pi^2}\,\frac{p_1}{E_{\rm CM}} \,d\Omega_1}~.
\label{tfps3}
\ee\vs\nin
Let us compute now the cross section in the CM frame. It is 
\vs\be
\frac{d\sigma}{d\Omega} =
\frac{1}{2E_A\,2E_B}\,\frac{1}{|v_A^z-v_B^z|}\,\frac{p_1}{16\pi^2\,E_{\rm
    CM}}\,|\cA|^2~,
\label{xsec1}
\ee\vs\nin
where the solid angle refers to the final states particles, and $z$ is
the direction of the incoming $A$ particle. 

\nin
 If we now consider the relative velocity we have 
\vs\be
|v_A^z-v_B^z| = \left|\frac{p_A^z}{E_A}-\frac{p_B^z}{E_B}\right| ~.
\ee\vs\nin
If we now consider the simplified  case $m_A=m_B=m_1=m_2=m$, we have 
\vs\be
|v_A^z-v_B^z| = \frac{2}{E_{\rm CM}}\,\left|p_A^z-(-p_A^z)\right| =
\frac{4p_A}{E_{\rm CM}} = \frac{4p_1}{E_{\rm CM}}~,
\label{relvel} 
\ee\vs\nin
Then, we arrive at a final expression for the angular distribution for
scattering in the CM of two particles into two particles, all of the
same mass $m$: 
\vs\be
\left(\frac{d\sigma}{d\Omega}\right)_{\rm CM} =
\frac{1}{64\pi^2}\,\frac{1}{E_{\rm CM}^2}\,\left|\cA\right|^2~.
\label{xsec2}
\ee\vs\nin 

\subsubsubsection{Decay rate of an unstable particle}

\nin
If instead of considering the transition probability per unit time
from a two-particle initial state we start with a state of one
particle, we are computing the decay rate for the process $A\to 1\dots
n$, for the decay of a particle $A$ to $n$ particles in the final
state. The derivation is just straightforward and the result is the
differential decay probability per unit time given by
\vs\be
d\Gamma =
\frac{1}{2m_A}\,\prod_{f=1}^n\left(\frac{d^3p_f}{(2\pi)^3}\,\frac{1}{2E_f}\right)\,(2\pi)^4\,\delta^{(4)}\left(P_A-\sum_f
    P_f\right)\,|\cA|^2~,
\label{dgamma}
\ee\vs\nin
where the factor of $2m_A$ comes from using $2E_A$ in the rest frame
of the decaying particle, and $\cA$ is the amplitude for the decay
process. For a given decay channel (i.e. a given final state), the
integral gives the so-called partial width of $A$ into that channel
\vs\be
\Gamma(A\to f_1) = \int d\Gamma(A\to f_1)~.
\ee\vs\nin 
The total width of $A$ is a property of the particle and corresponds
to the sum     of the partial widths into all the available channels into
which $A$ can possibly decay
\vs\be
\Gamma_A \equiv \sum_i \Gamma(A\to f_i)~. 
\label{totalwidth}
\ee\vs\nin 
 The lifetime of tha particle is then the inverse of the total decay
 rate or total width. Decay rates have units of energy, thus if we
 want the lifetime in seconds we can use 
\vs\be
\tau_A = \frac{\hbar}{\Gamma_A}~.
\ee\vs\nin
For instance, if we initially have a given number of particles of type
$A$, at a later time $t$ we have
\vs\be
N(t) = N(0) \, e^{-t/\tau_A}~.
\ee\vs\nin
The lifetime also determines the typical displacement of a particle
produced before it decays. This is 
\vs\be
c\,\tau_A\,\gamma~,
\ee\vs\nin
where $c$ is the speed of light, and $\gamma$ is the relativistic
factor. 

\nin
Finally, the propagation of an unstable particle is affected by its
decays. We will show later in the course that the propagator of a
particle with open decay channels gets modified to be
\vs\be
\frac{i}{p^2 -m_A^2 -i\Gamma_A m_A}~,
\label{unstableprop}
\ee\vs\nin
where we considered a scalar propagator and $p$ is the four-momentum
of $A$. We will derive (\ref{unstableprop}) in the context of
renormalization and see that the new term appears as a consequence of
an imaginary shift in the pole of the propagator that arises due to
the existence of open decay channels for $A$. As a result, unstable
particles appear in cross sections for processes that are mediated by
them as resonances of widths characterized by $\Gamma_A$. This is the
reason why these particles are called resonances, and also why the
total decay rate $\Gamma_A$ is called the particle width.

\subsection{Gauge theories}
Here we  introduce vector fields. Although it is
generically possible to write the action for a theory with such
fields, it turns out that these generic theories are not well defined
unless the vector fields are gauge fields, i.e. vector fields
associated with a local symmetry. We will eventually show this
relation further along our course. For now, let us introduce gauge
fields as a consequence of gauge invariance. We will start with a
fermion theory so as to derive quantum electrodynamics. 

\subsubsection{Gauge invariance} 

\nin
Let us consider the Lagrangian for a free fermion of mass $m$
\vs\be
\cL = \bar{\psi}(i\dsl\del -m)\psi~.
\label{lfree}
\ee\vs\nin
This is invariant under the {\em global}  $U(1)$
transformation\footnote{A unitary transformation determined by one parameter.}
defined by
\vs\bear
\psi(x) &\longrightarrow &e^{i\alpha}\,\psi(x)~,\nonumber\\
\bar{\psi}(x) &\longrightarrow &e^{-i\alpha}\,\bar{\psi}(x)~,\label{globalu1transf}
\eear\vs\nin
where $\alpha$ is a real constant. The conserved charge associated
with these symmetry transformations is fermion number: $+1$ for
fermions, $-1$ for antifermions.

\nin
But what if we want {\em local} $U(1)$ invariance, i.e. what if
$\alpha=\alpha(x)$ is a function of the spacetime position ? The local
transformation now reads
\vs\bear
\psi(x) &\longrightarrow &e^{i\alpha(x)}\,\psi(x)~,\nonumber\\
\bar{\psi}(x) &\longrightarrow &e^{-i\alpha(x)}\,\bar{\psi}(x)~,\label{localu1transf}
\eear\vs\nin
which leads to a transformation of the Lagrangian as
\vs\be
\cL\to \cL' =  \cL = \bar{\psi}(i\dsl\del -m)\psi
-\del_\mu\alpha(x)\,\bar{\psi}\gamma^\mu\psi \,\not= \cL~.
\label{lpisnotl}
\ee\vs\nin
From (\ref{lpisnotl}) we see that the local or {\em gauge}
transformation (\ref{localu1transf}) does not leave the Lagrangian
(\ref{lfree}) invariant. In order to obtain a theory invariant under
these local transformations we will need to add a new field that also
transforms in some way that depends on $\alpha(x)$ and whose
transformation cancels the extra term that appears in
(\ref{lpisnotl}). One way to do this is to define a covariant
derivative on $\psi(x)$, a generalization of the normal derivative.
We write
\vs\be
\cL = \bar{\psi}\left(i\dslcv D-m\right)\psi~,
\label{lginv1}
\ee\vs\nin
where we defined the covariant derivative $D_\mu \psi(x)$ so that it
must transform as the field $\psi(x)$ in order for (\ref{lginv1}) to
be invariant, i.e. under the transformations (\ref{localu1transf}) it
must transform as 
\vs\be
D_\mu\psi(x) \longrightarrow e^{i\alpha(x)} \,D_\mu\psi(x)~.
\label{covdertransf} 
\ee\vs\nin  
Clearly, we can see that if (\ref{covdertransf}) is satisfied at the
same time as (\ref{localu1transf}) then (\ref{lginv1}) is invariant. 
Next, we write the covariant derivative $D_\mu\psi(x)$ by introducing a
vector field as 
\vs\be
D_\mu\psi(x) \equiv \left(\del_\mu +ieA_\mu(x)\right)\,\psi(x)~,
\label{covderdef}
\ee\vs\nin
where $e$ is a constant. Then, it can be verified that in order for
the covariant derivative defined in (\ref{covderdef}) to satisfy
(\ref{covdertransf}) the vector field $A_\mu(x)$ must transform as
\vs\be
A_\mu(x) \longrightarrow A_\mu(x) -\frac{1}{e}\,\del_\mu\alpha(x)~.
\label{atransf} 
\ee\vs\nin
We notice in passing that the vector field $A^\mu(x)$ must be
real. This is a consequence of the fact that the gauge parameter
$\alpha(x)$ is real.
Thus, to summarize, the theory in (\ref{lginv1}) is invariant under
the gauge or local $U(1)$  transformations
\vs\bear
\psi(x) &\longrightarrow &e^{i\alpha(x)}\,\psi(x)~,\nonumber\\
\bar{\psi}(x) &\longrightarrow &e^{-i\alpha(x)}\,\bar{\psi}(x)~,\nonumber\\
A_\mu(x) &\longrightarrow & A_\mu(x) -\frac{1}{e}\,\del_\mu\alpha(x)~,
\label{gaugetransf}
\eear\vs\nin
with the covariant derivative defined by (\ref{covderdef}). Finally,
if the gauge field $A_\mu(x)$ is to be a dynamical degree of freedom,
we need appropriate quadratic terms in it, i.e. a kinetic term and a
mass term. 
A kinetic term that is trivially invariant under the transformations
(\ref{atransf}) is built from the contraction of 
\vs\be
F_{\mu\nu} \equiv \del_\mu A_\nu - \del_\nu A_\mu~,
\label{fmunudef1}
\ee\vs\nin
with itself, since the  tensor $F_{\mu\nu}$ is invariant. Furthermore,
a mass term for $A_\mu(x)$ must be something like
\vs\be
m_A^2\,A_\mu A^\mu~. 
\ee\vs\nin
But since this is clearly not gauge invariant, we must assume that
$m_A=0$. Thus a gauge field must have zero mass in order to respect
gauge invariance. Although there are exceptions to this statement,
they all correspond to the case when the mass is generated dynamically
via a scalar field coupled to $A_\mu(x)$ obtaining a non-zero vacuum
expectation value. We will study this case in the second part of this
course. For now, gauge invariance means zero mass for the gauge
fields.
Then, the complete theory that is  $U(1)$ gauge invariant is 
\vs\bear
\cL &=& \bar{\psi}\left(i\dslcv D-m\right)\psi
-\frac{1}{4}F_{\mu\nu}F^{\mu\nu} \nonumber\\
&=& \bar{\psi}\left(i\dsl\del-m\right)\psi -eA_\mu\bar{\psi}\gamma^\mu\psi
-\frac{1}{4}F_{\mu\nu}F^{\mu\nu} ~,
\label{lginvwitha}
\eear\vs\nin
where in the last equality we can see that the gauge field $A_\mu(x)$
interacts with the fermion current with a coupling $e$.
The factor of $-1/4$ in front of the gauge field kinetic term is a
convenient choice of normalization which results in  $F_{\mu\nu}$
being the electromagnetic stress tensor in the case of quantum
electrodynamics (QED).
In fact, this Lagrangian is the basis for QED, where
$\psi(x)$ is the charged electron field and $A_\mu(x)$ is identified
with the photon.  The next step in order to obtain QED as a quantum
field theory would be to quantize the gauge field $A_\mu(x)$. 

\subsubsection{Gauge fields and quantization}

\nin
From the Lagrangian for the gauge fields
\vs\be
\cL = -\frac{1}{4}F_{\mu\nu}F^{\mu\nu}~,
\label{lagphoton}
\ee\vs\nin
we can derive the equations of motion (Euler-Lagrange)
\vs\be
\del^2 A^\mu -\del^\mu\left(\del_\nu A^\nu\right) = 0~,
\label{eom1}
\ee\vs\nin
As usual in the classical case, if we choose the Lorentz condition 
\vs\be
\del_\mu A^\mu =0~,
\label{lorentz}
\ee\vs\nin
we obtain 
\vs\be
\del^2 A^\mu =0 ~.
\label{kgforphoton}
\ee\vs\nin
Thus, imposing the Lorentz condition (\ref{lorentz}) gives us a simple
equation with plane wave solutions, a massless Klein-Gordon equation
for each component of the four-vector $A^\mu(x)$.
Naively, we would then expand $A ^\mu(x)$ in these solutions and
quantize by imposing commutation relations between $A^\mu(x)$ and its
conjugate momentum $\pi^\mu(x)$. From (\ref{lagphoton}) we obtain the form of the
conjugate momentum as
\vs\be
\pi^\mu(x) = \frac{\del\cL}{\del(\del_0A_\mu)} = F^{\mu0}~.
\label{pimu}
\ee\vs\nin
However, from (\ref{pimu}) it is clear that there is a problem with
the time component of $\pi^\mu(x)$ coming from the fact that
$F^{\mu\nu}$ is antisymmetric. We have that
\vs\be
\pi^0(x) = 0~,
\ee\vs\nin
 meaning that it will not be possible to impose a quantization
 condition on $A^0(x)$. We can get around this by adding a term to the
 Lagrangian as
\vs\be
\cL = -\frac{1}{4}F_{\mu\nu}F^{\mu\nu} -c(\del_\mu A^\mu)^2~, 
\ee\vs\nin 
where $c$ is a arbitrary real constant.  
Now the equations of motion are
\vs\be
\del^2 A^\mu +(c-1)\del^\mu(\del_\nu A^\nu) =0~.
\label{eom2}
\ee\vs\nin
We can see that there are two ways of  obtaining  (\ref{kgforphoton}): either by using the Lorentz
condition or by choosing $c=1$. But now the second choice also allows us to
define a non-zero conjugate momentum of $A^0(x)$ since now
\vs\be
\pi^\mu(x) = F^{\mu 0} -c g^{\mu 0} (\del_\nu A^\nu)~,
\ee\vs\nin 
which results in 
\vs\be
\pi^0(x) = -c (\del_\nu A^\nu)~.
\ee\vs\nin
So choosing $c=1$ allows us to carry out the canonical quantization
procedure. This is called the Feynman gauge. But, as we will see
below, the upshot is that now we
will have non-physical degrees of freedom.

\nin
To proceed with the quantization, we start by expanding the field
$A^\mu(x)$ in momentum space. 
The most general solution to (\ref{kgforphoton}) can be written as
\vs\be
A_\mu(x) = \int\frac{d^3
  k}{(2\pi)^3\sqrt{2E_k}}\,\sum_{\lambda=0}^3\left\{ a_k^{(\lambda)}
  \epsilon_\mu^{(\lambda)} e^{-ik\cdot x} + 
a_k^{\dagger(\lambda)}
  \epsilon_\mu^{*(\lambda)} e^{+ik\cdot x} \right\}~,
\label{aexpansion} 
\ee\vs\nin 
where we have used the fact that $A^\mu(x)$ must be a real field, and
the $\epsilon_\mu^{(\lambda)}$ for $\lambda=0,1,2,3$ form a 
basis for a general expansion of any four-vector, the so-called
polarization vectors.  
If we could use the Lorentz condition (\ref{lorentz}) we could
eliminate one of the polarizations through
\vs\be
k^\mu\epsilon^{(\lambda)}_\mu = 0 ~,
\ee\vs\nin
In particular, using gauge invariance we can always
eliminate the polarization with time components. This is 
desirable for the quantization procedure given that in its presence
there appear negative norm states. To see this let us guess the form
of 
$
\langle 0| T A_\mu(x) A_\nu(y) | 0\rangle~,
$
which should be the gauge boson propagator. Since each component of
$A_\mu(x)$ obeys the massless Klein-Gordon equation all we lack to
write it is to guess its tensor form: it should be an isotropic second
rank tensor. Let  us try $g_{\mu\nu}$. We then write
\vs\be
\langle 0| T A_\mu(x) A_\nu(y) | 0\rangle = \int
\frac{d^4q}{(2\pi)^4}\,\frac{-i g_{\mu\nu}}{q^2 +i\epsilon}
  e^{-iq\cdot (x-y)}~.
\label{photon2point}
\ee\vs\nin
 To understand the sign choice we notice that doing the contour
 integral in $q_0$ we obtain
\vs\be
\langle 0| T A_\mu(x) A_\nu(y) | 0\rangle = \int
\frac{d^3q}{(2\pi)^3}\,\frac{- g_{\mu\nu}}{2E_q} \, e^{-iq\cdot
  (x-y)}~.
\label{taa1}
\ee\vs\nin
If we now take $x\to y$ (but with the limit $x_0\to y_0$ from the
positive side) and take $\mu=\nu$, then the quantity in (\ref{taa1}) becomes
the norm of the state
\vs\be
A_\mu(x) | 0\rangle~. 
\label{state1}
\ee\vs\nin
We want  states associated with the physical polarizations of real
photons,  which must be spatial in nature, e.g. $A_i(x)$ for $i=1,2$,  to have positive norm. This forces us to
choose the minus sign in front of $g_{\mu\nu}$ in
(\ref{photon2point}).  But at the same time this means that the sate 
\vs\be
A_0(x)|0\rangle~,
\ee\vs\nin
must have negative norm. This sounds troublesome. However, as we mentioned above, this polarization does not correspond to a physical degree of freedom. Both the temporal as well as the longitudinal components of $A_\mu(x)$ are not physical. They do not correspond to an asymptotic state (a real photon) satisfying $q_\mu q^\mu=0$, with $q^\mu$ being the photon momentum. 
One way to see this intuitively is to consider a process where two conserved currents, $j^\mu_A(x)$ and $j^\nu_B(x)$ interact exchanging a gauge boson (photon). Each of these currents is conserved and made up by the some fermion charged under the gauge symmetry, i.e $j^\mu_A(x) = \bar\psi_A(x)
\gamma^\mu\psi_A(x)$, etc. Then the amplitude can be schematically written as
\begin{eqnarray}
    A &\sim& \int d^4x\,  j^\mu_A(x) \,\frac{-i g_{\mu\nu}}{q^2+i\epsilon}\, j^\nu_B(x) \nonumber\\
    &=& \int d^4x\,\left\{ \frac{j^1_Aj^1_B + j^2_Aj^2_B}{q^2+i\epsilon} +\frac{j^3_A j^3_B -j^0_Aj^0_B}{q^2+i\epsilon}~\right\}.
    \label{currentscatt}
\end{eqnarray}
If we choose the longitudinal direction to be the $\hat z=\hat 3$ direction, the the photon momentum is   
\begin{equation}
    q^\mu = (q_0,0,0,|\mbf{q}|)~.
\end{equation}
Current conservation then implies, for both A and B currents, 
\begin{equation}
\del_\mu j^\mu(x)=0 \to q_\mu j^\mu = q_0 j^0 -|\mbf{q}| j^3 = 0~, 
\label{jconserv2}
\end{equation}
which means that if the photon is nearly real, i.e. if $q^2\simeq 0$ and $q_0 \simeq |\mbf{q}|$, then $j^3\simeq j^0$ and the longitudinal and temporal terms of the currents cancel in the second term of (\ref{currentscatt}). So we see that, for a real photon only the transverse polarizations contribute. 
However, the unphysical polarizations cannot be neglected when considering virtual photons. It is straightforward to see this by replacing $j^3_A$ and $j^3_B$ in (\ref{currentscatt}) by using (\ref{jconserv2}). Then, the amplitude can be seen to be
\begin{equation}
    A \sim \int d^4x\,\left\{ \frac{j^1_Aj^1_B + j^2_Aj^2_B}{q^2+i\epsilon} +\frac{j^0_A j^0_B}{|\mbf{q}|^2}\right\}~,
    \label{currentscat2}
\end{equation}
which shows a transverse contribution and a second contributions that corresponds to the {\em instantaneous} Coulomb potential, entirely given by the unphysical components. 

Thus, the photon propagator defined in (\ref{photon2point}) is consistent with current conservation and therefore gauge invariance. However, it corresponds to a particular gauge choice, called the Feynman gauge. A more formal and general derivation of the gauge boson  propagator can be performed in the functional integral approach. But making use of a trick due to Fadeev and Popov, it is possible to obtain the gauge boson propagator as 
\begin{equation}
    D_{F\mu\nu}(x-y) = \int\frac{d^4q}{(2\pi)^4}\,\hat{D}_{F\mu\nu}(q)
\, e^{-iq\cdot (x-y)}~,
\end{equation}
with the momentum space propagator given by 
\begin{equation}
    \hat{D}_{F\mu\nu}(q) = -\frac{i}{q^2}\left[ g_{\mu\nu}
  -\left(1-\xi\right) \frac{q_\mu q_\nu}{q^2}\right]~.
\end{equation}
This is the gauge boson propagator in the so-called $R_\xi$ gauge, for
arbitrary values of $\xi$. Choosing $\xi$ we fix the gauge. For
instance, with the $\xi=1$ corresponds to the Feynman gauge, and  we obtain the propagator of
(\ref{photon2point}). But in many cases other choices may be more
convenient. The choice $\xi=0$ is called the Landau gauge.

\subsection{Non-abelian gauge theories}

\subsubsection{Lie algebras and non-abelian symmetries}

\nin
Non-abelian gauge theories are based on non-abelian continuous
groups. These are defined by the fact that they include elements that
can be continuously deformed into the identity. For them then we have
that
\vs\be
g \in G  /
\ee\vs\nin
the we can write
\vs\be
g(\alpha) = 1 + i\alpha^a t^a + \cO(\alpha^2)~,
\label{gexp1}
\ee\vs\nin
where the $\alpha^a$'s are infinitesimally small real parameters, summation
over the index $a$ is understood and the $t^a$ are called the
generators of the group $G$. The definition (\ref{gexp1}) implies
\vs\be
g(0)=1~.
\label{iddef}
\ee\vs\nin
If $g(\alpha)$ is {\em unitary}  then the $t^a$
must be a set of linearly independent {\em hermitian}
operators. Groups defined by these properties 
are called Lie groups. 

\nin
In order to obtain the defining property of Lie groups (its algebra)
we start by defining the group's multiplication. The multiplication of
two elements of the group results in another element of $G$:
\vs\be
g(\alpha)\,g(\beta) = g(\xi)~,
\label{mult1}
\ee\vs\nin
where the real parameters of the product element satisfy
\vs\be
\xi^a = f(\alpha^a,\beta^a)~,
\label{xias}
\ee\vs\nin
with $f$ a continuously differentiable function of the $\alpha^a$'s
and  the $\beta^a$'s.
We can conclude various things about $f$. For instance,
\vs\be
f(\alpha^a,0) = \alpha^a~,
\label{betazero}
\ee\vs\nin
and similarly for $\alpha=0$. 
On the other hand, if in (\ref{mult1}) we have that
\vs\be
g(\beta) = g^{-1}(\alpha)~,
\label{theinverse}
\ee\vs\nin
then it must be that
\vs\be
f(\alpha,\beta)=0~.
\label{fiszero}
\ee\vs\nin
Armed with this knowledge we are going to compute the following
quadruple multiplication:
\vs\be
g(\alpha)\,g(\beta)\,g^{-1}(\alpha)\,g^{-1}(\beta) = g(\xi)~.
\label{quad1}
\ee\vs\nin
We will first focus on the left hand side of (\ref{quad1}). This is
given by
\vs\be
(1+i\alpha^a t^a+\cdots) \,(1+i\beta^b t^b+\cdots)\,(1-i\alpha^c
t^c+\cdots)\,(1-i\beta^d t^d+\cdots)~,
\label{lhs1}
\ee\vs\nin
from which we can see that the terms linear in $\alpha$ and $\beta$ cancel.
Multiplying the first order parameters and keeping only up to second
order products we have
\vs\bear
1 -\alpha^a\beta^b t^a t^b + \alpha^a\alpha^c t^a t^c + \alpha^a\beta^d
t^a t^d + \beta^d\alpha^c t^b t^c + \beta^b\beta^d t^b t^d -\alpha^c
\beta^d t^c t^d +\cdots~,
\label{lhs2}
\eear\vs\nin
where the dots include the second order terms in the expansions of the
$g$'s and they will also contain second order products of $\alpha$'s
and $\beta$'s which are not explicitly written in (\ref{lhs2}). In
fact, it is easy to see that  the {\em third} and {\em sixth} terms in
(\ref{lhs2})  actually are cancelled by them.
Then the left hand side of (\ref{quad1}) up to leading order in the
infinitesimal paramenters $\alpha$ and $\beta$ is given by
\vs\be
1+ \beta^b \alpha^c\,[t^b,t^c] +\cdots ~,
\label{lhs3}
\ee\vs\nin
where $[t^b,t^c] =t^bt^c-t^ct^b$ is the commutator of the generators.

\nin
Now let us consider the right hand side of (\ref{quad1}). We know that
\vs\be
\xi = f(\alpha,\beta)~.
\label{xiofab}
\ee\vs\nin
Then the  most general expansion of $\xi$ in terms of $\alpha$ and
$\beta$ is given by 
\vs\be
\xi^e = A^e + B^{ef} \alpha^f + \tilde{B}^{ef} \beta^f +
C^{efg}\alpha^f\beta^g +\tilde{C}^{efg}\alpha^f\alpha^g
+\hat{C}^{efg}\beta^f\beta^g +\cdots~,
\ee\vs\nin
where $A^e$, $B^{ef}$, $\tilde{B}^{ef}$, $C^{efg}$, $\tilde{C}^{efg}$
and $\hat{C}^{efg}$ are arbitrary real coefficients, and the dots
correspond to terms with more than two infinitesimal parameters.
However, since using (\ref{iddef}), (\ref{theinverse}) and (\ref{fiszero}) we
know that the function in (\ref{xiofab}) satisfies
\vs\be
f(\alpha,0)=f(0,\beta)=0~,
\label{zeroiszero}
\ee\vs\nin
we immediately conclude that
\vs\be
A^e=B^{ef}=\tilde{B}^{ef}=\tilde{C}^{efg}=\hat{C}^{efg}=0~.
\label{abczero}
\ee\vs\nin
Then we conclude that
\vs\be
\xi^e = C^{efg}\,\alpha^f\beta^g + \cdots~,
\label{xisosimple}
\ee\vs\nin
and therefore
\vs\bear
g(\xi) &=& 1+ i\xi^e t^e +\cdots\nonumber\\
&~&\label{gofxi}\\
&=&1+i C^{efg} \,\alpha^f\,\beta^g\,t^e+\cdots~.
\eear\vs\nin
We can now equate this with our result for the left hand side
(\ref{lhs3}). We then conclude that the commutator of the generators
must satisfy
\vs\be\boxed{
[t^b,t^c] = i\,C^{bce} \,t^e~}~.
\label{algebra1}
\ee\vs\nin
The expression above is the defining property of the group $G$ and is
called the algebra of the group. The set of constants $C^{bce}$ are
called structure constants and vary from one group to another.

\nin
Finally, the structure constants in (\ref{algebra1}) satisfy an
identity that is derived from the following cyclic  property of
commutators:
\vs\be
[t^a,[t^b,t^c]]+[t^b,[t^c,t^a]]+[t^c,[t^a,t^b]] = 0~.
\ee\vs\nin
Using (\ref{algebra1}) and the equation above we arrive at
\vs\be\boxed{
C^{ade}\,C^{bcd} + C^{bde}\,C^{cad}+C^{cde}\,C^{abd} =0~}~,
\label{jacobi}
\ee\vs\nin
which is the Jacobi identity for the structure constants.

\subsubsection{Classification of Lie algebras}

\nin
For the applications we are manly interested in here, we focus on
unitary transformations on a finite number of fields. These can be
represented by a finite number of hermitian operators. When the number
of of generators is finite we say that the group is {\em
  compact}. If one of the generators commutes with all
others, then it generates a $U(1)$ subgroup. If the algebra does not
contain such a $U(1)$ factor is called {\em semi-simple}.
Furthermore,  if it does not contain at least two sets of
generators whose members commute with the ones from the other set,
then the algebra is called {\em simple}. The most general Lie algebra
can be expressed as a direct sum of simple algebras plus $U(1)$
abelian factors.

\nin
The restriction that the algebra be compact and simple results in the three
so called classical groups, plus five exceptional groups. Here we will
not talk about the exceptional groups ($G_2$, $F_4$, $E_6$, $E_7$ and
$E_8$) although some of them have found applications, for instance  in
attempts to build model of the  unification of all fundamental interactions.  In fact we will mostly
concentrate on $SU(N)$, which is relevant in many applications such
as, for instance, the description of gauge theories in
the standard model of particle physics. The other classical
groups, $SO(N)$ and $Sp(N)$ have been also used 
in many applications. 

\vs
\nin
\underline{$SU(N)$}: Unitary transformations of $N$-dimensional
vectors.

\nin
If $u$ and $v$ are $N$-dimensional vectors, a linear transformation on
them defined by
\vs\be
u\to U \, u~,\qquad\qquad v\to U\, v~,
\label{uvtransf}
\ee\vs\nin
is a unitary transformation if it preserves the product
\vs\be
u^\dagger \, v~.
\label{proddef}
\ee\vs\nin
This is satisfied if
\vs\be
U^\dagger = U^{-1}~.
\label{uisunitary}
\ee\vs\nin
These transformations defined in this way also include the
multiplication by an overall  phase:
\vs\be
u\to e^{i\alpha}\,u~.
\label{phase}
\ee\vs\nin
But the transformation above corresponds to an example of a $U(1)$
factor. If we want our algebra to be {\em simple}, we should remove
it. We do this by requiring that
\vs\be
{\rm det} U =1~.
\label{detuone}
\ee\vs\nin
This requirement removes the phase transformation in (\ref{phase})
since we have
\vs\be
U = e^{i H}~,
\label{hdef}
\ee\vs\nin
where $H$ must be hermitian due to (\ref{uisunitary}). The unit
determinant constraint (\ref{detuone}) means that
\vs\be
\Tr\left[ H\right] =0~,
\label{traceless}
\ee\vs\nin
excluding the $U(1)$ transformation in (\ref{phase}). Without this
exclusion we would have $U(N)=SU(N)\times U(1)$. 
The
generators of $SU(N)$ are represented by $N^2-1$ $N\times N$
traceless matrices. Of these, $N-1$ are diagonal, which define the
rank of the group. As mentioned earlier, $SU(N)$ gauged groups figure
prominently in the standard model of particle physics, where the
interactions  are
described by  the
gauge group $SU(3)\times SU(2)\times U(1)$, where the first factor
refers to strong interactions and the last two to the electroweak
ones.  

\vs\vs\vs
\nin
\underline{$SO(N)$}: Orthogonal transformations on $N$-dimensional
vectors.

\nin
It is defined
as the unitary transformations that preserve the scalar product of any
two $N$ dimensional vectors
\vs\be
u\cdot v = u_a\,\delta_{ab}\, v_b~.
\label{scalarprod}
\ee\vs\nin
This is just the group of rotations in $N$ dimensions, but we need to
exclude the reflection so that (\ref{detuone}) is satisfied. Otherwise
we would have $O(N)$, which is not a simple group.
The number of generators is
\vs\be
\frac{N(N-1)}{2}~,
\label{songens}
\ee\vs\nin
which is the number of independent angles in $N$ dimensions.

\nin
$SO(N)$ gauge
theories have been used in extensions of the standard model, such as
for example $SO(10)$ grand unification models.   The are also often
used as spontaneously broken global symmetries in models where the
Higgs boson is composite.

\vs\vs\vs
\nin
\underline{$Sp(N)$}: Symplectic  transformations on $N$-dimensional
vectors.

\nin
These transformations preserve the anti-symmetric product of $N$
dimensional vectors
\vs\be
u\cdot v = u_a \,\epsilon_{ab} \,v_b~,
\label{asprod}
\ee\vs\nin
with
\vs\be
\epsilon = \left(
  \ba{cc}
  0&1\\
  -1 &0\ea\right)~.
\label{eab}
\ee\vs\nin
The groups has 
\vs\be
\frac{N(N+1)}{2}~,
\label{spngens}
\ee\vs\nin
generators, that means that it is represented by this number of 
$N\times N$ unitary matrices. 

\subsubsection{Representations}

\nin
A representation is a  realization of the multiplication of group
elements by using matrices. That is
\vs\be
a \, b  = c\qquad\to \qquad M(a) \, M(b) = M(c)~,
\label{repdef}
\ee\vs\nin
where $M(a)$, $M(b)$ and $M(c)$ are matrices. A representation is said
to be {\em reducible}  if it can be written in diagonal block form,
that is as
\vs\be
M(a) = \left(\ba{ccc}
  M_1(a) &\bf{0} &\bf{0}\\
  \bf{0} & M_2(a) &\bf{0}\\
  \bf{0} &\bf{0} &M_3(a)\ea\right)~.
\label{rrep}
\ee\vs\nin
A reducible representation is the direct sum of irreducible
representations (irreps). 

\nin
The dimension of  representation $r$, $d(r)$, is the dimension of the
vector space in which the matrices $M(a)$ act.  Irreps can be used to
have matrices representing the generators of the group, $t^a$. We
denote these matrices as $t^a_r$.  To fix their normalization we
define the trace of the product as
\vs\be
\Tr[t^a_r\, t^b_r] \equiv D^{ab}~,
\label{traceinrep}
\ee\vs\nin
which satisfies $D^{ab}>0$ if the $t^a_r$ are hermitian. We can always
choose a basis for the matrices $t^a_r$ such that
\vs\be
D^{ab}\propto \delta^{ab}~,
\label{isdelta}
\ee\vs\nin
meaning that
\vs\be
\Tr[t^a_r\, t^b_r] = C(r)\,\delta^{ab}~,
\label{crdef}
\ee\vs\nin
with $C(r)$ a constant that depends on the particular representation
$r$.

\nin
Expressing the generators by the $t^a_r$, we may write the algebra of
the Lie group as
\vs\be
[t^a_r,t^b_r] = i\,f^{abc}\,t^c_r~,
\label{fdef}
\ee\vs\nin
where the $f^{abc}$ are the structure constants (which we called
$C^{abc}$ before). Making use of (\ref{crdef}) and (\ref{fdef}) we can
write the structure constants as
\vs\be
f^{abc} = \frac{-i}{C(r)}\, \Tr[[t^a_r,t^b_r]\,t^c_r]~.
\label{fastrace}
\ee\vs\nin
Expanding the commutator and the trace it is straightforward to show
that (\ref{fastrace}) implies that $f^{abc}$ is totally anti-symmetric
under the exchange of the group indices $a$, $b$ and $c$.

\vs\nin
\underline{\em Complex Conjugate Representation}

\nin
For each irrep $r$ we can define a {\em complex conjugate}
representation $\bar r$. For instance, if we have a field $\phi$
undergoing an infinitesimal transformation we write
\vs\be
\phi \to (1+i\alpha^a \,t^a_r )\,\phi~.
\label{fitransf}
\ee\vs\nin
Then, the complex conjugate of the field transforms as
\vs\be
\phi^* \to (1-i\alpha^a \,(t^a_r )^*)\,\phi^*~.
\label{ficc}
\ee\vs\nin
Then, the generators of the complex conjugate representation are
defined as
\vs\be
t^a_{\bar r} = -(t^a_r)^* = -(t^a)^T~,
\label{ccrepgen}
\ee\vs\nin
where the last equality is a consequence of $t^a_r$ being hermitian.
There are cases when the complex conjugate representation $\bar r$
is equivalent with $r$.  This is the case if a unitary transformation
$U$ exists such that
\vs\be
t^a_{\bar r} = U\,t^a_r\,U^\dagger~.
\label{risreal}
\ee\vs\nin
Then we say that the representation $r$ is {\em real}.

\vs\nin
\underline{\em Adjoint Representation}

\nin
The generators of the adjoint representation $G$  are defined by the
structure constants $f^{abc}$ by 
\vs\be
\left(t^b_G\right)_{ac} \equiv i\,f^{abc} ~.
\label{adrepdef}
\ee\vs\nin
It is straightforward to verify that they satisfy the algebra, that is
that
\vs\be
[t^b_G,t^c_G]_{ae} = i f^{bcd}\, \left(t^d_G\right)_{ae}~,
\label{adjalg}
\ee\vs\nin
which is in fact the Jacobi identity (\ref{jacobi}). Since the
structure constants $f^{abc}$ are real, we can see that the generators
of the adjoint representation satisfy
\vs\be
t^a_G = -(t^a_G)^*~,
\label{gisreal}
\ee\vs\nin
which means that he adjoint representation is real. The dimension of
the adjoint representations, $d(G)$ is given by the number of
generators of the group, e.g. $N^2-1$ for $SU(N)$, etc.

\vs\nin
\underline{Casimir Operator}

\nin
The operator defined by
\vs\be
T^2 \equiv t^a\,t^a~,
\label{casimirdef}
\ee\vs\nin
is called the Casimir operator and it has the property that it
commutes with all the generators of the group. That is,
\vs\be
[T^2,t^a] =0~.
\label{t2tazero}
\ee\vs\nin
The most well known example is the operator for the total angular
momentum squared, $J^2$, which commutes with all the components of
$\vec{J}$. In a given irrep $r$ the Casimir is given by a constant:
\vs\be
t^a_r\,t^a_r = C_2(r)\,1/~,
\label{c2def}
\ee\vs\nin
where $1$ is the identity in $d(r)\times d(r)$
dimensions. Here we defined $C_2(r)$, the quadratic Casimir operator
of the representation $r$. For the particular case of the adjoint
representation, we have
\vs\be
\left(t^c\right)_{ad}\,\left(t^c\right)_{bd} = f^{acd}\,f^{bcd} =
C_2(G) \,\delta^{ab}~.
\label{casadjoint}
\ee\vs\nin
For a given representation $r$ it is possible to relate the Casimir
$C_2(r)$ with $C(r)$. To see this we start from (\ref{crdef}). We have
that if we multiply it by $\delta^{ab}$ on each side we arrive at 
\vs\bear
\delta^{ab} \,\Tr[t^a_r\,t^b_r] = C(r)\, \delta^{ab}\,\delta^{ab}
\label{relation}
\eear\vs\nin
The product of the two deltas in the right hand side above gives the
number of generators, which we can write as $d(G)$, the dimension of
the adjoint representation $G$. But inserting the factor of
$\delta^{ab}$  on the right
hand side of (\ref{relation}) inside the trace, we obtain the trace of
(\ref{c2def}). Noticing that  $\Tr[1] = d(r)$ we arrive at
the useful relation
\vs\be\boxed{
d(r)\,C_2(r) = d(G)\,C(r)~}~.
\ee\vs\nin
We are now ready to tackle gauge symmetries based on non-abelian
groups.

\subsubsection{Gauge invariance and geometry} 

\nin
We consider here the generalization of the concept of gauge invariance
when the gauge group $G$ is non-abelian. Below, we will see what this
means by presenting the basics of non-abelian group theory. We will
also study  the physical consequences of non-abelian gauge
invariance. But before we do all that, we will take another look at
abeliang gauge theory, i.e. when $G=U(1)$, by thinking about gauge
invariance in a geometric way.

\nin
We consider the action of a $U(1)$ local symmetry transformation on a
fermion field $\psi(x)$. It is given by
\vs\be
\psi(x) \to \psi'(x)= e^{i\alpha(x)}\,\psi(x)~.
\label{gt1}
\ee\vs\nin
As we well know, terms in the Lagrangian that do no contain
derivatives are trivially invariant under (\ref{gt1}). For instance,
the fermion mass term transforms as
\vs\be
m\bar\psi\psi \to m\bar\psi' \psi' = m \bar\psi \psi~.
\ee\vs\nin
However, terms containing derivatives are not invariant. Let us study
in detail how the problem arises. We write the  derivative
by using a direction in spacetime defined by a four-vector $n_\mu$,
such that 
\vs\be
n^\mu\del_\mu\psi(x) = \lim_{\epsilon\to 0}\,
\frac{1}{\epsilon}\,\left[\psi(x+\epsilon n) - \psi(x)\right]~,
\label{defder1}
\ee\vs\nin
where the argument of the first term on the left hand side must be
understood as
\vs\be
x_\mu + \epsilon n_\mu = x_\mu + \Delta x_\mu~.
\label{xplusdeltax}
\ee\vs\nin
But the fields $\psi(x+\epsilon n)$ and $\psi(x)$ have {\em different}
gauge transformations as clearly seen from (\ref{gt1}). The fact that
they are evaluated in different spacetime points means that the gauge
parameters of their transformations are different,
i.e. $\alpha(x+\epsilon n)$ and $\alpha(x)$.
This translates in $\del_\mu\psi(x)$ not having a well defined gauge
transformation. 

\nin
The situation is similar to what happens in general relativity when we
want to compare two objects with non-trivial transformation
properties, e.g. vectors or spinors, at two different positions in
spacetime. For instance, if the objects being compared are two vectors,
then part of the variation comes from the fact that the curvature will
change the orientation of a vector as we move it from one point to
another. But we are interested in the {\em intrinsic} variation due to
some dynamical effect. For this purpose we define a {\em parallel
  transport}.  Our case is no different. 

\nin
We define the scalar function
\vs\be
U(y,x)~,
\label{udef1}
\ee\vs\nin
depending on two spacetime points $x$ and $y$ in such as way that it
transforms under the $U(1)$ gauge symmetry as
\vs\be
U(y,x) \to e^{i\alpha(y)}\,U(y,x)\,e^{-i\alpha(x)}~.
\label{ugt1}
\ee\vs\nin
We call $U(y,x)$ a comparator. 
This clearly means that $U(y,y)=1$. Also, it means that
\vs\be
U(y,x)\,\psi(x) \to e^{i\alpha(y)}\, U(y,x) \,\psi(x)~.
\label{upsigt1}
\ee\vs\nin
Thus, the product of the comparator times the field in $x$, transforms
as an object located in $y$. We can use this to define a new
derivative as
\vs\be
n^\mu D_\mu \equiv \lim_{\epsilon\to 0}\,\frac{1}{\epsilon} \,
\left[\psi(x+\epsilon n) - U(x+\epsilon n, x)\,\psi(x)\right]~,
\label{covderdef1}
\ee\vs\nin
so that the two terms being subtracted transform in the same way under
the gauge symmetry.  This is the case given that under a $U(1)$ gauge
transformation 
\vs\bear
\psi(x+\epsilon n) &\to & e^{i\alpha(x+\epsilon n)}\,\psi(x+\epsilon
n)\nonumber\\
&~&\label{sametransf}\\
U(x+\epsilon n,x) \,\psi(x) &\to & e^{i\alpha(x+\epsilon n)} \,
U(x+\epsilon n,x)\, \psi(x)~.\nonumber
\eear\vs\nin
Based on the definition of the covariant derivative in
(\ref{covderdef1}) we can recover the familiar form of
$D_\mu\psi(x)$. For this purpose, we first expand the comparator at
leading order in $\epsilon$ as
\vs\be
U(x+\epsilon n,x) = 1 - i \epsilon n^\mu\,A_\mu(x) + \cO(\epsilon^2)~,
\label{uexp1}
\ee\vs\nin
where the linear term in the expansion must depend also on the
direction $n^\mu$, but then this Lorentz index must be contracted with
a four-vector that generally depends on $x$, which we call
$A_\mu(x)$. Implicit in the form of the expansion we used in
(\ref{uexp1}) is the assumption that the comparator can be written as a
phase, since the normalization  can always be absorbed in
redefinitions of the fields, here $\psi(x)$. Replacing (\ref{uexp1})
in (\ref{covderdef1}) we have
\vs\bear
n^\mu\,D_\mu\,\psi(x) &=& \lim_{\epsilon\to 0} \,
\frac{1}{\epsilon}\,\left[\psi(x+\epsilon n) - \psi(x) +i\epsilon
  n^\mu A_\mu(x)\right]~\nonumber\\
& ~&\label{covder3}\\
&=& \lim_{\epsilon\to 0} \, \frac{1}{\epsilon}\,\left[\psi(x+\epsilon
  n) -\psi(x)\right] + in^\mu A_\mu(x)~,\nonumber
\eear\vs\nin
where we neglected terms of higher order in $\epsilon$ since they do
not contribute when taking the limit $\epsilon\to 0$. The first term
above is just the normal derivative as defined in (\ref{defder1}), so
we obtain
\vs\be
D_\mu\psi(x) = \del_\mu\psi(x) + iA_\mu(x)\psi(x)~,
\label{covderdef2}
\ee\vs\nin
which is of course the usual definition of the covariant derivative. 
The vector field $A_\mu(x)$ will also transform under the gauge
symmetry. To extract its transformation law, we need to look at the
expansion of the transformation of the comparator $U(y,x)$ which
defines $A_\mu(x)$. This is,
\vs\bear
U(x+\epsilon n,x) &\to & e^{i\alpha(x+\epsilon n)} \,U(x+\epsilon
  n,x)\,e^{-i\alpha(x)}~\nonumber\\
&~&\nonumber\\
1-i\epsilon\, n^\mu A_\mu(x) +\cdots &\to & (1+i\alpha(x+\epsilon n)
+\cdots)\,(1-i\epsilon\, n^\mu
A_\mu(x)+\cdots)\,(1-i\alpha(x)+\cdots)\nonumber\\
&~&\nonumber\\
&\to &1+ i(\alpha(x+\epsilon n)-\alpha(x)) -i\epsilon\, n^\mu A_\mu(x)
+\cdots~,
\label{atransf1}
\eear\vs\nin
where the dots indicate both higher orders in $\epsilon$ and in the
$\alpha$'s. We point out that we are not using an infinitesimal
$\alpha(x)$, but that the higher orders terms in $\alpha$ actually
identically cancel. 
Dividing both sides of (\ref{atransf1}) by $\epsilon$ and taking the
limit for $\epsilon\to 0$ we obtain
\vs\be
A_\mu(x) \to A_\mu(x) -\del_\mu\alpha(x)~,
\label{atransf2}
\ee\vs\nin
as expected. Combining (\ref{atransf2}) and (\ref{covderdef2}) one can
easily verify that the covariant derivative transforms as 
\vs\be
D_\mu\psi(x) \to e^{i\alpha(x)}\,D_\mu\psi(x)~,
\label{covdertransf2}
\ee\vs\nin 
which guarantees that all terms in the Lagrangian are now gauge
invariant if the covariant derivative replaces the normal
derivative. That is, the first term in  
\vs\be
\cL = \bar\psi i\gamma^\mu D_\mu\psi -m\bar\psi\psi ~,
\label{gaugedirac}
\ee\vs\nin
is $U(1)$ gauge invariant since the covariant derivative of $\psi(x)$ transforms as
the field $\psi(x)$. 

\nin
A final question is the definition of a kinetic term for the {\em
  connection} field $A_\mu(x)$. Here, we will make use of a method that,
although appears too complicated for the abelian case, it will be very
useful when applied to non-abelian gauge theories later. 
What we are after is a term that depends quadratically on derivatives
of  $A_\mu(x)$. What we will start with is the following differential
operator applied to the fermion field:
\vs\be
[D_\mu, D_\nu] \,\psi(x) ~.
\label{dmudnudef}
\ee\vs\nin
 This is the commutator of the covariant derivatives applied
 $\psi(x)$. Using (\ref{covdertransf2}) is easy to verify that  (\ref{dmudnudef}) 
 transforms like the field, that is
\vs\be
[D_\mu, D_\nu] \,\psi(x) \to e^{i\alpha(x)}\, [D_\mu, D_\nu] \,\psi(x)
~.
\label{dmunutransf}
\ee\vs\nin
This can be interpreted as a transformation rule for the commutator:
\vs\be
[D_\mu, D_\nu]  \to e^{i\alpha(x)}\, [D_\mu, D_\nu] \,
e^{-i\alpha(x)}~.
\label{covdercomute}
\ee\vs\nin
On the other hand, we can explicitly compute the commutator by using (\ref{covderdef2}). 
This is
\vs\bear
[D_\mu,D_\nu]\,\psi(x) &=& [\del_\mu+iA_\mu,
\del_\nu+iA_\nu]\,\psi(x)\nonumber\\
&~&\nonumber\\
&=& i\,\left(\del_\mu A_\nu- \del_\nu A_\mu\right)\,\psi(x)~,
\label{comm1}
\eear\vs\nin
which reveals that the commutator of the covariant derivatives is
itself {\em not} a differential operator.

\nin
We then define 
\vs\be
[D_\mu,D_\nu]\equiv i\,F_{\mu\nu}~,
\label{fmunudef}
\ee\vs\nin
which is clearly gauge invariant, since the commutator transformation
rule (\ref{covdercomute}) implies 
\vs\be
F_{\mu\nu} \to e^{i\alpha(x)}\, F_{\mu\nu}\,
e^{-i\alpha(x)} = F_{\mu\nu}~.
\ee\vs\nin
This can be alternatively seen from
(\ref{dmunutransf}) in combination with  the field transformation
(\ref{gt1}), since the commutator is not a differential operator. 
Then, two powers of $F_{\mu\nu}$ would give us what we want for a
gauge field kinetic term.

\nin
This
concludes our rederivation of the $U(1)$ gauge invariant Lagrangian 
\vs\be
\cL = \bar\psi(i\gamma^\mu D_\mu -m)\psi -\frac{1}{4} F_{\mu\nu}
F^{\mu\nu} + \cdots~,
\label{u1gilag}
\ee\vs\nin
where the factor of $-1/4$ is necessary to recover the electromagnetic
strength tensor in the classical limit, and the dots denote possible
gauge invariant higher dimensional (non-renormalizable) terms.

\subsubsection{Non-abelian gauge groups}

\nin
We will now follow the same geometric procedure we applied for a
$U(1)$ gauge theory for the case of non-abelian groups. 
We first consider the case of $G=SU(2)$ and later generalize our results
for arbitrary non-abelian groups. $SU(2)$ is isomorphic with $SO(3)$
the group of rotations in 3 dimensions, so it should be familiar 
from the study of angular momentum in quantum mechanics.
The elements of $SU(2)$ are unitary matrices which we write as 
\vs\be
g(x) = e^{i\alpha^a(x) t^a}~,
\label{gdef}
\ee\vs\nin
where $t^a$ are the generators (three of them from $2^2-1$),  which are given in terms of the Pauli
matrices by
\vs\be
t^a =\frac{\sigma^a}{2}~,
\label{tissigma}
\ee\vs\nin
with 
\vs\be
\sigma^1 = \left(\ba{cc} 0&1\\1&0\ea\right)~, 
\quad
\sigma^2 = \left(\ba{cc} 0&-i\\i&0\ea\right)~, 
\quad 
\sigma^3 = \left(\ba{cc} 1&0\\0&-1\ea\right)~.
\ee\vs\nin
As we see from (\ref{gdef}), there are $3$ coefficient functions of
$x$, $\alpha^1(x)$, $\alpha^2(x)$ and $\alpha^3(x)$, so that the
exponent is the most general $x$ dependent expansion of the
generators. Let us consider, just as in the previous section for the
$U(1)$ case, the transformation of a fermion field under a $SU(2)$
gauge group. 
This is given by
\vs\be
\psi(x) \to \psi'(x) = e^{i\alpha^a(x) t^a}\,\psi(x) = g(x)\,\psi(x)~.
\label{fermionsu2}
\ee\vs\nin
If a fermion field does transform as in (\ref{fermionsu2}) this
implies that it has an $SU(2)$ internal index. Depending on the
representation under which they transform they will be different {\em
  multiplets}. The {\em fundamental } representation correspond to
using (\ref{tissigma}) and implies that the fermion field is an
$SU(2)$ {\em doublet} 
\vs\be
\psi(x) = \left(\ba{c}  \psi_1(x) \\~\\ \psi_2(x)\ea\right)~,
\label{doublet}
\ee\vs\nin  
which means there are two fermions. The local transformation
(\ref{fermionsu2}) mixes these two components. 

\nin
 We are now in a position to define the covariant
derivative. Just as before, we define the comparator $U(y,x)$ with the
gauge transformation property
\vs\be
U(y,x)\to g(y)\,U(y,x)\,g^\dagger(x)~.
\label{uinsu2}
\ee\vs\nin
and just as for the $U(1)$ case before, the covariant derivative has
the geometric definition
\vs\be
n^\mu D_\mu \psi(x) = \lim_{\epsilon\to 0}
\,\frac{1}{\epsilon}\,\left[\psi(x+\epsilon n) -U(x+\epsilon
n,x)\,\psi(x)\right]~.
\label{coversu2}
\ee\vs\nin
Noticing that 
\vs\be
U(y,y) = \mbf{1}~,
\label{uyyisid}
\ee\vs\nin
the identity in $2\times 2$ matrices, we can expand $U(y,x)$ around
this considering infinitesimal gauge transformations
$\alpha^a(x) \sim \cO(\epsilon)$. The most general expansion to leading order is
\vs\be
U(x+\epsilon n,x)  = \mbf{1} + i g \epsilon n^\mu A^a_\mu(x)\,t^a +
\cO(\epsilon^2)~,
\label{uexp4}
\ee\vs\nin
where we included a factor $g$, the coupling, and the Lorentz index in
$n^\mu$ is contracted by the fields $A^a_\mu(x)$, where the index $a$
contracts with the one in the generator. This reflects the fact that
the  most general expansion is a linear combination of the 3 Pauli
matrices, meaning that now we will have 3 gauge fields, $A^1_\mu(x)$,
$A^2_\mu(x)$ and $A^3_\mu(x)$.  Then, we have 
\vs\bear
n^\mu D_\mu\psi(x) &=& \lim_{\epsilon\to
  0}\,\frac{1}{\epsilon}\,\left[\psi(x+\epsilon n) - U(x+\epsilon
  n,x)\,\psi(x)\right]~\nonumber\\
&~&\label{nacovder1}\\
&=& n^\mu\del_\mu\psi(x) - ig n^\mu\,A^a_\mu(x)\,t^a\psi(x)~,
\nonumber
\eear\vs\nin
which results in the covariant derivative 
\vs\be
D_\mu\psi(x) = \left(\del_\mu - ig\,A^a_\mu(x)\,t^a\right)\psi(x)~.
\label{nacovder2}
\ee\vs\nin
For the case at hand, i.e. $G=SU(2)$ the generators in
(\ref{nacovder2}) are one half of the Pauli
matrices. This is the covariant derivative acting on a fermion $\psi$
that transforms under the $SU(2)$ gauge group as in (\ref{fermionsu2}).
As we will see below, this determines the interactions of fermions
with the $SU(2)$ gauge bosons $A^a_\mu(x)$. 

\nin
The next step is to obtain the gauge transformations for the gauge
fields. Once again, to do this we consider the infinitesimal gauge transformation of the
comparator. Using (\ref{uinsu2}) this is given by
\vs\bear
U(x+\epsilon n,x) &\to & g(x+\epsilon n)\,U(x+\epsilon
n,x)\,g^\dagger(x)\nonumber\\
&~&\label{uinf1}\\
\mbf{1} + i g \,\epsilon n^\mu A^a_\mu(x)\,t^a  &\to &
g(x+\epsilon n)\, \left(\mbf{1} + i g \epsilon n^\mu
  A^a_\mu(x)\,t^a \right) g^\dagger(x)~,\nonumber
\eear\vs\nin 
where in the second line we use the expansion in (\ref{uexp4}). We
notice that
\vs\bear
g(x+\epsilon n)\,g(x) &=& \left[\left(\mbf{1} + \epsilon
    n^\mu\frac{\del}{\del x^\mu}+
    \cO(\epsilon^2)\right)g(x)\right]\,g^\dagger(x)\nonumber\\
&~& \label{ggdagger}\\
&=& \mbf{1} + \epsilon\,n^\mu\del_\mu(g(x))\,g^\dagger\nonumber~.
\eear\vs\nin
Replacing the equation above in (\ref{uinf1}), we have that
\vs\be
A^a_\mu(x)\,t^a \to g(x)\,\left(A^a_\mu(x)\,t^a\right)\,g^\dagger(x)
-\frac{i}{g}\,(\del_\mu g(x))\,g^\dagger(x)~.
\label{aatransf1}
\ee\vs\nin 
If we now define the gauge field matrix
\vs\be
A_\mu(x) \equiv A^a_\mu(x) \,t^a~,
\label{amatrixdef}
\ee\vs\nin
we can rewrite (\ref{aatransf1}) as
\vs\be\boxed{
A_\mu(a) \to g(x)\left( A_\mu(x)
  +\frac{i}{g}\,\del_\mu\right)\,g^\dagger(x)~}~,
\label{amatrixtransf}
\ee\vs\nin
where we have used the fact that $g^\dagger g=g g^\dagger =
\mbf{1}$ in order to make the replacement
\vs\be
\del_\mu (g(x))\,g^\dagger(x) =- g(x)\del_\mu g^\dagger(x)~.
\ee\vs\nin 
The gauge transformation  of the matrix gauge field
(\ref{amatrixtransf}) is actually valid for any non-abelian gauge
group, not just $SU(2)$, as long as $g(x)$ is a group element expressed
in terms of the generators $t^a$ as in (\ref{gdef}). We can also
recover the {\em abelian} gauge field transformation (\ref{atransf2}) if we 
replace $t^a$ by the identity and $\alpha^a(x)$ is just
$\alpha(x)$. However this is deceiving since there are new
contributions that appear exclusively in the non-abelian case. To see
this in the gauge field transformation, we consider an infinitesimal
gauge transformation with 
\vs\be
g(x) = \mbf{1} + i\,\alpha^a(x)\,t^a+\cdots
\label{ginfini}
\ee\vs\nin
where the dots denote terms higher in powers of
$\alpha^a(x)$. Replacing (\ref{ginfini}) in (\ref{amatrixtransf}) we
arrive at
\vs\be
A^a_\mu(x)\,t^a \to A^a_\mu(x)\,t^a +
\frac{1}{g}\,\del_\mu\alpha^a(x)\,t^a + i\,\left[\alpha^a(x)\,t^a,
  A^b_\mu(x)\,t^b\right] +\cdots~.
\label{atransfinf}
\ee\vs\nin
The first two terms in (\ref{atransfinf}) are analogous to what we
find in the abelian case. But the third term  is only present in non
abelian gauge groups since it is proportional to the commutator of two 
generators. We will see below that this non commutativity has
important physical consequences.

\nin
With the definition of the covariant derivative in (\ref{nacovder2})
and the gauge field transformation (\ref{amatrixtransf}) we can prove
that the fermion kinetic term given by 
\vs\be
\bar\psi \gamma^\mu D_\mu\psi~,
\label{ferkin1} 
\ee\vs\nin
is invariant under the gauge transformations (\ref{fermionsu2}). This
means that under these gauge transformations
\vs\be
D_\mu\psi(x) \to g(x)\,D_\mu\psi(x)~,
\label{dmupsitransf}
\ee\vs\nin 
must be satisfied. This can be explicitly verified just by
substitution. 

\nin
The final step, just as in the abelian case considered earlier, is to
obtain the kinetic term for the gauge fields. Following the steps
taken there, we need to compute 
\vs\be
[D_\mu,D_\nu]\psi(x)~.
\label{comute1}
\ee\vs\nin
Using the matrix notation (\ref{amatrixdef}) and replacing the
explicit form of the covariant derivative (\ref{nacovder2}) in
(\ref{comute1}) we obtain
\vs\be
[D_\mu,D_\nu]\psi(x) = -i g\,\left(\del_\mu A_\mu -\del_\nu
  A_\mu\right)\psi(x) - g^2\,\left[A_\mu,A_\nu\right]\psi(x)~.
\label{comute2}
\ee\vs\nin 
Once again, just as for the abelian case, we see that the commutator
in (\ref{comute1}) is not a differential operator. But unlike for the
abelian case, there is a new term proportional to the commutator
\vs\be
[A_\mu,A_\nu] = A^a_\mu A^b_\nu\,[t^a,t^b]~.
\label{amuanucomute}
\ee\vs\nin 
Defining the gauge field strength (matrix) by
\vs\be
[D_\mu,D_\nu]\psi(x) \equiv -i g\,F_{\mu\nu}\,\psi(x)~,
\label{nafmunu1}
\ee\vs\nin 
we have that 
\vs\be
F_{\mu\nu} = \del_\mu A_\nu -\del_\nu A_\mu - ig\,[A_\mu,A_\nu]~,
\label{nafmunu2}
\ee\vs\nin
which can be expressed in gauge field components using
(\ref{amatrixdef}) to give
\vs\be
 F_{\mu\nu} = \left(\del_\mu A^a_\nu -\del_\nu A^a_\mu\right)\,t^a
 -ig\,A^a_\mu A^b_\nu\,[t^a,t^b]~.
\label{nafmunu3}
\ee\vs\nin
Defining the gauge field strength $F^a_{\mu\nu}$ by
\vs\be
F_{\mu\nu}\equiv F^a_{\mu\nu}\,t^a~,
\label{fmunuadef}
\ee\vs\nin
and writing the commutator out in terms of the structure constants we
arrive at
\vs\be\boxed{
F^a_{\mu\nu} = \del_\mu A^a_\nu - \del_\nu A^a_\mu + g f^{abc} A^b_\mu
A^c_\nu~}~,
\label{fmunua2}
\ee\vs\nin
which is the non-abelian gauge field strength in all generality. For
instance for and $SU(2)$ gauge theory we have 
 \vs\be
F^a_{\mu\nu} = \del_\mu A^a_\nu - \del_\nu A^a_\mu + g \epsilon^{abc} A^b_\mu
A^c_\nu~,
\label{fmunuainsu2}
\ee\vs\nin
since the structure constants are given by the epsilon tensor
$\epsilon^{abc}$.

\nin
Now, given (\ref{dmupsitransf}), we know that the commutator acting on the
fermion field transforms as
\vs\be
[D_\mu,D_\nu]\psi(x) \to g(x)\, [D_\mu,D_\nu]\psi(x),
\label{comutetransf}
\ee\vs\nin
which results in the gauge  transformation for the commutator
\vs\be
[D_\mu,D_\nu] \to g(x) \, [D_\mu,D_\nu]\, g^\dagger(x)~.
\label{dmudnuto}
\ee\vs\nin
Then, using (\ref{nafmunu1}), we obtain the gauge transformation for
the matrix $F_{\mu\nu}$ :
\vs\be
F_{\mu\nu} \to g(x)\,F_{\mu\nu}\,g^\dagger(x)~.
\label{nafmunutransf}
\ee\vs\nin 
We can use this information to guess the form of the gauge invariant
kinetic term. From (\ref{nafmunutransf}), we see that $F_{\mu\nu}$ is
not gauge invariant, unlike what happens in the abelian case. Then,
although 
\vs\be
F_{\mu\nu} F^{\mu\nu} \to g(x)\,F_{\mu\nu} F^{\mu\nu}\,g^\dagger(x)~,
\ee\vs\nin
is not gauge invariant, its trace actually is. Then, we have 
\vs\bear
\Tr[F_{\mu\nu} F^{\mu\nu}] &=& F^a_{\mu\nu}\, F^{b\mu\nu}\, \Tr[t^a
t^b]\nonumber\\
&~&\label{traceoffmunu2}\\
&=&  F^a_{\mu\nu}\, F^{b\mu\nu}\,\frac{\delta^{ab}}{2}~,
\nonumber
\eear\vs\nin
so the form of the kinetic term that corresponds to the abelian
normalization is
\vs\be\boxed{
-\frac{1}{2} \Tr[F_{\mu\nu}\,F^{\mu\nu}] = -\frac{1}{4}
F^a_{\mu\nu}\,F^{a\mu\nu}~}~.
\label{nagkinterm}
\ee\vs\nin
Although at first the form of the gauge kinetic term above looks just
like a simple sum of the kinetic terms of the individual gauge bosons
(for $a=1,\dots,N^2-1$), this is deceiving. When plugging in the
explicit form of $F^a_{\mu\nu}$ from (\ref{fmunua2}) we see that
(\ref{nagkinterm}) not only leads to terms quadratic in the
derivatives of each of the fields, but also to interactions among the
gauge fields: there will be a triple interaction and a quartic
one. This is a crucial feature of  non-abelian gauge theories: the
gauge bosons interact with each other, whereas this is not the case
for the gauge bosons  of the abelian $U(1)$, e.g. the photons. This
will have very important consequences, from the behavior of scattering
amplitudes to the renormalization group flow.

\subsubsection{Feynman rules in non-abelian gauge theories}

\nin
Here we  press on with non-abelian gauge theories by
deriving their Feynman rules. However, before we can safely apply  them
to compute scattering amplitudes in perturbation theory and,
specially before we can study the renormalization of  these gauge
theories, we will see at the end of this lecture that there is
something missing. In order to solve this problem, we will have to be
careful in quantizing non-abelian gauge theories, as we will do in
the next lecture. 

\nin
We start by considering a generic a theory of a fermion that transforms as 
\vs\be
\psi(x)\to g(x)\,\psi(x) = e^{i\alpha^a(x) t^a}\,\psi(x)~,
\label{ftransfgen}
\ee\vs\nin
under a generic non-abelian gauge symmetry. The Lagrangian of the
theory is then 
\vs\be
\cL = \bar\psi\left(i \dsl D -m\right)\psi -\frac{1}{4} F^a_{\mu\nu}
F^{a\mu\nu}~,
\label{lag1}
\ee\vs\nin
where the covariant is given by
\vs\be
D_\mu\psi(x) = (\del_\mu-ig\,A^a_\mu(x)\,t^a)\,\psi(x)~,
\label{covderfr}
\ee\vs\nin 
and the $t^a$ are the generators of the gauge group $G$ written in the
appropriate representation. The non-abelian field strength is 
\vs\be
F^a_{\mu\nu} = \del_\mu A^a_\nu -\del_\nu A^a_\mu + g f^{abc} A^b_\mu
A^c_\nu~.
\label{famunufr}
\ee\vs\nin
As we saw earlier, this means that there will be interactions terms in
the gauge field ``kinetic term'', the last one in (\ref{lag1}). Thus,
for the purpose of deriving all the Feynman rules it is convenient to
split the Lagrangian in (\ref{lag1}) into a truly free Lagrangian and
interacting terms. We define 
\vs\be
\cL = \cL_0 + \cL_{\rm int.} 
\label{lagsplit}
\ee\vs\nin 
where the free Lagrangian is now
\vs\be
\cL_0 \equiv \bar\psi (i\dsl\del -m)\psi -\frac{1}{4} \left(\del_\mu
  A^a_\nu -\del_\nu A^a_\mu\right)\left(\del^\mu A^{a\nu} - \del^\nu
  A^{a\mu}\right)~.
\label{freelag1} 
\ee\vs\nin
On the other hand, the interaction part of the Lagrangian defined in
(\ref{lagsplit}) can be itself separated into three terms given by
\vs\be
\cL_{\rm int.} = \cL_{\rm int.}^f + \cL_{\rm int.}^{3G} + \cL_{\rm
  int.}^{4G}~,
\label{lintsplit}
\ee\vs\nin
denoting the interactions of gauge bosons with fermions, 
\vs\be
\cL_{\rm int.}^f = g A^a_\mu \bar\psi\gamma^\mu t^a\psi~,
\label{lintfermions}
\ee\vs\nin
the triple
gauge boson interaction 
\vs\be
\cL_{\rm int.}^{3G} = -g f^{abc} \del^\mu A^{a\nu} A^b_\mu A^c_\nu~,
\label{lint3g}
\ee\vs\nin
and the quartic one
\vs\be
\cL_{\rm int.}^{4G} = -\frac{1}{4} g^2 f^{abc} f^{ade} A^b_\mu A^c_\nu
A^{d\mu} A^{e\nu}~,
\label{lint4g}
\ee\vs\nin
 respectively. It is now straightforward to derive the Feynman rules
 from (\ref{lintfermions}), (\ref{lint3g}) and (\ref{lint4g}). 

\nin
We start with the fermion interaction. The Feynman rule is very
similar to that of QED, but with the addition of the gauge group
generator. This is shown in the figure below:
\vs
\begin{figure}[h]
\hspace*{1.0in}\includegraphics[width=0.3\textwidth]{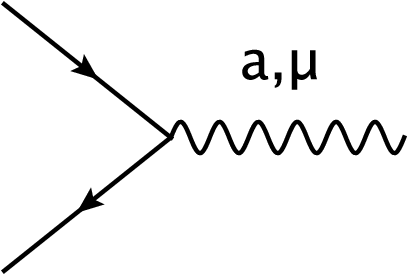}
\end{figure}
\nin
\vspace*{-1.25in}
\be
\qquad\qquad\qquad\qquad\qquad=\qquad i \,g \gamma^\mu\,t^a\nonumber
\ee
\vs\vs\vs\vs\vs\vs\vs\vs\vs
\nin
\nin    
Next, we consider the triple gauge boson interaction in
(\ref{lint3g}). Here we have to be more careful with the momentum
flow since it involves a derivative on one of the gauge fields. To
obtain the Feynman rule from $i \cL_{\rm int.}^{3G}$ we need to
contract it with all possible combinations of the state
\vs\be
| k,\epsilon(k);\, p,\epsilon(p);\, q,\epsilon(q)\rangle~.
\label{state3g}
\ee\vs\nin
There are $3!$ such contractions. For instance, if we contract the
gauge boson of momentum $k$ with $\del^\mu A^{a\nu}$, the one with
momentum $p$ with $A^b_\mu$ and the one with momentum $q$ with
$A^c_\nu$,  we obtain the following
contribution to the Feynman rule
\vs\be
-i g f^{abc} (-ik^\nu) g^{\mu\rho}~.
\label{frcontrib}
\ee\vs\nin
This corresponds to the last term in the Feynman rule shown in
the figure below.
All possible 6 contractions result in the Feynman rule
shown there. 
\vs
\begin{figure}[h]
\includegraphics[width=0.3\textwidth]{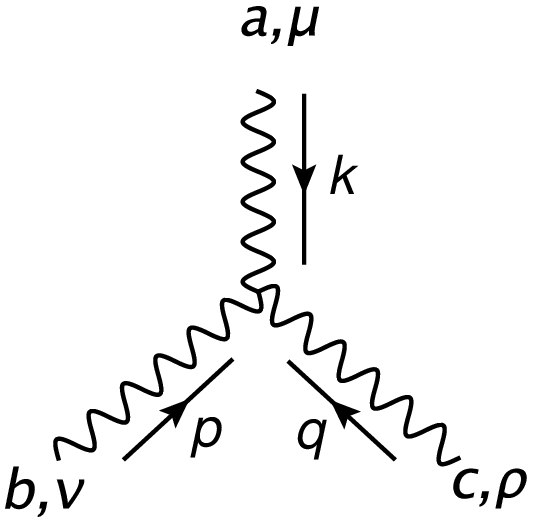}
\end{figure}
\nin
\vspace*{-1.25in}
\be
\qquad\qquad\qquad = \qquad  g\,f^{abc} \left[ g^{\mu\nu}(k-p)^\rho +
    g^{\nu\rho}(p-q)^\mu + g^{\rho\mu}(q-k)^\nu\right]\nonumber
\ee
\vs\vs\vs\vs\vs\vs\vs\vs\vs\vs\vs\vs\vs\vs\vs\vs\vs
\nin
Finally, we derive the Feynman rule for the quartic interaction from
(\ref{lint4g}).
coming from the product of the last term in
$G_{\mu\nu}^a$ with the similar term in $G^{a\mu\nu}$. This is given
by
\vs
\begin{figure}[h]
\includegraphics[width=0.3\textwidth]{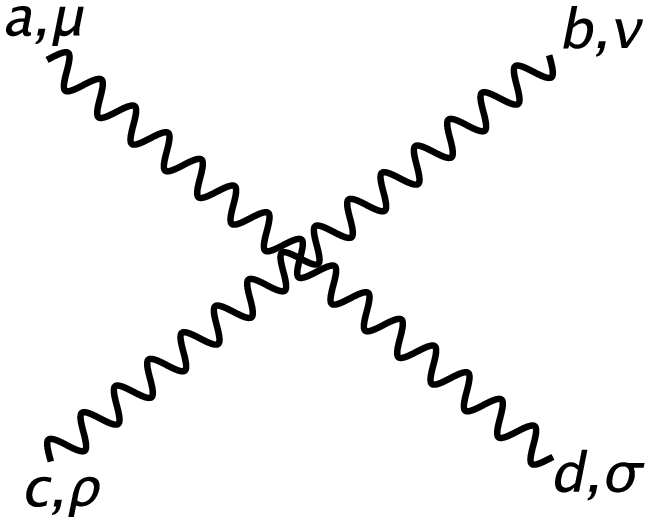}
\end{figure}
\nin
\vspace*{-1.3in}
\bear
\qquad\qquad\qquad= \qquad && -i g^2\,\left[  f^{abe} f^{cde}\left(
    g^{\mu\rho}g^{\nu\sigma}-g^{\mu\sigma}g^{\nu\rho}\right)
  \right.\nonumber\\
& +& \left. f^{ace} f^{bde}\left(
  g^{\mu\nu}g^{\rho\sigma}-g^{\mu\sigma}g^{\nu\rho}\right) \right.\nonumber\\
 &+& \left. f^{ade} f^{bce}\left(
  g^{\mu\nu}g^{\rho\sigma}-g^{\mu\rho}g^{\nu\sigma}\right) \right]\nonumber
\eear
\vs\vs\vs\vs\vs\vs\vs\vs\vs\vs\vs\vs\vs
\nin
Notice that, although this last Feynman rule starts at order $g^2$, it cannot be
considered of a higher order in perturbation theory than the other
two. What matter is the computation of the amplitude o a given process
to the desired order in $g$. For instance, if we wish to compute the
leading order contributions to the scattering of two gauge bosons
going to two gauge bosons, we see that  the second Feynman rule can be
used to form contributions with two vertices and one gauge boson
propagator. These are of order $g^2$. On the other hand, the last
Feynman rule is a contribution to the amplitude in and on itself. So
all the leading order contributions to this process are of the same order, $g^2$.

\section{The electroweak Standard Model}
\label{sec:lecture2}

The standard model (SM) of particle physics is first and foremost a
gauge theory. It is described by the product of three groups,
$SU(2)\times SU(2)\times U(1)$. Two of them non-abelian and one
abelian. Most commonly this is written as
\begin{equation}
SU(3)_c \times SU(2)_L\times U(1)_Y~,\label{smgaugegroup}
\end{equation}
where the subscript $c$ in the first factor stands for ``color'', the
$L$ in the second stands for ``left'' and the $Y$ in the third factor
refers to hypercharge. The group $SU(3)_c$ describes the interactions
of quarks with the gauge fields called {\em gluons}. These are the
degrees of freedom and interactions relevant at energies above the
$O(1)$~GeV scale, where the theory of the strong interactions is 
quantum chromodynamics (QCD). This theory and its applications in
various topics in particle physics are the subject of the lectures by
Giulia Zanderighi~\cite{giulia}. Here, we concentrate on the other two factors in
(\ref{smgaugegroup}),
\begin{equation}
SU(2)_L\times U(1)_Y~,\label{electroweak}
\end{equation}
which we call the electroweak standard model (EWSM). This will be the
subject of the rest of these lectures.

The EWSM is built from experimental observations, coupled to our
understanding of gauge theories. All SM fermions transform under the 
gauge theory in (\ref{electroweak}). In the next section we briefly
review how is it that we know this.

\subsection{Building the electroweak Standard Model}
\label{sec:building}

Let us review the main evidences leading to the gauge structure of the
electroweak theory.
\begin{itemize}
\item \underline{Weak Interactions (Charged)}: Weak decays, such as
  $\beta$ decays $n\to p\,e^- \bar\nu_e$ or
  $\mu^-\to\nu_\mu\,\bar\nu_e\,e^-$ among many others, are mediated by
  {\em charged} currents. Let us look at the case of muon decay. It is
  very well described by a four fermion interaction, i.e. with a  non
  renormalizable coupling $G_F$, the Fermi constant. In fact, all
  other weak interactions can be described in this way with the same
  Fermi constant ( to a very good approximation, more later). The
  relevant Fermi Lagrangian is 
  \vs\be
\cL_{\rm Fermi} = -4\,\frac{G_F}{\sqrt{2}}\,\bigl(\bar\mu_L\gamma_\mu
  \nu_L\bigr)\,\bigl(\bar e_L \gamma^\mu \nu_e\bigr)~,
  \label{fermi1}
  \ee\vs\nin
  where we already included the fact that the charged weak interactions
  only involve {\em left handed fermions}. That is, the
  phenomenologically built Fermi Lagrangian above tells us that the
  weak decay of a muon is described by the product of two {\em charged
    vector currents} coupling only left handed fermions. 
  The fact that only left handed
  fermions participate in the charged weak interactions is an
  experimentally established fact, observed in {\em all charged weak
    interactions}. This is done by a variety of experimental
  techniques. For instance,  in the case of muon decay, the angular
  distribution of the outgoing electron is very different if this is
  left or right handed. Precise measurements (performed over decades
  of increasingly accurate experiments) have concluded that the
  outgoing electron is left handed only. The different couplings
  involving left  and right handed fermions require {\em parity
    violation}. Moreover, the charged weak interactions require {\em
    maximal parity violation}: only one handedness participate.  
Now, if we  assume that the non renormalizable four fermion interaction is the
  result of integrating out a gauge boson with a renormalizable interaction, this would point to the
  need of $2$ charged gauge bosons.This is schematically shown in
  Figure~\ref{fig:203}. Assuming that $m_\mu \ll M_W$, we integrate
  out the massive vector gauge boson to obtain
  \vs\be
\frac{G_F}{\sqrt{2}} = \frac{g^2}{8 M_W^2}~,
  \ee\vs\nin
  where $g$ is the renormalizable coupling of the gauge bosons to
  fermions in diagram $(b)$.  The charged vector gauge bosons, $W^\pm$ were
  discovered in the 1980s and studied with great detail ever since.

 \vs
\begin{figure}
  \begin{center}
    \includegraphics{Figures/fig_20-3.eps}
\end{center}
\caption{Diagram  $(a)$ is the Feyman diagrams associated with the four
  fermion Fermi Lagrangian (\ref{fermi1}). Diagram $(b)$ shows the
  corresponding exchange of a massive charged gauge boson, $W_\mu^\pm$. }
\label{fig:203}
\end{figure}
\nin

\item \underline{Weak Neutral Currents}: In addition to the charged
  currents described by (\ref{fermi1}), we have known since
  experimental evidence first appeared in the 1970s, that there are
  also {\em  weak neutral currents}. These were first observed by
  neutrino scattering off nucleons. Normally, the charged currents
  would result in $\nu_e\,N \to e^- \,N'$, with $N$ and $N'$ protons
  and neutrons. This is just a crossed diagram of $\beta$ decay. But
  the reaction $\nu\,N\to \nu\,N$ was also observed. Many other
  reactions involving neutral currents have been observed since
  then. They also violate parity. However, they do not do so
  maximally. This means that the neutral currents, or the vector gauge
  boson that we need to integrate out to obtain them at low energies,
  couple differently to left and right handed fermions but, unlike the
  charged currents, the do couple to right handed fermions.
  The neutral vector gauge boson, $Z^0$, was also discovered in the
  1980s and its properties studied with great precision. 

  \item \underline{Electromagnetism}: Of course, we know that the
    electromagnetic interactions are described by a quantum field
    theory, QED, mediated by a neutral {\em massless} vector gauge
    boson, the photon. One important feature to remember is that the
    photon coupling in QED is {\em parity invariant}. No parity
    violation is present in QED.

  \end{itemize}

  The elements described above suggest that we need: $4$ gauge bosons
  for a unified description of the weak and electromagnetic
  interactions. Three of them appear to be massive: the $W^\pm$ and the
  $Z^0$. One, the photon, must
  remain massless. The SM gauge
  group is then $G=SU(2)\times U(1)$ which matches the number of gauge
  bosons. However, we know that two of these only couple to left
  handed fermions. whereas the massive neutral one 
  couples differently to left and right handed fermions. Finally, the
  photon must remain massless and its couplings parity invariant. The
  choice of gauge group is then
  \vs\be\boxed{
  G = SU(2)_L\times U(1)_Y~},
  \label{smgroup}
  \ee\vs\nin
  where the three gauge bosons couple to left handed fermions only,
  and the $U(1)_Y$ is {\em not identified} with the $U(1)_{\rm EM}$, the
  abelian gauge symmetry responsible for electromagnetism. As we will
  see below, two of the $SU(2)_L$ gauge bosons will result in the
  $W^\pm_\mu$. On the other hand to obtain the $Z^0$.

\subsection{The electroweak gauge theory}
\label{sec:ewgaugetheory}

The EWSM is a {\em chiral gauge theory}. As we discussed in the
previous section, this means that in general the gauge fields do not
couple equally to left and righ handed fermion chiralities.  
The fact that the gauge group is  $SU(2)_L\times U(1)_Y$ tells us the
transformation properties of left and right handed fermions
under a given gauge transformation. For instance, the left handed
fermion fields transform as
\begin{equation}
\psi_L(x)\to e^{i\alpha^a(x) \frac{\sigma^a}{2}} \,e^{i\beta(x) Y_{\psi_L}}\,\psi_L(x)~,\label{lhgaugetrans}
\end{equation}
where $\sigma^a$ ($a=1,2,3$) are the Pauli matrices, which are twice
the generators of $SU(2)$, and $Y_{\psi_L}$ is the {\em hypercharge}
of the fermion $\psi_L$. Here, $\alpha^a(x)$ is the arbitrary gauge
parameter corresponding to an $SU(2)_L$ transformation (one per
generator $\sigma^a/2$), whereas $\beta(x)$ is the arbitrary gauge
parameter corresponding to the $U(1)_Y$ gauge transformation, both acting
on the  left handed fermion. On the other hand, a right handed fermion
would transform as
\begin{equation}
 \psi_R(x)\to ,e^{i\beta(x) Y_{\psi_R}}\,\psi_R(x)~,\label{rhgaugetrans}
\end{equation}
where $Y_{\psi_R}$ is the right handed fermion hypercharge. As we
discussed in the previous lecture, for each generator in a gauge group
there is a gauge parameter {\em function}. The EWSM gauge group has
four generators so the gauge transformations introduce the four
functions of spacetime $\alpha^1(x)$, $\alpha^2(x)$, $\alpha^3(x)$ and
$\beta(x)$. This means that we need to introduce four gauge bosons for
the theory to be invariant under local $SU(2)_L\times U(1)_Y$
transformations. Then, the covariant derivative acting on left handed
fermion fields is given by
\begin{equation}
D_\mu\psi_L(x) = \bigl(\del_\mu -igA^a_\mu t^a -i g' Y_{\psi_L} B_\mu
\bigr) \psi_L(x)~,
\label{lhfcd1}
\end{equation}
where $A^a_\mu(x)$ are the three $SU(2)_L$ gauge bosons, $B(x)$ is the
$U(1)_Y$ hypercharge gauge boson, and $g$ and $g'$ are the
corresponding (dimensionless) gauge couplings.  On the
other hand, since right handed fermions do not feel the $SU(2)_L$
interaction, their covariant derivative is given by
\vs\be
D_\mu\psi_R(x) = \bigl(\del_\mu -i g' Y_{\psi_R} B_\mu
\bigr) \psi_R(x)~,
\label{rhfcd}
\ee\vs\nin
with $Y_{\psi_R}$ its hypercharge.

Next, we have to see how to accommodate all the SM fermions in {\em
  representations} of $SU(2)_L\times U(1)_Y$. Starting with left
handed fermions, since they transform under $SU(2)_L$ they must carry
a non-abelian gauge group index. We can see this from the expression
for the covariant derivative in (\ref{lhfcd1}): the covariant
derivative here must be a $2\times 2$ matrix since one of the terms is
an $SU(2)$ generator. The other two terms must be thought of as
implicitly  multiplied by the identity matrix. I.e. writing
explicitly the $SU(2)_L$ indices, we have
\begin{equation}
(D_\mu)_{ij}\psi_j(x)~,
\end{equation}  
where $j=1,2$. Thus, left handed fermions are {\em doublets} of
$SU(2)_L$. In the SM there are two types of left handed doublets:
lepton and quark doublets. For instance, for the first generation
these are
\begin{equation}
L=\left(\ba{c} \nu_{e L}\\ e^-_L\ea\right)~, \qquad\qquad Q=\left(\ba{c} u_L
  \\ d_L\ea\right)~,
\end{equation}
and similarly for the second and third generations. Notice that the
$SU(2)_L$ covariant derivative in (ref{lhcd1}) is applied to the
doublets $L(x)$ and $Q(x)$ as a whole. This means that the
hypercharges quantum numbers $Y_L$ and $Y_Q$ apply to the doublets,
not just the individual components. For instance, in $D_\mu L(x)$, the
hypercharge {\em matrix} acting on $L(x)$ is 
\begin{equation}
\left(\ba{cc} Y_L &0\\0& Y_L\ea\right)~.
\end{equation}  
Moving on to the right handed fermions, since they are {\em singlets}
under $SU(2)_L$ (they only  transform under $U(1)_Y$), they just have
their own hypercharge assignment. For instance, $e^-_R$ has
hypercharge $Y_{e^-_R}$, $u_R$ has $Y_{u_R}$, etc.

Now that we know how to accommodate fermions in representations of the
EW gauge group $SU(2)_L\times U(1)_Y$ we can address a problem of the
electroweak gauge theory: masses. We know that fermions have
masses. If we write the mass term of a generic fermion of mass $m$  this is
\begin{equation}
  m\bar\psi\psi = m \bar\psi_L\psi_R + {\rm h.c.}~,
  \label{massterm}
\end{equation}  
where $h.c.$ stands for ``hermitian conjugate.
But if we subject the mass term to an $SU(2)_L\times U(1)_Y$ gauge
transformations in (\ref{lhgaugetrans}) and (\ref{rhgaugetrans})
\begin{equation}
  \bar\psi_L\psi_R \to \bar\psi_L \,e^{-i\alpha^a(x)t^a}\,e^{-i\beta(x)Y_{\psi_L}}\,e^{i\beta(x)Y_{\psi_R}}\psi_R\not=\bar\psi_L\psi_R~,
\end{equation}
we see that it is not invariant. The $\bar\psi_L$ transformation
is not balanced since $\psi_R$ does not transforms under $SU(2)_L$,
and also $Y_{\psi_L}\not= Y_{\psi_R}$. So we conclude that fermion
masses are forbidden by EW gauge invariance.

Nest, we can consider the electroweak gauge boson sector.  The kinetic
terms for the $SU(2)_L$ and $U(1)_Y$ gauge boson fields are
\begin{equation}
{\cal L}_{\rm GB} = -\frac{1}{4} F^a_{\mu\nu} F^{a\mu\nu}
-\frac{1}{4} B_{\mu\nu} B^{\mu\nu}~,
\label{lgaugebosons}
\end{equation}
where $F^a_{\mu\nu}$ and $B_{\mu\nu}$ are the $SU(2)_L$ and $U(1)_Y$
field strengths respectively. Absent in this gauge boson Lagrangian
are gauge boson mass term just as
\begin{equation}
M_B^2 B_\mu B^\mu \qquad\quad {\rm or}\qquad\quad M^2_{A^a} A^a_\mu A^{a\mu}~,
\end{equation}
are not invariant under the gauge transformations
\begin{equation}
  B_\mu(x) \to B_\mu(x) +\frac{1}{g'} \partial_\mu\beta(x)~,
\end{equation}
and
\begin{equation}
A^a_\mu(x)\,t^a \to g(x)\,\left(A^a_\mu(x)\,t^a\right)\,g^\dagger(x)
-\frac{i}{g}\,(\del_\mu g(x))\,g^\dagger(x)~,
\label{aatransf2}
\end{equation}
where in the last expression
\begin{equation}
  g(x) = e^{i\alpha^a(x) t^a}~.
\end{equation}
Thus, we arrive at the conclusion that neither fermions nor gauge
bosons can have masses in the EWSM due to gauge invariance. But we
know that all fermions and some of the EW gauge bosons are massive! 
The solution of this problem requires that we introduce a new concept:
the {\em spontaneous} breaking of a gauge symmetry.

\subsection{The origin of mass in the electroweak Standard Model}
\label{sec:mass}

To solve the problem of mass in the EWSM we need to implement the
Anderson-Brout-Englert-Higgs (ABEH) mechanism. This is what is at play when a
gauge theory like the EWSM is {\em spontaneously broken}. Then masses
are generated out of gauge invariant operators, unlike the mass terms
for fermion and gauge bosons in the previous section, which constitute
an {\em explicit} breaking of the gauge symmetry. In order to apply
the ABEH mechanism to the case of the EWSM we need to consider in
turn: 1) The Spontaneous Breaking of a {\em global} symmetry and
Goldstone's theorem  and  2) The Spontaneous Breaking of a gauge or
local symmetry, the case of the SM. We will go through these two in turn.

\subsubsection{Spontaneous breaking of a global symmetry} 

\nin
Noether's theorem tells us that for each continuous symmetry in the
Lagrangian $\cL(\phi,\del_\mu\phi)$  there is a conserved current
$J^\mu$, i.e.\footnote{Here we go back to relativistic notation and
  Minkowski space.}
\vs\be
\del_\mu J^\mu = 0~.
\label{jconserved1}
\ee\vs\nin 
We can restate this by saying that the charge associated with this symmetry
\vs\be
Q = \int d^3x \,J^0~,
\label{chargedef}
\ee\vs\nin
is conserved. This is easily checked by computing
\vs\be
\frac{d Q}{dt} = \int d^3x\,\del_0 J^0 = \int d^3x \,\grad\cdot\vf J =
\int_{S_{\infty}} d\vf s\cdot\vf  J =0~,
\label{dqdtzero} 
\ee\vs\nin
where in the last step we assume there are no sources at infinity.

\nin
Now, in the presence of a continuous symmetry, quantum states
transform under the symmetry as
\vs\be
|\psi\rangle \to e^{i\alpha Q}\,|\psi\rangle ~,
\label{psitransf1}
\ee\vs\nin 
where $\alpha$ is a real constant, i.e. a continuous parameter. In
particular, if the ground state is invariant under the symmetry this
means that
\vs\be
|0\rangle \to e^{i\alpha Q} |0\rangle = |0\rangle~,
\label{gsinvariant}
\ee\vs\nin
with the last equality implying 
\vs\be
Q |0\rangle =0~.
\label{qkillvacuum}
\ee\vs\nin
In other words, if the ground state is invariant under a continuous
symmetry the associated charge $Q$ annihilates it. This is the normal
realization of a symmetry. 

\nin
But if 
\vs\be
Q|0\rangle \not=0~,
\label{gsnotinvariant}
\ee\vs\nin
then this means that 
\vs\be
|0\rangle \to e^{i\alpha Q} |0\rangle  \equiv |\alpha\rangle\not=|0\rangle~,
\label{alphadef1}
\ee\vs\nin
where we defined the states $|\alpha\rangle$ by the continuous 
parameter of the transformation connecting it to the ground state. 
In general, this is the situation when a symmetry is broken. But it is
possible to have (\ref{gsnotinvariant}) and still have a conserved
charge. In other words to have 
\vs\be
\frac{dQ}{dt} = 0~.
\label{dqdtstillzero}
\ee\vs\nin
Having both (\ref{gsnotinvariant}) and (\ref{dqdtstillzero}) satisfied
at the same time corresponds to what we call spontaneous symmetry
breaking (SSB): the charge is still conserved, but the ground state is not
invariant under a symmetry transformation. 
\vs\be\boxed{
\left(Q|0\rangle \not=0, \quad \frac{dQ}{dt} =0\right)\Rightarrow
{\rm SSB}}~.
\label{thisisssb}
\ee\vs\nin
For instance, this is what happens in a ferromagnet below a critical
temperature. The free energy
\vs\be
F = E-TS~,
\label{freeenergy}
\ee\vs\nin
can be  minimized, at high temperature,  by increasing the entropy $S$. So
at high $T$ disorder rules. 
However, below a critical temperature, the free energy would be
minimized by 
minimizing $E$, which is achieved by aligning the
interacting spins, resulting in a macroscopic magnetization. This is
an ordered phase. But since the magnetization picks a direction in
space it corresponds to the spontaneous 
breaking the symmetry of the system, i.e. $O(3)$. 

\nin
Since the charge is conserved we have that $[H,Q]=0$. Then, 
given a Hamiltonian $H$ acting on a state $|\alpha\rangle$ connected
to the ground state, we can write
\vs\bear
H|\alpha\rangle &=& H e^{i\alpha Q}|0\rangle = e^{i\alpha Q} H|0\rangle
= E_0 e^{i\alpha Q}|0\rangle\nonumber\\
&=& E_0 |\alpha\rangle~.
\label{eofalpha}
\eear\vs\nin
So we conclude that (\ref{thisisssb}) results in a continuous family
of degenerate states $|\alpha\rangle$ with the same energy of the
ground state, $E_0$. Going from the ground state $|0\rangle$ to the
$|\alpha\rangle$ states costs no energy. These are the gapless states
characteristic of SSB. They are the Nambu-Goldstone modes. In a
relativistic quantum field theory they correspond to massless
particles, as we will see in the following example.

\nin
\underline{Spontaneous Breaking of a Global $U(1)$ Symmetry}

\nin
We will consider a complex scalar field, the simplest systems to
illustrate the spontaneous breaking of a global symmetry and the
appearance  of massless particles. This is the relativistic version of
the  superfluid. The Lagrangian is 
\vs\be
\cL = \frac{1}{2} \del_\mu\phi^* \del^\mu\phi
-\frac{1}{2}\mu^2\phi^*\phi -\frac{\lambda}{4}
\left(\phi^*\phi\right)^2~.
\label{cxscalar1}
\ee\vs\nin
As we well know, $\cL$ is invariant under the $U(1)$ symmetry
transformations
\vs\be
\phi(x)\to e^{i\alpha} \phi(x)~,\qquad \phi^*(x) \to e^{-i\alpha}
\phi^*(x)~,
\label{u1forphi}
\ee\vs\nin
where $\alpha$ is a real constant. Here the $U(1)$ symmetry is
equivalent (isomorphic)  to a rotation in the complex plane defined by
\vs\be
\phi(x) = \phi_1(x) + i\phi_2(x)~,\qquad \phi^*(x) = \phi_1(x)
-i\phi_2(x)~,
\label{realimg}
\ee\vs\nin
where $\phi_{1,2}(x)$ are real scalar fields. Then we see that  $U(1) \simeq O(2)$. For instance, had we started with a purely
real field $\phi(x) = \phi_1(x)$, i.e. $\phi_2(x)=0$, the $U(1)$
transformations (\ref{u1forphi}) would result in 
\vs\be
\phi(x) = \phi_1(x) \to \cos\alpha \phi_1(x) + i \sin\alpha\phi_1(x)~,
\label{rotation}
\ee\vs\nin
as illustrated in Figure~\ref{fig91} below.
\vs
\begin{figure}[h]
  \begin{center}
    \includegraphics[width=3in]{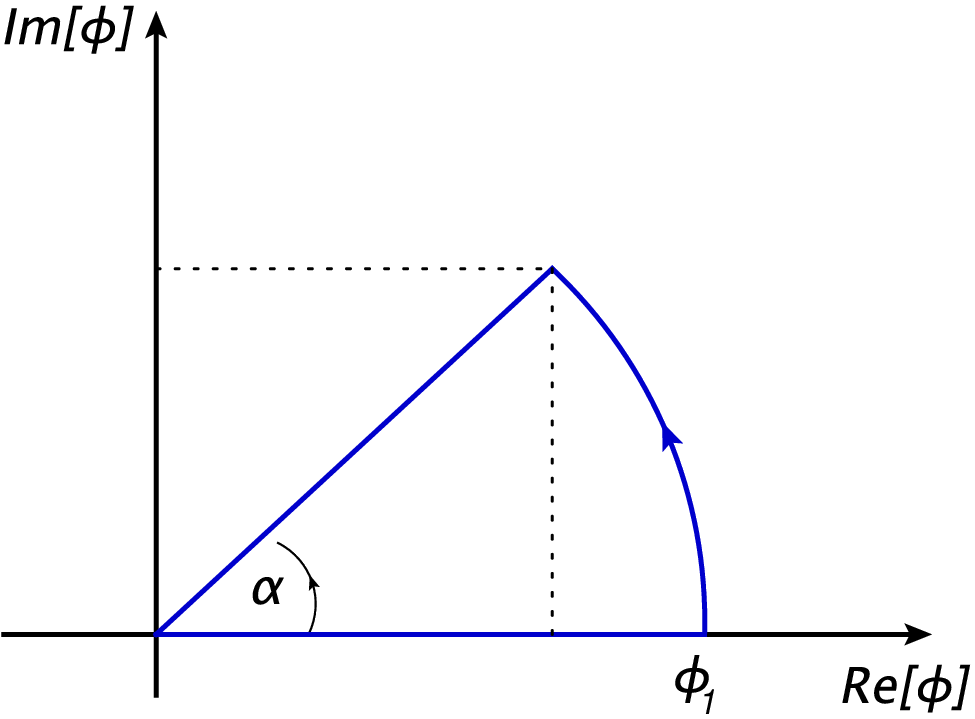}
\end{center}
\caption{
The $U(1)$ rotation $\phi\to e^{i\alpha}\phi$ for an initially real
field. 
}
\label{fig91}
\end{figure}
\nin   
We now consider the (classical) potential
\vs\be
V = \frac{1}{2}\mu^2\phi^*\phi +
\frac{\lambda}{4}\left(\phi^*\phi\right)^2~.
\label{potential1}
\ee\vs\nin    
For $\mu^2>0$ $V$ has a minimum at $(\phi^*\phi)_0=0$. On the other
hand, if $\mu^2<0$ there is a non trivial minimum for $\lambda>0$
resulting from the competition of the first and second terms in
(\ref{potential1}). Redefining 
\vs\be
\mu^2 \equiv -m^2~,
\label{mutomdef}
\ee\vs\nin
with $m^2>0$, the minimum of the potential now is
\vs\be
\left(\phi^*\phi\right)_0 = \frac{m^2}{\lambda} \equiv v^2~.
\label{vevdef1}
\ee\vs\nin
Here $v^2$ is the expectation value of the $\phi^*\phi$ operator in
the ground state, i.e.
\vs\be
\langle 0| \phi^*\phi |0\rangle =v^2~.
\label{vdef2} 
\ee\vs\nin
The potential looks just as the one for the superfluid case in the
previous lecture, shown  in Figure~8.1. The projection onto the
$(\phi_1,\phi_2)$ plane is shown in Figure~\ref{fig92} below. 
\vs
\begin{figure}[h]
\begin{center}
 \includegraphics[width=3in]{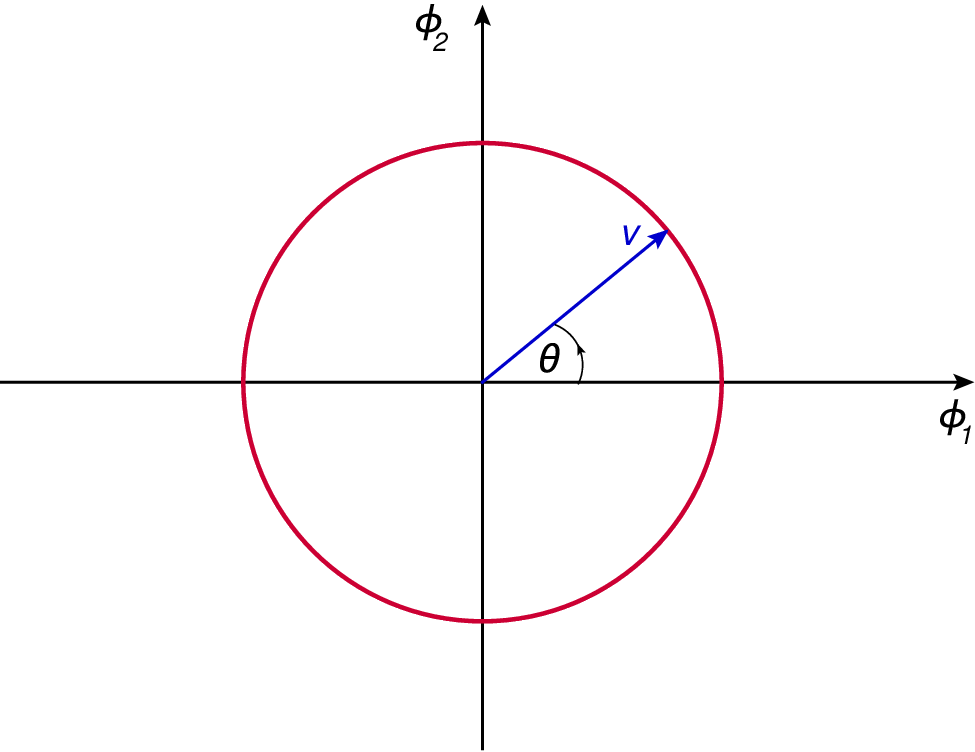}
\end{center}
\caption{
The red circle represents the locus points of the minimum of the
potential (\ref{potential1}) for $\mu^2<0$. The radius is $v$, a real
number. The phase is not determined by the minimization.  
}\label{fig92}
\end{figure}
\nin    
The radius is fixed through
\vs\be
(\phi^*\phi)_0 =v^2 = \phi_1^2 +\phi_2^2~,
\label{circle1}
\ee\vs\nin
but the phase is undetermined. We need to fix it in order to choose a
ground state to expand around. Any choice should be equivalent
\vs\bear
\langle\phi_1\rangle &=&v\qquad\qquad\langle\phi_2\rangle =0\nonumber\\
\langle\phi_1\rangle &=&\frac{v}{\sqrt{2}}  \qquad\quad\langle\phi_2\rangle
  =\frac{v}{\sqrt{2}}\nonumber\\
\vdots &~&\qquad\qquad \vdots \nonumber\\
\langle\phi_1\rangle& =&0 \qquad\qquad\langle\phi_2\rangle =v\nonumber~.
\eear\vs\nin
This particular choice is what constitutes spontaneous symmetry
breaking. We need to fix the phase $\theta=\theta_0$ arbitrarily in
order to expand around {\em this}  ground state. For instance, let us
choose the first line above, i.e. $\langle\phi_1\rangle =v$, and
$\langle\phi_2\rangle=0$. This allows us to expand the field $\phi(x)$
around this ground state as
\vs\be
\phi(x) = v + \eta(x) + i\xi(x)~,
\label{expansion1}
\ee\vs\nin
where $\eta(x)$ and $\xi(x)$ are {\em real} scalar fields statisfying
\vs\be
\langle 0 | \eta(x)|0\rangle =0,\qquad\qquad \langle 0|
\xi(x)|0\rangle =0~.
\label{zerovevs}
\ee\vs\nin
This obviously corresponds to $\phi_1(x)=v+i\eta(x)$ and
$\phi_2(x)=\xi(x)$. We can now rewrite the Lagrangian (\ref{cxscalar1})
in terms of $\eta(x)$ and $\xi(x)$. This is
\vs\bear
\cL &=& \frac{1}{2}\del_\mu\eta\del^\mu\eta
  +\frac{1}{2}\del_\mu\xi\del^\mu\xi +\frac{1}{2}m^2
  \left(v+\eta-i\xi\right)\left(v+\eta+i\xi\right) \nonumber\\
&~&-\frac{\lambda}{4}\left[\left(v+\eta-i\xi\right)\left(v+\eta+i\xi\right)\right]^2~,
\label{linetaxi1}
\eear\vs\nin
where we used  (\ref{mutomdef}).  Using (\ref{vevdef1}) and focusing
on the terms quadratic in the fields, we obtain
\vs\be
\cL = \frac{1}{2}\del_\mu\eta\del^\mu\eta
  +\frac{1}{2}\del_\mu\xi\del^\mu\xi -m^2\eta^2 + {\rm
    ~~interactions}~.
\label{noximass1}
\ee\vs\nin
So we see that when we expand around the ground state defined by
(\ref{expansion1}) we end up with a theory of a real scalar field with
mass ($\eta$) and a massless state $\xi$. That is 
\vs\be
m_{\eta} = \sqrt{2} m,\qquad\qquad m_\xi =0~.
\ee\vs\nin
This result is a reflection of Goldstone's theorem: a spontaneously
broken continuous symmetry, here a $U(1)$, results in massless
states. Notice that the result would be exactly the same had we chosen
any other angle in Figure~\ref{fig92} instead of $\theta=0$. One
simple way to check this is to use a different parametrization of
$\phi(x)$. We write
\vs\be
\phi(x) \equiv \left[v + h(x)\right] e^{i\pi(x)} ~,
\label{radial1}
\ee\vs\nin 
where $h(x)$ and $\pi(x)$ are real scalar fields, also satisfying 
\vs\be
\langle 0|h(x)|0\rangle = 0,\qquad\qquad \langle 0|\pi(x)|0\rangle =0~.
\ee\vs\nin
Then from (\ref{radial1}) it is pretty obvious that $\pi(x)$ does not
enter in the potential, and therefore will not have a mass term. It is
very simple to obtain the Lagrangian (\ref{cxscalar1}) in terms of
$h(x)$ and $\pi(x)$ using (\ref{radial1}). This is 
\vs\be
\cL =\frac{1}{2} \del_\mu h\del^\mu h +
\frac{1}{2}\del_\mu\pi\del^\mu\pi -m^2 h^2 + {\rm ~~interactions}~,
\label{nopimass1}
\ee\vs\nin
which is exactly the same theory as the one in (\ref{noximass1}),
i.e. a massive state with $m_h=\sqrt{2}m$ and a massless particle,
here the $\pi(x)$.

To understand more intuitively the appearance of the massless state it
is helpful to look at the possible excitations of the potential, as
illustrated in Figure~\ref{hpotential}.
\begin{figure}[h]
\begin{center}
 \includegraphics[width=3in]{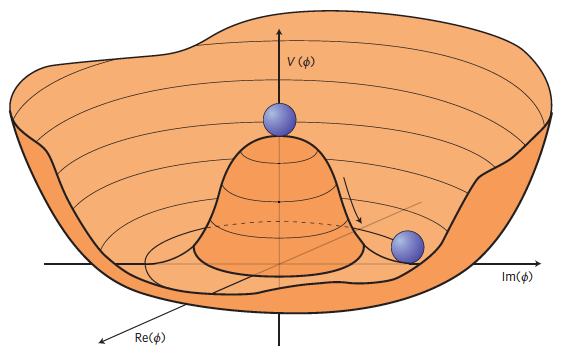}
\end{center}
\caption{
  The scalar potential. There are two types of independent excitations
  about the minimum: the radial excitation implies a cost of energy
  since results in a larger value of $V(\phi)$ than the minimum. The
  excitation along the circle cost no energy and so it corresponds to
  a massless state.
}\label{hpotential}
\end{figure}

\nin
We can see that in order to obtain the particle states we must expand
about the minimum of the potential. But there are two independent
(orthogonal)  directions we can choose. If the expand in the radial
direction, no matter how small the fluctuation it will cost
energy. This fluctuation corresponds to the massive field $h(x)$. On
the other hand, if we expand about the minimum along the circle, this
has no energy cost since all the points in the circle have the same
energy as the minimum we picked arbitrarily. This is the massless
fluctuation $\pi(x)$, the Nambu-Godstone bosons.

We will later see a derivation of Goldsone's theorem that is more
geared towards quantum field theory. We will see that there will be a
NGB for each {\em broken } symmetry generator, i.e. for each spontaneously
broken symmetry.

\subsubsection{Spontaneous breaking of a gauge symmetry}
\label{sec:ssbgauge}

\nin
We have seen that the spontaneous breaking of a continuous symmetry
results in the presence of massless states in the spectrum, the
Nambu--Goldstone Bosons (NGB). We have seen this in particular for a
$U(1)$ global symmetry where the potential was such that the ground
state was not $U(1)$ invariant. In that case, the NGB corresponded to
the degeneracy of the ground state, i.e. it was the fluctuation going
around the degenerate minimum and as such it corresponded to a
massless state. We will see later that this picture generalizes for
non-abelian global continuous symmetries so that the number of NGBs
corresponds to the number of degenerate directions in group space,
i.e. the number of broken generators. 

Before we go into non-abelian symmetries, we 
will consider the situation when the $U(1)$ symmetry studied earlier
is gauged. That is, is a local $U(1)$ symmetry such as for example in
QED. As we will soon see, the consequences for the spectrum when the
spontaneously broken symmetry is gauged are drastic.
We start with the Lagrangian of a scalar field charged under a gauged
$U(1)$ symmetry just as QED. This is given by
\vs\be
\cL = \frac{1}{2} (D_\mu\phi)^* D^\mu\phi -V(\phi^*\phi) -\frac{1}{4}
F_{\mu\nu}F^{\mu\nu}~,
\label{lag2} 
\ee\vs\nin 
where the covariant derivative is defined by
\vs\be
D_\mu \phi = (\del_\mu +ieA_\mu)\phi~,
\label{covderdef3}
\ee\vs\nin
and the scalar and gauge field transformations under the $U(1)$ gauge
symmetry are 
\vs\bear
\phi(x) &\to & e^{i\alpha(x)} \phi(x) \nonumber\\
&~&\label{fieldtransf1}\\
A_\mu(x) &\to & A_\mu(x) -\frac{1}{e}\del_\mu\alpha(x)~.\nonumber
\eear\vs\nin
Finally, the gauge field $A_\mu(x)$ has a kinetic term given by the
square of the  gauge invariant field strength as usual
\vs\be
F_{\mu\nu} = \del_\mu A_\nu - \del_\nu A_\mu~.
\label{fmunudef2}
\ee\vs\nin 
With (\ref{covderdef3}), (\ref{fieldtransf1}) and (\ref{fmunudef2}) the
Lagrangian in (\ref{lag2}) is clearly gauge invariant. 

\nin
In order to implement spontaneous breaking we choose the potential as
\vs\be
V(\phi^*\phi) = \frac{1}{2}\mu^2\phi^*\phi +
\frac{\lambda}{4}\left(\phi^*\phi\right)^2~,
\label{pot1}
\ee\vs\nin
which is the same form we used for the breaking fo the global $U(1)$
and corresponds to the only renormalizable terms allowed by the
symmetry in four spacetime dimensions. What follows next pertaining
the minimum of the potential is identical to what we saw for the
global symmetry case. If $\mu^2>0$ the minimum of $V$ in (\ref{pot1})
is $\phi=0$. However if $\mu^2<0$ then we rewrite the potential as
\vs\be
V(\phi^*\phi) = -\frac{1}{2}m^2\phi^*\phi +
\frac{\lambda}{4}\left(\phi^*\phi\right)^2~,
\label{pot2}
\ee\vs\nin
where we have defined the positive constant $m^2=-\mu^2$. As before,
in this case the minimum is now given by the solution of 
\vs\be
-\frac{1}{2}m^2 +\frac{\lambda}{2} (\phi^*\phi)_0 =0~,
\label{mineq}
\ee\vs\nin
which results in 
\vs\be
(\phi^*\phi)_0 = \langle 0|\phi^*\phi|0\rangle =
\frac{m^2}{\lambda}\equiv v^2~.
\label{vevsquared}
\ee\vs\nin
Choosing the value of the field to be real at the minimum, we use the
expansion 
\vs\be
\phi(x) = v + \eta(x) + i\xi(x)~,
\label{phiexp1}
\ee\vs\nin
such that the physical real fields satisfy 
\vs\be
\langle 0| \eta(x)|0\rangle = \langle 0|\xi(x)|0\rangle =0~.
\ee\vs\nin
Just as we expect, writing the potential in terms of $\eta(x)$ and
$\xi(x)$ 
\vs\be
V(\phi^*\phi) = V((v^2+\eta(x)^2) + \xi(x)^2)~,
\ee\vs\nin
allows us to identify the spectrum which is given by
\vs\bear
m_\eta &=& \sqrt{2} m = \sqrt{2\lambda} v\nonumber\\
& ~&\label{spectrum}\\
m_\xi&=& 0~.\nonumber
\eear\vs\nin
Thus, we identify $\xi(x)$ with the massless NGB. The difference with
respect to the SSB of a global $U(1)$ comes in when we look at what
happens in the scalar kinetic term. 
This is 
\vs\bear
\frac{1}{2} (D_\mu\phi)^*D^\mu\phi &=&
\frac{1}{2}\del_\mu\eta\del^\mu\eta +
\frac{1}{2}\del_\mu\xi\del^\mu\xi +\frac{1}{2} \,e^2\,v^2\,A_\mu A^\mu
\nonumber\\
& ~& \label{kinscal1}\\
&&+ \,e\, v\, A_\mu\,\del^\mu\xi + \cdots~,
\nonumber
\eear\vs\nin
where we have explicitly written the terms quadratic in the fields,
and the dots denote interactions terms that are cubic or quadratic in
them. Besides the kinetic terms for $\eta(x)$ and $\xi(x)$ we notice
two terms. The first one is an apparent gauge boson mass term. It
implies that the gauge boson has acquired a mass given by
\vs\be
m_A = e\,v~.
\label{amass1}
\ee\vs\nin
However, this does not mean that the gauge symmetry is not been
respected. In fact, all we have done with respect to the (\ref{lag2} )
is to expand the theory around the ground state in terms of fields
that have zero expectation values there. In other words, we just
performed a change of variables. However, the fact the we are
expanding the theory around a minimum that {\em does not}  respect the
symmetry is resulting in a mass for the gauge boson. This means that
the gauge symmetry has been {\em spontaneously} broken. But since we
have not added any terms that violated explicitly the $U(1)$ gauge
symmetry, the symmetry {\em has not} been {\em explicitly} broken and
therefore currents and charges must still be conserved. 

\nin
The second notable aspect in (\ref{kinscal1}) is the term mixing the
gauge boson with the $\xi(x)$ field, the would-be NGB. Having a term
like this, i.e. non-diagonal two-point function, implies that we have
to include a Feynman diagram as the one in
Figure~\ref{fig131}. Although in principle there is no problem with
having a non-diagonal Feynman rule such as this as long as we always
remember to include it, it is interesting to see how to diagonalize it
and what are the consequences of doing that. The idea is to choose a
gauge for $A_\mu(x)$ such that we can cancel this term once we go to
the new gauge. The theory has to be physically equivalent to the one
with (\ref{kinscal1}). Choosing a specific gauge corresponds to
choosing a scalar function $\alpha(x)$ in the gauge transformations
(\ref{fieldtransf1}). In particular, if we choose
\vs
\begin{figure}
\begin{center}
\includegraphics[width=4in]{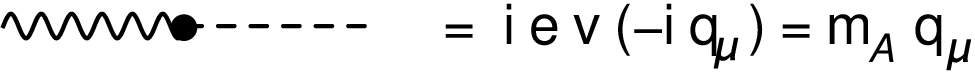}
\end{center}
\caption{
Feynman rule for the non-diagonal contribution to the two-point
function in (\ref{kinscal1}).
}
\label{fig131}
\end{figure}
\vs\nin    

\vs\be
\alpha(x) = -\frac{1}{v}\,\xi(x)~,
\label{alphachoice}
\ee\vs\nin
we then have the gauge transformation
\vs\be
A_\mu(x) \to A'_\mu(x) = A_\mu(x) +\frac{1}{e v} \xi(x)~. 
\label{achoice}
\ee\vs\nin
Replacing $A_\mu(x)$ in terms of $A'_\mu(x)$ and $\xi(x)$ in
(\ref{kinscal1}) we have
\vs\bear
\frac{1}{2}(D_\mu\phi)^*D^\mu\phi &=& \frac{1}{2}\del_\mu\eta\del^\mu\eta +
\frac{1}{2}\del_\mu\xi\del^\mu\xi +
\frac{1}{2}e^2\,v^2\,\left(A'_\mu-\frac{1}{e v}\del_\mu\xi\right) \left(A'^\mu-\frac{1}{e v}\del^\mu\xi\right)
\nonumber\\
&~&\label{kinscal2}\\
&&+ \, e\,v\,\left(A'_\mu-\frac{1}{e v}\del_\mu\xi\right) \del^\mu\xi
+ \cdots~,\nonumber, 
\eear\vs\nin
Carefully collecting all the terms in (\ref{kinscal2}) we arrive at
the surprisingly simple expression for the scalar kinetic term:
\vs\be
\frac{1}{2}(D_\mu\phi)^*D^\mu\phi =
\frac{1}{2}\del_\mu\eta\del^\mu\eta + \frac{1}{2}\,e^2\,v^2\,A'_\mu
A'^\mu + \cdots~.
\label{kinscal3}
\ee\vs\nin
We see that the gauge boson mass term is still the same as
before. However, the $\xi(x)$ field, the massless field that we thought
would be the NGB is now gone. Its kinetic term is gone and, as we will
see later, no term with $\xi(x)$ remains in the Lagrangian after this
gauge transformation. So the would-be NGB is not! When a degree of
freedom disappears from the theory just by performing a gauge
transformation, we say that this is not a physical degree of
freedom. This particular gauge without the NGB $\xi(x)$ is called the
{\em unitary gauge}, since it exposes the actual degrees of freedom of
the theory: a real scalar field $\eta(x)$ with mass $m_{\eta} =
\sqrt{2} m$ and a gauge boson with mass $m_A =e v$. In fact if we count
degrees of freedom before and after we expanded around the non-trivial
ground state, we see that before we had {\em two real scalar fields},
and {\em two degrees of freedom } corresponding to the two helicities
of a massless gauge boson, for a total of {\em four degrees of
  freedom}. But after we expanded around the ground state, we have
{\em one real scalar field}, plus {\em three polarizations} for the
now massive gauge boson, again a total of {\em four degrees of freedom}.   
It is in this sense that sometimes we say that when a gauge symmetry
is spontaneously broken, the NGB is {\em ``eaten''} by the gauge boson
to become its longitudinal polarization. This statement can be made
more precise through the {\em equivalence theorem}, which says that in
processes at energies much larger than $v$ (so that it does not matter
that the expectation value of the field is not zero in the ground
state) computing any observable by using the theory with a massive
gauge boson should yield the same result as using the theory with a
massless gauge boson and a massless NGB, up to corrections that go
like $v^2/E^2$, where $E$ is the characteristic energy scale of the
process in question. We will come back to  the equivalence theorem
later on when we consider the spontaneous breaking of non-abelian
gauge symmetries.

\nin
There is another, perhaps more direct, way to see that the NGB can be
{\em gauged away}, i.e. it disappears from the theory by performing a
gauge transformation. For this purpose, it is advantageous to
parameterize the scalar field not in terms of  real and imaginary
parts, but of modulus and phase. We write
\vs\be
\phi(x) = e^{i\pi(x)/f}\,\left(v+\sigma(x)\right)~,
\label{expform1}
\ee\vs\nin
where we see that this automatically satisfies (\ref{vevsquared}). We
have two real scalar fields, just as before. One is the modulus field
$\sigma(x)$ and the other one is the phase field $\pi(x)$. The scale
$f$ is defined so that the argument of the exponent is
dimensionless. To fix $f$ we demand that the $\pi(x)$ field has a
canonically normalized kinetic term, i.e. we impose 
it be
\vs\be
\frac{1}{2}\del_\mu\pi\del^\mu\pi~.
\label{picanon}
\ee\vs\nin
This fixes 
\vs\be
f =v~,
\label{feqv}
\ee\vs\nin
so that we have 
\vs\be
\phi(x) = e^{i\pi(x)/v}\,\left(v+\sigma(x)\right)~,
\label{expform2}
\ee\vs\nin
instead of (\ref{phiexp1}).
From the form above, it is immediately clear that $\pi(x)$ will not
appear in the potential. In fact, this is given by
\vs\be
V(\phi^*\phi) = -\frac{m^2}{2}\left[v+\sigma(x)\right]^2 +
\frac{\lambda}{4}\left[v+\sigma(x)\right]^4~.
\label{nopiinv}
\ee\vs\nin
From this form above we see that $\sigma(x)$ is the massive real
scalar field with 
\vs\be
m_\sigma = \sqrt{2\lambda} \,v~,
\label{sigmamass}
\ee\vs\nin
just as before.
This also means that $\pi(x)$ cannot get a mass, i.e.
\vs\be
m_\pi = 0 ~,
\label{pimass}
\ee\vs\nin
 and therefore is the
NGB. In fact, it will only appear in the Lagrangian in derivative form
since it is the only way it will come down from the exponentials
before these annihilate in the kinetic scalar term.  

\nin
From the parameterization (\ref{expform2}) it is also obvious how to
remove $\pi(x)$ by means of a gauge transformation. Clearly, choosing
the gauge transformation
\vs\be
\phi(x) \to \phi'(x)=e^{-i\pi(x)/v}\,\phi(x)~,
\label{gtnongb}
\ee\vs\nin
results in 
\vs\be
\phi'(x) = \left[v+\sigma(x)\right]~.
\label{nongbinphi}
\ee\vs\nin
Of course, the gauge transformation (\ref{gtnongb}) is the same we
introduced earlier in (\ref{alphachoice}) only substituting $\pi(x)$
for $\xi(x)$, and it therefore results in the same transformation for
the gauge fields as in (\ref{achoice}). Therefore, our conclusions are
exactly the same as the ones we derived by using (\ref{phiexp1}) as
the field expansion: there is a massive gauge boson field with mass
$m_A=e\,v$ and a massive reals scalar with mass given by
(\ref{sigmamass}). 

\nin
We finally comment on the meaning of spontaneously breaking a gauge
symmetry. Specifically, we want to address the point that although the
gauge boson has acquired a mass, the gauge symmetry is still present. 
To show this, let us go back to the gauge where we have both the gauge
boson and the NGB. We want to compute the gauge boson two-point
function  at tree level. In particular we want to consider the effect
of spontaneous symmetry breaking. We will need to use the Feynman rule
illustrated in Figure~\ref{fig131}. The calculation is illustrated in
Figure~\ref{fig132}. 
In addition to the tree-level gauge boson propagator, there are two
new terms contributing: the gauge boson mass insertion and
the massless NGB pole. They are
\vs\bear
i\delta\Pi_{\mu\nu} &=& i m_A^2 g_{\mu\nu} + m_A q_\mu\,\frac{i}{q^2} \,m_A
(-q_\nu)\nonumber\\
&~&\label{pimunu}\\
&=& i m_A^2\left(g_{\mu\nu} -\frac{q_\mu q_\nu}{q^2}\right)~.\nonumber
\eear\vs\nin
\vs
\begin{figure}
\begin{center}
\includegraphics[width=5in]{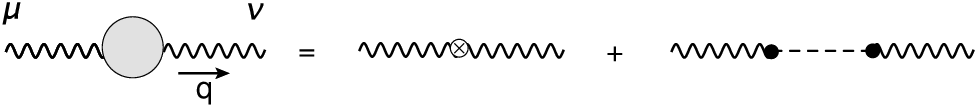}
\end{center}
\caption{
New contributions to the gauge boson two-point function at tree level
in the presence of spontaneous symmetry breaking.
The first diagram is the gauge boson mass term insertion. The second one
corresponds to the massless NGB contribution.}
\label{fig132}
\end{figure}
\vs\nin    
In the first line in (\ref{pimunu})  we used the gauge boson--NGB
mixing Feynman rule of Figure~\ref{fig131}. The result is that the new
additions to the 
two-point function result to be  actually transverse. That is, we have
that 
\vs\be
q^\mu \delta\Pi_{\mu\nu} = 0~,
\label{stilltransverse}
\ee\vs\nin
so that the two-point function remains transverse, therefore
respecting the Ward identities. Since the Ward identities are
equivalent to current conservation, we conclude that the gauge
symmetry is still preserved, even in the presence of the gauge boson
mass term. We can see that this required the presence of the NGB
pole. Just having the gauge boson mass term would have resulted in a
non-transverse contribution to the two-point function, and an explicit
violation of the gauge symmetry. So having a gauge boson mass is
compatible with gauge invariance as long as it is the result of
spontaneous symmetry breaking.

\subsubsection{Spontaneous breaking of non-abelian global symmetries}

Before we can finally go into the application of the ABEH mechanism to
the EWSM, we need to generalize the spontaneous breaking to the cases
of non-abelian symmetries, both global and gauged. 
We start with the simpler case of the global symmetry and we will restate Goldstone's
theorem in a more general way so as to include different symmetry
breaking patterns, which will result in a different number of
Nambu--Goldstone Bosons (NGBs). Then we will consider the spontaneous
breaking of non-abelian gauge symmetries, i.e. the most general
version of the ABEH mechanism.

We start with the Lagrangian for a scalar field $\phi$,
\vs\be
\cL = \del_\mu\phi^\dagger \del^\mu\phi -\frac{\mu^2}{2}
\phi^\dagger\phi -\frac{\lambda}{4}\Bigl(\phi^\dagger\phi\Bigr)^2~.
\label{lagscalar1}
\ee\vs\nin
The Lagrangian above is invariant under the transformation
\vs\be
\phi(x) \to e^{i \alpha^a\,t^a}\,\phi(x)~,
\label{globalt1}
\ee\vs\nin
where the $t^a$ are the generators of the non-abelian group $G$, and
the arbitrary parameters $\alpha^a$ are constants.
Here the scalar field $\phi(x)$ must carry a group index in order for
(\ref{globalt1}) to make sense. 
\nin
We say the  symmetry is spontaneously broken  if
we have
\vs\be
\mu^2=-m^2 < 0~,
\ee\vs\nin
then the potential has a non trivial minimum at
\vs\be
\Bigl(\phi^\dagger\phi\Bigr)_0 =\langle\phi^\dagger \phi\rangle =
\frac{m^2}{\lambda}\equiv v^2~.
\label{minimum1}
\ee\vs\nin
However, we need to ask {\em how} is the symmetry spontaneously
broken. In other words, Spontaneous Symmetry Breaking (SSB) means that
the value of the field at the minimum, let us call it the vacuum
expectation value (VEV) of the field $\langle\phi\rangle$ , is not invariant under the
symmetry transformation (\ref{globalt1}). That is,
\vs\be
\langle\phi\rangle \to e^{i\alpha^a t^a}\,\langle\phi\rangle = \Bigl(1
+ i\alpha^a t^a +\cdots\Bigr)\,\langle\phi\rangle~,
\ee\vs\nin
can be either equal to $\langle\phi\rangle$ or not. 
This tells us that if 
\vs\be
t^a\langle\phi\rangle =0~,
\label{unbroken1}
\ee\vs\nin
the ground state is invariant under the action of the symmetry ({\em
  unbroken symmetry directions }), whereas if
\vs\be
t^a\langle\phi\rangle \not=0~,
\label{broken1}
\ee\vs\nin
the ground state is not invariant ({\em broken symmetry
  directions}). We see that some of the generators will annihilate the
ground state $\langle\phi\rangle$, such as in (\ref{unbroken1}),  whereas
others will not.
In the first case, these directions in group space will correspond to
preserved or unbroken symmetries. Therefore, there should not be
massless NGBs associated with them. On the other hand, if the
situation is such as in (\ref{broken1}), then the ground state is not
invariant under the symmetry transformations {\em defined by these
  generators}. These directions in group space defined {\em broken
  directions or generators} and there should be a massless NGB
associated with each of them. Thus, as we will see in more detail
below, the number of NGB will correspond to the total number of
generators of G, minus the number of unbroken generators, i.e. the
number of {\em broken generators}.

\vskip0.5in
\nin
\underline{Example: $SU(2)$}

\nin
As a first example, let us consider the case where the symmetry
transformations are those associated with the group $G=SU(2)$. The
{\em three} generators of $SU(2)$ are
\vs\be
t^a = \frac{\sigma^a}{2}~,
\label{gensofsu2}
\ee\vs\nin
with ${\sigma^a}$ the three Pauli matrices. This means that the scalar
fields appearing in the Lagrangian (\ref{lagscalar1}) are {\em
  doublets} of $SU(2)$, i.e. we can represent them by a column vector
\vs\be
\phi(x) = \left(\ba{c}
  \phi_1(x)\\
  ~\\
  \phi_2(x)\ea\right)
\label{fundrep}~,
\ee\vs\nin
and that the symmetry transformation can be written as\footnote{We put
  the group indices in the fields upstairs for future notation
  simplicity. There is no actual meaning to them being ``up'' or
  ``down'' indices, but the summation convention still holds.}
\vs\be
\phi^i(x) =\Bigl(\delta^{ij}  + i \alpha^a
t^a_{ij}+\cdots\Bigr)\,\phi^j(x)~,
\label{globalt2}
\ee\vs\nin
where $i,j=1,2$ are the group indices for the scalar field in the
fundamental representation. We now need to {\em choose}  the vacuum
$\langle\phi\rangle$. This is typically informed by either the
physical system we want to describe or by the result we want to get. 
Let us choose
\vs\be
\langle \phi\rangle = \left(\ba{c} 0\\
  v\ea\right)~.
\label{vev1}
\ee\vs\nin
Clearly this satisfies (\ref{minimum1}). This choice corresponds to
having
\vs\bear
\langle {\rm Re}[\phi_1]\rangle &=&0\qquad\qquad\langle {\rm Im}[\phi_1]\rangle=0\nonumber\\
\langle {\rm  Re}[\phi_2]\rangle &=&v\qquad\qquad \langle
{\rm Im}[\phi_2]\rangle=0~,
\label{vevsofphis1}
\eear\vs\nin
in (\ref{fundrep}). We can now test what generators annihilate the
vacuum (\ref{vev1}) and which ones do not. We have
\vs\be
t^1\,\langle\phi\rangle = \frac{1}{2}\left(\ba{cc}
  0&1\\
  1&0\ea\right)\,\left(\ba{c}0\\
  v\ea\right) = \frac{1}{2}\,\left(\ba{c} v\\
  0\ea\right) \not = \left(\ba{c}0\\
0\ea\right) ~.
\ee\vs\nin
Similarly, we have
\vs\be
t^2\,\langle\phi\rangle = \frac{1}{2}\left(\ba{cc}
  0&-i\\
  i&0\ea\right)\,\left(\ba{c}0\\
  v\ea\right) = \frac{1}{2}\,\left(\ba{c} -i v\\
  0\ea\right) \not = \left(\ba{c}0\\
0\ea\right) ~,
\ee\vs\nin
and
\vs\be
t^3\,\langle\phi\rangle = \frac{1}{2}\left(\ba{cc}
  1&0\\
  0&-1\ea\right)\,\left(\ba{c}0\\
  v\ea\right) = \frac{1}{2}\,\left(\ba{c} 0\\
  -v\ea\right) \not = \left(\ba{c}0\\
0\ea\right) ~.
\ee\vs\nin
So we conclude that with the choice of vacuum (\ref{vev1}), all $SU2)$
generators are broken. This means that all the continuous symmetry
transformations generated by (\ref{globalt1}) change the chosen vacuum
$\langle \phi\rangle$. Thus, Goldstone's theorem predicts there must
be {\em three} massless NGBs. In order to explicitly see who are
these NGBs, we write the Lagrangian (\ref{lagscalar1}) in terms of the real
scalar degrees of freedom as in 
\vs\be
\phi(x) = \left(\ba{c} {\rm Re}[\phi_1(x)] + i \,{\rm Im}[\phi_1(x)]\\
  ~\\
  v+ {\rm Re}[\phi_2(x)] + i\, {\rm Im}[\phi_2(x)]\ea\right)~,
\label{phiexp}
\ee\vs\nin
which amounts to expanding
about the vacuum (\ref{vev1}) as long as (\ref{vevsofphis1}) is
satisfied. Substituting in (\ref{lagscalar1}) we will find that there
are three massless states, namely, ${\rm Re}[\phi_1(x)]$ , ${\rm
  Im}[\phi_1(x)]$ and ${\rm Im}[\phi_2(x)]$, and that there is a
massive state corresponding to ${\rm Re}[\phi_2(x)]$ with a mass given
by $m$. this looks very similar to what we obtain in the
abelian case, of course. Also analogously to the abelian case, we could
have parameterized $\phi(x)$ as in
\vs\be
\phi(x) = e^{i\pi^a(x) t^a/f}\,\left(\ba{c}0\\
  v+c\,\sigma(x)\ea\right)~,
\label{exppar1}
\ee\vs\nin
where $\sigma(x) $ and $\pi^a(x)$, with $a=1,2,3$ are real scalar
fields, and the scale $f$ and the constant $c$ are to be determined so
as to obtain canonically normalized kinetic terms for them in
$\cL$. In fact, choosing
\vs\be
f=\frac{v}{\sqrt{2}}, \qquad\qquad c=\frac{1}{\sqrt{2}}~,
\ee\vs\nin
we arrive at
\vs\be
\cL = \frac{1}{2}\del^\mu\sigma\del_\mu\sigma
+\frac{1}{2}\del^\mu\pi^a\del_\mu\pi^a  -\frac{m^2}{2}
\Bigl(v+\frac{\sigma(x)}{\sqrt{2}}\Bigr)^2
+\frac{\lambda}{4}\, \Bigl(v+\frac{\sigma(x)}{\sqrt{2}}\Bigr)^4~,
\label{lagsigpi1}
\ee\vs\nin
from which we see that the three  $\pi^a(x)$ fields are massless and
are therefore the NGBs. Furthermore, using $m^2=\lambda v^2$, we can
extract
\vs\be
m_\sigma=m=\lambda\,v
\ee\vs\nin
The choice of vacuum $\langle\phi\rangle$ resulting in this spectrum
could have been different. For instance, we could have chosen
\vs\be
\langle\phi\rangle = \left(\ba{c}v\\0\ea\right)~.
\ee\vs\nin
But it is easy to see that this choice is equivalent to (\ref{vev1}),
and that it would result in an identical real scalar
spectrum. Similarly,
the apparently different vacuum
\vs\be
\langle\phi\rangle = \frac{1}{\sqrt{2}}\left(\ba{c}v\\v\ea\right)~,
\ee\vs\nin
results in the same spectrum.
All these vacuum choices spontaneously break $SU(2)$
{\em completely}, i.e. there are not symmetry transformations that
respect these vacua. Below we will see an example of partial
spontaneous symmetry breaking.

\subsubsubsection{Goldstone theorem revisited}

\nin
We now can reformulate Goldstone theorem for the case of the
spontaneous breaking of the global non-abelian symmetry. 
We go back to considering the infinitesimal transformation
(\ref{globalt2}), but we rewrite it as
\vs\be
\phi^i \to \phi^i + \Delta^i(\phi)~,
\label{globalt3}
\ee\vs\nin
where we defined
\vs\be
\Delta^i(\phi) \equiv i \alpha^a\,(t^a)_{ij}\phi^j~.
\label{deltadef1}
\ee\vs\nin
If the potential has a non trivial minimum  at $\Phi^i(x)=\phi^i_0$, then it is satisfied that
\vs\be
\frac{\del V(\phi^i)}{\del\phi^i}\Bigr|_{\phi_0} =0~.
\label{der1iszero}
\ee\vs\nin
We can then expand the potential around the minimum as
\vs\be
V(\phi^i) = V(\phi^i_0)
+\frac{1}{2}\,\Bigl(\phi^i-\phi^i_0\bigr)\,\Bigl(\phi^j-\phi^j_0\Bigr)\,\frac{\del^2
  V}{\del\phi^i\del\phi^j}\Bigr|_{\phi_0}
+\cdots~,
\label{taylorexp1}
\ee\vs\nin
where the first derivative term is omitted in light of
(\ref{der1iszero}). The second derivative term in (\ref{taylorexp1})
Above defines a matrix with units of square masses:
\vs\be
M^2_{ij}\equiv \frac{\del^2
  V}{\del\phi^i\del\phi^j}\Bigr|_{\phi_0} \geq 0~.
\label{msqdef1}
\ee\vs\nin
where the last inequality results from the fact that  $\phi^0$ is a
minimum. $M^2_{ij}$ is the mass squared matrix.
We are now in the position to state Goldstone's theorem in this
context.
\vs\vs\vs\vs\nin
\underline{\bf Theorem}: \\
``For each symmetry of the Lagrangian that {\em is
  not } a symmetry of the vacuum $\phi_0$, there is a zero eigenvalue
of $M^2_{ij}$ .''

\vs\vs\vs\nin
\underline{\bf Proof}: \\
The infinitesimal symmetry transformation in (\ref{globalt3}) leaves the Lagrangian
invariant. In particular, it also leaves the potential invariant,
i.e.
\vs\be
V(\phi^i) = V\Bigl(\phi^i +\Delta^i(\phi)\Bigr)~.
\label{visinvariant}
\ee\vs\nin
Expanding the right hand side of (\ref{visinvariant}) and keeping
only terms leading in $\Delta^i(\phi)$, we can write
\vs\be
V(\phi^i) = V(\phi^i)  +
\Delta^i(\phi)\,\frac{\del V(\phi^i)}{\del\phi^i} ~,
  \label{expandv1}
\ee\vs\nin
which, to be satisfied  requires that
\vs\be
\Delta^i(\phi)\,\frac{\del V(\phi)}{\del\phi^i}=0~.
\label{vanish1}
\ee\vs\nin
To make this result useful, we take a derivative on both sides and
specified for $\phi^i=\phi^i_0$, i.e. we evaluate all the expression
at the minimum of the potential. We obtain
\vs\be
\frac{\del\Delta^i(\phi)}{\del\phi^j}\Bigr|_{\phi_0}\,\frac{\del
  V(\phi)}{\del\phi^i}\Bigr|_{\phi_0} +
    \Delta^i(\phi_0)\,\frac{\del^2
      V(\phi)}{\del\phi^j\del\phi^i}\Bigr|_{\phi_0} =0~.
 \ee\vs\nin

      But by virtue of (\ref{der1iszero}), the first term above
      vanishes, leaving us with 
      \vs\be\boxed{
\Delta^i(\phi_0)\,\,\,\frac{\del^2
  V(\phi)}{\del\phi^j\del\phi^i}\Bigr|_{\phi_0} =0}~.
\label{thisiszero}
      \ee\vs\nin
      There are two ways to satisfy (\ref{thisiszero}):

      \begin{enumerate}
      \item $\Delta^i(\phi_0) = 0$. \\
        But this means that, under a symmetry
  transformation, the vacuum is invariant, since according to
  (\ref{globalt3}) this results in
  \vs\be
\phi^i_0\to\phi^i_0~.
\ee\vs\nin

\item $\Delta^i(\phi_o)\not=0$. \\
  This requires that the second derivative factor in (\ref{thisiszero})
  must vanish, i.e.
  \vs\be
M^2_{ij}=0~.
\ee\vs\nin
We then conclude that for each symmetry transformation that {\em does not
  leave the vacuum invariant} there must be a zero eigenvalue of the
mass squared matrix $M^2_{ij}$. QED.
\end{enumerate}

\subsubsection{Spontaneous breaking of non-abelian gauge symmetries}

\nin
We will now consider the case when the spontaneously broken non
abelian symmetry is gauged. As we saw for the case of abelian gauge
symmetry, the spontaneous breaking of the symmetry will be realized in
the sense of the ABEH mechanism, i.e. the NGBs would not be
in the physical spectrum, and the gauge bosons associated with the
{\em broken } generators will acquire mass.
We will derive these results carefully in what follows.

\nin
We consider a Lagrangian invariant under the gauge transformations
\vs\be
\phi(x)\to e^{i\alpha^a(x)\,t^a}\,\phi(x)~,
\label{gauget1}
\ee\vs\nin
where $t^a$ are the generators of the group $G$, and the gauge fields
transform as they should. 
If we consider infinitesimal gauge transformations and write out the
field $\phi(x)$ in its groups components, we have 
\vs\be
\phi_i(x) \to \bigl(\delta_{ij} + i\alpha^a(x)
\,(t^a)_{ij}\bigr)\,\phi_j(x)
\label{gauget2}
\ee\vs\nin
In general, we consider representations where the $\phi_i(x)$ fields
in (\ref{gauget2}) are  complex. But for the purpose of our next
derivation, it would be advantageous to consider their real
components. So if the original representation had dimension $n$, we
now have $2n$ components in the real fields $\phi_i(x)$. If this is
the case, then the generators in (\ref{gauget2}) {\em must be
  imaginary}, since the $\alpha^a(x)$ are real parameter
functions. This means we can write them as
\vs\be
t^a_{ij} = i \,T^a_{ij}~,
\label{realts}
\ee\vs\nin
where the $T^a_{ij}$ are real. Also, since the $t^a$ are hermitian, we
have
\vs\be
\bigl(t^a_{ij}\bigr)^\dagger = t^a_{ij}~,
\ee\vs\nin
we see that
\vs\be
T^a_{ij} = -T^a_{ji}~,
\ee\vs\nin
so the $T^a$ are antisymmetric. In general, the Lagrangian of the
gauge invariant 
theory for a scalar field in terms of the real scalar degrees of
freedom would be\footnote{Here we concentrate on the
  scalar sector of $\cL$ since it is here that SSB of the gauge
  symmetry arises. We can imagine adding fermion terms to $\cL$
  coupling them both to the gauge bosons through the covariant
  derivative, as well as Yukawa couplings between the fermions and the
  scalars. Of course, all these terms must also respect gauge invariance.} 
\vs\be
\cL = \frac{1}{2} \Bigl(D_\mu\phi_i\Bigr)\Bigl(D^\mu\phi_i\Bigr)
-V\bigl(\phi_i\bigr)~,
\label{lagrealfis}
\ee\vs\nin
where the repeated $i$ indices are summed. We can write the covariant
derivatives above as
\vs\be
D_\mu\phi(x) = \bigl(\del_\mu -ig A_\mu^a(x) t^a\bigr)\,\phi(x) = \bigl(\del_\mu + g
A^a_\mu(x) T^a\bigr)\,\phi(x)~,
\label{realcovder}
\ee\vs\nin
where we omitted the group indices for the fields and the
generators. We are interested in the situation when the potential in
(\ref{lagrealfis}) induces spontaneous symmetry breaking. To see how
this affects the gauge boson spectrum we must examine in detail the
scalar kinetic term:
\vs\bear
\frac{1}{2}\bigl(D_\mu\phi_i\bigr)\,\bigl(D^\mu\phi_i\bigr)
&=&\frac{1}{2}\del_\mu\phi_i\,\del^\mu\phi_i + \frac{1}{2} g^2 A^a_\mu
A^{b\mu} \bigl(T^a\phi\bigr)_i \bigl(T^b\phi\bigr)_i\nonumber\\
~\label{kinterm1}\\
&+& g A^a_\mu \bigl(T^a\phi\bigr)_i \,\del^\mu\phi_i~,
\nonumber
\eear\vs\nin
where we used the notation
\vs\be
\bigl(T^a\phi\bigr)_i = T^a_{ij} \phi_j~, 
\ee\vs\nin
and as usual repeated group indices $i,j$ are summed.
If the potential $V(\phi_i)$ has a non trivial minimum then,
the vacuum expectation value (VEV) of the fields $\phi_i$ at the
minimum is
\vs\be
\langle 0 |\phi_i|0\rangle =\langle\phi_i\rangle \equiv
\bigl(\phi_0\bigr)_i~,
\label{vevedef2}
\ee\vs\nin
which says that we are singling out directions in field space which
may have non trivial VEVs. Then the terms in $\cL$ quadratic in the gauge boson
fields, i.e. the gauge boson mass terms, can be readily read off
(\ref{kinterm1}): 
\vs\be
\cL_{m} = \frac{1}{2}  M^2_{ab}\,A^a_\mu A^{b\mu}~,
\label{gbmass1}
\ee\vs\nin
where the gauge boson mass matrix is defined by
\vs\be
M^2_{ab} \equiv
g^2\,\bigl(T^a\phi_0\bigr)_i\,\bigl(T^b\phi_0\bigr)_i~.
\label{massmatrixdef}
\ee\vs\nin
Since the $T^a$'s are real, the non zero eigenvalues of $M^2_{abv}$
are definite positive.  We can clearly see now that if
\vs\be
T^a\phi_0 = 0,
\label{stillgood}
\ee\vs\nin
then the associated gauge boson $A^a_\mu$ remains massless. That is,
the{\em  unbroken}  generators, which as we saw in the previous lecture, {\em
  do not have NGBs associated with them}, do not result in a mass term
for the corresponding gauge boson. On the other hand, if
\vs\be
T^a\phi_0\not=0~,
\label{notgood}
\ee\vs\nin
then we see that this results in a gauge boson mass term. The
generators satisfying (\ref{notgood}) are of course the {\em broken
  generators} which result in massless NGBs. However, just as we saw
for the abelian case, these NGBs can be removed from the spectrum by a
gauge transformation. To see how this works we consider the last term
in (\ref{kinterm1}), the mixing term. This is
\vs\be
\cL_{\rm mix.} = g A^a_\mu \bigl(T^a\phi_0\bigr)_i\,\del^\mu\phi_i~.
\label{lmix}
\ee\vs\nin
Thus, we see that if the associated generator is broken,
i.e. (\ref{notgood}) is satisfied, then there is mixing of the
corresponding gauge boson with the massless $\phi_i$ fields, the
NGBs. It is clear that, just as in the abelian case,  we can eliminate
this term by a suitable gauge transformation on $A^a_\mu$. This would
still leave the mass term unchanged, but would  completely eliminate
the NGBs mixing in (\ref{lmix}) from the spectrum. But even if we
leave the NGBs in the spectrum, and we still have to deal with the
mixing term (\ref{lmix}), we can still see that the gauge boson two
point function remains transverse, a sign that gauge invariance is
still respected despite the appearance of a gauge boson mass. This is
depicted in Figure~\ref{fig:201}.
\vs
\begin{figure}
\begin{center}
  \includegraphics[width=6in]{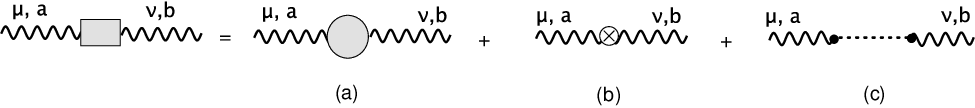}
 \end{center}    
\caption{Contributions to the gauge boson two point function in the
  presence of spontaneous gauge symmetry breaking. Diagram $(a)$
  includes the tree level as well as loop diagrams, all of which are
  transverse contributions. Diagram $(b)$ is the contribution from the
  gauge boson mass term. Diagram $(c)$ depicts the contribution from
  the massless NGBs.
}
\label{fig:201}
\end{figure}
\nin
In order to obtain diagram $(c)$ we need to derive the Feynman rule
resulting from the mixing term $\cL_{\rm mix}$ (\ref{lmix}). In
momentum space this becomes
\vs\vs\vs\vs
\begin{center}
\begin{tabular}{c c} 
  \includegraphics[width=2in]{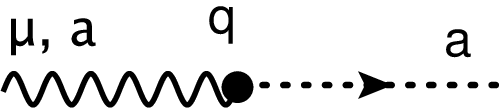}
  &
    $\qquad\quad=  g \bigl(T^a \phi_0\bigr)_i\,q^\mu~,$
\end{tabular}
\end{center}
\vs\vs\vs\vs
\nin
where the NGB momentum is flowing out of the vertex (its sign changes
if it is flowing into the vertex). The contributions to diagram $(a)$
are transverse as they come from either the leading order propagator
or the loop corrections to it, both already shown to be transverse.
Then the two point function for the gauge boson in the presence of
spontaneous symmetry breaking is
\vs\bear
\Pi_{\mu\nu} &=& \Pi^{(a)}_{\mu\nu} + i M^2_{ab} \,g_{\mu\nu} + g
\bigl(T^a\phi_0\bigr)_i\,q_\mu\,\frac{i\delta_{ab}}{q^2}\, g
\bigl(T^b\phi_0\bigr)_i\,(-q_\nu\bigr)~\nonumber\\
&=& \Pi^{(a)}_{\mu\nu} + i M^2_{ab}\bigl(g_{\mu\nu} -\frac{q_\mu\,a_\nu}{q^2}\bigr)~,
\label{istransverse}
\eear\vs\nin
where to obtain the second line we used (\ref{massmatrixdef}). Then,
just as we saw for the abelian case, we see that the gauge boson two
point function is transverse even in the presence of gauge boson masses. 

\nin
\underline{ Example: $SU(2)$}

\nin
In this first example we gauge the $SU(2)$ of the first example in the
previous lecture.  The Lagrangian
\vs\be
\cL = \bigl(D_\mu\phi\bigr)^\dagger\,D^\mu\phi -V(\phi^\dagger\phi)
-\frac{1}{4}F^a_{\mu\nu} F^{a\mu\nu}~,
\label{lagsu2}
\ee\vs\nin
with the covariant derivative on the scalar field is\footnote{We have
  gone back to complex scalar fields for the remaining of the lecture.}
\vs\be
D_\mu\phi(x) =\bigl(\del_\mu - ig A^a_\mu(x) t^a\bigr)\,\phi(x)~,
\label{covder2}
\ee\vs\nin
where the $SU(2)$ generators are given in terms of the Pauli matrices as
\vs\be
t^a=\frac{\sigma^a}{2}~,
\ee\vs\nin
with $a=1,2,3$. Since they transform according to
\vs\be
\phi(x)_j\to e^{i\alpha^a(x) t^a_{jk}}\,\phi_k(x)~,
\ee\vs\nin
with $j,k=1,2$, 
then that are {\em doublets} of $SU(2)$. Since each of the $\phi_j(x)$
are complex scalar fields, we have {\em four} real scalar degrees of
freedom.
We will consider the vacuum
\vs\be
\langle\phi\rangle = \left(\ba{c} 0\\
  \frac{v}{\sqrt{2}}\ea\right)~,
\label{su2vac}
\ee\vs\nin
such that, as required by imposing a non trivial minimum, we have
\vs\be
\langle\phi^\dagger\phi\rangle =\frac{v^2}{2}~,
\label{phi2vac}
\ee\vs\nin
where the factor of $2$ above is chosen for convenience. We are
particularly interested in the gauge boson mass terms. These can be
readily obtained by substituting the vacuum value of the field in the
kinetic term. This is
\vs\bear
\cL_{\rm m}&=&
\bigl(D_\mu\langle\phi\rangle\bigr)^\dagger\,D^\mu\langle\phi\rangle~,\nonumber\\
&=& \frac{g^2}{2}\,A^a_\mu A^{b\mu}\,\bigl(0\quad v\bigr)\,t^a t^b \left(\ba{c}0\\v\ea\right)~, 
\label{gaugemass2}
\eear\vs\nin
 where we used (\ref{su2vac}) in the second line. But for the case of
 $SU(2)$ we can use the fact that
 \vs\be
\bigl\{\sigma^a,\sigma^b\bigr\} = 2\delta^{ab}~,
\ee\vs\nin
which translates into
\vs\be
\bigl\{t^a,t^b\bigr\} = \frac{1}{2} \delta^{ab}~,.
\label{su2acom}
\ee\vs\nin
Then, if we write
\vs\bear
A^a_\mu \,A^{b\mu} \,t^a\, t^b &=& \frac{1}{2} \,A^a_\mu \,A^{b\mu} \,t^a\, t^b  +
\frac{1}{2} \,A^b_\mu\, A^{a\mu} \,t^b \,t^a \nonumber\\
&=& \frac{1}{2} \, A^a_\mu\,A^{b\mu} \, \bigl\{t^a,t^b\bigr\}=
\frac{1}{4}\,A^a_\mu\,A^{a\mu}~,
\label{su2trick}
\eear\vs\nin
where in the last equality we used (\ref{su2acom}). Then we obtain
\vs\be
\cL_{\rm m} = \frac{1}{8} g^2\,v^2\,A^a_\mu\,A^{a\mu}~,
\label{lmass2}
\ee\vs\nin
which results in a gauge boson mass of
\vs\be
M_A = \frac{g\,v}{2}~.
\ee\vs\nin
Notice that {\em all three} gauge bosons obtain this same mass. It is
interesting to compare this result with what we obtained in the
previous lecture for the spontaneous breaking of a {\em global }
$SU(2)$ symmetry using the same vacuum as in (\ref{su2vac}). 
In that case, we saw that all generators were broken, i.e. there are
three massless NGBs in the spectrum and the $SU(2)$ is completely
(spontaneously) broken in the sense that none of its generators leaves
the vacuum invariant. In the case here, where the $SU(2)$ symmetry is
gauged, we see that  all three gauge bosons get masses. This is in
fact the same phenomenon: none of the gauge symmetry leaves the
$SU(2)$ vacuum (\ref{su2vac}) invariant. However, the end result is
three massive gauge bosons, not three massless NGBs. We argued in our
general considerations above that, just as for  the abelian case
before, the NGBs can be removed by a gauge transformation. Let us see
how this can be implemented.
\nin
We consider the following parameterization of the $SU(2)$ doublet scalar field:
\vs\be
\phi(x) = e^{i\pi^a(x) t^a/v}\,\left(\ba{c}0\\
  ~\\
  \frac{v+\sigma(x)}{\sqrt{2}}\ea\right)~,
\label{phiparam2}
\ee\vs\nin
where $\sigma(x)$ and $\pi^a(x)$ with $a=1,2,3$ are real scalar
fields satisfying
\vs\be
\langle\sigma(x)\rangle=0=\langle\pi^a(x)\rangle~,
\ee\vs\nin
so that this choice of parameterization is consistent with the vacuum
(\ref{su2vac}). Clearly, the potential will not depend on the
$\pi^a(x)$ fields
\vs\be
V(\phi^\dagger\phi)= -\frac{m^2}{2}\,\phi^\dagger\phi
+\frac{\lambda}{2}\,\bigl(\phi^\dagger\phi\bigr)^2~,
\label{pot3}
\ee\vs\nin
The minimization results in\footnote{Notice the different factor in
  the denominator of the second term. This is due to the factor of
  $\sqrt{2}$ in the definition of the vacuum.} 
\vs\be
\langle\phi^\dagger\phi\rangle = \frac{m^2}{2\lambda} ~,
\ee\vs\nin
which results in
\vs\be
v^2=\frac{m^2}{\lambda}~.
\ee\vs\nin
Replacing this in the potential (\ref{pot3}) we obtain
\vs\be
m_\sigma=\sqrt{2\,\lambda}\,v~.
\ee\vs\nin
And of course, the implicit result of having
\vs\be
m_{\pi^1}=m_{\pi^2}=m_{\pi^3}=0~.
\ee\vs\nin
But how do we get rid of the massless NGBs ? If we define the
following gauge transformation
\vs\be
U(x)\equiv e^{-i\pi^a(x) t^a/v}
\ee\vs\nin
under which the fields transform as
\vs\bear
\phi(x) &\to &  \phi'(x) = U(x) \,\phi(x) = \left(\ba{c}0\\
  ~\\
  \frac{v+\sigma(x)}{\sqrt{2}}\ea\right)~,\nonumber\\
~\label{ugauge1}\\
A_\mu &\to & A'_\mu = U(x)\,A_\mu \,U^{-1}(x)
-\frac{i}{g}\,\bigl(\del_\mu U(x)\bigr)\,U^{-1}(x)~,\nonumber
\eear\vs\nin
where we used the notation $A_\mu=A^a_\mu t^a$. It is clear from the
first transformation above, that $\phi'(x)$ does not depend on the
$\pi^a(x)$ fields. Thus, the gauge transformation (\ref{ugauge1}) has
removed them from the spectrum completely. However, the number of
degrees of freedom is the same in both gauges. We had three transverse
gauge bosons (i.e. 6 degrees of freedom)  and four real scalar fields. In this new gauge we have
three massive gauge bosons (i.e. 9 degrees of freedom) plus one real
scalar, $\sigma(x)$. The total number of degrees of freedom is always
the same. The gauge were the NGBs disappear of the spectrum is
called the {\em unitary gauge}.

\subsubsection{The ABEH mechanism in the electroweak Standard Model}
\label{sec:higgsmechinsm}

In order to apply what we learned in the previous section to the EWSM,
we have to introduce a scalar field in to it.   We must define the
representation of $SU(2)_L\times U(1)_Y$ for this new field. 
 We consider a scalar field $\Phi$ in the fundamental representation
of $SU(2)_L$ and with assignment of hypercharge $U(1)_Y$,
  \vs\be
  Y_{\Phi}=1/2~.
  \ee\vs\nin
  That the scalar is in the fundamental representation of $SU(2)_L$
  means that it is a scalar {\em doublet}, dubbed the Higgs doublet. It can be written as
  \vs\be
  \Phi = \left(\ba{c}\phi^+\\\phi^0\ea\right)~,
  \label{hdoublet}
  \ee\vs\nin
  where $\phi ^+$ and $\phi^0$ are complex scalar fields, resulting in
  four real scalar degrees of freedom\footnote{At this point, the
    labels ``$+$'' and ``$0$'' are just arbitrary, since we have not
    even defined electric charges But these labels will be consistent in
    the future, after we have done this.}.    Under a $SU(2)_L\times
  U(1)_Y$ gauge transformation, the Higgs doublet transforms as
  \vs\be
  \Phi(x)\to e^{i \alpha^a(x) t^a}\,e^{i \beta(x) Y_{\Phi}},\Phi(x)~,
  \label{hdtransf}
  \ee\vs\nin
  where $t^a$ are the $SU(2)_L$ generators (i.e. Pauli matrices
  divided by $2$), $\alpha^a(x)$ are the three
  $SU(2)_L$  gauge
  parameters, $\beta(x)$ is the $U(1)_Y$ gauge parameter, and it is
  understood that the $U(1)_Y$ factor of the gauge transformation
  contains a factor of the identity $I_{2\times 2}$ after the
  hypercharge $Y_{\Phi}$. Thus, the covariant derivative on $\Phi$ is
  given by
  \vs\be
D_\mu\Phi(x) = \Bigl(\del_\mu -ig A^a_\mu(x) t^a -ig' B_\mu(x)
Y_{\Phi} I_{2\times 2}\Bigr)\,\Phi(x)~.
\label{covderphi}
\ee\vs\nin
Here, $A^a_\mu(x)$ is the $SU(2)_L$ gauge boson, $B_\mu(x)$ the
$U(1)_Y$ gauge boson, and $g$ and $g'$ are their corresponding
couplings. The Lagrangian of  the scalar and gauge sectors of the SM is then
\vs\be
\cL = \bigl(D_\mu\Phi\bigr)^\dagger D^\mu\Phi - V(\Phi^\dagger \Phi)
-\frac{1}{4} F^a_{\mu\nu} F^{a\mu\nu} -\frac{1}{4} B_{\mu\nu}
B^{\mu\nu}~,
\label{smlag1}
\ee\vs\nin
where $F^a_{\mu\nu}$ is the usual $SU(2)$ field strength built out of
the gauge fields $A^a_\mu(x)$ and $B_{\mu\nu}$ is the $U(1)_Y$ field
strength given by the abelian expression
\vs\be
B_{\mu\nu} = \del_\mu B_\nu(x) -\del_\nu B_\mu(x) ~.
\label{bmunu2}
\ee\vs\nin
As usual, we consider the potential
\vs\be
V(\Phi^\dagger\Phi) = -m^2\bigl(\Phi^\dagger\Phi\bigr) + \lambda
\bigl(\Phi^\dagger\Phi\bigr)^2~,
\label{smpotential}
\ee\vs\nin
which is minimized for
\vs\be
\langle\Phi^\dagger\Phi\rangle = \frac{m^2}{2\lambda}\equiv
\frac{v^2}{2}~.
\label{vmindef}
\ee\vs\nin
In order to fulfil this, we choose the vacuum
\vs\be
\langle\Phi\rangle = \left(\ba{c}0\\
  \frac{v}{\sqrt{2}}\ea\right)~.
\label{smvac}
\ee\vs\nin
Just as in the previous examples of SSB of non-abelian gauge
symmetries, the next question is what is the symmetry breaking
pattern, i.e. what gauge bosons get what masses, if any. In
particular, we want one of the four gauge bosons in $G$ to remain
massless after imposing the vacuum $\langle\Phi\rangle$ in
(\ref{smvac}). This means that there must be a generator or, in this
case, a linear combination of generators of $G$ that annihilates
$\langle\Phi\rangle$, leaving the vacuum invariant under a $G$
transformation. This combination of generators must be associated with
the massless photon in $U(1)_{\rm EM}$, the remnant gauge group after
the spontaneous breaking. One trick to identify this combination of
generators is to consider the gauge transformation defined by
\vs\bear
\alpha^1(x)=\alpha^2(x)=0\nonumber\\
\alpha^3(x)=\beta(x)~.
\label{t3plusygauge}
\eear\vs\nin
The exponent in the gauge transformation has the form
\vs\bear
i\alpha^3(x) t^3 + i \beta(x) Y_\Phi I_{2\times2} &=& i
\frac{\beta(x)}{2} \Bigl[\left(\ba{cc}1&0\\0&-1\ea\right) +
\left(\ba{cc}1&0\\0&1\ea\right)\Bigr]\nonumber\\
&=&\frac{i\beta(x)}{2}\left(\ba{cc}1&0\\0&0\ea\right)~.
\label{charge1}
\eear\vs\nin
Then we see that this combination
\vs\be\boxed{
\bigl(t^3+Y_\Phi\bigr)\,\langle\Phi\rangle =0~},
\label{killsvac}
\ee\vs\nin
indeed annihilates the vacuum, leaving it invariant. 
Thus, we suspect that this linear combination of $SU(2)_L\times
U(1)_Y$ generators must be associated with the massless photon. We
will come back to this point later.
\\
\nin
We now go to extract the gauge boson mass terms from the scalar
kinetic term in (\ref{smlag1}). This is
\vs\bear
\cL_{\rm m} &=& \bigl(D_\mu\langle\Phi\rangle\bigr)^\dagger
D^\mu\langle\Phi\rangle\nonumber\\
&=& \frac{1}{2}\,(0\quad v)\,\bigl(g A^a_\mu t^a
+ g' Y_\Phi B_\mu \bigr) \bigl(g A^{b\mu} t^b
+ g' Y_\Phi B^\mu \bigr)\,\left(\ba{c}0\\v\ea\right)~.
\label{gbmass2}
\eear\vs\nin
For the product of the two $SU(2)$ factors we will use the trick in
(\ref{su2trick}). Then, the only terms we need to be careful about are
the mixed ones: one $SU(2)$ times one $U(1)_Y$ contribution. There are
two of them, and each has the form
\vs\be
\frac{1}{2}\,(0\quad v)\, g\,g' \frac{\sigma^3}{2}
Y_\Phi\,\left(\ba{c}0\\v\ea\right)
= -\frac{1}{2} \frac{v^2}{4}\,g\,g'
\,A^3_\mu\,B^\mu~,
\label{su2u1mix}
\ee\vs\nin
where in the second equality we used $Y_\Phi=1/2$.
We then have
\vs\be
\cL_{\rm m} = \frac{1}{2}\,\frac{v^2}{4}\Bigl\{g^2A^1_\mu A^{1\mu}+g^2
A^2_\mu A^{2\mu} +g^2A^3_\mu A^{3\mu} + g'^2 B_\mu B^\mu -2g g'
A^3_\mu B^\mu\Bigr\}~.
\label{gbmass3}
\ee\vs\nin
From this expression we can clearly ee that $A^1_\mu$ and $A^2_\mu$
acquire masses just as we saw in the pure $SU(2)$ example. It will be
later  convenient  to define the linear combinations
\vs\be
W^\pm_\mu\equiv \frac{A^1_\mu\mp iA^2_\mu}{\sqrt{2}}~,
\label{wpmdef}
\ee\vs\nin
which allows us to write the first two terms in (\ref{gbmass3}) as
\vs\be
\cL_{\rm m}^W = \frac{g^2\,v^2}{4}\,W^+_\mu W^{-\mu}~.
\label{wmassterm}
\ee\vs\nin
These two states have masses
\vs\be
M_W=\frac{g\,v}{2}~.
\label{wmass1}
\ee\vs\nin
On the other hand, the
fact that $A^3_\mu$ and $B_\mu$ have a mixing term prevents us from
reading off masses. We need to rotate these states to go to a bases
without mixing, a diagonal basis. In order to clarify what needs to be
done, we can write the last three terms in (\ref{gbmass3} in matrix
form
\vs\be
\cL_{\rm m}^{\rm neutral} = \frac{1}{2}\frac{v^2}{4}
(A^3_\mu\quad B_\mu)\,\left(\ba{cc} g^2&-g\,g'\\
  -g\,g'&g'^2\ea\right) \,\left(\ba{c}A^{3\mu}\\B^\mu\ea\right)~,
\label{neutralmass1}
\ee\vs\nin
where the task is to find the eigenvalues and eigenstates of the
matrix above. It is clear that one of the eigenvalues is zero, since
the determinant vanishes. Then the squared masses of the physical neutral
gauge bosons are
\vs\bear
M^2_\gamma&=&0\nonumber\\
~\label{neutralmass2}\\
M^2_Z &=&\frac{v^2}{4}\,(g^2+g'^2)~\nonumber 
\eear\vs\nin
The eigenstates in terms of $A^3_\mu$ and $B_\mu$, the original $SU(2)_L$ and
$U(1)_Y$ gauge bosons respectively, are
\vs\bear\boxed{
A_\mu \equiv \frac{1}{\sqrt{g^2+g'^2}}\,\bigl(g' A^3_\mu + g
B_\mu\bigr)~}\label{photondef1}\\
~\nonumber\\
\boxed{Z_\mu\equiv \frac{1}{\sqrt{g^2+g'^2}}\,\bigl(g A^3_\mu - g'
B_\mu\bigr)~}.\label{zdef1}
\eear\vs\nin
Alternatively, we could have obtained the same result by defining an
orthogonal rotation matrix to diagonalize the interactions above.
That is,  rotating the states by
\vs\be
\left(\ba{c}Z_\mu\\A_\mu\ea\right)=
\left(\ba{cc}\cos\theta_W & -\sin\theta_W\\
  \sin\theta_W&\cos\theta_W\ea\right)\,\left(\ba{c}
  A^3_\mu\\B_\mu\ea\right)~,
\label{wangledef1}
\ee\vs\nin
results in diagonal neutral interactions if we have
\vs\be
\cos\theta_W\equiv \frac{g}{\sqrt{g^2+g'^2}},\qquad
\sin\theta_W\equiv \frac{g'}{\sqrt{g^2+g'^2}},
\label{wangledf2}
\ee\vs\nin
where $\theta_W$ is called the Weinberg angle. It is useful to
invert (\ref{wangledef1}) to obtain
\vs\bear
A^3_\mu &=& \sin\theta_W A_\mu +\cos\theta_W Z_\mu\label{a3asaz}\\
~\nonumber\\
B_\mu &=& \cos\theta_W A_\mu -\sin\theta_W Z_\mu~.\label{basaz}
\eear\vs\nin
Using these expressions for $A^3_\mu$ and $B_\mu$ we can replace them
in the covariant derivative acting on the scalar doublet $\Phi$. Their
contribution fo $D_\mu$ is
\vs\bear
-i g A^3_\mu t^3 - i g' Y_\Phi B_\mu &=&
-iA_\mu\bigl(g\sin\theta_Wt^3 +  g'\cos\theta_W
Y_\Phi\bigr) 
- i\bigl(g\cos\theta_W t^3 - g'\sin\theta_W Y_\Phi\bigr)\,Z_\mu
\nonumber\\
~\label{photonandzcpl}\\
&=&- i g\sin\theta_W \Bigl(t^3+Y_\Phi\Bigr) A_\mu 
-i\frac{g}{\cos\theta_W}\Bigl(t^3 -(t^3+Y_\Phi)\sin^2\theta_W\Bigr) Z_\mu~,
\nonumber
\eear\vs\nin
where it is always understood that the hypercharge $Y_\Phi$ is always
multiplied by the identity, and in the last identity we used the fact
that
\vs\be
g' \cos\theta_W = g \sin\theta_W~,
\ee\vs\nin
and trigonometric identities. We can conclude that is $A_\mu$  is to
be identified with the photon field, then its coupling must be $e$
times the charged of the particle it is coupling to (e.g. $-1$ for an
electron.
Thus we must impose that
\vs\be\boxed{
  e= g\,\sin\theta_W~},
\label{qedcdef1}
\ee\vs\nin
and that the charge operator,  acting here on the field $\Phi$ coupled to $A_\mu$ is
defined as
\vs\be\boxed{
Q = t^3 + Y_\Phi~}.
\label{emchargedef}
\ee\vs\nin
Then we can read the photon coupling to the doublet scalar field $\Phi$ from
\vs\be
-i\, e\, A_\mu \,Q \,\Phi(x) = -i \,e \,A_\mu \,Q
\,\left(\ba{c}\phi^+\\\phi^0\ea\right) ~.
\ee\vs\nin
Substituting $Y_\Phi=1/2$ we have
\vs\be
Q \,\left(\ba{c}\phi^+\\\phi^0\ea\right) = \left(\ba{cc}1&0\\
  0&0\ea\right) \,\left(\ba{c}\phi^+\\\phi^0\ea\right) = \,\left(\ba{c}\phi^+\\0\ea\right) ~,
\ee\vs\nin
which tells us that the top complex field in the scalar doublet has
charge equal to $1$ (in units of $e$, the proton charge), whereas the
bottom component has zero charge, justifying our choice of labels.
On the other hand, we see that fixing $Q$ to be the electromagnetic
charge operator, completely fixes the couplings of $Z_\mu$ to the
scalar $\Phi$. This is now,  from (\ref{photonandzcpl}),
\vs\be
-i \frac{g}{\cos\theta_W} Z_\mu \Bigl(t^3
-Q\sin^2\theta_W\Bigr)\,\Phi~.
\label{zcpltophi}
\ee\vs\nin
We will see below that the choice of fixing the $A_\mu$ couplings to
be those of electromagnetism, fixes completely the $Z_\mu$ couplings to all
fermions, giving a wealth of predictions.

\subsubsection{Gauge couplings of fermions}
\label{sec:gaugefermion}

\nin
The SM is a {\em chiral gauge theory}, i.e. its gauge couplings differ
for different chiralities.
To extract the left handed fermion gauge couplings, we look at the covariant derivative
\vs\be
D_\mu\psi_L = \bigl(\del_\mu -igA^a_\mu t^a -i g' Y_{\psi_L} B_\mu
\bigr) \psi_L~,
\label{lhfcd2}
\ee\vs\nin
where $Y_{\psi_L}$ is the left handed fermion hypercharge. On the
other hand, since right handed fermions do not feel the $SU(2)_L$
interaction, their covariant derivative is given by
\vs\be
D_\mu\psi_R = \bigl(\del_\mu -i g' Y_{\psi_R} B_\mu
\bigr) \psi_R~,
\label{rhfcd1}
\ee\vs\nin
with $Y_{\psi_R}$ its hypercharge.
Using the covariant derivatives above, we can extract the neutral and
charged couplings. We start with the neutral couplings, which in terms
of the gauge boson mass eigenstates are the couplings to the photon
and the $Z$.

\vs\vs\nin
\underline{Neutral Couplings}: 
From (\ref{lhfcd2}), the neutral gauge couplings of a left handed fermions
are 
\vs\bear
\bigl(-ig t^3 A^3_\mu-ig' Y_{\psi_L} B_\mu\bigr)\psi_L &=&
ig\sin\theta_W\bigl(t^3+Y_{\psi_L}\bigr) \,A_\mu\,\psi_L \nonumber\\
&-&i\frac{\bigl(g^2 t^3-i g'^2 Y_{\psi_L}\bigr)}{\sqrt{g^2+g'^2}} \,Z_\mu \,\psi_L~,
  \label{neutralf1}
  \eear\vs\nin
  where on the right hand side we made use of (\ref{a3asaz}) and
  (\ref{basaz}).
  Now, we know that the photon coupling should be
  \vs\be
-ie \,Q_{\psi_L}~,
\ee\vs\nin
with $Q_{\psi_L}$ the fermion electric charge operator. Thus, we must
identify
\vs\be
Q_{\psi_L} = t^3+ Y_{\psi_L}~,
\label{fermioncharge1}
\ee\vs\nin
as the fermion charge. We can use our knowledge of the fermion charges
to fix their hypercharges. As an example, let us consider the left
handed lepton doublet. For the lightest family, this is written in the notation 
\vs\be
L = \left(\ba{c}\nu_{eL}\\e^-_L\ea\right)~.
\label{lepdoublet1}
\ee\vs\nin
The action of $t^3$ on $L$ is
\vs\bear
t^3\,L&=& \left(\ba{cc}1/2&0\\
  0&-1/2\ea\right)\, \left(\ba{c}\nu_{eL}\\e^-_L\ea\right)\nonumber\\
~\nonumber\\
&=&\left(\ba{c}(1/2)\,\nu_{e_L}\\(-1/2)\, e^-_L\ea\right) \equiv \left(\ba{c}
  t^3_{\nu_{e_L}}\,\nu_{e_L}\\t^3_{e_L}\, e^-_L\ea\right)~,
\label{t3onl}
\eear\vs\nin
where in the last equality we defined $t^3_{\nu_{e_L}}=1/2$ and
$t^3_{e_L}$ as the eigenvalues of the operator $t^3$ associated to the
electron neutrino and the electron. Then, we have
\vs\be
Q_{L}\,L = \left(\ba{cc} 1/2+Y_L&0\\
  -1/2+Y_L &0\ea\right)\, \left(\ba{c}\nu_{eL}\\e^-_L\ea\right) =
\left(\ba{c}(1/2+Y_L)\,\nu_{e_L} \\
  (-1/2+Y_L) \, e_L\ea\right)~.
\label{qonl} 
\ee\vs\nin
But we know that the eigenvalue of the charge operator applied to the
neutrino must be zero, as well as that the eigenvalue of the electron
must be $-1$. Thus, we obtain the hypercharge of the left handed lepton
doublet
\vs\be\boxed{
  Y_L=-\frac{1}{2}~},
  \label{lhlddhyp}
\ee\vs\nin
which is fixed to give us the correct electric charges for the members
of the doublet $L$.
\nin
We can do the same with the right handed fermions. These, however do
not have $t^3$ in the covariant derivative (see (\ref{rhfcd1})
). Then, for $e^{-}_R$, the right handed electron, we have that
$t^3_{e_R}=0$, which means that, since 
\vs\be
Q_{e^-_R}=  -1~,
\ee\vs\nin
then the right handed electron's hypercharge is equal  to it:
\vs\be\boxed{
Y_{e^-_R} =-1~}.
\ee\vs\nin
Similarly, the right handed electron neutrino has zero electric
charge, which results in
\vs\be\boxed{
Y_{\nu_R}=0~} .
\ee\vs\nin
Now that we fixed all the lepton hypercharges by imposing that they
have the QED couplings to the photon, we can extract their couplings
to the $Z$ as predictions of the electroweak SM.  From
(\ref{neutralf1}) we have
\vs\bear
-i \bigl(g\,\cos\theta_W \,t^3 - g'\,\sin\theta_W \,Y_\psi\bigr) Z_\mu\,
\psi
&=& -i \frac{g}{\cos\theta_W}\,\bigl( \cos^2\theta_W\,t^3 -
\sin^2\theta_W\,Y_\psi\bigr)\,Z_\mu\,\psi\nonumber\\
~\nonumber\\
&=& -i\,\frac{g}{\cos\theta_W} \,\bigl(t^3 -
\sin^2\theta_W\,Q_\psi\bigr)\,Z_\mu\,\psi~,
\label{neutralfcoup2}
\eear\vs\nin
where the initial expressions makes use of $\cos\theta_W $ and
$\sin\theta_W$ in terms of $g$ and $g'$, in the first equality we used
that $\tan\theta_W=g'/g$ and, in the final equality, we  used that in
general $Q_\psi=t^2+Y_\psi$, independently of the fermion chirality, as
long as we generalize (\ref{fermioncharge1}) for right handed fermions
using $t^3_{\psi_R}=0$. For instance, from (\ref{neutralfcoup2}) we can read off the
lepton couplings of the $Z$ boson. These are,
\vs\bear
\nu_{e_L} : &&\qquad
-i\frac{g}{\cos\theta_W}\,\Bigl(\frac{1}{2}\Bigr)\nonumber\\
  e^-_L: &&\qquad  -i\frac{g}{\cos\theta_W}\,\Bigl(-\frac{1}{2}
  +\sin^2\theta_W\Bigr)\nonumber\\
  ~\label{lzcoup1}\\
  e^-_R: &&\qquad -i\frac{g}{\cos\theta_W}\,\Bigl(
  \sin^2\theta_W\Bigr)\nonumber\\
  \nu_{e_R}: &&\qquad 0\nonumber~.
\eear\vs\nin
From the couplings above, we see that every lepton has a different
predicted coupling to the $Z$. These are, of course, three level
predictions. Measurements of these $Z$ couplings have been performed
with subpercent precision for a long time, and the SM predictions for
the fermion gauge couplings have
passed the tests every time. Another, interesting point, is that right
handed neutrinos have {\em no gauge couplings} in the SM: no $Z$
coupling, certainly no electric charge and no QCD couplings. Thus,
from the point of view of the SM, the right handed neutrino need not
exist. 

\vs\vs\nin
\underline {Charged Couplings}:

\nin
We complete here the derivation of the gauge couplings of leptons by extracting their
charged couplings.These come from the $SU(2)_L$ gauge couplings, as we
see from
\vs\be
-i g\bigl(A^1_\mu t^1 + A^2_\mu t^2\bigr) =
-i\frac{g}{\sqrt{2}}\,\left(\ba{cc}0& W^+_\mu\\
  W^-_\mu&0\ea\right)~,
\label{chargedgauge2}
\ee\vs\nin
which then involve only left handed fermions. 
Then, from the gauge part of the left handed doublet kinetic term 
\vs\be
\cL_{L} = \bar L i\dslcv D L ~,
\ee\vs\nin
we obtain their charged couplings 
\vs\bear
\cL_L^{\rm ch.} &=& (\bar\nu_{e_L}\quad \bar e_L)\,\gamma^\mu
\frac{g}{\sqrt{2}}\,
\left(\ba{cc}0& W^+_\mu\\
  W^-_\mu&0\ea\right)\,\left(\ba{c}\nu_{e_L}\\e_L\ea\right)\nonumber\\
~\label{chargedcoup2}\\
&=& \frac{g}{\sqrt{2}}\,\Bigl\{\bar\nu_{e_L}\gamma^\mu e_L \, W^+_\mu
+ \bar e_L\gamma^\mu \nu_{e_L}\,W^-_\mu\Bigr\}~,
\nonumber
\eear\vs\nin
where we can see that, as required by hermicity, the second term is the hermitian conjugate of
the first. The Fermi Lagrangian can be obtained from $\cL_L^{\rm ch.}$
by integrating out the $W^\pm$ gauge bosons.

\nin
We now briefly comment on the electroweak gauge couplings of quarks.
Just as for leptons, we concentrate on the first family. The left
handed quark doublet is
\vs\be
q_L = \left(\ba{c}u_L\\d_L\ea\right)~,
\label{qldef1}
\ee\vs\nin
We know  that, independently of helicity, the charges of the up and
down quarks are $Q_u=+2/3$ and $Q_d=-1/3$. Then we have
\vs\be
Q_{q_L} = \Bigl(t^3+ Y_{q_L}\Bigr) = \left(\ba{cc}
  +2/3&0\\0&-1/3\ea\right)~,
\label{qlcharge1}
\ee\vs\nin
which results in
\vs\be\boxed{
  Y_{q_L}=\frac{1}{6}~}~.
\label{qlhyp}
\ee\vs\nin
The hypercharge assignments for the right handed quarks are again
trivial and given by the quark electric charges. We have
\vs\be\boxed{
Y_{u_R} = +\frac{2}{3} ~,\qquad Y_{d_R} = -\frac{1}{3}~}~.
\label{rhqhyp}
\ee\vs\nin
With these hypercharge assignments we can now write the quark
couplings to the $Z$.  Using
(\ref{neutralfcoup2}) we obtain
\vs\bear
u_L: && \qquad -i\frac{g}{\cos\theta_W}\,\Bigl(\frac{1}{2}
-\sin^2\theta_W\frac{2}{3}\Bigr)~\nonumber\\
d_L: && \qquad -i\frac{g}{\cos\theta_W}\,\Bigl(-\frac{1}{2} +
\sin^2\theta_W\frac{1}{3}\Bigr)\nonumber\\
u_R: && \qquad
-i\frac{g}{\cos\theta_W}\,\Bigl(-\sin^2\theta_W\frac{2}{3}\Bigr)\nonumber\\
d_R: &&\qquad -i\frac{g}{\cos\theta_W}\,\Bigl(\sin^2\theta_W\frac{1}{3}\Bigr)~.
\label{qzcoup1}
\eear\vs\nin
Once again, we see that each type of quark has a different coupling to
the $Z$. All of these predictions have been tested with great
precision, confirming the SM even beyond leading order.

\nin
The charged gauged couplings of left handed quarks are trivial to obtain: they are
dictated by $SU(2)_L$ gauge symmetry and therefore there must be the
same as those of the left handed leptons in (\ref{chargedcoup2}). So we have
\vs\be
\cL_{q}^{\rm ch.} = \frac{g}{\sqrt{2}}\,\Bigl\{\bar u_L\gamma^\mu d_L \, W^+_\mu
+ \bar d_L\gamma^\mu u_L\,W^-_\mu\Bigr\}~.
\label{qchargedcoup}
\ee\vs\nin

\subsubsection{Fermion masses} 

\nin
We have seen that SSB leads to masses for same of the gauge bosons,
preserving gauge invariance. We now direct our attention to fermion
masses. In principle these terms
\vs\be
\cL_{\rm fm} = m_\psi \bar \psi_L \psi_R +{\rm h.c.} ~,
\ee\vs\nin
are forbidden by $SU(2)L\times U(1)_Y$ gauge invariance since thy are
not invariant under
\vs\bear
\psi_L &\to & e^{i\alpha^a(x) t^a}\, e^{i\beta(x)
  Y_{\psi_L}}\,\psi_L\nonumber\\
~\label{fgaugetrans}
\psi_R &\to & e^{i\beta(x)
  Y_{\psi_R}}\,\psi_R ~.\nonumber
\eear\vs\nin
But the operator
\vs\be
\bar\psi_L\, \Phi \,\psi_R ~,
\ee\vs\nin
is clearly invariant under the $SU(2)_L$ gauge transformations, and it
would be $U(1)_Y$ invariant if\vs\be
-Y_{\psi_L} + Y_\Phi + Y_{\psi_R} =0~.
\label{hyperiszero}
\ee\vs\nin
Since $Y_\Phi=1/2$, this form of the operator will work for the down
type quarks and charged leptons. For instance, since $Y_L=-1/2$ and
$Y_{e_R}=-1$, the operator
\vs\be
-\cL_{m_e}=
\lambda_e \,\bar L \,\Phi \,e_R~ + {\rm h.c.},
\label{emass1}
\ee\vs\nin
is gauge invariant since the hypercharges satisfy (\ref{hyperiszero}).
In (\ref{emass1}) we defined the dimensionless coupling $\lambda_e$
which will result in a Yukawa coupling of electrons to the Higgs
boson. To see this, we write $\Phi(x)$ in the unitary gauge, so that
\vs\bear
-\cL_{m_e} &=&\lambda_e\, (\bar\nu_{e_L}\quad \bar e_L)\, \left(\ba{c}0\\
  ~\\
  \frac{v+h(x)}{\sqrt{2}}\ea\right)\, e_R  + {\rm h.c.} \nonumber\\
&=&\lambda_e \frac{v}{\sqrt{2}}\,\bar e_L e_R +
\lambda_e\,\frac{1}{\sqrt{2}}\,h(x)\,\bar e_L e_R + {\rm h.c.}~,
\label{emass2}
\eear\vs\nin
where the first term is the electron mass term resulting in
\vs\be
m_e = \lambda_e\,\frac{v}{\sqrt{2}}~,
\label{emass3}
\ee\vs\nin
and the second term is the Yukawa interaction of the electron and the
Higgs boson $h(x)$. We can rewrite (\ref{emass2}) as
\vs\be
-\cL_{\rm m_e} = 
m_e\,\bar e_L e_R +
\frac{m_e}{v}\,h(x)\,\bar e_L e_R + {\rm h.c.}~,
\label{emass4}
\ee\vs\nin
from which we can see that the electron couples to the Higgs boson
with a strength equal to its mass in units of the Higgs VEV
$v$. Similarly, for quarks we have that the operator
\vs\be
-\cL_{m_d}=
\lambda_e \,\bar q_L \,\Phi \,d_R~ + {\rm h.c.},
\label{dmass1}
\ee\vs\nin
is gauge invariant since $Y_{q_L}=1/6$ and $Y_{d_R}=-1/3$ satisfy (\ref{hyperiszero}).
Them we obtain
\vs\be
-\cL_{\rm m_d} = 
m_d\,\bar d_L d_R +
\frac{m_d}{v}\,h(x)\,\bar d_L d_R + {\rm h.c.}~, 
\label{dmass2}
\ee\vs\nin
and where the down quark mass was defined as
\vs\be
m_d = \lambda_d\,\frac{v}{\sqrt{2}}~.
\label{dmass3}
\ee\vs\nin
As we can see, it will be always the case that fermions couple to the
Higgs boson with the strength $m_\psi/v$. Thus, the heavier the
fermion, the stronger its coupling to the Higgs.

\nin
Finally, in order to have gauge invariant operators with up type right handed
quarks we need to use the operator
\vs\be
-\cL_{\rm m_u} =  \lambda_u \, \bar q_L \tilde\Phi\,u_R + {\rm h.c.}~,
\label{umass1}
\ee\vs\nin
where we defined
\vs\be
\tilde\Phi(x) = i\sigma^2\,\Phi(x)^* = \left(\ba{c}
  \frac{v+h(x)}{\sqrt{2}}\\
  ~\\0\ea\right)~,
\label{tildephidef}
\ee\vs\nin
where in the last equality we are using the unitary gauge. It is
straightforward\footnote{Only need to use that $\sigma^2 \sigma^2=1$,
  and that $\sigma^2 (\sigma^a)^* \sigma^2 = -\sigma^a$.} to prove that $\tilde\Phi(x)$ is an $SU(2)_L$ doublet
with $Y_{\tilde\Phi}=-1/2$, which is what we need so as to make the
operator in (\ref{umass1}) invariant under $U(1)_Y$. Then we have
\vs\be
-\cL_{\rm m_u} = 
m_u\,\bar u_L u_R +
\frac{m_u}{v}\,h(x)\,\bar u_L u_R + {\rm h.c.}~, 
\label{umass2}
\ee\vs\nin
with
\vs\be
m_u = \lambda_u\,\frac{v}{\sqrt{2}}~.
\ee\vs\nin
The fermion Yukawa couplings are parameters of the SM. In fact, since
there are three families of quarks their Yuakawa couplings are in
general a non diagonal three by three matrix. This fact has important
experimental consequences. On the other hand, we could imagine having
something similar if we introduce a right handed neutrino. This
however, might be beyond the SM, since this state does not have any
SM gauge quantum numbers.
\nin
Overall, the SM is determined by the parameters $v, g, g' $ and
$\sin\theta_W$ in the electroweak gauge sector, plus all the Yukawa
couplings in the fermion sector leading to all the observed fermion
masses and mixings.

\subsubsection{Fermion mixing}
\label{sec:fermionmix}

In the previous section, we considered the fermion masses arising from
Yukawa couplings assuming only one generation of fermions. But instead
of (\ref{emass1}), (\ref{dmass1}) and (\ref{umass1}), the most general
interactions of fermions with the Higgs doublet can be written as
\begin{equation}
-{\cal L}_{HF} = \lambda_u^{ij} \,\bar q_{L,i}\tilde\Phi u_{R,j}+
\lambda_d^{ij} \,\bar q_{L,i}\Phi d_{R,j} + \lambda_\ell^{ij} \,\bar
\ell_{L,i}\Phi \ell_{R,j}~,
\label{yukawa3gen}
\end{equation}  
where  $(i,j)=1,2,3$ are generation indices, we denote the quark
and lepton three generation doublets as $q_{L,i}$ and $\ell_{L,i}$
respectively, and similarly with the right handed fermions $u_{R,i}$,
$d_{R,i}$ and $\ell_{R,i}$.  The Yukawa couplings now are $3\times 3$
matrices in flavor space: $\lambda_u^{ij}$, $\lambda_d^{ij} $ and
$\lambda_e^{ij} $.  These matrices are generally non diagonal and
complex. Therefore, so are the mass matrices
\begin{equation}
M_u^{ij} = \lambda_u^{ij} \,\frac{v}{\sqrt{2}}, \qquad M_d^{ij} =
\lambda_d^{ij} \,\frac{v}{\sqrt{2}}, \qquad M_\ell^{ij} =
\lambda_\ell^{ij} \,\frac{v}{\sqrt{2}}~.
\label{massmatrices}
\end{equation}
These matrices need to be diagonalized by unitary transformation on
the fermion fields. For instance, for the up quark mass matrix we want
\begin{equation}
M_u^{\rm diag.} = \left(\begin{array}{ccc}m_u & 0 & 0\\0 & m_c & 0\\0 & 0 & m_t\end{array}\right), \dots~,
\end{equation}
where the eigenvalues above are the physical (real) masses of the up
type quarks,  and similarly for $M_d^{\rm diag.}$ and $M_\ell^{\rm diag.}$.

  We now concentrate on the quark sector. A similar procedure can be
  followed in the lepton sector~\cite{renata}.  The quark mass terms
  before diagonalization are
  \begin{equation}
    -{\cal L}_{\rm mass} = \bar u_L^i \,M_u^{ij}\,u_R^j
    + \bar d_L^i \,M_d^{ij}\,d_R^j  + {\rm h.c.}~,
  \end{equation}
To obtain diagonal mass matrices, we define four unitary
transformations acting separately on left and right handed up and down
type quarks. These are
\begin{eqnarray}
u_L \to S_L^u\,u_L &&\qquad\qquad u_R\to S_R^u\,u_R \nonumber\\
  \qquad d_L \to S_L^d\,d_L &&\qquad\qquad d_R\to S_R^d\,d_R~.
\label{unitarys}                               
\end{eqnarray}
We choose these quark field unitary transformations such that they
satisfy
\begin{equation}
 M_u^{\rm diag.} =(S_L^u)^\dagger \,M_u\, S_R^u \qquad {\rm and }
 \qquad  M_d^{\rm diag.} =(S_L^d)^\dagger \,M_d\, S_R^d~.
\end{equation}
At the same time that 
the quark field rotations above not diagonalize the mass matrices, it
also does so with the Yukawa 
couplings of the Higgs bosons to fermions, which are diagonal and in
fact given by
\begin{equation}
\frac{m_f}{v}. 
\end{equation}
However, we should rotate the quark fields appearing in the vector
currents, both neutral and charged.

\nin Let us first consider the {\bf neutral currents}. 
Since vector currents do not
change chirality, we always have
\begin{equation}
\bar u_L\gamma^\mu u_L  \qquad {\rm or}\qquad \bar u_R\gamma^\mu u_R ~,
\end{equation}
or alternatively,
\begin{equation}
\bar d_L\gamma^\mu d_L  \qquad {\rm or}\qquad \bar d_R\gamma^\mu d_R ~.
\end{equation}
But these currents are clearly invariant under the unitary
transformations in (\ref{unitarys}), since they involve the product of
a unitary transformation and its hermitian adjoint, i.e. its inverse. 
We then conclude that {\bf in the SM there are no flavor changing neutral
currents (FCNC) at leading order in perturbation
theory}.\footnote{FCNC can be generated in the SM at one loop
order. We will see this briefly below, but in much more  detail in~
\cite{matthias} .}

\nin We now consider the quark {\bf charged currents} . Their
contribution to the Lagrangian is given by

\begin{equation}
{\cal L}_{\rm ch.}=\frac{g}{\sqrt{2}}\,\bar u_L^i\gamma^\mu d_L^i \, W^+_\mu
+ {\rm h.c.}~,
\end{equation}
where the repeated flavor index is summed over, and the fields above
are those before the diagonalization of the mass matrices. But once we
applied the different field transfomations on $u_L^i$ and $d_L^i$
defined in (\ref{unitarys}), the charged current becomes
\begin{eqnarray}
{\cal L}_{\rm ch} &\to& \frac{g}{\sqrt{2}} ((S_L^u)^\dagger
                        S_L^d)_{ij} \,\bar u_L^i\gamma^\mu d_L^j \,
                        W^+_\mu  + {\rm h.c.} \nonumber\\
  &=&\frac{g}{\sqrt{2}} (V_{\rm CKM})_{ij} \,\bar u_L^i\gamma^\mu d_L^j \, W^+_\mu
      + {\rm h.c.}~,
      \label{charged}
\end{eqnarray}
where we defined the Cabibbo-Kobayashi-Maskawa (CKM) matrix as
\begin{equation}
V_{\rm CKM}\equiv (S_L^u)^\dagger\, S_L^d~.
\end{equation}
The CKM matrix is non-diagonal which results in generation-changing
charged currents. As an example, Figure~\ref{fig:ckmvertices} shows
the possible charge current vertices involving an up quark. Not only
can it go to a first generation anti-down quark, but also --thanks to
the CKM matrix being non-diagonal -- it can go to an anti-strange or
an anti-bottom quarks. As indicated in (\ref{charged}), each of these
vertices is accompanied by a factor of the corresponding CKM matrix
element: in this case  $V_{ud}$, $V_{us}$ and $V_{ub}$.    
\begin{figure}
  \begin{center}
\includegraphics[width=5in]{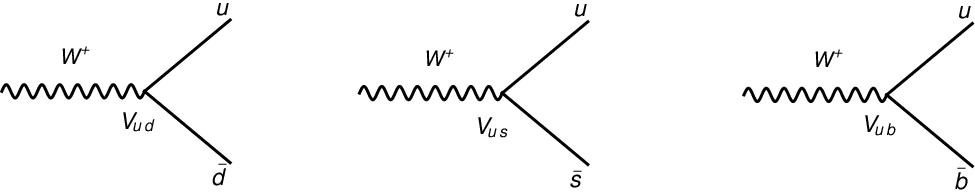}
\caption{Charged current vertices with an up quark. In addition to the
generation-conserving vertex with a down quark, the CKM matrix allows
the generation-changing vertices with the strange and bottom quarks.}
\label{fig:ckmvertices}
\end{center}
\end{figure}
The conjugate vertices involving a $W^-$ and a $\bar u$ quark, are
multiplied by the complex conjugate of the CKM matrix elements
mentioned above. Of course, similar vertices can be obtained for the
other up-type quarks, $c$ and $t$.

These charged current vertices and the CKM matrix are at the heart of
a wealth of phenomena that we typically call  quark ``flavor
physics''. Not only are behind the typical (unsuppressed, leading
order) decays of heavier quarks, but also enter crucially in the loop
generated FCNC in the SM, such $b\to s\gamma$, as well as $b\to
s\ell^+\ell^-$, $K^0-\bar{K^0}$, $D^0-\bar{D^0}$ and $B^0-\bar{B^0}$
mixing among others.

\nin Of all the possible phases in $V_{\rm CKM}$ all but one can be
removed by fields redefinitions. This leads to the phenomenon of CP
violation, which was first observed in kaon decays in 1964, and was
further observed in $B$ meson decays, leading to a precise mapping of
the CKM matrix elements and phase structure. 
A detailed presentation of these topics can be
found in \cite{matthias}. A similar application to leptons is in the
lectures  of Ref. \cite{renata}.

\section{Testing the electroweak Standard Model}
\label{sec:testing}

Now that we know how all the particles in the electroweak SM couple to
each other, we can turn to testing the SM. In this lecture we 
review the past, present and future tests of the various sectors of the
SM that consolidated our understanding of particle physics in the last
decades. We divide this in three distinct parts: testing the couplings
of fermions to gauge bosons, the gauge boson self-couplings and
finally the Higgs couplings to all particles in the SM. However, due
to the high precision  the experimental tests  have achieved, we need
to match this with theoretical precision. This requires that, in many
cases we need to go beyond leading order calculations in order to make
predictions in the EWSM that can be meaningfully tested by these
experiments.  This forces us to introduce one more aspect of the
quantum field theory tool box: renormalization. We start with a brief
summary of renormalization and its applications to some of the
electroweak observables of interest. Then we move to the tests of the
electroweak SM.

\subsection{Renormalization}
\label{sec:renorma}

\nin
Virtual processes in quantum field theory will modify the parameters
of a theory, i.e the parameters in the Lagrangian. In perturbation
theory these contributions are ordered by an expansion parameter,
typically a coupling constant, in order to have a controlled
approximation. For instance, in a theory with a real scalar with the
Lagrangian given by
\vs\be
\cL = \frac{1}{2}\,\del_\mu\phi\del^\mu\phi -\frac{1}{2}\,m^2\,\phi^2
-\frac{\lambda}{4!}\phi^4~,
\label{phi4191}
\ee\vs\nin 
the two-point function to order $\lambda$ admits the one-particle
irreducible diagrams (1PI) shown in Figure~\ref{fig:191}.
\vs
\begin{figure}[h]
\begin{center}
    \includegraphics[width=3in]{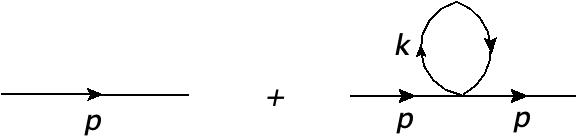}
\end{center}
\caption{
1PI diagrams contributing to the two-point function in 
the theory with Lagrangian (\ref{phi4191}), to order $\lambda$.
}
\label{fig:191}
\end{figure}
\vs\nin      
The first diagram is the free propagator. The second one gives a
contribution to the two-point function that must be integrated over
the undetermined four-momentum $k$, and is
\vs\be
\frac{(-i\lambda)}{2}\,\int\frac{d^4k}{(2\pi)^4}\,\frac{i}{k^2-m^2+i\epsilon}~,
\label{loop1}
\ee\vs\nin 
where the factor of two is due to the symmetry of the diagram. The
need for the integration is a consequence of the momentum conservation
at the vertex and is consistent with the quantum mechanical character
of the computation: all possible values of the four-momentum $k$
contribute to the amplitude, such as we saw in the first lecture when
deriving the Feynman rules. 
The contribution form (\ref{loop1}) will result in a shift of the
two-point function. It will change the position of the pole of the
propagator through a shift $\delta m^2$ in the parameter $m^2$ in
(\ref{phi4191}), and will change the residue at the pole. The latter
will be absorbed by a redefinition of the field $\phi(x)$ itself.

\nin
In addition to shifting the parameters of the theory entering in the
two-point function, 
the one-loop diagram of Figure~\ref{fig:191} diverges for large values
of the momentum. This is a consequence of the fact that the momentum
integration is not limited. This is an example of an ultra-violet (UV)
or high momentum divergence.\footnote{There also infra-red (IR)
  divergences, or low momentum divergences. We will focus here solely
  on UV divergences.}  We can also think of the UV divergence as a
consequence of taking a distance to zero. It is interesting to look
closely at the UV limit of the integral in (\ref{loop1}). To this
effect, we define the Euclidean
four-momentum by
$
k_0 \to ik_4\Longrightarrow \quad k^2=k_0^2-\mbf{k}^2 = -k_4^2
-\mbf{k}^2\equiv -k_E^2~,
$
such that now the integral in (\ref{loop1}) can be written in terms of
the 4D Euclidean momentum $k_E$ as
\vs\be
\frac{(-i\lambda)}{2}\,\int\frac{d^4k_E}{(2\pi)^4}\,\frac{1}{k_E^2+m^2}
=\frac{(-i\lambda)}{2}\,\int\frac{dk_E\,k_E^3 \,d\Omega_E}{(2\pi)^4}\,\frac{1}{k_E^2+m^2}~,
\ee\vs\nin
where the 4D solid angle is $\Omega_E = 2\pi^2$.\footnote{You may need
  to think a bit about this. We will derive a general expression later
  on.} Finally, the
remaining Euclidean momentum integral can be cutoff at some value
$\Lambda$ giving 
\vs\be
\frac{(-i\lambda)}{16\pi^2}\,\int_0^\Lambda\frac{dk_E\,k_E^3}{k_E^2+m^2}\simeq
-i\frac{\lambda}{32\pi^2}\,\Lambda^2 +\dots~,
\label{quaddiv19}
\ee\vs\nin
where the dots denote terms diverging with less than two powers of
$\Lambda$, or terms that are finite after the limit $\Lambda\to\infty$
is taken. In this example, we say that this quantity is quadratically
sensitive to the UV cutoff $\Lambda$.

\nin
Similarly, the 1PI contributions to the four-point function up to
order $\lambda^2$ include loop diagrams that result in a quantum
correction of the sort given by
\vs\bear
&&\frac{-i\lambda)^2}{2}\int \frac{d^4k}{(2\pi)^4}
\frac{i}{k^2-m^2}\,\frac{i}{(p-k)^2-m^2}~,\nonumber\\
&=&\frac{i\lambda^2}{16\pi^2}\int_0^\Lambda \frac{dk_E
  k_E^3}{\left(k_E^2+m^2\right) \left(p-k_E)^2+m^2\right)} ~,\nonumber\\
&\simeq&\frac{i\lambda^2}{16\pi^2}\ln\Lambda^2 + \dots~,
\eear\vs\nin
which is logarithmically sensitive to the UC cutoff $\Lambda$. The UV
sensitivity is smaller since there is one extra propagator with
respect to (\ref{loop1}). This will result in shifts to the coupling
$\lambda$.
UV divergences like these are always present in relativistic quantum
field theory.  They come from the fact that undetermined momenta can
be as large as possible, or the distance between any two positions in
spacetime can be made as small as possible. Although their presence
requires care, it is still possible to define the changes in the
theory due to the quantum corrections in loop diagrams. The process of
regularizing divergences is part of the renormalization procedure. 
Renormalization redefines all the parameters of a theory in the
presence of interactions. That is, as in our example, redefinitions of
$m$, $\lambda$ and the field $\phi(x)$ itself. 
In what follows, we describe a well defined method for absorbing the
UV sensitivity in loops into {\em counterterms}.

\subsubsection{Renormalization by counterterms}
\label{sec:ctrenorma}

\nin
Starting from the Lagrangian for a real scalar field with quartic
self-interactions
\vs\be
\cL = \frac{1}{2}\del_\mu\phi_0 \del^\mu\phi_0 -\frac{1}{2}m_0^2
\phi_0^2 -\frac{\lambda_0}{4!}\phi_0^4~,
\label{unrenlphi421}
\ee\vs\nin  
with unrenormalized parameters $m_0^2$, $\lambda_0$ and the
unrenormalized field $\phi_0$, we defined the renormalized parameters
$m^2,\lambda$ and $\phi$. First, we start by the field, just as we did
in the previous lecture.
\vs\be
\phi = Z_\phi^{-1/2}\,\phi_0~.
\label{renphi21}
\ee\vs\nin
Then, we rewrite (\ref{unrenlphi421}) replacing the renormalized field
$\phi$ for $\phi_0$ to obtain
\vs\be
\cL = \frac{1}{2}Z_\phi\del_\mu\phi \del^\mu\phi -\frac{1}{2}m_0^2Z_\phi
\phi^2 -\frac{\lambda_0}{4!}Z_\phi^2\phi^4~,
\label{renlagphi21}
\ee\vs\nin  
We now define
\vs\bear
\delta Z_\phi&\equiv& Z_\phi -1\nonumber\\
\delta m^2 &\equiv & m_0^2Z_\phi -m^2\nonumber\\
\delta\lambda &\equiv & \lambda_0 Z_\phi^2 -\lambda~,\label{ctdef1}
\eear\vs\nin
which we can rewrite in a more convenient way as 
\vs\bear
Z_\phi&=& 1+\delta Z_\phi \nonumber\\
 m_0^2 Z_\phi &= & m^2 +\delta m^2\nonumber\\
\lambda_0 Z_\phi^2&= & \lambda +\delta\lambda~.\label{ctdef2}
\eear\vs\nin
Replacing (\ref{ctdef2}) in (\ref{renlagphi21}) we obtain
\vs\bear
\cL &=& \frac{1}{2}\del_\mu\phi\del^\mu\phi -\frac{1}{2}m^2\phi^2
-\frac{\lambda}{4!}\phi^4\nonumber\\
&+&  \frac{1}{2}\delta Z_\phi\,\del_\mu\phi\del^\mu\phi
-\frac{1}{2}\delta m^2\phi^2
-\frac{\delta\lambda}{4!}\phi^4~,
\label{lagwct1}
\eear\vs\nin
where we see that the first line is the renormalized Lagrangian
whereas the second line is what we will call the counterterms. These
new terms will result in  new Feynman rules for the theory and will
cancel divergencies in the renormalized theory. 
We have seen (see lecture 19) that for this theory the degree of
divergence of diagrams is given by
\vs\be
D =4-\sum_f E_f (s_f+1)
\ee\vs\nin
where $E_f$ is the number of external lines of the field type $f$ in
the diagram and here $s_f=0$ for a scalar field. This meant that there
are divergences in the two-point function ($E_f=2\Rightarrow$ $D=2$)
and in the four-point function ($E_f=4\Rightarrow$ $D=0$). The
divergences in the two-point function affect the terms in $\cL$
quadratic in the fields and there will be cancelled by the
counterterms $\delta Z_\phi$ and $\delta m^2$, whereas the ones in the four-point function 
impact the quartic term and are canceled by $\delta \lambda$.  The
cancellation takes place at a given order in the perturbative expansion
in the coupling constant $\lambda$. In order to define the physical
parameters we need to impose renormalization conditions. To compute a
given process up to some order in perturbation theory we need to use
the Feynman rules that include the counterterms. These new
contributions will ensure that the cancelation takes place in every
process.

\subsubsubsection{Counterterm Feynman rules}  

\nin
The Feynman rules of the theory in terms of
renormalized parameters are shown below, and derived
from the first line in (\ref{lagwct1}).
In addition to the tree-level Feynman rule,  we now need to derive new
rules from the second line. This results in 
\vs
\begin{minipage}{0.3\textwidth}
\vskip0.75cm
\hspace*{2cm}\includegraphics[width=1in]{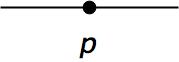}
\end{minipage}%
\begin{minipage}{0.7\textwidth}
\vs\be
i\left(\delta Z_\phi p^2-\delta m^2\right)~,
\label{propct1}
\ee
\end{minipage}
\vs
\vs\vs\vs
\begin{minipage}{0.3\textwidth}
\vskip0.75cm
\hspace*{2.5cm}\includegraphics[width=1in]{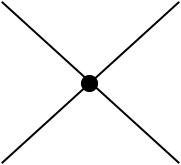}
\end{minipage}%
\begin{minipage}{0.7\textwidth}
\vs\be
-i\delta\lambda~,
\label{couplingct1}
\ee
\end{minipage}
\vs\vs\nin
where the dots indicate the insertion of the counterterm. To
understand the form of the counterterm for the two-point function we
should imagine inserting it as one more 1PI contribution to
$-i\Sigma(p^2)$ in the summed propagator, as we did in lecture
20. With the form (\ref{propct1}) the propagator now would be
\vs\be
\frac{i}{p^2 -m^2 -\Sigma_\ell(p^2) +\delta Z_\phi\, p^2 -\delta
  m^2}~,
\label{deltafwithcts}
\ee\vs\nin 
where $-i\Sigma_\ell(p^2)$ is the sum of the actual loop contributions
to the two-point function.  Notice that since the mass squared in the
propagator is already the renormalized mass, the divergences in
$-i\Sigma_\ell(p^2)$ will now be canceled exclusively by the
counterterms $\delta Z_\phi$ and $\delta m^2$.  
To implement the program of renormalization by counterterms, we
compute any desired amplitude up to the desired order in $\lambda$,
including all the counterterms. Divergent integrals must be regulated,
i.e. expressed in terms of an appropriate regulator that respects the
symmetries of the theory. In the next lectures we will specify
regularization procedures. But the regulator is typically either an
euclidean  momentum  cut off $\Lambda$, or some other parameter that
exposes the divergences in some limit. The answer of the calculation
initially depends on the couterterms $\delta Z_\phi$, $\delta m^2$ and
$\delta\lambda$.  These are fixed by imposing {\em renormalization
  conditions} that result in the cancellation of divergences. The
resulting expression is then independent of the regulator. This
procedure removes all divergences in a renormalizable theory.

\subsubsubsection{Fixing $\delta Z_\phi$ and $\delta m^2$}

\nin
These counterterms are fixed by the renormalization of the two-point
function. The 1PI diagrams that need to be summed in order to obtain
the     propagator (\ref{deltafwithcts})  now include the counterterm
contribution, as shown in Figure~\ref{fig:213}, where we show the 1PI
up to $\cO(\lambda)$. In addition to the one-loop diagram we now have
to consider the couterterm contribution to the two-point function as
in (\ref{propct1}) .
\begin{figure}[h]
\begin{center}
\includegraphics[width=4in]{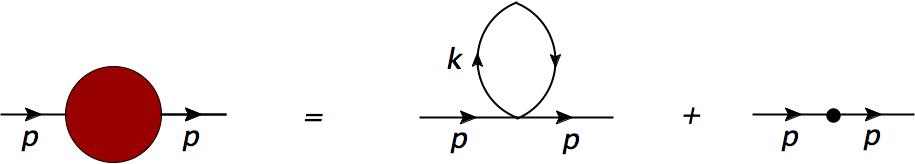}
\end{center}
\caption{
The  1PI  diagrams contributing to the two-point
function to $\cO(\lambda)$. The last diagram corresponds to the
counterterm in (\ref{propct1}).
}
\label{fig:213}
\end{figure}
\nin   
The sum of the two diagrams is
\vs\be
-i\Sigma(p^2) =
\frac{(-i\lambda)}{2}\int\frac{d^4k}{(2\pi)^4}\,\frac{i}{k^2-m^2+i\epsilon}  +
i\left(\delta Z_\phi p^2-\delta m^2\right)~.
\label{sigmatot1}
\ee\vs\nin
We will impose the renormalization conditions on the propagators
\vs\be
\Delta_F(p) = \frac{i}{p^2-m^2-\Sigma(p^2)}~,
\label{propwsigma}
\ee\vs\nin
we now we use the renormalized mass parameter  $m^2$ from the
Lagrangian in  (\ref{lagwct1}) and $\Sigma(p^2)$ has the expansion
\vs\be
\Sigma(p^2) = \Sigma(m^2) + (p^2-m^2)\,\Sigma'(m^2) +
\tilde{\Sigma}(p^2)~,
\label{sigmaexp} 
\ee\vs\nin
where the first two terms are divergent, but the last is not.
Now, the renormalization conditions are a little different than
before
because here we are adding the contributions of the loop plus those of
the counterterms and ger the {\em renormalized } propagator. 
This means that the renormalization condition now should leave $m^2$
as the pole {\em and} the residue should be unity times $i$, since the field is
already renormalized. This translates into the conditions 
\vs\be
\Sigma(m^2) =0, \qquad\qquad \Sigma'(m^2) =0,
\label{renconds}
\ee\vs\nin
with the first condition ensuring that $m^2$ is the pole of the
propagator, whereas the second one leads to the desired residue of
$i$. We can see from (\ref{sigmatot1}) that,  since the loop integral does not contain any $p^2$ dependence, 
the second renormalization condition in (\ref{renconds}) leads to 
$\delta Z_\phi=0$. However, this is only the case at this order in
$\lambda$. In fact, going to $\cO(\lambda^2)$ there will be such
dependence in the integral, leading to the more accurate statement
\vs\be\boxed{
\delta Z_\phi = 0 + \cO(\lambda^2)}~.
\label{deltazphi}
\ee\vs\nin 
Finally, we may use the first condition in (\ref{renconds}) in
(\ref{sigmatot1}) to obtain the mass squared counterterm
\vs\be\boxed{
\delta m^2 = -\frac{\lambda}{2}\int \frac{d^4k}{(2\pi)^4)}
\,\frac{i}{k^2-m^2+i\epsilon}}~.
\label{dm21}
\ee\vs\nin 
To actually compute $\delta m^2$ we will need to regulate the
integral above. We will do this in detail in the next two lectures. In any case, the answer will not depend on the details of
the regularization procedure.

\subsubsubsection{Fixing $\delta\lambda$ through the four-point function}  
\vs
\nin
The renormalization of the four-point function leads to the fixing of
the coupling counterterm $\delta\lambda$. In this case the first loop
corrections will introduce a momentum dependence absent at leading
order. Let us consider a scattering process in the $\phi^4$
theory up to one loop. The relevant diagrams are shown in
Figure~\ref{fig:214}.
\begin{figure}[h]
\begin{center}
\includegraphics[width=5in]{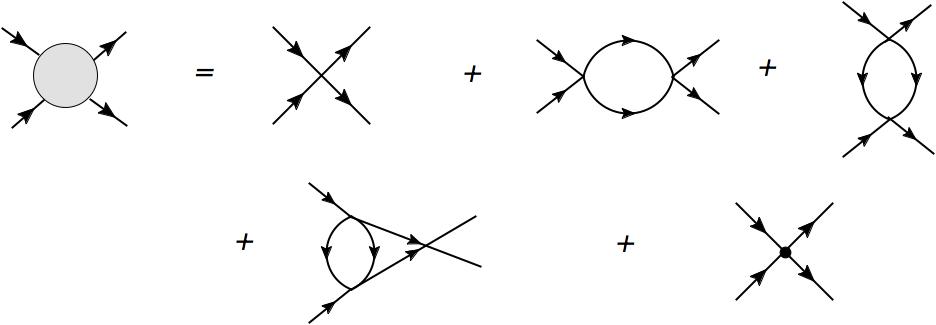}
\end{center}
\caption{
The  diagrams contributing to the four-point
amplitude. The leading order, i.e. $\cO(\lambda)$ diagram is followed
by the three possible $\cO(\lambda^2)$ 1PI diagrams.  The last diagram is the counterterm $\delta\lambda$. 
}
\label{fig:214}
\end{figure}
\nin   
 The amplitude for scattering two scalars of momenta $p_1$ and $p_2$
 into two scalars of momenta $p_3$ and $p_4$ is
\vs\be
i{\cal A}(p_1,p_2\to p_3,p_4) = -i\lambda + \Gamma(s) + \Gamma(t) +
\Gamma(u) -i\delta\lambda~,
\label{scaamp1}
\ee\vs\nin
where the Mandelstam variables are  $s=(p_1+p_2)^2$, $t=(p_3-p_1)^2$
and $u=(p_4-p_1)^2$. Since the loop diagrams introduce kinematic
dependence, we need once again to choose a point in order to impose
the renormalization condition on the four-point function. This time we
choose the zero-momentum condition, i.e.
\vs\be
s_0=4m^2,\qquad t_0=0,\qquad u_0=0~,
\label{kinpoint}
\ee\vs\nin 
which corresponds to $p_1=p_2=(m,\mbf{0})$. We then impose the
renormalization condition
\vs\be
i{\cal A}(s_0,t_0,u_0) = -i\lambda~,
\label{fprencond}
\ee\vs\nin
which results in 
\vs\be\boxed{
\delta_\lambda = -i\left(\Gamma(4m^2) + 2\Gamma(0)\right)}~.
\label{deltalam21}
\ee\vs\nin
We can then rewrite the amplitude as 
\vs\be
i{\cal A}(s,t,u) = -i\lambda + \tilde{\Gamma}(s) + \tilde{\Gamma}(t)
+\tilde{\Gamma}(u)~,
\label{renamp21} 
\ee\vs\nin
where the $\tilde{\Gamma}$'s are finite and satisfy
$\tilde{\Gamma}(s_0)=\tilde{\Gamma}(t_0)=\tilde{\Gamma}(u_0)=0$.
The amplitude in (\ref{renamp21}) is expressed in terms of the
renormalized coupling $\lambda$ and it has a well defined kinematic
dependence acquired at order $\lambda^2$ through the finite parts of
the loop diagrams. 

\nin
We see in this example the full extent of the renormalization
procedure. The renormalization condition is used in order to remove the
UV sensitivity of the parameter $\lambda$. But this is done at one
specific, arbitrarily chosen, kinematic point defined by
(\ref{kinpoint}).  Once this is done, the amplitude and its {\em
  dependance} on $s$, $t$ and $u$ are obtained. This is a physical
result: the amplitude up to one loop in perturbation theory contains
the physical kinematic dependance induced by the quantum
corrections. The renormalization condition fixes the value of the
amplitude at (\ref{kinpoint}) and removes the UV sensitivity in the
process. The same is true of the other parameters of the theory. Thus,
despite the appearance  of divergences, the renormalization procedure
yields physically meaningful predictions in perturbation theory coming
from the quantum corrections.

\subsection{Electroweak precision constraints and fermion couplings to gauge bosons}
\label{sec:fermions2gauge}

We start by considering the low energy charged and neutral current
effective Lagrangians. The weak charged current Fermi effective Lagrangian
in (\ref{fermi1}) can be generalized as  as
\be
{\cal L}_{\rm ch.} = - \frac{G_F}{\sqrt{2}}   \bar f\gamma_\mu\left(
  1-\gamma_5\right)f\,\bar f'\gamma^\mu\left(1-\gamma_5\right)f' ~.
\label{eq:charged}
\ee
On the other hand, the weak neutral current results from integrating
out the $Z^0$ and is given by
\be
{\cal L}_{\rm nt.} = - \rho_0\frac{G_F}{\sqrt{2}}   \bar f\gamma_\mu\left(
  g^{(f)}_{v,0}-g^{(f)}_{a,0}\gamma_5\right)f\,\bar f'\gamma^\mu\left(g^{(f')}_{v,0}-g^{(f')}_{a,0}\gamma_5\right)f' ~~,
\label{eq:neutral}
\ee
where the $0$ subscript denotes the unrenormalized or tree-level
quantities, and the vector and axial-vector couplings are obtained
from the left and right handed couplings to the $Z^0$ in
Section~\ref{sec:gaugefermion} and are given by
\be
g^{(f)}_{v,0} = t^{3}_f -2 Q_f s^2_{W,0} \qquad g^{(f)}_{a,0}
= t^3_f~,
\label{eq:vectoraxial}
\ee
where $t^3_f$ is the eignevalue of the third component of isospin for
the fermion $f$. E.g. $t^3_\nu =+1/2$ , $t^3_{e^-}=-1/2$, etc. Notice
that we defined the tree-level Weinberg angle above, in anticipation
of the renormalization processs. Finally, in (\ref{eq:neutral}) we
defined the tree-level $\rho$ parameter as
\be
\rho_0 \equiv \frac{1}{c^2_{W,0}} \frac{M_W^2}{M_Z^2}~,
\label{rhodef}
\ee
which measures the relative strengths of the weak neutral to the charged
effective Lagrangians (\ref{eq:neutral}) and (\ref{eq:charged}). In
the SM, the tree-level value is $\rho_0^{\rm SM}=1$. However, this is
not necessarily the case in extensions beyond the SM, and it certainly
is not the case in the SM beyond tree-level. In particular, since the
measurements of electroweak observables has achieved such a large
precision, it become necessary to go beyond tree level in order to
compare experimental values with the SM predictions. 

\nin
As a first step, let us write the {\em renormalized} vector and
axial-vector couplings as
\be
g^{(f)}_{v,0}\to g^{(f)}_v = \sqrt{\rho_f} \,\left(t^3_f -2 \kappa_f
  \,s^2_W\,Q^{(f)}\right)~,\qquad\qquad
  g^{(f)}_{a,0}\to g^{(f)}_a = \sqrt{\rho_f} \,t^3_f~,
  \label{rengvga}
\ee
where we defined the non-universal factors $\rho_f$ in such a way that
they absorb the renormalized universal overall factor of
$\rho_0\to\rho$ but also allows for non-universal vertex corrections
specific to $f$. Also defined in (\ref{rengvga}) above is the factor
$\kappa_f$, which is unity at tree level, but when considering quantum
corrections it changes the relationship between the two terms in
$g^{(f)}_v$. As we will see below, it is possible to re-interpret
$\kappa_f$ as a renormalization of the effective weak angle. The
corrections defined by (\ref{rengvga}) affecting the $Z\to f\bar f$
couplings  are some of the leading electroweak corrections. Others are
the running of the QED coupling $\alpha(\mu)$, as well as the Fermi
constant $G_F$. They all are integral part of the precision
electroweak constraints. 

\nin
There have been a large number of tests of the electroweak
interactions at relatively low energies over the years. These include
deep inelastic neutrino scattering, atomic parity violation in cesium, 
as well as polarized M\"oller scattering. Although these measurements
were able to probe neutral currents with some precision, 
the most precise tests have come from the 
neutral current interactions at the $Z^0$ pole, both at LEP at CERN and at the SLD at
SLAC. 
This is due to the very large statistics achieved at the$Z^0$  pole, where
the $e+e^- \to Z^0\to f\bar f$ cross section is more than three orders
of magnitude larger than that of photon exchange. 

\nin
In order to analyse the experiments at the $Z$ pole it has become
customary to use the so-called {\em effective description} of the
renormalized vector and axial-vector couplings in
(\ref{rengvga}). This is defined by
\be
\bar g_v^{(f)} = \sqrt{\rho_f}\,\left(t^3_f -2\,\bar
  s^2_f\,Q^{(f)}\right)~,
\quad\quad
\bar g_a^{(f)} = \sqrt{\rho_f}\, t^3_f~,
\label{effgvga}
\ee
where the bars on top refers to the use of the effective
renormalization scheme, which uses $\mu=M_Z$. In this scheme, the
effective renormalized weak mixing angle depends on the fermion flavor
$f$ and its defined as
\be
\bar s^2_f \equiv  \kappa_f s^2_W~,
\label{defbarsw}
\ee
which can be extracted by measuring $\bar g_v/\bar g_a$. This is done
by measuring asymmetries at the $Z$ pole. In particular it is
convenient to define the fermion $f$ asymmetry parameter
\be
{\cal A}_f \equiv 2\frac{\bar g_v^{(f)}\,\bar g_a^{(f)}}{\bar
  g_v^{(f)2} + \bar g_a^{(f)2}} = 2\frac{\bar g_v^{(f)}/\bar g_a^{(f)}
}{1+\left(\bar g_v^{(f)}/\bar g_a^{(f)}\right)^2 }~,
\label{asymparam}
\ee
which can be extracted from the angular data. We start by defining
the forward and backward cross sections
\be
\sigma_F =2\pi\int_0^1 d\cos\theta
\frac{d\sigma}{d\Omega}~,\qquad\qquad
\sigma_B =2\pi\int_{-1}^0 d\cos\theta
\frac{d\sigma}{d\Omega}~. 
\label{fbsigma}
\ee
Also useful in the context of $e^+e^- \to f \bar f$ at the $Z^0$ pole
are the cross sections $\sigma_L$ and $\sigma_R$ denoting the case of
a left-handed and right-handed electron beam, respectively. Then,
defining the asymmetries at the $Z^0$ pole we can have:
\be
A_{FB}\equiv \frac{\sigma_F -\sigma_B}{\sigma_F+\sigma_B}~,
\label{afbdef}
\ee
which is the forward-backward asymmetry;
\be
A_{LR} \equiv \frac{\sigma_L-\sigma_R}{\sigma_L+\sigma_R}~,
\label{alrdef}
\ee
which defines the left-right asymmetry.
For instance, in the presence of electron polarization ${\cal P}_e$ we
have that
\be
A_{FB}^{(f)} = \frac{3}{4}\,{\cal A}_f \,\frac{{\cal A}_e +{\cal
    P}_e}{1+{\cal A}_e {\cal P}_e}~,
\label{afbdef2}
\ee
and
\be
A_{LR} = {\cal A}_e \,{\cal P}_e~.
\label{alrdef2}
\ee

But before we go into the details of the electroweak precision data
and the derived constraints on the SM, we must discuss the various
possible definitions of the weak mixing angle. This variety appears
when going beyond tree level and reflects the various possible
renormalization conditions imposed. We have already introduced the
{\em effective } mixing angle, $\bar s_f^2$ in (\ref{defbarsw}). 
It can be directly determined by experimentally measuring the vector
to axial ratio $\bar g_v/\bar g_a$ in asymmetries, such as
$A_{FB}^{(f)}$ in (\ref{afbdef2}) and (\ref{asymparam}).
Alternatively, we can define the modified minimal substruction scheme ($\overline{\rm MS}$) weak mixing angle, $\hat
s_W^2$, a renormalization scale dependent quantity, as
\be
\hat s_W^2(q^2) \equiv \frac{e^2(q^2)}{g^2(q^2)}~,
\label{defmsbarsw}
\ee
and that can be implemented by using dimensional regularization for
the renormalization of the couplings above. Finally, we can also
define the {\em on-shell} weak mixing angle as
\be
s^2_W \equiv 1- \frac{M_W^2}{M_Z^2}~,
\label{defonshellsw}
\ee
which is given in terms of the physical masses of the $W$ and the $Z$
and is therefore directly determined experimentally from the
measurements of $M_W $ and $M_Z$. All these definitions of the weak
mixing angle agree at tree-level. It is possible to relate the
different weak mixing angles at a given order in perturbation
theory. For instance, we have
\be
\bar s_\ell^2 =\hat\kappa_\ell\,\hat s^2_W(M_Z^2)
\simeq \hat s^2_W(M_Z^2) + 0.00032~, 
\ee
where the $\overline{\rm MS})$ weak mixing angle is evaluated at the
$Z$ pole, $q^2=M_Z^2$. The great experimental precision obtained at
the $Z$ pole both at LEP 1 and at SLD requires great precision in the
loop corrections. Here, $\bar s_\ell^2$ must be computed at full
two-loop precision, as well as partial higher orders. Extracting the
ratios of vector to axial vector couplings from asymmetry measurements,  and the sum of their squares from the decay widths, i.e. 
\be
\Gamma(Z\to f\bar f) = \eta_f\,\frac{N_c}{6\pi}\,\frac{G_F
  M_Z^3}{\sqrt{2}}\left(\bar g_v^2 +\bar g_a^2\right)~,
\label{ztoffbar}
\ee
where $N_c=3$ and $\eta_f=\delta_{\rm QCD}$ if $f$ is a quark, or both are
unity otherwise, and
\be
\delta_{\rm QCD} = 1 +\frac{\alpha_s(M_Z^2)}{\pi} +
1.41\left(\frac{\alpha_s(M_Z^2)}{\pi}\right)^2 +\dots \simeq 1.04,
\ee
it is possible to measure the effective couplings for all SM
fermions.
 For instance , in
Figure~\ref{fig:gvga} we see the results of the measurements at LEP
and SLC at the Z pole for the lepton couplings. In the figure, we see
the combination of the three measurements assuming lepton universality
in the black contour. Also shown, is the SM best value show in the
black cross and according to the definition of the effective couplings
in (\ref{effgvga}).
\begin{figure}[h]
\begin{center}
\includegraphics[width=6in]{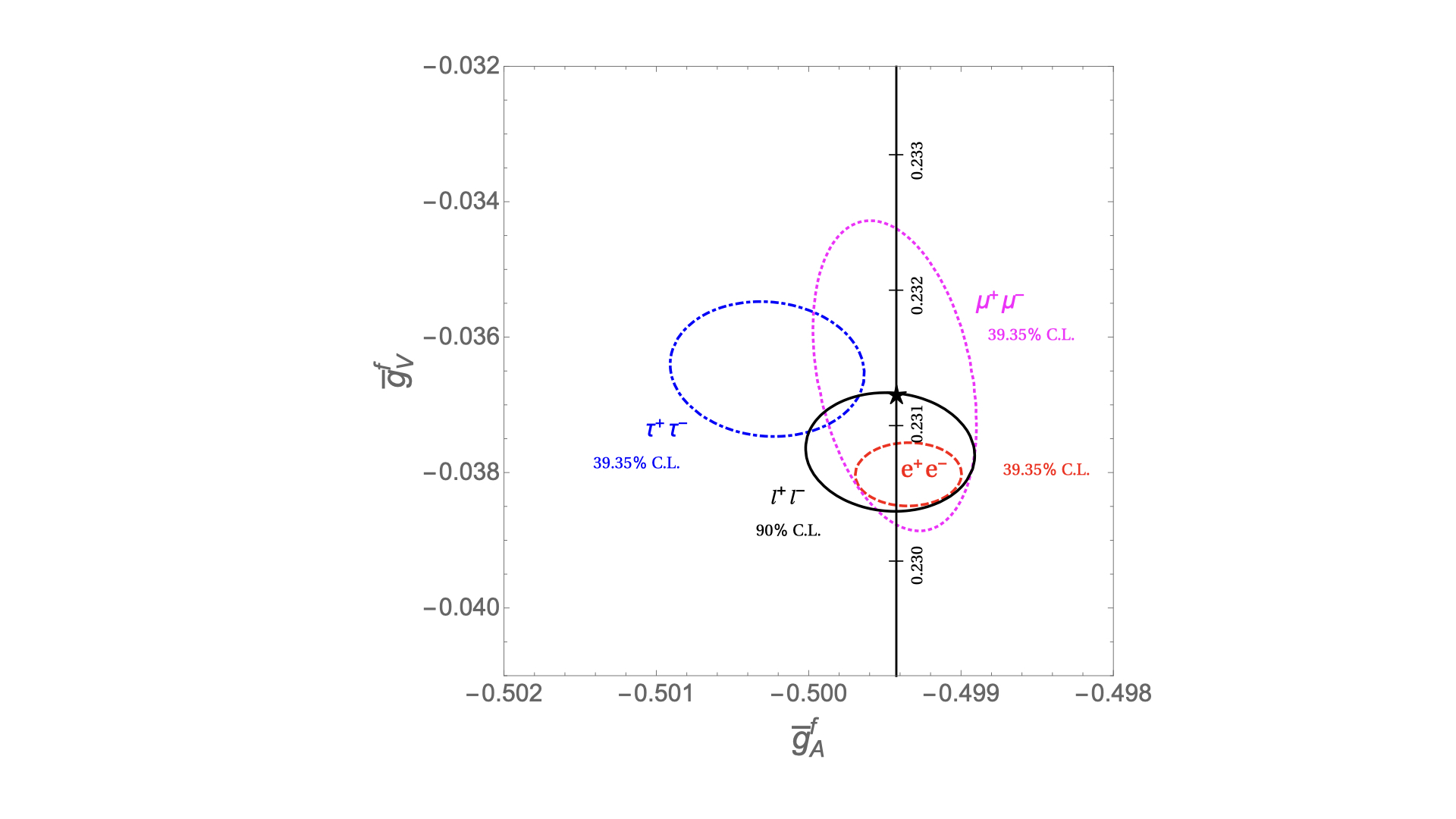}
\end{center}
\caption{$1 \sigma (39.35 {\rm C.L.})$ contours for the Z pole
  observables $\bar g^f_v$ and $\bar g^f_a$, for $f=e,\mu,\tau$
  obtained at LEP and at SLC, compared to the SM expectation as a
  function of $\hat s^2_Z$ , with the best value ($\hat
  s^2_Z=0.23122$) indicated. Also, in black, is the $90 \%$
  C.L. allowed region when assuming lepton universality. From \cite{rpp}. 
}
\label{fig:gvga}
\end{figure}
\nin   
Notice the agreement with the Z pole data requires the higher order
calculations named earlier. Many more tests of the gauge couplings to
fermions have been performed since, most notably at hadron colliders
such as the Tevatron and the LHC. From the figure, we can see that the
agreement in the effective coupling is at the sub-percent level. So
we can conclude that the electroweak gauge couplings to fermions in
the SM are tested with a great level of precision.
\nin
\begin{figure}[h]
\begin{center}
\includegraphics[width=6in]{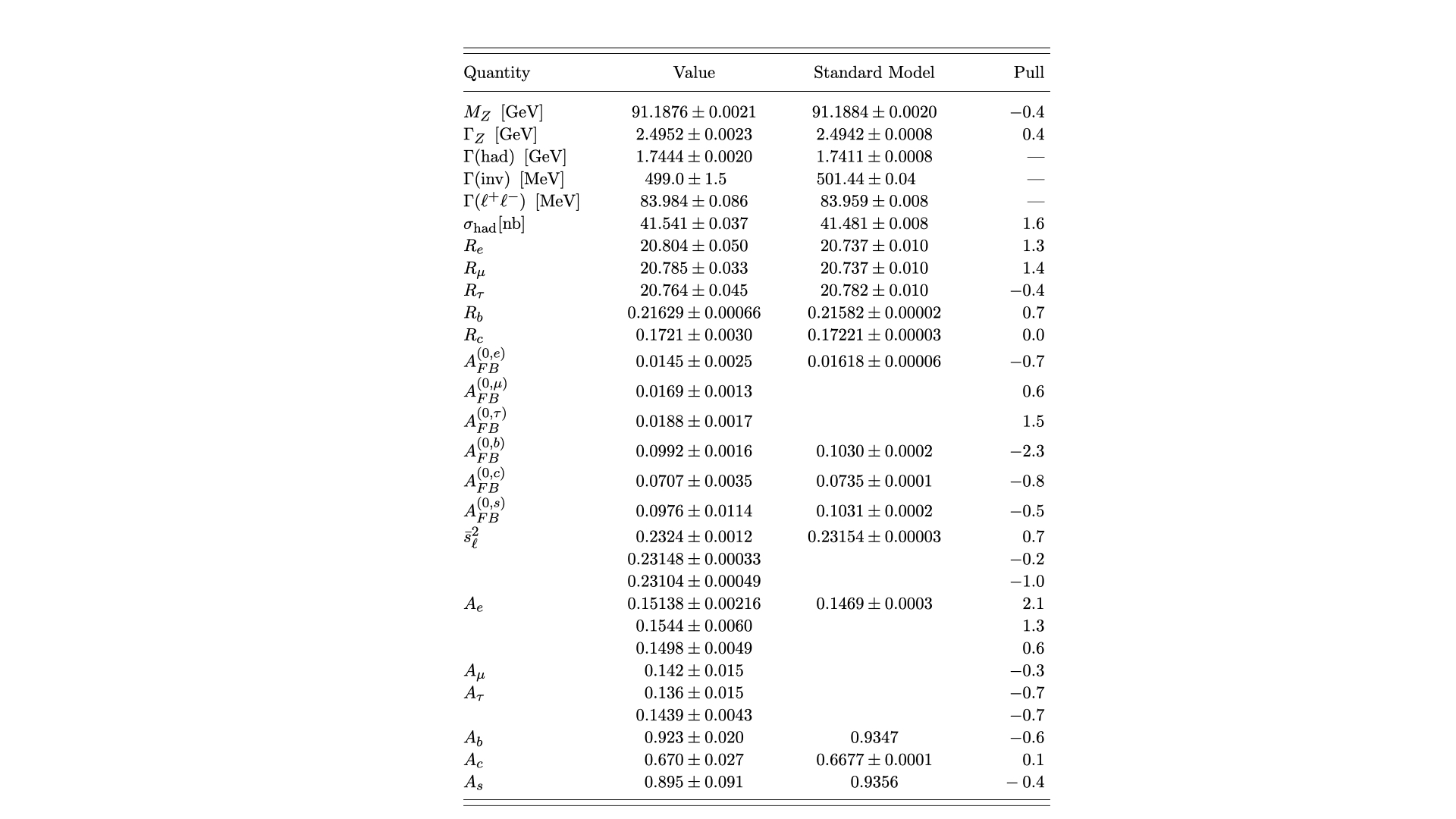}
\end{center}
\caption{ Fit of electroweak observables at the Z Pole. From \cite{rpp}. 
}
\label{fig:ewpo}
\end{figure}
\nin

To show the extent of this success, we close this section showing the
SM fit for electroweak observables at the Z pole,
Figure~\ref{fig:ewpo}.  The electroweak sector has three fundamental
parameters, i.e. $g$, $g'$ and $v$. However, it is advantageous to
use a combination of these that is measured with the greatest
precision. These are: $M_Z$, $G_F$ and $\alpha$. In addition, it is
necessary to incorporate the dependence of observables of the Higgs
mass $m_h$, the quark masses and mixings, and the strong coupling
$\alpha_s$, all entering through radiative corrections. In general we have the parameter set given by
\be
\{p\}\equiv \{\alpha, M_Z,G_F, \alpha_s,\lambda, m_h, m_t, \cdots\}~,
\label{smparameters}~.
\ee
\nin
As
we already discussed, simple tree level relations that only involve
$\alpha$, $G_F$ and $M_Z$, such as the $W$ mass 
\be
M_W = \cos\theta_W\,M_Z~,
\ee
or the Weinberg angle 
\be
\sin 2 \theta_W = \left(\frac{2\sqrt{2}\pi\alpha}{G_F M_Z^2}\right)^{1/2}~,  
\ee
will now be affected by loop corrections. These, in effect, will make
all the floating parameters depend of all others. That is, we can
write for some observable ${\cal O}_i$
\be
{\cal O}_i^{\rm theory}(\{p\}) = {\cal O}_i^{\rm tree-level}(\{\alpha,
G_F, M_Z\}) + \delta {\cal O}_i(\{p\})~,
\label{fitof}
\ee
where the $\delta{\cal O}_i$ are the loop corrections. Measuring a
large number of electroweak observables with large enough precision to
be sensitive to the loop corrections we can extract information on all
parameters.

\nin
The strategy
is to have some parameters that are kept fixed and others are let
float in the fit.
We consider as {\bf fixed parameters}: $\alpha(M_Z)$, measured at low
energies and then evolved to $M_Z$; $G_F$, as measured from the muon
lifetime; and the fermion masses (originally with the exception of
$m_t$). The rest of the parameter set is let to float in the fit.
Thus, for each observable in Table in (\ref{fig:ewpo}), there is an
experimental measurement 
 and, next to it, the SM prediction resulting from the fit including
 all the parameter set in (\ref{fitof}). Finally, the ``pulls'' are
 computed using
 \be
{\rm pull} = \frac{{\cal O}^{\rm th.} - {\cal O}^{\rm
    exp.}}{\sigma^{\rm th.}}~,
 \ee
 where $\sigma^{\rm th.}$ are the errors in ${\cal O}^{\rm th.}$.
We see that the agreement of the SM fit with experiment is very good
with a high level of precision.

\nin
The electroweak data is so precise that allows for a determination of
$M_Z$, $m_h$, $m_t$ and $\alpha_s(M_Z)$. Although the Higgs boson mass
enters only logarithmically in electroweak loop corrections, it is
possible to obtain
\be
m_h = 90^{+17}_{-16} ~{\rm GeV}~,
\ee
once the kinematic constraints from the LHC are removed. All this
information can be seen in Figure~\ref{fig:mtvsmh} below.
\begin{figure}[h]
\begin{center}
\includegraphics[width=5.5in]{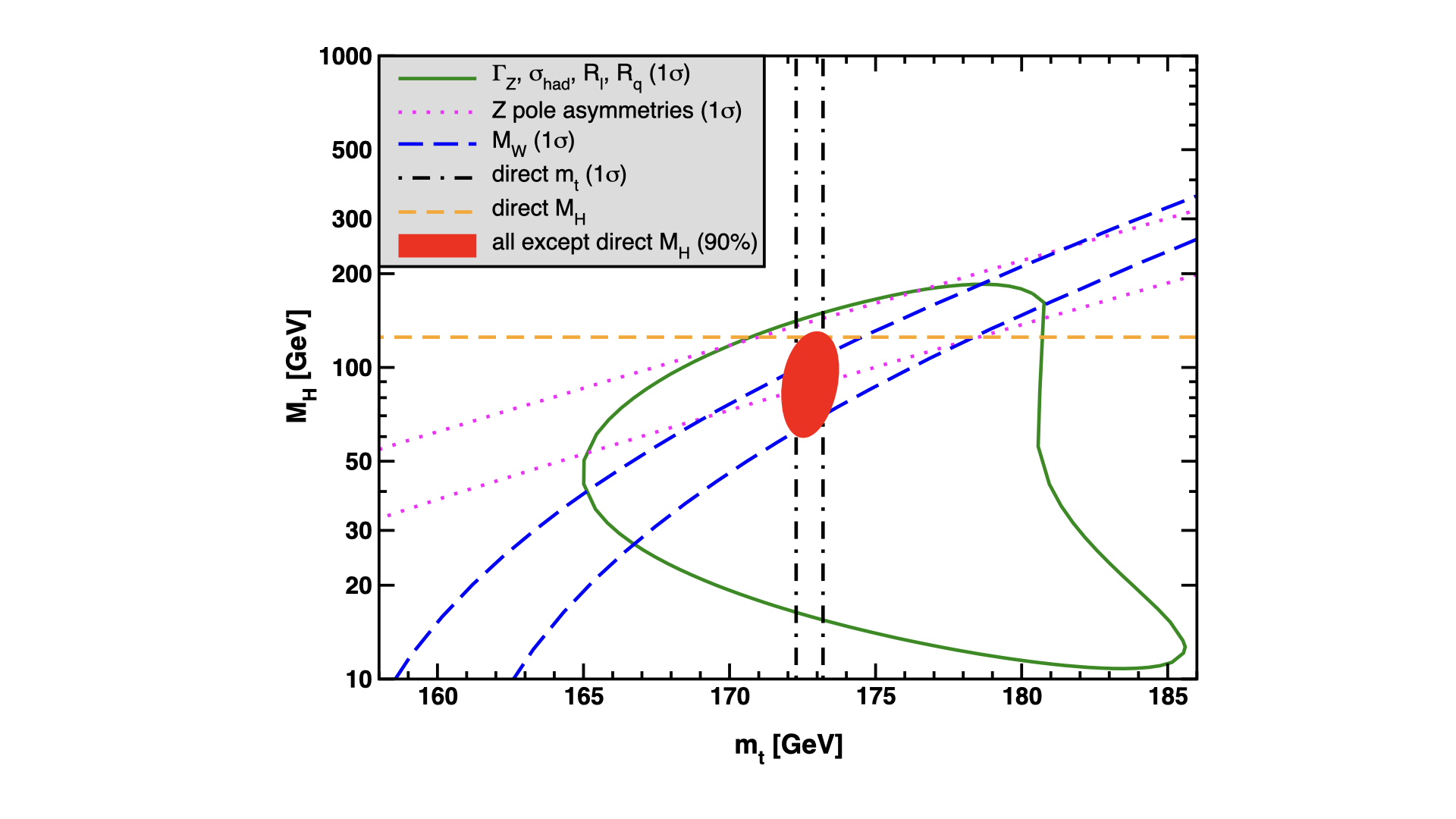}
\end{center}
\caption{Fit result and one-standard-deviation ($39.35\%$ for the
  closed contours and 68\% for the others) uncertainties in $m_h$ as a
  function of $m_t$ for various inputs, and the $90\%$ CL region
  allowed by all data. $\alpha_s(MZ) = 0.1187$ is assumed except for
  the fits including the Z lineshape.
  From \cite{rpp}. 
}
\label{fig:mtvsmh}
\end{figure}

\nin
Finally, precision electroweak data like these can be used to
constrain new physics beyond the SM. The model independent approach to
constrain new physics in precision data is to make use of the effective
field theory framework. The Lagrangian of the SM contains only
renormalizable (i.e. dimension 4) operators. But higher dimensional
operators (HDO) coming from physics BSM at higher energy can modify
the physics at the electroweak scale. The systematic expansion of the
electroweak SM as an effective field theory (EFT) in terms of HDOs can
be schematically written as~\cite{Shepherd:2022rsg,Falkowski:2023hsg}
\be
{\cal L}_{\rm SMEFT} = {\cal L}_{\rm SM}~ + \sum_i
\frac{c_i}{\Lambda^2} {\cal O}_i^{d=6} + \sum_j\frac{c_j}{\Lambda^4}
{\cal O}_j^{d=8} +\cdots~,
\label{smeft}
\ee
where the ${\cal O}_j^{d=n}$ are the dimension $n$ operators,
$\Lambda$ is the UV scale where the physics integrated out resides and
gives rise to the HDOs, and the coefficients $c_j$ are in principle
unknown and depend on the UV physics.  This so
called SMEFT is a road map for using high precision data to constrain
new physics BSM coming from higher energy scales. This will be an
important part of the program of the HL-LHC and even beyond in a
scenario where the energy frontier remains at around $14~$TeV.
Already it is possible to put bounds on the coefficient of the
dimension 6 operators~\cite{Ellis:2014jta}, although much more data
will be necessary for a tighter set of constraints \cite{deBlas:2022ofj}. One of the reasons
is that, even if we restrict ourselves to dimension 6 operators, there
are 59 of them~\cite{Grzadkowski:2010es}. Is even possible that for
some observables dimension 8 operators, of which there are thousands)
might be necessary. All of these is beyond the scope of these
lectures. But we can give a glimpse of this procedure by  selecting a
couple of dimension 6 operators, which are quite well known and very
well constrained by electroweak precision data. These are
\be
{\cal O}_S=H^\dagger \sigma^i H A^i_{\mu\nu} B^{\mu\nu}  \qquad\qquad
{\cal O}_T= |H^\dagger D_\mu H|^2~,
\label{stdef}
\ee
where the $\sigma^i$ are the Pauli matrices, $A^i_{\mu\nu}$ are the
$SU(2)_L$ gauge field strength and $B_{\mu\nu}$ is the $U(1)_Y$ gauge
field strength. Once the Higgs field $H$ is replaced by its VEV, the
operator ${\cal O}_S$ induces kinetic mixing between $A^3_\mu$ and
$B_\mu$, absent in the SM\footnote{The mass mixing between $A^3_\mu$
  and $B_\mu$, which leads to the need to diagonalize the neutral
  gauge boson mass matrix, is of a different character.}. On the other
hand, the operator ${\cal O}_T$ induces a shift on the $Z^0$ mass, but
none on the $W^\pm$ mass. This is then a new physics contribution to the $\rho$
parameter defined in (\ref{rhodef}) as a tree level relation. The SM
contributions to $\rho$ have been both computed and measured with
great precision, so the coefficient of this operator is greatly
constrained by electroweak data. These type of corrections are called {\em oblique}, since
they really are only corrections to the gauge two point functions,
ignoring the corrections to vertices. The rationale behind considering
these type of corrections alone in a fit is that maybe the new physics
states (e.g. heavy fermions or scalars) have electroweak quantum
numbers so they would induce loop corrections to the electroweak gauge
boson two point functions as picture in Figure~\ref{fig:oblique}.
\begin{figure}[h]
\begin{center}
\includegraphics[width=2.0in]{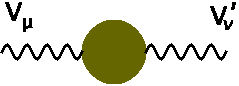}
\end{center}
\caption{Oblique corrections to the electroweak gauge boson two point
  functions. They can be one loop contributions from new fermions or
  scalars carrying electroweak quantum numbers and contribute to the
  coefficient of the operators ${\cal O}_S$ and ${\cal O}_T$, among
  others.   
}
\label{fig:oblique}
\end{figure}
\nin
The corrections arising from these two dimension six operators give
rise to {\em universal} electroweak corrections, in the sense that
they will appear in all electroweak amplitudes independently of the
identity of the external fermions. So the data constraining them could
be coming from muon decay, $Z$ pole observables such as the $Z\to
f\bar f$ widths to hadrons or leptons, asymmetries, etc.

\nin
Another important aspect of these quantities, i.e. $c_S$ and $c_T$, is
that they are {\em finite}, since there are no counterterms to absorbe
divergences coming from loops that would have this form. So, even if
individual loop diagrams contributing to either $c_S$ and $c_T$ could
be divergent, the sum of all of them must give a finite answer. So
these are indeed measurable effects of quantum corrections to the
EWSM. Originally ~\cite{Peskin:1991sw} these two parameters where defined as $S$ and $T$, as
given by
\be
S\equiv \frac{4 s_W c_W v^2}{\alpha}\,c_S~,\qquad \qquad T\equiv
-\frac{v^2}{2\alpha}\,c_T~,
\label{stdefs}
\ee
 where the Weinberg angle and $\alpha$ are to be evaluated at the weak
 scale, and $v\simeq 246~$GeV. For instance, adding these dimension
 six operators to
 the SM Lagrangian  ${\cal L}_{\rm SM}$, we can add their
 contributions to the predictions for ${\cal O}_i^{\rm th.}$ in
 (\ref{fitof}) through the corrections in $\delta {\cal O}_i^{\rm
   th.}$, where the operators ${\cal O}_i$  are the dimension four SM
 operators affected by the shifts induced by the dimension 6 operators
 ${\cal O}_{S,T}$. For instance the $W$ mass is shifted as~\cite{rpp}
 \be
M_W^2 = M_{W,{\rm SM}}^2\frac{1}{1-G_F M_{W,\rm SM}^2 \,S /2\sqrt{2} \pi}~,
\label{mw2st}
 \ee
 whereas the $Z$ mass is given by
 \be
M_Z^2 = M_{Z,{\rm SM}} \,\frac{1-\alpha(M_Z) T}{1-G_F M_{Z,\rm SM}^2
  \,S /2\sqrt{2} \pi}~,
\label{mz2st}
 \ee
 and similarly for all observables in the fit.  For any {\em neutral
 current}  amplitude $A_i$, we would have
 \be
 A_i = A_{i,{\rm SM}}\,\frac{1}{1-\alpha(M_Z) \,T}~,
   \label{neutralampt}
 \ee
   where the $A_{i,{\rm SM}}$ are the corresponding SM amplitudes. 
 Then, adding $S$ and
 $T$ as floating parameters in the fit, one obtains~\cite{rpp}
 \be
 S = 0.02 \pm 0.10~,\qquad \quad T = 0.07\pm 0.12~.
 \label{stfit}
 \ee
 We see from the results above that $S$ and $T$, and therefore the
 coefficients $c_S$ and $c_T$ corresponding to the dimension 6 oblique
 operators  defined in (\ref{stdef}), are consistent with zero. This
 constitutes a very important constraint to possible extensions of the
 SM, which typically  generate non zero values of these
 parameters. Since the SM predictions in the expressions above are
 computed to two loop accuracy, any increased precision in electroweak
 precision observables tightens the bounds on new physics.

 \nin
 Many extensions of the SM have been severely constrained by
 electroweak precision observables such as $S$ and $T$. These continue
 to be one the most important bounds on extensions of the SM. In order
 to test a  BSM model against these measurements, one needs to
 consider the quantum corrections to the electroweak gauge bosons as
 depicted in Figure~\ref{fig:oblique}. The most general form of the
 gauge boson two point function is
 \be
\Pi^{\mu\nu}_{VV'}(q^2) = \Pi_{VV'}(q^2) g^{\mu\nu} +
\Sigma_{VV'}(q^2) q^\mu q^\nu~,
\label{tpfundef}
 \ee
where $q^\mu$ is the momentum going through the gauge boson line. The
second term in (\ref{tpfundef}) can be safely neglected since either
the gauge boson is coupled to a conserved current or its effects are
suppressed if the external particles have small
masses. Since we can assume that the scale of new physics giving rise
to these corrections to the SM come from some energy scale $\Lambda$
such that  $\Lambda^2\gg q^2$, then we can expand the $\Pi_{VV'}(q^2)$
functions around $q^2=0$ and keep only the first terms as in
\be
\Pi_{VV'}(q^2) \simeq \Pi_{VV'}(0) + q^2\,\Pi'_{VV'}(0) + \cdots~,
\label{piexp}
\ee
where $\Pi'_{VV'}$ denotes the derivative with respect to $q^2$. 
So, in principle, we have 8 quantities we need: ($\Pi_{\gamma\gamma}, \Pi_{\gamma
  Z}, \Pi_{ZZ}, \Pi_{WW}, \Pi'_{\gamma\gamma}, \dots$).  However, from
the renormalization conditions on the electric charge (e.g. from QED)
we know that
\be
\Pi_{\gamma\gamma}(0) = 0~,\qquad\qquad \Pi_{\gamma Z}(0)=0~.
\ee
Then we are down to 6 quantities. But another 3 can be absorbed in to the
renormalization of $\alpha$, $G_F$ and $M_Z$ as shifts defined by
\be
\frac{\delta\alpha}{\alpha} = -\Pi'_{\gamma\gamma}(0), \qquad
    \frac{\delta G_F}{G_F} = \Pi_{WW}(0)~,\qquad \frac{\delta
      M_Z^2}{M_Z^2} = - \Pi'_{ZZ}(0)~,
    \label{parshifts}
\ee
The remaining 3 parameters then must be accounting for the loop
corrections coming from new physics. These are~\cite{Peskin:1991sw}
the Peskin-Takeuchi parameters $S$, $T$ and $U$. While in this
formalism $S$ and $T$ 
can be matched to  the coefficients of dimension 6 operators, in this
case  ${\cal O}_S$ and ${\cal O}_T$, On the other hand, the third one,
$U$, would correspond to a dimension 8 operator in the SMEFT, and this
is the reason why in BSM models it typically gives no important
contributions. The $S$ and $T$ parameters can be defined  in terms of
the gauge boson two point functions as~\cite{Peskin:1991sw,rpp}
\be
\alpha T \equiv \frac{\Pi_{WW}(0)}{M_W^2} -
\frac{\Pi_{ZZ}(0)}{M_Z^2}~,
\label{tdef}
\ee
and
\be
\frac{\alpha}{4 \hat s_W^2 \hat c_W^2}\,S\equiv \frac{\Pi_{ZZ}(M_Z^2)
  -\Pi_{ZZ}(0)}{M_Z^2} - \frac{\hat c_W^2 -\hat s_W^2}{\hat c_W \hat
  s_W} \frac{\Pi_{Z\gamma}(M_Z^2)}{M_Z^2} -
  \frac{\Pi_{\gamma\gamma}(M_Z^2)}{M_Z^2}~,
    \label{sdef}
    \ee
Given a BSM theory, if it contains fermions and/or scalars charged
under the electroweak gauge group, it is possible to compute the loop
contributions to $S$ and $T$ directly, resulting on tight constraints
on the masses and couplings of the new particles. 

\nin
 Finally, to make clear contact with the dimension 6 operators defined
 earlier, we can express the gauge boson two point functions in the
 basis before electroweak symmetry breaking, i.e. in terms of the
 $SU(2)_L$ and $U(1)_Y$ gauge bosons. Then, we define the $\Pi_{11},
 \Pi_{22}, \Pi_{33}, \Pi_{YY}$ and $\Pi_{3Y}$ vacuum polarization
 functions of $q^2$ in terms of which we can write
 \be
T = \frac{4\pi}{\hat s_W^2 \hat c_W^2}\,\frac{\Pi_{11} +
  \Pi_{22} - \Pi_{33}}{M_Z^2}~,
\label{tredef}
\ee
and
\be
S = -16\pi \frac{\Pi_{3Y}(M_Z^2) - \Pi_{3Y}(0) }{M_Z^2} = -16\pi
\frac{\Pi'_{3Y}(0)}{M_Z^2}~.
\label{sredef}
\ee
Writing $T$ and $S$ in this way it is easier to understand their
physical significance. The $T$ parameter measures the breaking of the
isospin symmetry present in the EWSM~\footnote{This so called
  custodial symmetry is a remnant of the electroweak symmetry breaking
and is an accidental global symmetry in the EWSM.}, which is the
difference between the (identical)
$(11,22)$ components and the $33$ component. So, for instance, if there
is a new heavy $SU(2)L$ doublet $(U ~D)^T$, then the $T$ parameter
measures the different contribution from the charged loop containing
both $U$ and $D$ to the neutral loops containing either $U$ or D. This
contribution goes like $(M_U^2 -M_D^2)$ is this is a chiral doublet
and like $\ln( M_U/M_D)$ if is a vector-like one.  
Also new scalars in various representations can contribute to $T$.
\nin
On the other hand, the $S$ parameter clearly measures the amount of
kinetic mixing (as opposed to mass mixing) between the $A^3_\mu$ and
the $B_\mu$ gauge bosons, as evidenced by the presence of the $q^2$
derivative of $\Pi_{3Y}$. For instance, early on the $S$ parameter was used
used to exclude heavy chiral fermions. More recently, the
contributions of resonances in composite Higgs models, gives rise to
an important contribution to $S$ putting pressure on the mass scale of
these models vector resonance masses. 
   
\subsection{Gauge boson self couplings}
\label{sec:selfgauge}

In addition to testing the gauge boson couplings to fermions as seen
in the previous section, a crucial test of the electroweak theory is
its non-abelian character. Recalling the form of the electroweak pure
gauge boson sector:
\be
{\cal L}_{\rm GB} = -\frac{1}{4}\,F^a_{\mu\nu} F^{a\mu\nu} -\frac{1}{4}
B_{\mu\nu} B^{\mu\nu} ~,
\label{lgb_2}
\ee\vs\nin
where
\be
F^a_{\mu\nu} = \partial_\mu A^a_\nu -\partial_\nu A^a_\mu + g
\epsilon^{abc} A^b_\mu A^c_\nu~,
\label{fmunu}
\ee
is the $SU(2)_L$ gauge field strength and
\be
B_{\mu\nu} = \partial_\mu B_\nu - \partial_\nu B_\nu~,
\label{bmunu}
\ee
is the $U(1)_Y$ one, we can then derive the self-couplings of the
electroweak gauge bosons. We can immediately see that the $SU(2)_L$
term in (\ref{lgb_2}) will result in triple as well as quartic gauge
boson couplings. This is a direct consequence of the non-abelian
nature of the $SU(2)_L$ electroweak sector. In order to test this
experimentally however, we need to rewrite (\ref{lgb_2}) in terms of
the mass eigenstate gauge bosons. So we again make the transformation
from the $(A^a,B)$ basis to the  $(W^\pm, Z^0, \gamma)$ basis.  We
will concentrate on triple gauge boson  couplings (TGC).  A general
form of their interactions can be schematically written as
\be
{\cal L}_{\rm WWV} = i\,g_{\rm WWV} \left[\left(W^\dagger_{\mu\nu}
    W^\mu - W_{\mu\nu} W^{\mu}\right)\,V^\nu + W_\mu^\dagger
  W_\nu\,V^{\mu\nu}\right]~,
\label{tgc_1}
\ee
with $V=\gamma, Z^0$, and we defined the tensors
\be
W_{\mu\nu} = \partial_\mu W_\nu -\partial_\nu W_\mu~, \qquad {\rm
  and}\qquad
V_{\mu\nu} = \partial_\mu V_\nu - \partial_\nu V_\mu~.
\ee\nin
In the SM, we have
\be
g_{WW\gamma}= -e \qquad  {\rm and} \quad g_{WWZ} = -e \cot\theta_W~.
\label{gsmvalues}
\ee\vs\nin
The first experimental tests of these TGC were performed at LEP II in
the early 1990s through $W^+ W^-$ pair production. The corresponding
diagrams are shown in Figure~\ref{fig:wwprod}.
\nin
\begin{figure}[h]
\begin{center}
\includegraphics[width=4in]{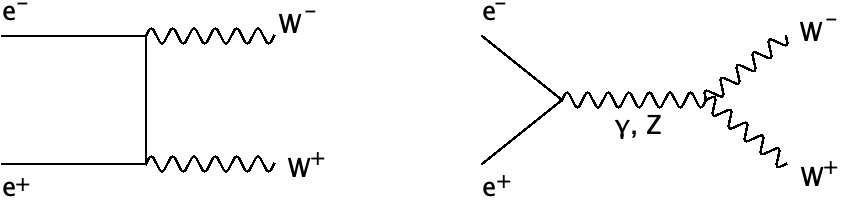}
\end{center}
\caption{Tree-level diagrams for $W$ pair production . The second type
 of diagrams are described by (\ref{tgc_1})   
}
\label{fig:wwprod}
\end{figure}
\nin
In Figure~\ref{fig:lep2} below, we see the early data testing the
electroweak TGCs. Already with these data it was clear that the triple
gauge boson couplings must be included in the calculations in order to have agreement with
the experiment. Even if one argues that the $\gamma W ^+ W^-$ TGC does
not really test the non-abelian nature of the electroweak gauge sector
since it is just the coupling of a charged particle to the photon as
expected in QED, the presence of the $Z^0 W^+ W^-$ contribution to $W$
pair production is necessary to bring agreement with data. Current
data are much more constraining. 
\nin
\begin{figure}[h]
\begin{center}
\includegraphics[width=6in]{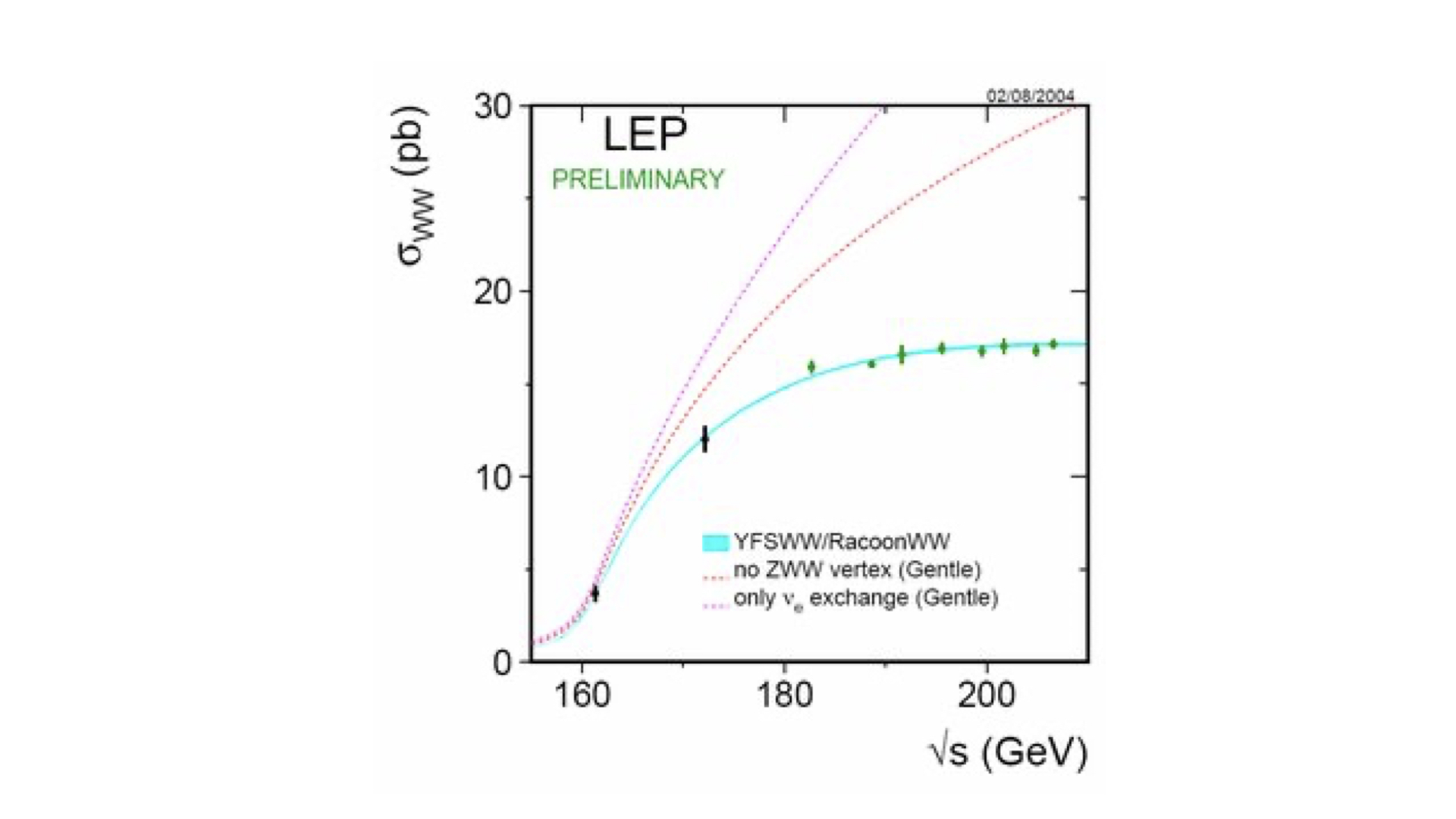}
\end{center}
\caption{Early LEP II data on $W$ pair production: testing the non-abelian
  nature of the electroweak sector. For the data to agree with the SM
  prediction, all diagrams in Figure~\ref{fig:wwprod} must be
  considered, including the $Z^0 W^+ W^-$.
}
\label{fig:lep2}
\end{figure}
\nin
In order to compare with more modern data, we first define an
effective Lagrangian for the TGCs that allows for anomalous deviations
from the SM, although still maintaining parity and charge conjugation
invariance. This is conventionally written as
\be
{\cal L}_{\rm WWV} = i g_{WWV}\left[ g_1^V\left(W^\dagger_{\mu\nu}
    W^\mu - W_{\mu\nu} W^{\mu}\right)\,V^\nu + \kappa_V W^\dagger_\mu
  W_\nu V^{\mu\nu} +i\frac{\lambda_V}{M_W^2}\,W^\dagger_{\rho\mu}
    W^\mu_\nu V^{\nu\rho}\right]~,
    \label{tgc_2}
\ee
where $g_{WWV}$ is still given by (\ref{gsmvalues}) and, in the SM we
have
\be
g_1^V =1 \qquad \kappa_V =1\qquad \lambda_V =0~.
\ee
The introduction of the last term in (\ref{tgc_2}) corresponds to a
higher dimensional operator, as seen by the appearance of an energy
squared in the denominator, here chosen to be $M_W^2$.  If we further
impose gauge invariance, the couplings defined in (\ref{tgc_2}) are
constrained to satisfy
\be
\lambda_\gamma = \lambda_Z, \qquad \kappa_Z =g_1^Z-(\kappa_\gamma-1)\tan^2\theta_W~.
\ee
Various experiments have constraints these TGC over the years. In order
to compare with them, it is customary to define quantities that are
zero in the SM:
\be
\Delta g_1^Z\equiv g_1^Z -1, \qquad \Delta\kappa_Z\equiv
\kappa_Z-1,\qquad \Delta\kappa_\gamma \equiv \kappa_\gamma-1~,
\ee
in addition to  $\lambda_\gamma$ and $\lambda_Z$.
\nin
\begin{figure}[h]
\begin{center}
\includegraphics[width=7in]{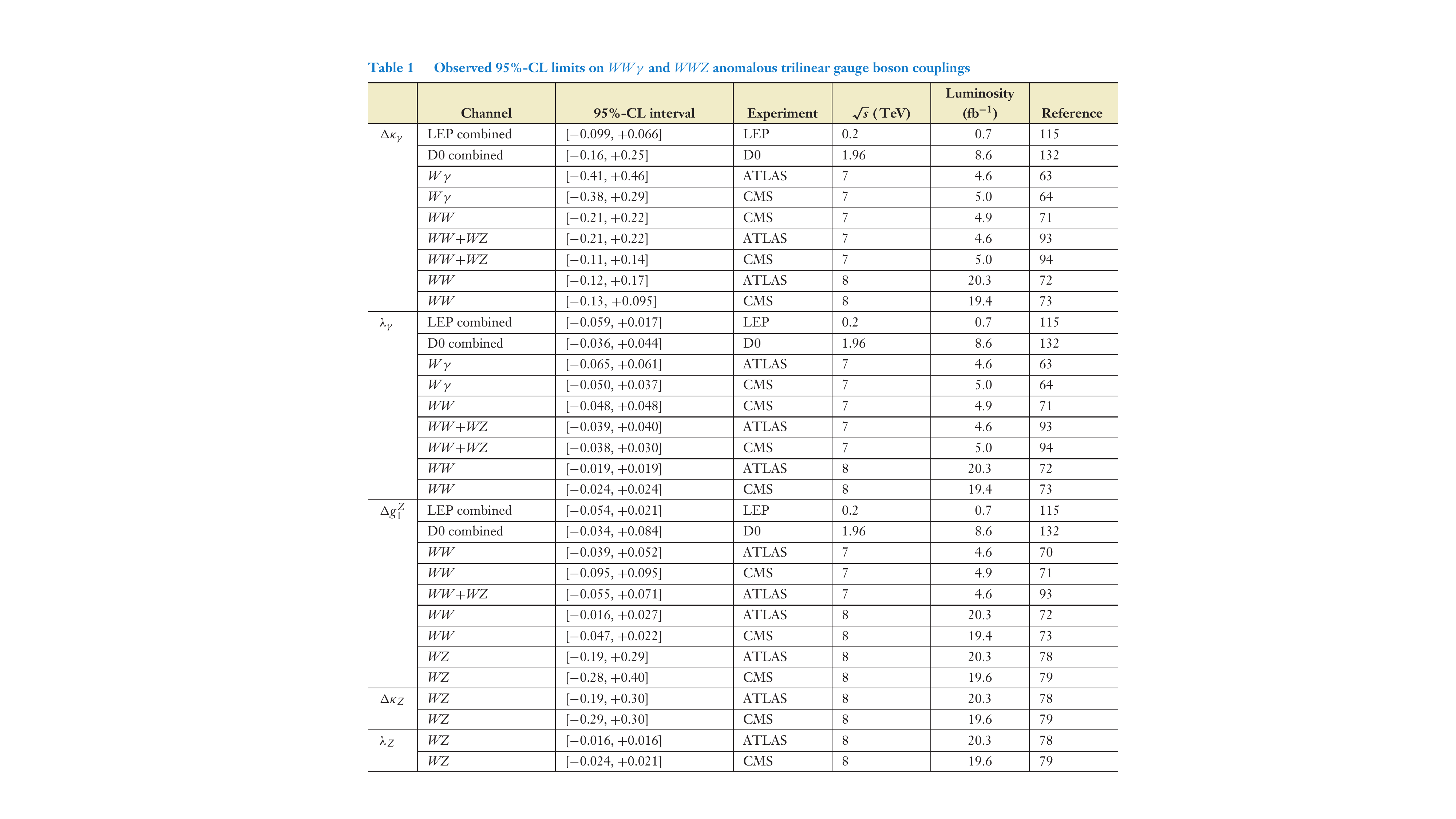}
\end{center}
\caption{Measurements of TGC at various experiments. From
  \cite{tgcbounds}. 
}
\label{fig:tgc_bounds}
\end{figure}
\nin
We can see that all TGC measurements are consistent with the SM
within experimental errors. 

\subsection{Higgs boson couplings}
\label{sec:higgscouplings}
\nin
The Lagrangian for the EWSM is schematically given by
\be
{\cal L}_{\rm EW} = (D_\mu\Phi)^\dagger D^\mu\Phi -V(\Phi^\dagger\Phi)
+ {\cal L}_{\rm HF} + {\cal L}_{\rm GB} + {\cal L}_{\rm GF}~,
\label{lagewsm}
\ee
where ${\cal L}_{\rm GF}$ contains the interactions of fermions with
gauge bosons, ${\cal L}_{\rm GB}$ contains just the gauge bosons
including their TGC and quartic self-interactions and ${\cal L}_{\rm  HF}$ contains the fermion Yukawa couplings to the Higgs bosons which
will be discussed below. Working in the unitary gauge with
\be
\Phi(x) = \left(\ba{c} 0\\ \frac{v+h(x)}{\sqrt{2}} \ea\right)~,
\label{ugauge}
\ee
we can read off the couplings to gauge bosons from the first term in
(\ref{lagewsm}). For instance, for the Higgs couplings to $W$'s, these can be written as
\be
{\cal L}_{\rm hWW} = \left[ g_{\rm hWW}\,h  +\frac{g_{\rm hhWW}}{2!}\,
  h^2\right] W^{+\mu} W^-_\mu~,
\label{higgs2w}
\ee
where we defined
\be
g_{\rm hWW}=\frac{2 M_W^2}{v},\qquad g_{\rm hhWW} = \frac{2M_W^2}{v^2}~.
\label{h2wdefs}
\ee
Analogously, we can obtain the couplings of the Higgs boson to the
$Z$:
\be
{\cal L}_{\rm hZZ} = \left[ \frac{g_{\rm hZZ}}{2!}\,h  +\frac{g_{\rm hhWW}}{(2!)^2}\,
  h^2\right] Z^{\mu} Z_\mu~,
\label{higgs2z}
\ee
with
\be
g_{\rm hZZ}=\frac{2 M_Z^2}{v},\qquad g_{\rm hhZZ} = \frac{2M_Z^2}{v^2}~.
\label{h2zdefs}
\ee
The way the couplings are defined above allows us to write them in
(\ref{higgs2w}) and (\ref{higgs2z}) with the explicit factors of $2!$
counting the number of identical particles in the vertex.

\nin
The triple vertices $g_{\rm hVV}$, with $V=(W^\pm, Z)$, have been
tested at the LHC with considerable precision. Defining the coupling
strengths normalized to the SM values
\be
\kappa_V = \frac{g^{\rm exp.}_{\rm hVV}}{g^{\rm SM}_{\rm hVV}}~.
\ee
We can see some recent results for $\kappa_W$ and $\kappa_Z$ in
Figure~\ref{fig:hcouplings1}.  The best measurements of $\kappa_W$ and
$\kappa_Z$ come from $pp\to h \to V V^*$, as well as indirectly from
the loop  $W^\pm$ contribution to $pp\to h \to \gamma\gamma$. 
\nin
\begin{figure}[h]
\begin{center}
\includegraphics[width=7in]{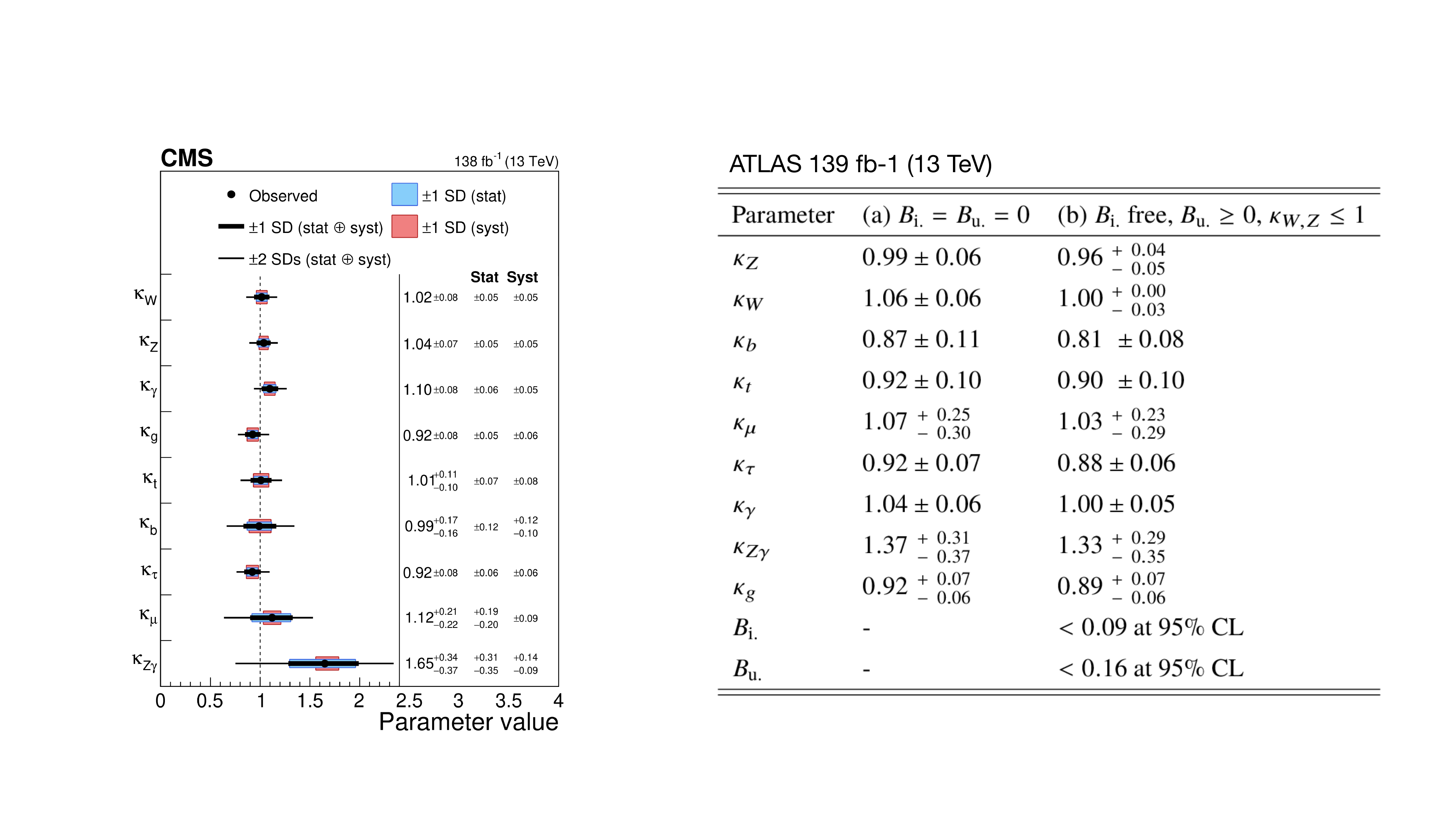}
\end{center}
\caption{Measurements of the Higgs boson couplings. From CMS (left)
  and ATLAS (right). In the latter, the left column assumes no
  invisible ($B_i=0$) or undetected ($B_u=0$) events, whereas in the
  right column these are allowed to float in the fit.
}
\label{fig:hcouplings1}
\end{figure}
\nin
We can see that the agreement with the SM predictions is quite
remarkable, already somewhat better than $10\%$ in the couplings. At
the moment, the LHC is not directly sensitive to the quartic couplings
$g_{\rm hhVV}$, which will require detailed understanding of double
Higgs production.

\nin
We move now to tests of the Higgs boson couplings to fermions. We go
back to the discussion of Section~\ref{sec:fermionmix}, and rewrite
(\ref{yukawa3gen}), i.e. the
  third term in (\ref{lagewsm}):
\be
-{\cal L}_{HF} =\lambda^{ij}_u\bar q_{L,i}\tilde\Phi u_{R,j} +
\lambda^{ij}_d \bar q_{L,i}\Phi d_{R,j} + \lambda_{\ell}^{ij}
\bar\ell_{L,i} \Phi \ell_{R,j}~,
\label{h2fermions}
\ee
where we remind ourselves that  $q_{L,i}$ is the quark $SU(2)_L$ doublet of generation $i$,
$u_{R,I}$ and $d_{R,i}$ are the corresponding right handed up and down
type quarks of generation $i$, and we denoted the $SU(2)_L$ lepton
doublet by $\ell_{L,i}$, and the right handed lepton ($SU(2)_L$
singlet) by $\ell_{R,i}$. The dimensionless Yukawa matrices
$\lambda_u$, $\lambda_d$ and $\lambda_\ell$ are parameters of the EWSM,
and are  generically complex and non diagonal in the basis where the
gauge interactions of fermions are diagonal, the gauge basis. As we
saw in Section~\ref{sec:fermionmix}, the mass matrices that result
from taking just the VEV of $\Phi$ in (\ref{h2fermions})
\be
M^{ij}_u, \qquad M^{ij}_d,~\qquad M^{ij}_\ell~,
\ee
are diagonalized by  bi-unitary transformations on the quark and
lepton fields. As a result, when writing the theory in terms of the
fermion mass eigenstates, the Higgs couplings to fermions will be 
automatically  diagonal and given by
\be
\lambda_f = \frac{m_f}{v}~,
\label{diagonalyuk}
\ee
where we see that the Yukawa coupling to a given fermion is generation
diagonal (as it should be so as to not result in tree level FCNCs! )
and is proportional to the fermion mass. Once again, we may define
$\kappa_f$ as the fermion Yukawa coupling normalized by the SM
prediction (\ref{diagonalyuk}). Although the top quark has the
strongest coupling to the Higgs, its measurement can only achieved
indirectly due to the fact that $h\to\bar  t t $ is kinematically
forbidden. The indirect measurement is performed thorough the
measurement of the Higgs production cross section, $\sigma(pp\to h)$
which is dominated by the gluon fusion channel. This, in turn is
dominated by the top quark loop, as is shown in Figure~\ref{fig:gg2h}.
\nin
\begin{figure}[h]
\begin{center}
\includegraphics[width=5in]{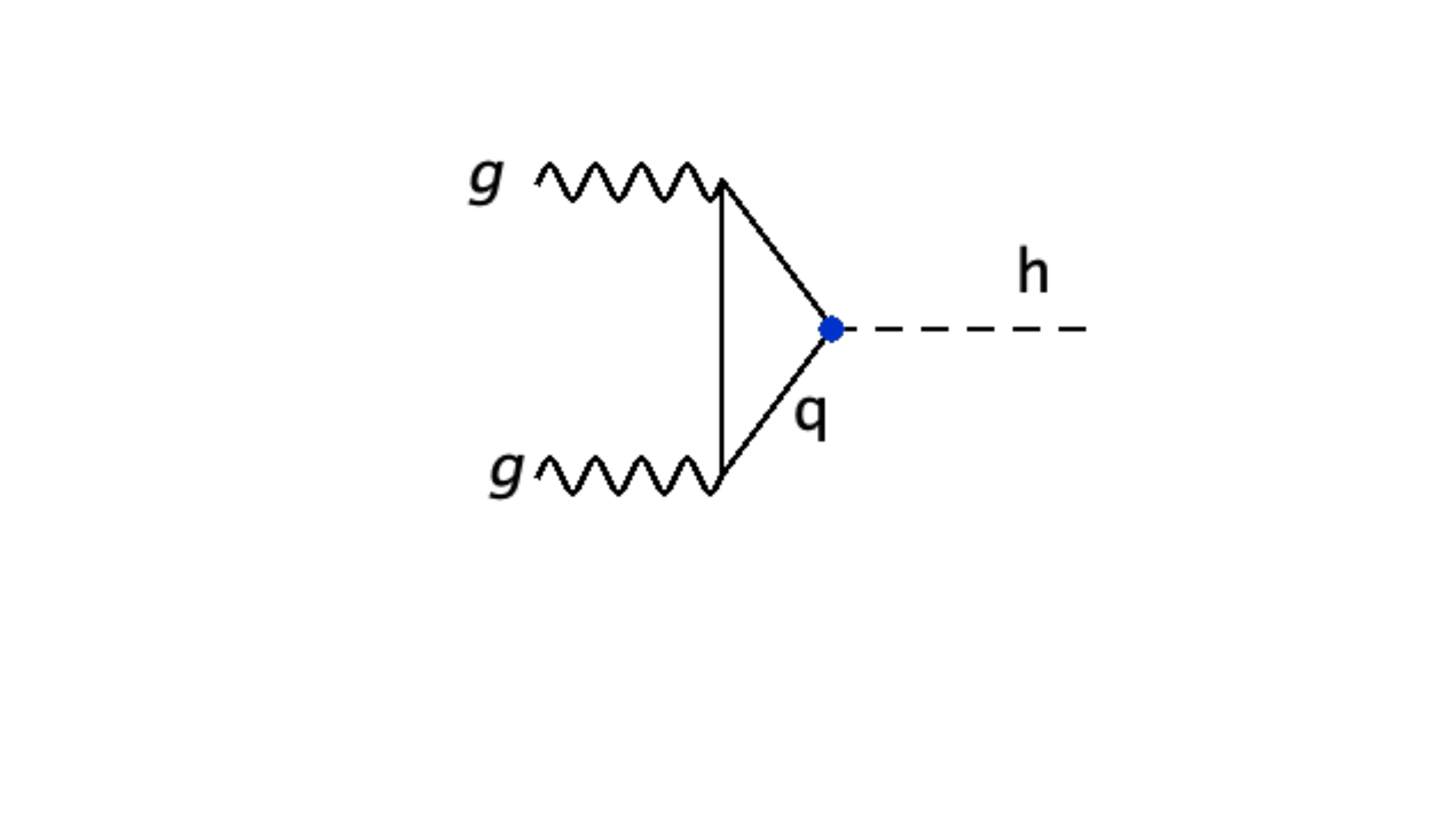}
\end{center}
\caption{Quark loop contributing to $gg \to h$. It is largely
  dominated by the top quark in the loop. 
}
\label{fig:gg2h}
\end{figure}
\nin
We see from Figure~\ref{fig:hcouplings1} that $\kappa_t$ is in
agreement with the SM value of $1$ within the error bars. The next
fermion with the largest coupling is the $b$ quark, which in fact
dominates the Higgs boson decays, with the largest branching
ratio. We,
see that $\kappa_b$ also agrees with the SM prediction. Despite the
$\bar b b$ mode being directly observable, the error in its
determination of $\kappa_b$ is similar to that of $\kappa_t$ since the
$b$ quark modes suffers from large backgrounds. Regarding couplings to
leptons, the LHC has achieved measurements of $h\to\tau^+\tau^-$ with
similar error bars. More recently, $h\to\mu^+\mu^-$ has been observed
but the errors in the determination of $\kappa_\mu$ are considerably
larger.
\nin
\begin{figure}[h]
\begin{center}
\includegraphics[width=8in]{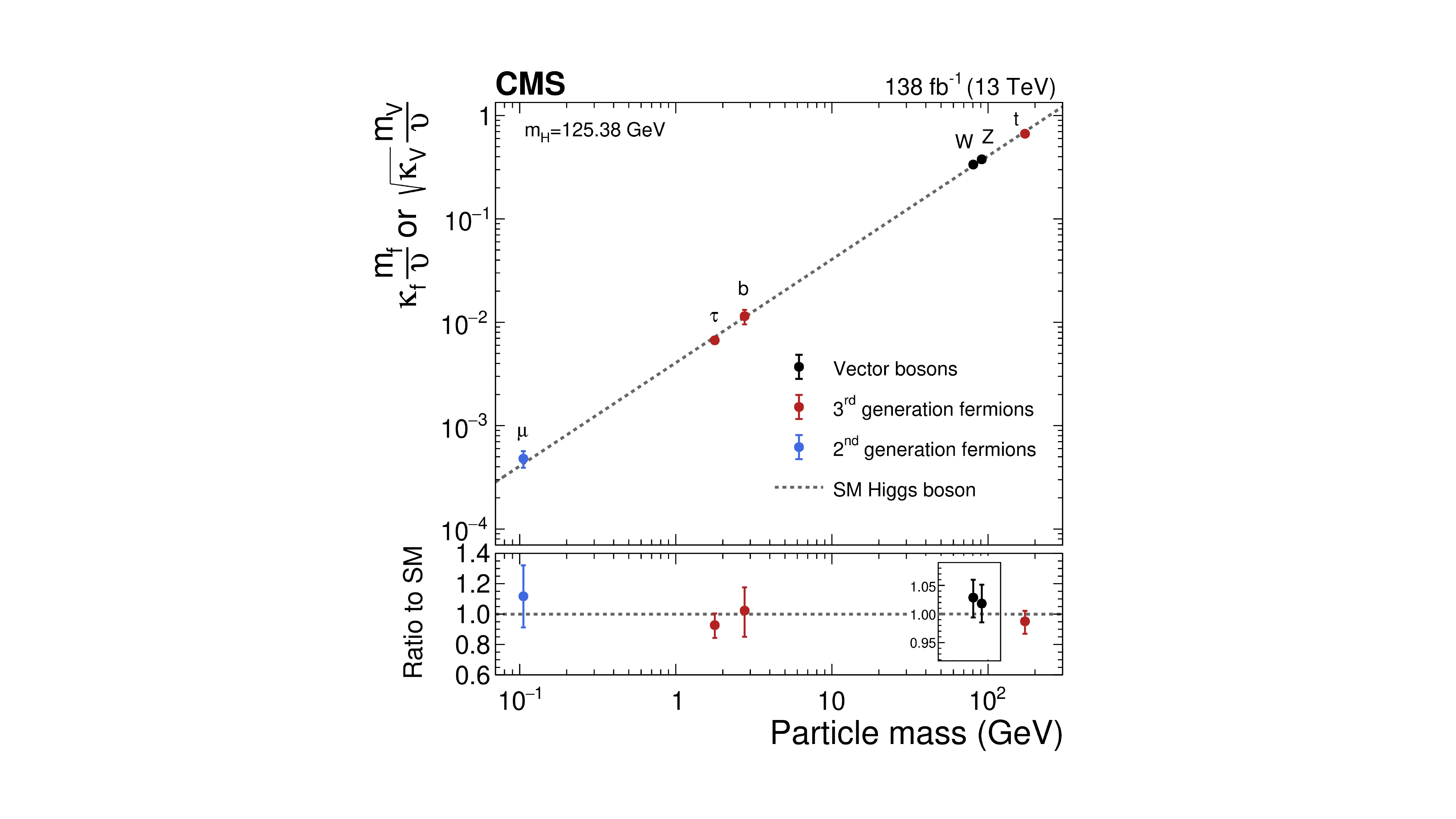}
\end{center}
\caption{Couplings of the Higgs boson to SM particles vs. the particle
  mass, as measured by
  the CMS collaboration. 
}
\label{fig:CMS_K_vs_mf}
\end{figure}
\nin
All of this information about the Higgs couplings to SM particles can
be seen summarized in Figure~\ref{fig:CMS_K_vs_mf}, where the
couplings as measured by the CMS collaboration are plotted versus the
fermion masses. We see that the agreement with the predictions of the
EWSM is excellent within the experimental errors.

\nin
Finally, we consider the Higgs boson self couplings. These come from
the Higgs potential:
\be
V(\Phi^\dagger\Phi)= -m^2\Phi^\dagger\Phi
+\lambda\left(\Phi^\dagger\Phi\right)^2~.
\label{hpot_2}
\ee
Using the unitary gauged form for $\Phi$ in (\ref{ugauge}), 
as well as expressing the Higgs VEV as
\be
v =\sqrt{\frac{m^2}{\lambda}}~, 
\ee
allows us to write the Higgs self interaction as
\be
{\cal L}_h = -\frac{1}{2} m_h^2 h^2 -\frac{g_{h^3}}{3!} h^3
  -\frac{g_{h^4}}{4!} h^4~,
  \label{hself}
  \ee
  where we defined the triple and quartic Higgs self couplings as
  \be
  g_{h^3} = 3\frac{m_h^2}{v}, \qquad g_{h^4} = 3\frac{m_h^2}{v^2}~,
  \label{hselfdefs}
  \ee
  In order to experimentally access these couplings we need double
  Higgs production data. Before we go into some details of double
  Higgs production, let us make clear why this is such a fundamental
  test of the EWSM and, in particular of the whole picture of
  electroweak symmetry breaking. To see this, let us recall that the
  Higgs mass is given by 
  \be
  m_h=\sqrt{2\lambda} v.
  \label{smmh}
\ee
  Thus, using
  $m_h\simeq 125~$GeV and $v\simeq 246~$GeV (from various electroweak
  precision measurements) we arrive at the SM prediction for the Higgs
  quartic coupling in the potential (\ref{hpot_2})
  \be
  \lambda\simeq 0.13~.
  \label{smlambda}
  \ee
  This is a value extracted from the Higgs mass measurements, plus our
  knowledge of the electroweak scale from electroweak data (e.g. muon
  decay, $M_W$ measurements, etc.). 
Thus, a fundamental test of the {\em shape}  of the Higgs potential,
is the direct measurement of the quartic coupling $\lambda$. If we
rewrite the triple and quartic Higgs self couplings in (\ref{hselfdefs}) using the SM
prediction (\ref{smmh}) we obtain
\be
g_{h^3} = 6 \lambda v, \qquad g_{h^4} = 6\lambda ~.
\label{hscinsm}
\ee
It is possible to measure $g_{h^3}$ in double Higgs production, so as
to experimentally test the SM prediction in (\ref{hscinsm}). The main
contribution to double Higgs production come from $gg\to hh$ as
illustrated in Figure~\ref{fig:doublehprod}.
\nin
\begin{figure}[h]
\begin{center}
\includegraphics[width=5in]{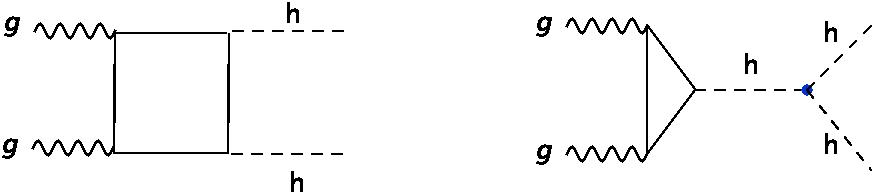}
\end{center}
\caption{One loop diagrams contributing to $gg\to hh$.
Only the diagram on the right is sensitive to the triple Higgs self
coupling $g_{h^3}$.
}
\label{fig:doublehprod}
\end{figure}
\nin
The two contributing diagrams interfere destructively. The box
diagram, which does not depend on $g_{h^3}$, dominates for the SM value
of the coupling. If we define
\be
\kappa_\lambda \equiv \frac{\lambda^{\rm exp.} }{\lambda^{\rm SM}},
\ee
the SM computations show that for values of $\kappa_\lambda$
sufficiently larger than unity (about $\kappa_\lambda>2.5$) or negative there could be an
enhancement in the double Higgs production cross
section~\cite{Frederix:2014hta}. The current status of searches for
Higgs pair production and bounds on the Higgs triple self coupling are
shown in Figure~\ref{fig:dihiggsexp}. Shown are the bounds on
$\kappa_\lambda$ as a function of $\kappa_t$. It is clear that this
are preliminary studies since the allowed values of $\kappa_\lambda$
when fixing all other couplings to the SM values, including $\kappa_t$
, span a huge interval, roughly $-5 \leq \kappa_\lambda\leq
10$. More meaningful constraints on $\kappa_\lambda$ will be available
with the HL-LHC.  For instance, simulations for the ATLAS experiment
in the HL-LHC with $3{\rm ab}^{-1}$ accumulated luminosity point to a
measurement of the SM value of $\lambda$ (i.e. $\kappa_\lambda=1$ of
about $3.2\sigma$ from a combination of channels~\cite{Mete:2022voi}. Although this will be quite an improvement over the
current situation, it is clear that a precision in $\kappa_\lambda$
comparable to the one attained in the other couplings will require to
go beyond the HL-LHC. We will comment on the importance of this
measurement in the next section.
\nin
\begin{figure}[h]
\begin{center}
  \includegraphics[width=3in]{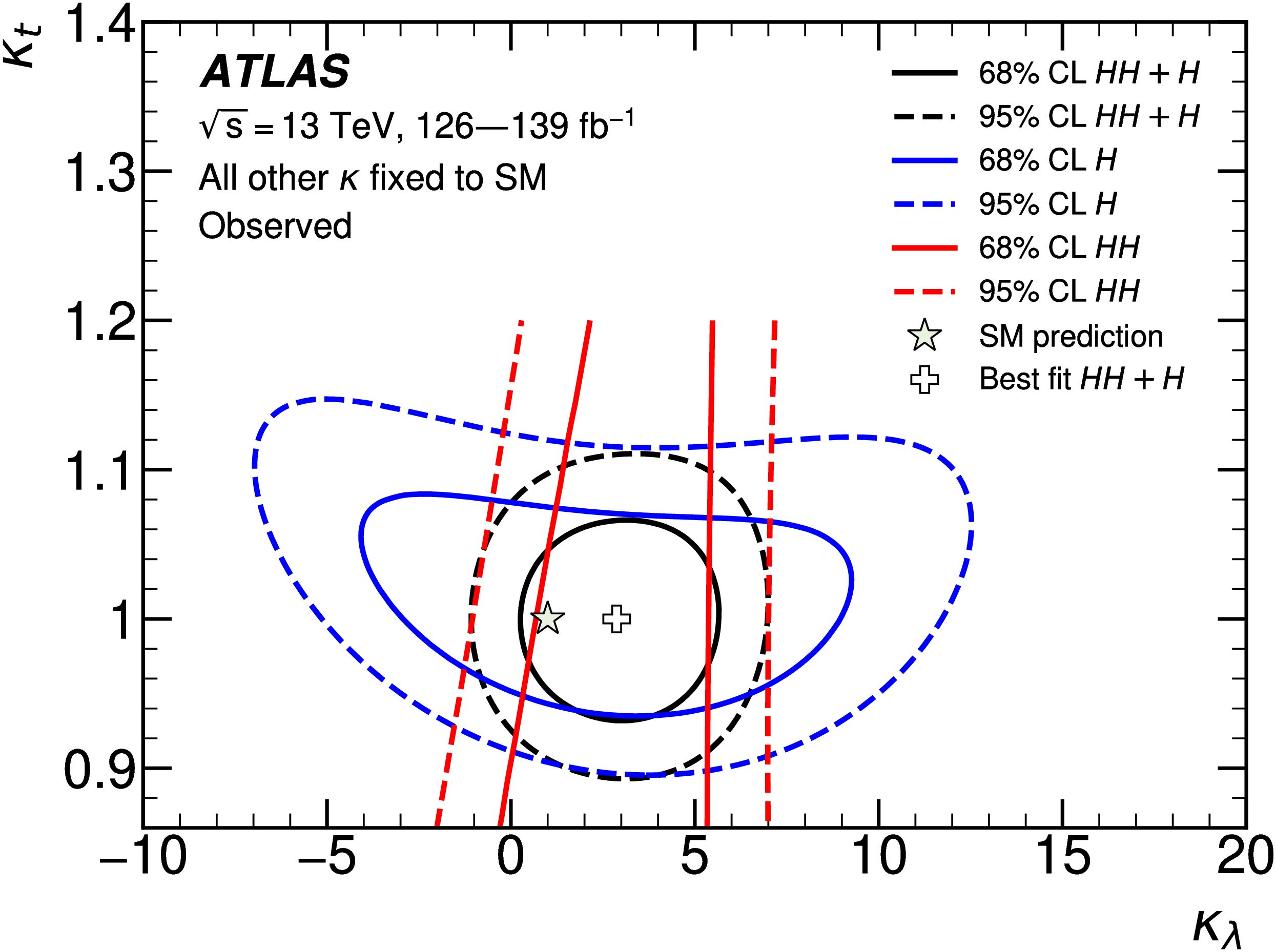}\hfill
  \includegraphics[width=3in]{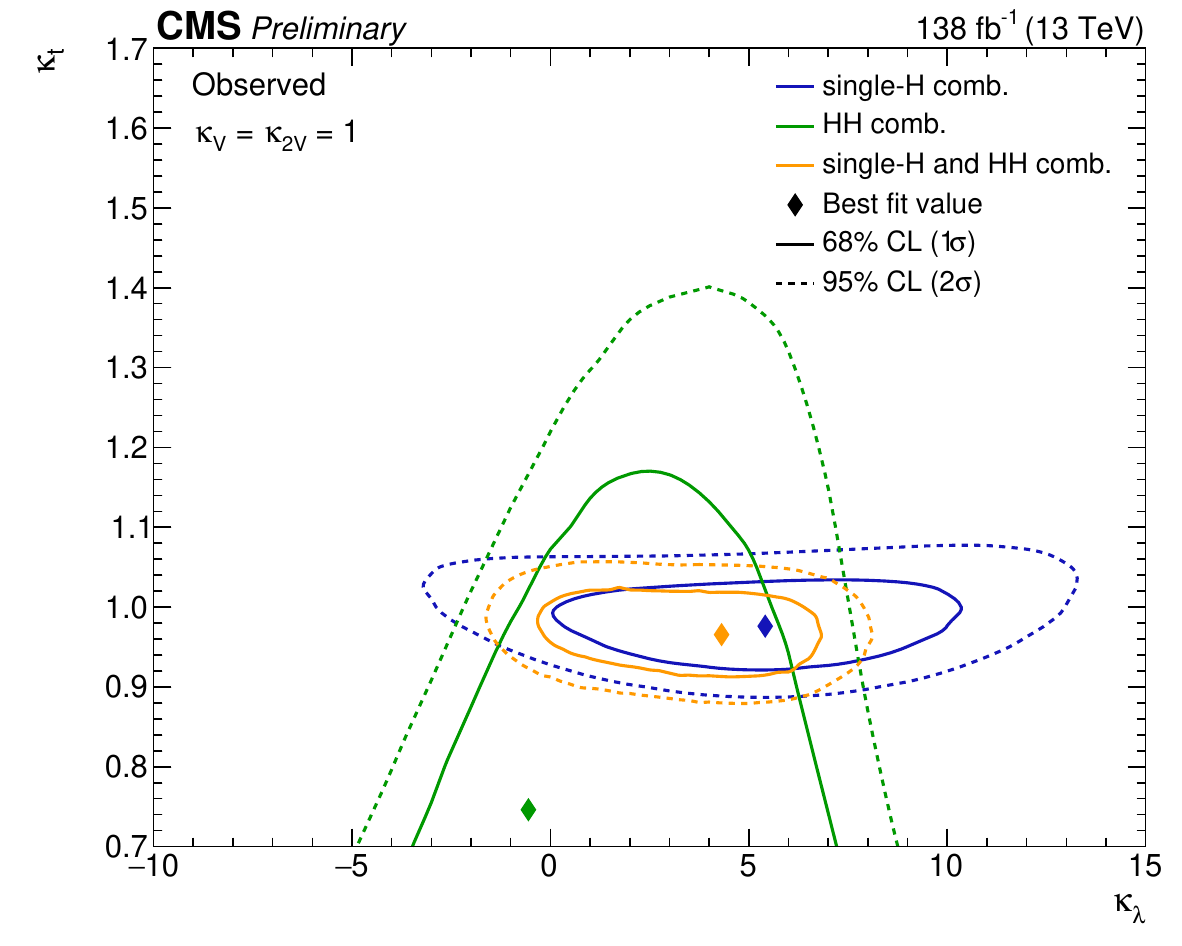}
\end{center}
\caption{Bounds on the Higgs self coupling, normalized to the SM
  prediction, vs the top Yukawa coupling $\kappa_t$. ATLAS result
  (left panel) from \cite{ATLAS:2022jtk}. CMS result (right panel)
  from \cite{CMS:2023cee}.
}
\label{fig:dihiggsexp}
\end{figure}
\nin

\section{Conclusions and outlook}
\label{sec:concout}
As we have seen in the previous sections, the EWSM is a quantum field
theory, a spontaneously broken gauge theory that describes all
available data so far used to test it. The $SU(2)_L\times U(1)_Y$
gauge theory, spontaneously broken to $U(1)_{\rm EM}$, When we add the
unbroken $SU(3)_c$ interaction (QCD), it describes with
great experimental success all the interactions of all elementary
particles known today.  The precision achieved in this description
varies. It is great for the interactions of fermions to gauge bosons
and the gauge boson self interactions. The interactions of the Higgs
boson with gauge bosons and fermions are being tested with increasing
precision. However, the Higgs self interactions are yet to be
experimentally observed. This is of great importance since it would
constitute a direct test of the form of the Higgs potential (more on
this below). The HL-LHC will begin to make this observation
possible. But it will be very short of a precise test of the Higgs
sector of the EWSM. This one of the main reasons why the high energy
physics community must consider options for future accelerators~\cite{futureaccel}.

\subsection{The electroweak Standard Model: Open questions}
\label{sec:smproblems}

\nin
Despite all of its successes, there are many open questions that are
not answered by the SM of particle physics. Some of these exist
independently of the theory, others are actually raised by it.
We first briefly mention some of the first type.  

\nin
{\bf Dark Matter}: It appears that more than $80\%$ of the
    matter in the universe does not behave as the matter described by
    the SM. All we know so far about it is that it gravitates.In fact,
    cosmological data are rather precise about the abundance of dark
    matter necessary to fit them. The SM cannot accommodate anything
    of the sort~\cite{dmrefs}. Extensions of the SM have can be
    proposed that would accommodate the correct dark matter
    abundance. Experimental bounds coming from direct and indirect
    detection 
    
\nin
    {\bf The Baryon--Anti-Baryon Asymmetry}: The asymmetry between
    the number of baryons and anti-baryons in the universe can be
    measure in terms of the number density of photons. This is
    \be
\eta = \frac{n_B - n_{\bar B}}{n\gamma}~.
    \ee
    Observations result in $\eta\simeq 10^{-10}$~\cite{basym}. Although this
    appears to be a small number, the problem for the SM is to explain
    why is not zero. The existence of  $\eta\not=0$ is incompatible
    with the SM.  The SM respects both baryon and lepton numbers in
    the form of accidental global $U(1)_{\rm B}$ and
    $U(1)_\ell$ symmetries in the Lagrangian. These global symmetries are however anomalous due to
    the existence of non trivial gauge field configurations associated
    with the non-abelian nature of the SM. Then, in principle, these
    anomalies could produce baryon violating processes. However, these
    processes are exponentially suppressed since their rate is
    essentially that of a tunnelling process and at zero temperature
    this is roughly suppressed as $e^{-1/\alpha}$, with $\alpha$ the
    QED coupling. The only hope to overcome this enormous suppression
    is to consider it at large enough temperatures such that they are
    unsuppressed due to thermal effects (going over the potential
    barrier). This is the situation expected to occur, in the
    cosmological history of the universe, around the temperature of
    the electroweak phase transition, $T_{\rm EW}\simeq 150~$GeV, the
    critical temperature for the vacuum to go from its symmetric value
    of $\langle\Phi\rangle =0$ to the broken phase value of
    $\langle\Phi\rangle=v/\sqrt{2}$. Thus, it looks like there is hope
    that one can explain $\eta\not=0$ in the SM. Unfortunately, 
    it has been known for some time that
    in order to generate the baryon asymmetry $\eta$ three conditions,
    called Sakharov's conditions,
    must be met: 1) Baryon number violation; 2) C and CP violation;
    and 3) Out of equilibrium dynamics. Although, as we just
    discussed, baryon number violation via the anomaly is present in
    the SM, the need of out of equilibrium dynamics requires the
    electroweak phase transition to be first order. This is not
    satisfied in the SM with the measured Higgs boson mass since it is
    too large and results in a smooth crossover (not even a second
    order phase transition). In addition, the second condition is only
    partially fulfilled in the SM, since the amount of CP violation is
    many orders of magnitude too small to be enough to produce the
    observed value of $\eta$. So it appears that, just as in the case
    for dark matter, an extension beyond the SM is needed to explain
    the baryon asymmetry.

\nin
    {\bf Dark Energy}: For about 25 years, we have known that the
      expansion of the universe is accelerating. The source of this is
      an energy density in the energy momentum tensor that does not
      behave like matter or radiation. It can be a constant (i.e. the
      cosmological constant), and the data is up to now consistent
      with this interpretation, or it can be a more complex effect,
      perhaps associated with a cosmic fluid. The cosmological
      standard model assumes that this dark energy (dark for lack of a
      better name) is indeed just the cosmological constant,
      $\Lambda_{\rm CC}$. Assuming this plus the correct abundance of {\em
        cold} dark matter, in addition to all the SM interactions for
      baryons, all the cosmological data can be fit rather well with
      what is called the $\Lambda_{\rm CDM}$
      model~\cite{cosmodata}. On the other hand, although the SM of
      particle physics can accommodate dark energy just by adding a
      cosmological constant in it, its value $\Lambda_{\rm CC}\simeq
      (10^{-3}eV)^4$, appears to be orders of magnitude smaller to
      what QFT would generically estimate. We will discuss this
      further below when we talk about other problems created by the
      SM. But the origin of this particular energy scale of dark
      energy is a mystery that cannot be ignored, since it represents
      about $70\%$ of the energy budget of the universe.

    \nin
    In addition to the points above, there are several questions
    that are actually raised by the SM itself.

    \nin
    {\bf The hierarchy of fermion masses}. The EWSM allows for fermion
    masses in a way that is consistent with the gauge theory
    $SU(2)_L\times U(1)_Y$ by introducing Yukawa couplings of the
    Higgs doublet and fermions which result in masses after
    electroweak symmetry breaking. But the resulting fermion Yukawa
    couplings are all over the place. For instance, the top Yukawa
    coupling is $\lambda_t \simeq O(1)$ whereas the up quark has a
    Yukawa coupling of $O(10^{-5})$. These two fermions have exactly
    the same SM quantum numbers. They only differ by this aspect. The
    same can be said about the electron Yukawa, $\lambda_e\simeq
    10^{-6}$, but the tau Yukawa is $\lambda_\tau\simeq 10^{-2}$. This
    is of course all consistent with the SM, but whay are there three
    generations of fermions ? And why do they have so greatly
    differing Yukawa couplings ?
    
    \nin
    {\bf The strong CP problem}. The
    simplest  way to state the problem is the fact that the gauge
    symmetry in QCD allows for a term like
    \be
    G^a_{\mu\nu} \tilde G^{a\mu\nu},
    \label{ggdual}
    \ee
    where $G^a_{\mu\nu}$ is the $SU(3)_c$ gluon field strength, and
    \be
\tilde G^{a\mu\nu} =\frac{1}{2} \epsilon^{\mu\nu\alpha\beta}G^a_{\alpha\beta}~,
    \ee
is called the dual field strength. The presence of this operator in
QCD would lead to CP violation in the strong interactions. The story
is a bit more nuanced and in fact this operator is related to the
anomalies mentioned earlier. In particular the chiral anomalies in QCD
require the presence of this operator, despite the fact that in
principle it can be written as a total derivative. The reason this
total derivative cannot be ignored once this term is integrated in all
of spacetime, as so often we do in QFT, is that it can be shown that
in non-abelian gauge theories 
\be
\int d^4x G^a_{\mu\nu} \tilde G^{a\mu\nu} \not =0~.
\ee
As a matter of fact, the integral is proportional to an integer
characterizing the vacuum of the theory. The true vacuum of QCD then
is a superposition of these vacua. 
As a result, this operator can and should be included in the QCD
action. Its coefficient is related to the arbitrary phase associated
to the true vacuum superposition, referred to as $\theta$.
Thus,
\be
{\cal L}_{\rm QCD} = {\cal L}_{\rm QCD}^{\theta=0} +\theta
\frac{\alpha_s}{4\pi^2}G^a_{\mu\nu} \tilde G^{a\mu\nu}~,
\label{qcdvacuum}
\ee
    where $\alpha_s$ is the QCD coupling strength. A final
    complication is the fact that chiral quark rotations are in fact
    equivalent to a shift in $\theta$. So the final coefficient is
    given by
    \be
\theta_{\rm phys.} = \theta -{\rm arg}\left({\rm det}\left[M\right]\right)~,
\ee
where $M$ is the original, non diagonal mass matrix. Thus, unless
there is at least one massless quark, in which chase is always possible
to choose the arbitrary chiral rotation parameter ($\alpha_L$ or
$\alpha_R$), then the value of the $\theta$ coefficient in
(\ref{qcdvacuum}) is physical. This implies that CP violation in the
strong interactions should be observed, a way to extract
$\theta_{\rm phys.}$. The leading effect is to generate an electric dipole 
of the neutron. Since this has not been observed we can put a bound:
\be
\theta_{\rm phys.}\leq 10^{-11}~.
\label{tethabound}
\ee
This is the strong CP problem: why is this dimensionless parameter of
the SM bound to be so small ? Once again, all possible answers require extending
the SM~\cite{cpstrong}.

\nin
 {\bf The origin of neutrino masses}.
As we have seen in Section~\ref{sec:gaugefermion}, the EWSM does not
include a right handed neutrino. On the other hand, we have plenty of
experimental evidence for the existence of neutrino
masses~\cite{nuexp}, however small. In principle, one could {\em add}
a right handed neutrino to the SM just so as to be able to write down
a gauge invariant operator as in (\ref{emass1}), which would look
like
\be
\lambda_{\nu_e} \bar\ell_L \tilde\Phi \nu_R~.
\label{numass1}
\ee
This would result in a neutrino mass, with a rather tiny Yukawa
coupling. But we already have a problem with the Yukawa couplings of the
other fermions. So this in and on itself is not a new problem. The
problem with (\ref{numass1} is that we added a new field, $\nu_R$ with
no SM quantum numbers just in order to generate a neutrino mass. Then,
building a {\em Dirac neutrino mass}  as in (\ref{numass1})  requires
extending the SM. Another possibility to accommodate neutrino masses
without the need to add a new field to the SM spectrum is to write an
operator containing only left handed neutrinos. This is
\be
\frac{c}{\Lambda}\left(\bar\ell_L \tilde\Phi\right)^2~.
\label{numass2}
\ee
where $c$ is an order one constant and $\Lambda$ is an energy scale
needed to make this term dimension four since the operator itself is
dimension five. Then the price we pay in order to write a neutrino
mass term just with the SM fields is to need a higher dimensional (non
renormalizable) operator, suppressed by the UV scale $\Lambda$. The
neutrino mass resulting from such operator (sometimes referred to as
Weinberg's operator) is
\be
m_\nu = \frac{c}{\sqrt{2}} \frac{v^2}{\Lambda}~.
\label{woperator}
\ee
This is a Majorana neutrino mass. There various extensions of the SM
that would result in this effect once the new particles are integrated
out. The most common models are seesaw models~\cite{seesaw}. But the
main message is that in order to obtain the operator in
(\ref{woperator}), we need to go beyond the SM, even if we insist in
only using SM fields. It is not yet know what the nature of the
neutrino mass is: Dirac or Majorana. This question can be settled in
the future, for
instance, in neutrinoless double beta day
experiments~\cite{ndoublebeta}. But what is already clear is that
neutrino masses require an extension of the SM.

\nin
{\bf The origin of the electroweak energy scale}:
If we write down the entire SM Lagrangian as the EW Lagrangian of
(\ref{lagewsm}) plus the QCD Lagrangian
\be
{\cal L}_{\rm SM} = {\cal L}_{\rm EW}+{\cal L}_{\rm QCD} ~,
\label{lagsm}
\ee
we would notice that among the dozens of terms there is {\em only one
energy scale} in the entire ${\cal L}_{\rm SM}$. This corresponds to
the coefficient of the quadratic term in the Higgs potential in
(\ref{hpot_2}), the mass scale that appears here as $-m^2$, gives rise
to the VEV of the Higgs field $\Phi(x)$ and all the masses of the SM
particles, including the Higgs boson mass
\be
m_h =\sqrt{2} m = \sqrt{2\lambda} v~.
\label{ewscale}
\ee
Using the measured value of the Higgs mass we have $m\simeq
89~$GeV. Where does this energy scale come from ? In the SM it is put
by hand in $V(\Phi^\dagger\Phi)$. It is true that the EWSM has a large
number of unexplained parameter, mostly in the form of Yukawa
couplings. But all of these are dimensionless. The one and only energy
scale in the SM is yet another unexplained quantity, but one rather
central in defining all the masses of all the elementary
particles. This is not to say that there are no other energy scales in
the low energy theory. For instance, in the QCD sector at low energy
confinement and chiral symmetry breaking lead to a spectrum of
hadrons. This happens at an energy scale $\Lambda_{\rm hadronic}\simeq
O(1)$~GeV, a scale that defines the hadron spectrum. However, this
scale can be understood  as {\em dynamically generated} by the
underlying QCD interactions of quarks and gluons: the QCD gauge
coupling becomes stronger at lower energies, so that eventually it
will be strong enough for spontaneous chiral symmetry breaking and
confinement at a scale called $\Lambda_{\rm QCD}\simeq$ few hundred
MeV. Thus, the hadronic scale is generated by a process called
dimensional transmutation, by which a {\em dimensionless} coupling generates
an energy scale when it gets very strong due to its running. Fermion
masses are not new scales, since they are all proportional to the
electroweak scale, multiplied by a dimensionless Yukawa coupling
(perhaps with the exception of the neutrino mass, but outside of the
SM). In the SM, the electroweak scale is the only scale put a priori
(by hand) in the theory. It is determined experimentally.

\noindent
In fact, the only other energy scales in the fundamental theory
describing particle physics and cosmology are the Planck scale,
\be
M_P=1.2~10^{19}~{\rm GeV}~,
\label{mplanck}
\ee and the cosmological constant/dark energy
density given by
which is
\be
\Lambda_{\rm CC}\simeq (10^{-3}~{\rm eV})^4.
\label{lamcc}
\ee
Are these three scales in (\ref{ewscale}), (\ref{mplanck} and (\ref{lamcc}) the only ones introduced  {\em ad hoc} in all
of the standard models of particle physics and cosmology, really
independent, or they are related to each other ? Since $M_P$ is a
scale associated with the extreme UV of the quantum field theory, the
scale at which quantum gravity effects become important, it appears
that this might be a more {\em fundamental } energy scale. Can the
other two, i.e. the electroweak scale and $\Lambda_{\rm CC}$, be
derived from it ? If this was the case, would there be any
experimentally accessible consequences, particularly just above the
electroweak scale ? So the mere origin of the Higgs mass scale is not
understood and it might lead to interesting phenomena if we explore
Higgs physics further.

\noindent
{\bf The Hierarchy Problem}: In addition to the question of the origin
of the electroweak scale $v\simeq 246~$GeV, the Higgs sector of the
EWSM posses a more formal question: the apparent lack of radiative
stability of this scale. Another way to state this problem, is to say
that a sector involving a fundamental scalar field, such as the Higgs
sector, is greatly sensitive to UV physics. To see what is meant by
this let us consider the one loop corrections to the Higgs boson mass
in the SM. These include loops of all SM fermions, as well as the
massive gauge bosons $W^\pm, Z^0$ as well as the Higgs boson
itself. (See Figure~\ref{fig:mhloops}.)
\nin
\begin{figure}[h]
\begin{center}
\includegraphics[width=5in]{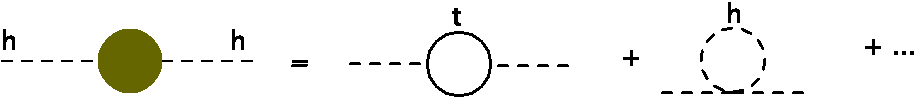}
\end{center}
\caption{One loop diagrams contributing to the quantum corrections to the Higgs
  mass. Show as examples are the top and the Higgs boson contributions.
}
\label{fig:mhloops}
\end{figure}
\vskip0.3in
\nin
The corrections to $m_h^2$ resulting from these one loop diagrams can
be generically written as
\be
\Delta m_h^2 = \frac{c}{16\pi^2}\,\Lambda^2 +\cdots~,
\label{quaddiv}
\ee
where $\Lambda$ is a momentum scale signifying the highest momentum
where the EWSM is valid, and $c$ is a constant that can be computed
and depends on the SM particles  going around the loop. For instance,
for a top quark, gauge bosons and the Higgs boson, respectively in the loop, we have
\be
c_{\rm top} = - 2 N_c y^2_t~, \qquad\qquad c_{\rm
  gauge}=g^2~,\qquad\qquad c_{\rm h} = \lambda^2~,
\label{ctop}
\ee
where $N_c=3$ is the number of colors,  $y_t$ is the top quark
Yukawa coupling to the Higgs, $g$ is a generic electroweak gauge
couplings and $\lambda$ is the Higgs self-coupling. . The dots in (\ref{quaddiv}) denote
either terms that depend logarithmically on the cutoff $\Lambda$,
i.e. proportional to $\ln \Lambda$, or terms that are finite in the
$\Lambda\to\infty$ limit. For a given value of $\Lambda$, it is clear
that the top quark loop will dominate. For instance, if the cutoff is
$\Lambda=10~$TeV, we have that the top loop contributes to $\Delta
m_h^2$ with about $(2~{\rm TeV})^2$, the gauge boson loops with $(700
  {\rm ~GeV})^2$, and the Higgs boson loop wiht about $(100 {\rm
   ~ GeV})^2)$. Then, the {\em  renormalization condition} that we need
  to impose on the physical Higgs boson mass is roughly:
  \be
m_{h,{\rm phys.}}^2 = m_0^2 - \left( 256 - 31 -0.7\right)\,(125~{\rm
    GeV})^2~,
  \label{mhrencon}
  \ee
  where $m_0$ is the unrenormalized Higgs mass. We see that the top
  quark loop already requires a fine tuning of the renormalization
  condition of more than 1 part in 100. The tuning only gets worse
  (quadratically) as we increase the cutoff $\Lambda$. To avoid this
  tuning, the cutoff should be as close to the electroweak scale as
  possible. This is the {\bf hierarchy problem}.

  \nin
  But is this really a problem ? After all, in QFT we are allowed to
  take the cutoff all the way to infinity, i.e. $\Lambda\to \infty$,
  since once the renormalization procedure is completed, no physical
  quantity should depend on it. Then, QFT does not have a hierarchy
  problem, even in theories where there is an elementary scalar field
  and its mass squared parameter is quadratically sensitive to UV
  scales. In fact, once the renormalization condition (\ref{mhrencon})
  is imposed, the Higgs mass evolves logarithmically with the energy
  scale
  \be
\frac{d m_h^2}{d \ln \mu^2} = \beta_{m_h}^{\rm SM}\,m_h^2~,
\label{mhlogevolution}
\ee
with the Higgs mass beta function to one loop given by
\be
\beta_{m_h}^{\rm SM} = \frac{1}{16\pi^2}\left(12\lambda+ 12y_t^2
  -(9g^2+3g'^2) +\cdots\right)~.
\ee
This logarithmic dependence of course is just the statement of  the fact that, after
renormalization, the evolution of physical parameters with the scale
$\mu$ corresponds to the re-scaling of energies/distances, which is
logarithmic. This logarithmic evolution of $m_h^2$ seems to belie the
problem of having {\em quadratic} sensitivity to UV scales. So, no
hierarchy problem then ?

\nin
It turns out that the problem resurfaces if  we have
heavy states, with masses well above the TeV scale,  coupled to the
Higgs. Let us consider as an example, a vector like fermion coupled to
the Higgs as in
\be
{\cal L} \supset y_N  \bar  L H N + M_N NN ~,
\label{rhneutrino}
\ee
where the vector-like mass $M_N$ can be arbitrarily large and we
coupled this singlet (or ``right handed neutrino'') to $H$ through the
lepton doublet $L$.  Independently of what this state does to neutrino
masses, one thing we can see is that it results in a threshold
correction to the RGE evolution of $m_h^2(\mu)$ in
(\ref{mhlogevolution}). This is given by
\be
\frac{d m_h^2}{d \ln \mu^2} |_{\rm threshold} = \frac{y_N^2}{16\pi^2}
\, M_N^2~.
\label{threshold}
\ee
The above correction represents a large jump in the logarithmic
evolution of the Higgs mass, so that for $\mu=\Lambda_{\rm UV} > M_N$
now we have a much larger value of $m_h$ that we would have otherwise
obtained by the SM RGE evolution. This quadratic (in $M_N$) jump is
one more reflection of the quadratic sensitivity of $m_h^2$ to the UV
scales. So when integrating out heavy scales, it would require a large
tuning in order to obtain the observed Higgs mass in the IR, very much
like what happened in (\ref{mhrencon}).

\nin
One could think to solve the
problem either by 1) forbidding any heavy particle to couple to the
Higgs in the UV, or 2) by imposing the renormalization condition on
$m_h^2$ only once we run the RGEs all the way up to the UV,  and
therefore know of all the
possible threshold corrections. However, either of these two ways
involves knowledge of the UV, which is not supposed to be necessary  
to define the theory in the IR ! So the UV sensitivity of the Higgs
sector is real and we have to live with it. At this point, we should
remind ourselves that the Higgs is the only particle in the SM for
which this problem arises. Fermion masses  are protected by chiral
symmetry, resulting in only a mild logarithmic dependence on the cutoff
$\Lambda$. Gauge boson masses are IR phenomena arising from (soft)
spontaneous symmetry breaking. They are not UV sensitive.  The Higgs
boson is unique in its role of introducing an {\em ad hoc} energy
scale in the SM, as well as having this scale (or its mass, which is
the same) very UV sensitive. 

\nin
The central question is then not whether the hierarchy problem exists
or not, but what does it imply for the scale of new physics. We used
to believe that it implied the existence of new physics at roughly the
$1~\rm TeV$ scale. The experimental absence of evidence for new
physics so far has turned this question into a more puzzling and
interesting one, nor less.  

\subsection{The EWSM and the future}
\label{sec:future}

We have seen that the EWSM is an extremely successful description of
the electroweak interactions. It is a spontaneously broken gauge
theory, $SU(2)\times U(1)_Y\to U(1)_{\rm EM}$ which has been tested
extensively over several decades. The couplings of gauge bosons to
fermions are the best tested ones, as detailed in
Section~\ref{sec:fermions2gauge}. Similarly, the gauge boson
self-couplings are the subject of increasing precision at the LHC. On
the other hand, the Higgs sector, introduced in order to trigger the
spontaneous breaking of the electroweak gauge theory, is the less
tested.
Although we have measured several of the Higgs boson couplings to
other SM particles, such as gauge bosons and the heavier fermions, it
remains the least precisely tested. In particular the Higgs potential,
introduced in an {\em ad hoc} to break the gauge symmetry in the
desired way, and introducing the {\em only dimensionfull } quantity
in the  theory, has not been tested. In fact, as seen in
Section~\ref{sec:higgscouplings}, the only parameter in the Higgs
potential in (\ref{hpot_2}) that we have had access to so far is
$m^2$. We extract this from the measurement of the Higgs boson mass by
making use of the relation (\ref{smmh}) between $m_h$, $v$ and
$\lambda$, the Higgs quartic coupling in (\ref{hpot_2}),
i.e. $m_h=\sqrt{2\lambda}\,v$, which results in $\lambda \simeq 0.13$
and
\be
m~\simeq 89~{\rm GeV}~.
\label{minsm}
\ee
But these values are obtained by making use of the minimization
procedure assuming the form of the potential in (\ref{hpot_2}). It
corresponds to the only two terms that are renormalizable and gauge
invariant. But we do not know if there are additional terms either
involving other fields or coming from higher dimensional
operators. For this purpose we need to measure the triple Higgs couple
$g_{h^3}$ with some precision in double Higgs production. This alone
would take a lot of data in the HL-LHC and it is not clear that would
be enough to settle the issue. To ``map'' the Higgs potential with
precision it might be necessary to go to a new experimental facility
such as a Higgs factory.

\nin
Still on the issue of the Higgs sector, there is the question of its
origin. As we mentioned earlier, this sector of the EWSM appears
for the specific purpose of breaking the gauge symmetry spontaneously
in the way it is observed experimentally. Although the discovery of
the Higgs boson has confirmed the Higgs mechanism beyond any doubt, it
is not clear where the Higgs sector comes from. In other physical
systems where a scalar degree of freedom is introduced to
spontaneously break a symmetry, it has turned out that the scalar or
scalars are collective excitations and not elementary fields. For
instance, we can describe superconductivity~\cite{Anderson:1963pc} by the Higgs mechanism,
but the Higgs is a fermion composite. In QCD at low energies, the
spontaneous breaking of chiral symmetry can be modeled as occurring
through the so called $\sigma$ model, where the only remnant light
degrees of freedom are the pNGB (e.g. the pions) whereas the $\sigma$
particle, which would be playing the role of the Higgs, is known to be
heavy and strongly coupled. Technicolor models~\cite{Hill:2002ap} from the 1970s and
1980s played with this analogy by postulating that the Higgs boson
would be heavy and strongly coupled, as well as a composite of
fermion/anti-fermion pairs. Clearly this is not the case in the EWSM,
since the Higgs seems to be weakly coupled $\lambda\simeq 0.13$, which
means is light.    
But what if instead of being the $\sigma$ the Higgs boson is a pNGB
just as the pions ? This would explain why is lighter than the new
physics scale and why is weakly coupled. This idea goes by the name of
Composite Higgs Models (CHM) ~\cite{Agashe:2004rs, Panico:2015jxa}: the Higgs
is a pNGB of the spontaneously broken global symmetry (just as chiral
symmetry in QCD). But what are the observable consequences of the
Higgs boson compositeness ?  First, in most CHM there are resonances,
both bosonic spin 1 and fermionic, that should be present at a scale
considerably above the electroweak scale, perhaps several TeV. So, as
it appears that the LHC has had not enough energy to produce them, we
should look for the effects of the new physics in deviations in the
Higgs behavior, particularly its couplings. Deviations in the Higgs
couplings with respect to the SM predictions are almost certain in
these models~\cite{Panico:2015jxa,Burdman:2014zta}, even momentum dependent ones~\cite{Bittar:2022wgb}. Thus, very precise
measurements in various different channels will be necessary to fully
test this hypothesis at the HL-LHC and perhaps beyond. 

\nin
Beyond the better understanding of the Higgs sector of the SM, we are
left with a number of fundamental questions that the SM does not
answer. Both theoretical and experimental exploration of these will be
a central part of particle physics in the next decades. The search for
particle dark matter will continue in direct~\cite{dmdirect} and indirect~\cite{dmindirect} detection
experiments, as well as at the LHC. New kinds of experiments are being
proposed to look for dark sectors in many different mass ranges from
the ones looked at so far.  Neutrino experiments such as DUNE~\cite{dune} and
HYPER-K~\cite{hyperk} will explore the neutrino question with great detail. 

\nin
The interaction of particle physics with astrophysics and cosmology
will continue through some of these questions. Will the
CMB~\cite{cmbfuture} data exclude any new relativistic degrees of
freedom through a very precise measurement of $N_{\rm eff.}$, the effective
number of neutrinos ? Will the precise determination of the dark
energy equation of state or the age of the universe, point  in the direction of new physics in the
cosmic history ? Many new gravitational wave detectors  will be
built. In particular, LISA~\cite{lisagw} will be sensitive to
gravitational wave signals from the electroweak phase transition. But
in the EWSM, the Higgs potential is unable to produce such
signal. Observation of it would point to new physics in the Higgs
potential (\ref{hpot_2}).

\nin
The EWSM is a great success of quantum field theory and experimental
ingenuity. But it leaves and/or creates enough open questions that the
future experimental and theoretical programs based on it
are very broad and increasingly exciting. 


\begin{flushleft}
\interlinepenalty=10000

\end{flushleft}

\end{document}